\documentclass[11pt,a4paper]{article}
\pdfoutput=1
\usepackage{amssymb,amsmath,amsfonts, mathtools, mathrsfs}
\usepackage[utf8]{inputenc} 
\usepackage[dvipsnames]{xcolor}

\newif\ifnatbibsort\natbibsorttrue

\relax

\ifnatbibsort\RequirePackage[numbers,sort&compress]{natbib}\else\RequirePackage[numbers,compress]{natbib}\fi
\RequirePackage[colorlinks=true
,urlcolor=blue
,anchorcolor=blue
,citecolor=blue
,filecolor=blue
,linkcolor=blue
,menucolor=blue
,pagecolor=blue
,linktocpage=true
,pdfproducer=medialab
,pdfa=true
]{hyperref}

\usepackage{graphicx}
\usepackage{caption}
\usepackage{tensor}
\usepackage{subfigure}
\usepackage{enumerate}
\usepackage{dsfont}
\usepackage{verbatim}
\usepackage{amsfonts,euscript,amssymb,stmaryrd,braket}
\usepackage{tikz}
\usetikzlibrary{arrows,decorations.markings,patterns}
\usepackage{slashed}

\def\clock{{\count0=\time
		\divide\count0 60
		\ifnum\count0<10 0\fi\the\count0
		\multiply\count0 -60 \advance\count0 \time
		:\ifnum\count0<10 0\fi \the\count0
}}
\newcommand{\timestamp}{{\small\vbox{\hbox{\tt\jobname.tex}
			\hbox{\the\day/\the\month/\the\year, \clock}}}}

\newcommand{\A}{\mathcal{A}}

\newcommand{\R}{\mathbb{R}}

\newcommand{\nn}{\nonumber}
\newcommand{\bea}{\begin{eqnarray}}
\newcommand{\eea}{\end{eqnarray}}

\newcommand{\be}{\begin{equation}}
\newcommand{\ee}{\end{equation}}

\makeatletter
\let\old@startsection=\@startsection
\let\oldl@section=\l@section
\renewcommand{\@startsection}[6]{\old@startsection{#1}{#2}{#3}{#4}{#5}{#6\mathversion{bold}}}
\renewcommand{\l@section}[2]{\oldl@section{\mathversion{bold}#1}{#2}}
\makeatother

\numberwithin{equation}{section}

\def \D {\pta}

\def \W {{\cal W}}
\def \V {{\cal V}}
\def \D {{\cal D}}

\usepackage{color}

\def\D{\mathcal{D}}

\def \RR {{\mathbb R}}

\def\ri {{\rm i}}
\def\rd {{\rm d}}
\def\e {{\rm e}}

\begin{document}
	\renewcommand{\thefootnote}{\arabic{footnote}}

	\overfullrule=0pt
	\parskip=2pt
	\parindent=12pt
	\headheight=0in \headsep=0in \topmargin=0in \oddsidemargin=0in

	\vspace{ -3cm} \thispagestyle{empty} \vspace{-1cm}
	\begin{flushright} 
		\footnotesize
		\textcolor{red}{\phantom{print-report}}
	\end{flushright}

\begin{center}
	\vspace{.0cm}

	{\Large\bf \mathversion{bold}
	Modular evolutions and causality 
	}
	\\
	\vspace{.25cm}
	\noindent
	{\Large\bf \mathversion{bold}
	in two-dimensional conformal field theory}


	\vskip  0.8cm
	{
		Dobrica Jovanovic$^{\,a}$, 
		Mihail Mintchev$^{\,b}$
		and Erik Tonni$^{\,c}$
	}
	\vskip  1.cm
	
	\small
	{\em
		$^{a}\,$Center for Theoretical Physics, Massachusetts Institute of Technology, \\
		77 Massachusetts Avenue, Cambridge, MA 02139, USA
		\vskip 0.05cm
		$^{b}\,$Dipartimento di Fisica, Universit\'a di Pisa and INFN Sezione di Pisa, \\
		largo Bruno Pontecorvo 3, 56127 Pisa, Italy
		\vskip 0.05cm
		$^{c}\,$SISSA and INFN Sezione di Trieste, via Bonomea 265, 34136, Trieste, Italy 
	}
	\normalsize

\end{center}

\vspace{0.3cm}
\begin{abstract} 

In two-dimensional conformal field theories (CFT) in Minkowski spacetime,
we study the spacetime distance between two events along two distinct modular trajectories.
When the spatial line is bipartite by a single interval, 
we consider both the ground state and the state
at finite different temperatures for the left and right moving excitations. 
For the free massless Dirac field in the ground state, 
the bipartition of the line given by the union of two disjoint intervals is also investigated. 
The modular flows corresponding to connected subsystems
preserve relativistic causality.
Locality along the modular flows of some fields is explored by evaluating their (anti-)commutators.
In particular, the bilocal nature of the modular Hamiltonian of two disjoint intervals for the massless Dirac field
provide multiple trajectories leading to  Dirac delta contributions in the (anti-)commutators
even when the initial points belong to different intervals, thus being spacelike separated. 

\end{abstract}

\newpage
\tableofcontents

\newpage
\section{Introduction}
\label{sec:intro}

The modular (entanglement) Hamiltonian $K$, 
introduced within the Tomita-Takesaki modular theory 
\cite{takesaki-book-70, Haag:1992hx, Brattelli2, Borchers:2000pv, takesaki-book-03},
is a crucial operator to investigate 
in order to understand the entanglement associated to the spatial bipartition of a quantum system.
Another important object in the Tomita-Takesaki modular theory is the modular conjugation, usually denoted by $J$.
The operator $K$, which depends both on the bipartition and on the state of the whole system,
generates a unitary evolution of the fields known as modular flow. 
In general $K$ is non-local and displays a very complicated structure;
hence its explicit form is known is very few cases.

For the sake of simplicity, we consider only two-dimensional relativistic field theories in the Minkowski spacetime 
${\mathbb M} = \R_x \times \R_t$, 
parameterised by the pair $(x,t)$,
where $x$ and $t$ are the spatial and temporal coordinates respectively. 
In this setup we investigate the deeply related
concepts of causality (the sign of the Minkowski spacetime distance
between two events) and locality of quantum fields whose evolution is generated by the entanglement associated to the bipartition of the system.

The first crucial result for $K$ has been obtained by Bisognano and Wichmann
\cite{Bisognano:1975ih, Bisognano:1976za}
for a generic local relativistic quantum field theory in its ground state
and the spatial bipartition of the real line in two halves, 
namely $\R_x = \R_- \cup \R_+$,
where $\R_+$ and $\R_-$ correspond to the positive and negative real numbers respectively. 
In this case, the operator $K$ coincides with the generator of Lorentz boosts in $\mathbb{M}$.
The corresponding modular flow has a well defined geometric action that allows identifying
modular trajectories for the local fields belonging 
either to the right Rindler wedge $\W_+ = \big\{ (x,t)  :  x \geqslant  |t| \big\}$ 
or to the left Rindler wedge $\W_- = \big\{ (x,t)  :  x \leqslant  |t| \big\}$,
depending on the initial position of the local field on $\R_x $.
Since the Lorentz boosts are isometries of the Minkowski metric,
this modular flow clearly preserves relativistic causality.
Also the modular conjugation $J$ has a geometric action 
(relating $\W_+$ and $\W_-$) given by the reflection with respect to the origin,
combined with the charge conjugation.

Conformal field theories (CFT) in two spacetime dimensions (see e.g. \cite{Luscher:1974ez,  Todorov:1978rf, Belavin:1984vu})
provide an important class of 
relativistic quantum field theories where some modular Hamiltonians have been found in explicit form. 
The most important example has been studied by Hislop and Longo \cite{Hislop:1981uh}
(see also \cite{Brunetti:1992zf, Casini:2011kv}) and consists of a free massless scalar in $\mathbb{M}$ and in its ground state, 
whose space $\R_x $ is bipartite by a finite interval $A_x$ and its complement $B_x$.
This case has been explored by combining the Bisognano-Wichmann result with
a conformal mapping sending e.g. $\R_+$ into $A_x$ and $\R_{-}$ into $B_x$.
The image of $\W_+$ under this mapping is the causal diamond $\D_A$ whose diagonal is the interval $A_x$
and the modular flow crossing $A_x$, which remains in  $\D_A$, preserves relativistic causality \cite{Fredenhagen:1984dc}.
The situation is more complicated for the image of $\W_-$ 
because in this domain the conformal map has singularities
and therefore induces singularities also in the flow generated by $K$ that crosses $B_x$. 
In this case, the analysis of relativistic causality along this flow is more subtle, 
as discussed in Sec.\,\ref{sec-1int-line-vacuum}.
The geometric action of the modular conjugation, relating the flows across $A_x$ and $B_x$,
and the two-point correlation functions of primaries evolving along the flow generated by $K$
are also described \cite{Haag:1992hx, Mintchev:2022fcp}. 
In Sec.\,\ref{sec-1int-line-vacuum} we also study
the properties of the spacetime support of the commutator of certain combinations of primary fields,  
discussing the relation between the flow generated by $K$ and the Tomita-Takesaki modular theory.
In two-dimensional CFT,  the modular Hamiltonian of the bipartition 
given  by an interval is known in a few other cases,  including one at finite temperature
\cite{Wong:2013gua, Cardy:2016fqc}.
In Sec.\,\ref{sec-1int-line-thermal}, 
the analysis of the spacetime distance between two points
along distinct modular trajectories in $\D_A$ is extended
to the case where the left and right moving  excitations have different temperatures, 
finding that relativistic causality is preserved.

In the second part of this manuscript (see Sec.\,\ref{sec-2int-Dirac}) we investigate 
the bipartition of the spatial real line $\R_x$ provided the union of two disjoint intervals $A_x \equiv A_{1,x} \cup A_{2,x}$ 
and its complement $B_x$ on the line. 
Considering a generic CFT on the line and in its ground state, 
while the expression of the modular Hamiltonian for a single interval holds for any CFT model, 
the modular Hamiltonian for the union of two disjoint intervals is model-dependent. 
For the massless Dirac field, this operator and the corresponding modular flow for the field
have been obtained by Casini and Huerta \cite{Casini:2009vk} 
within the framework developed by Peschel \cite{peschel-03} for the underlying lattice model,
while the modular correlators have been found in \cite{Longo:2009mn} 
(see also \cite{Hollands:2019hje}).
The crucial difference with respect to the single interval case is that 
this modular Hamiltonian for the union of two disjoint intervals is non-local because it involves a bilocal contribution. 
This term affects the modular flow of the massless Dirac field in a relevant way 
and, consequently, also its anti-commutator. 
In particular, the modular flow of each chiral component of the massless Dirac field 
is a mixing of two fields and it involves both the intervals, 
although at the beginning of the flow  the field is localised only in one interval \cite{Casini:2009vk}.
The spacetime distance between two points 
along two distinct modular trajectories describing the modular flow of the massless Dirac field is also explored, 
finding that relativistic causality is preserved along this evolution.
Nevertheless, because of the bilocal term in the modular Hamiltonian, 
some Dirac delta contribution occur in the anti-commutator of two modular flows of the massless Dirac field;
hence this modular evolution does not respect local commutativity.

In this paper we study  the causality properties of the evolution generated by $K$ in the conventional 
Minkowski spacetime $\mathbb{M}$, where only the infinitesimal special conformal transformations are well defined. 
In order  to implement the action of generic finite special conformal transformations, 
a compactification $\overline {\mathbb M}$ of ${\mathbb M}$ must be introduced. 
It is known \cite{Segal:1971aa, Luscher:1974ez, Segalbook, Todorov:1978rf, Brunetti:1992zf} 
that such compactification (often called Dirac-Weyl compactification) is given by
$\overline {\mathbb M} = ({\mathbb S} \times {\mathbb S} ) /{\mathbb Z}_2$,
where ${\mathbb S}$ is the unit circle. 
However, since this manifold is not causally orientable, 
the universal covering ${\widetilde {\mathbb M}}$ of $\overline {\mathbb M}$
given by ${\widetilde {\mathbb M}} = {\mathbb S}  \times  {\mathbb R}$ is employed
\cite{Segal:1971aa, Luscher:1974ez, Brunetti:1992zf}. 
Thus,  conformal invariance naturally leads to consider a spacetime with the geometry of a cylinder. 
It is worth mentioning  that, from the group theoretical point of view, 
the time $t_c$ on ${\widetilde {\mathbb M}}$ is associated to the conformal Hamiltonian $\frac{1}{2}(P_0 + K_0)$ 
rather than to the Hamiltonian $P_0$ in ${\mathbb M}$, 
where $K_0$ is the generator of the special conformal transformations. 
The physical impact of employing the conformal time $t_c$ 
in $3+1$ dimensional Minkowski spacetime has been explored in \cite{Segalbook, Todorov:1978rf}.

The outline of the paper has been mainly described above. 
Besides the sections already mentioned, 
some conclusions are drawn in Sec.\,\ref{sec-conclusions}
and in Appendices\,\ref{app-mappings}, \ref{app-derivative-Dirac-delta},
\ref{app-1int-circle-vacuum} and \ref{app-mod-flow-2int} 
some technical details and further discussions 
supporting the analyses in the main text are reported.

\section{Interval in the line, vacuum state}
\label{sec-1int-line-vacuum}


We consider a unitary CFT in $1+1$ Minkowski spacetime ${\mathbb M} = {\mathbb R}_x  \times  {\mathbb R}_t$. 
A generic point $(x,t) \in \mathbb{M}$ can be equivalently characterised by its light-cone coordinates 
\be 
u_\pm \equiv x \pm t 
\label{light-cone coordinates}
\ee 
along  the two independent chiral directions. 
A CFT on ${\mathbb M}$ is the tensor product of a sector of right (depending on $u_+$) and left (depending on $u_-$) moving excitations \cite{Luscher}.
Hence, it is convenient to study 
first the quantity of interest along a single chiral direction parameterised by $u \in \RR$
(the indices $\pm$ are often removed without ambiguity in order to enlighten the notation)
and then describe its complete expression in ${\mathbb M} $
by properly combining the results along $u=u_{+}$ and $u=u_{-}$.  
Denoting by $\phi_+(u_+)$ and $\phi_-(u_-)$ the generic right and left chiral primary field respectively, 
with conformal dimension $h_+$ and $h_-$ respectively,
all chiral primary fields can be treated simultaneously by adopting the notation $\phi(u)$ for the field and $h$ for its conformal dimension,
assuming that it corresponds to either $\phi_{+} (u_{+})$ or $\phi_{-} (u_{-})$,
with either $h=h_+$ or $h=h_-$ respectively. 
Unitarity implies that $h  >0$.

\subsection{Chiral modular evolution and modular conjugation} 
\label{sec-1int-chiral-dir-mod-evo}


A compact connected interval $A_x \subset {\mathbb R}_x $
defines a bipartition $A_x \cup B_x = {\mathbb R}_x$ of the system,
where  $B_x$ is  its complement in the spatial direction. 
We  denote by $\mathcal{D}_A \equiv \big\{ (u_+, u_-) \,|\, u_\pm \in A \big\} $ 
the causal diamond generated by $A_x$ 
(see the grey domain in Fig.\,\ref{fig:diamond-mod-hyper}),
whose orthogonal projections along the chiral axes $\RR_{u_\pm}$ (parameterised by $u_\pm$) 
define the interval  $A \equiv [a,b]$ and its complement $B$ along each chiral axis.
This uniquely identifies the corresponding spacetime domain 
$\mathcal{B}_A \equiv \big\{ (u_+, u_-) \,|\, u_\pm \in B \big\} $ 
(see the light blue region in Fig.\,\ref{fig:diamond-mod-hyper}).
In the limit $b \to +\infty$
the spacetime regions  $\mathcal{D}_A$ and $\mathcal{B}_A$
become the right and left Rindler wedges respectively,
while for $a \to -\infty$
they become the left and right Rindler wedges respectively.

The entanglement Hamiltonian $K$  
induced by the bipartition $A \cup B$ and associated with each chirality is
\cite{Hislop:1981uh, Casini:2011kv}
\be
\label{K-chiral-1int-line}
K =
\int_{-\infty}^{+\infty} V(u) \, T(u) \, \rd u
\,\equiv\,
K_{A} \otimes \boldsymbol{1}_B - \boldsymbol{1}_A \otimes K_{B} 
\ee
where $T(u)$ is  the chiral component of the energy density (i.e. either $T_+(u_+)$ or $T_-(u_-)$)
and the weight function reads
\be
\label{velocity_fund}
V(u) =2\pi \, \frac{(b-u)(u-a)}{b-a}=\frac{1}{w'(u)} 
\ee
being $w(u) $ defined as 
\be
\label{w_fund}
w(u) \equiv 
\left\{
\begin{array}{ll}
\displaystyle
\frac{1}{2\pi}\,\log \!\left( \frac{u-a}{b-u} \right)  \hspace{1cm}& u \in A
\\
\rule{0pt}{.9cm}
\displaystyle
\frac{1}{2\pi}\,\log \!\left( \frac{u-a}{u-b} \right)  & u \in B
\end{array}
\right.
\ee
whose inverse in each subsystem reads respectively
\be
\label{w_fund_inv}
w_A^{-1}(\zeta) \equiv \frac{a + b\, \e^{2\pi \zeta} }{ 1 +  \e^{2\pi \zeta}} 
\;\;\;\;\qquad\;\;\;
w_B^{-1}(\zeta) \equiv \frac{a - b\, \e^{2\pi \zeta} }{ 1 - \e^{2\pi \zeta}} 
\;\;\;\;\;\;\qquad\;\;\;\;\;\;
\zeta \in \RR\,.
\ee

The weight function (\ref{velocity_fund}) is obtained through a conformal map
from the full modular Hamiltonian of half line, 
found by Bisognano and Wichmann \cite{Bisognano:1975ih, Bisognano:1976za} 
for any unitary and Lorentz invariant quantum field theory,
as reviewed  in Appendix\;\ref{app-HL-map}.


The commutation properties of the energy-momentum tensor $T$  with any primary field $\phi$ with dimension $h$
allow to derive the evolution of $\phi$ generated by (\ref{K-chiral-1int-line}).
Denoting by $\tau \in \RR$ the modular evolution parameter, it reads
\be
\label{phi-mod-evo}
\phi(\tau, u)
\equiv \e^{\ri \tau K }\, \phi(u)\, \e^{-\ri \tau K}  
= \big[ \partial_{u} \xi(\tau, u) \big]^h \,\phi\big(\xi(\tau, u) \big)
\ee
where $\phi(u)$ defines the initial configuration and $\xi(\tau, u) $ 
can be written in terms of (\ref{w_fund}) and (\ref{w_fund_inv}) as follows
\cite{Hislop:1981uh, Brunetti:1992zf}
\be
\label{xi-map-fund}
\xi(\tau,u) 
=
\frac{(b-u)\,a+(u-a)\,b\,\mathrm{e}^{2\pi \tau}}{(b-u)+(u-a)\,\mathrm{e}^{2\pi \tau}}  
=
\left\{ \begin{array}{ll}
w_A^{-1}\big(w(u) + \tau\big) \hspace{1cm} & u \in A
\\
\rule{0pt}{.6cm}
w_B^{-1}\big(w(u) + \tau\big) \hspace{1cm} & u \in B\,.
\end{array}
\right.
\ee
At a given $\tau \in \RR$, the expression (\ref{xi-map-fund}) has the form
 $ u \mapsto \tfrac{ \mathsf{a} u + \mathsf{b} }{ \mathsf{c} u + \mathsf{d}  }$ 
 with $\mathsf{a} \mathsf{d} - \mathsf{b} \mathsf{c} = (b-a)^2 \,\e^{2\pi \tau}  $.

With a slight abuse of terminology, 
in the following  we call $K$  full modular Hamiltonian
and (\ref{phi-mod-evo}) modular evolution of $\phi$. 
Indeed, as explained in Sec.\,\ref{sec-tau-evolution}, 
the $K$-evolution generated by (\ref{K-chiral-1int-line}) 
can be related to the Tomita-Takesaki modular theory only for primaries with dimension $h \in \mathbb N$ and 
$h \in \mathbb N + \frac{1}{2}$. 
However, the correlation functions 
of (\ref{phi-mod-evo}) are well defined and satisfy the Kubo-Martin-Schwinger 
(KMS) condition for a generic $h>0$,
as discussed in Sec.\,\ref{sec-chiral-correlators}.

The expression (\ref{xi-map-fund}) describes the evolution of the initial point $u$ as $\tau \in \RR$;
indeed $\xi(0,u) = u$ and it satisfies 
\be
\label{der-xi}
\partial_\tau \xi(\tau,u) = V(\xi)
\;\;\;\qquad\;\;\;
\partial_u \xi(\tau,u) = \frac{V(\xi)}{V(u)}
\ee
where $V(u)$ is the weight function (\ref{velocity_fund}).
An interesting property of (\ref{xi-map-fund})
is that $\xi( -\tau , \xi(\tau,u) ) = u$,
which supports further the fact that $\xi(\tau,u)$ describes a proper evolution.
Moreover,  (\ref{xi-map-fund}) is invariant under the transformation that simultaneously 
exchanges $a$ and $b$ and replaces $\tau$ with $-\,\tau$.
The entangling points at $u=a$ at $u=b$, 
that characterise the bipartition along the chiral direction, do not evolve; indeed, 
$\xi(\tau,a) = a$ and $\xi(\tau,b) = b$ for any $\tau \in \RR$.

\begin{figure}[t!]
\vspace{-.5cm}
\hspace{1.3cm}
\includegraphics[width=.8\textwidth]{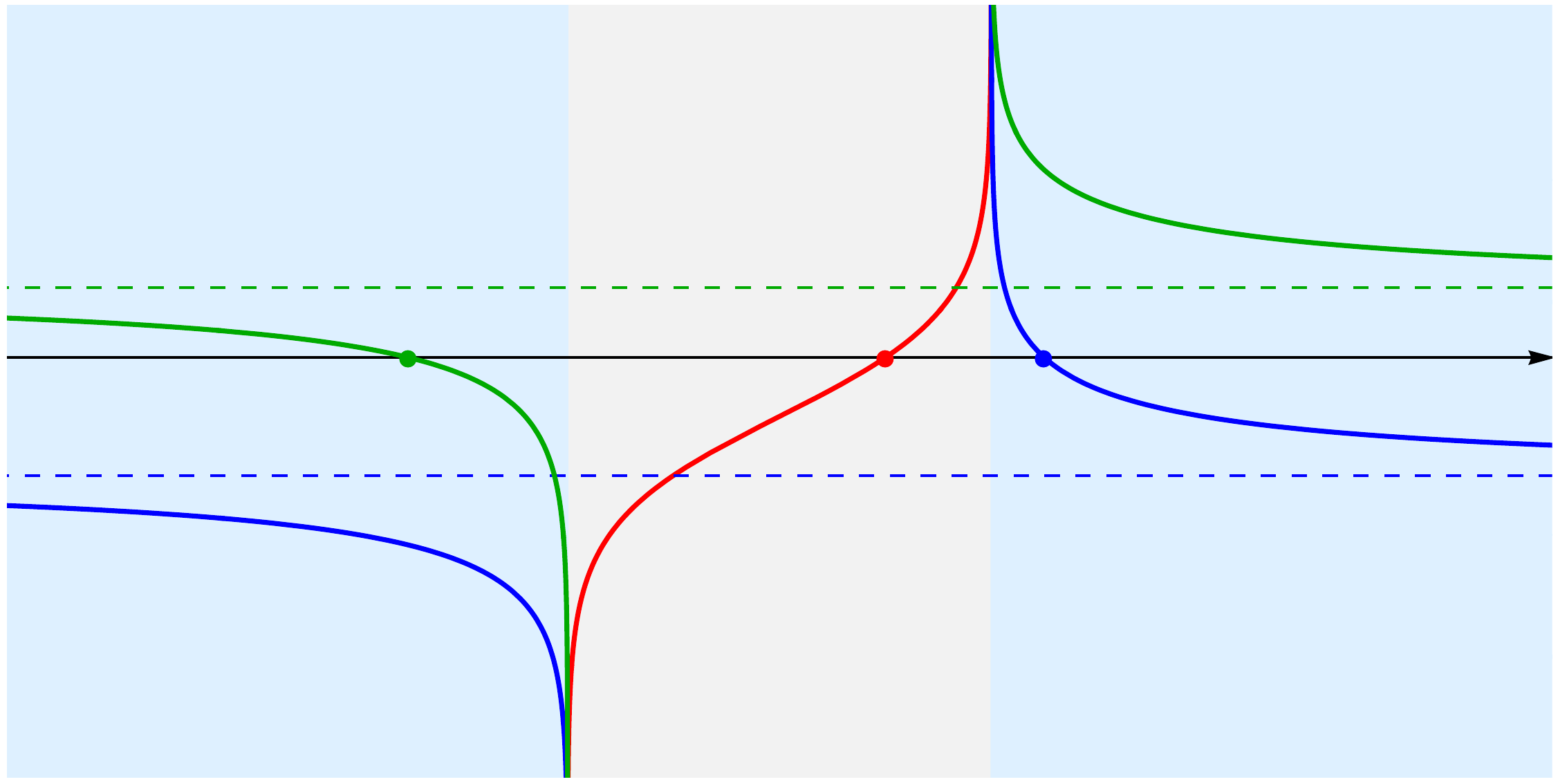}
\vspace{.2cm}
\caption{Modular evolutions along the chiral direction  in the plane $(\xi, \tau)$ given by (\ref{xi-map-fund}),
whose initial points correspond to the dots. 
Any horizontal dashed line is obtained from (\ref{tauB_def}) for the initial point in $B$ having the same color. 
}
\label{fig:1int-line-xi-v0}
\end{figure}

We remark that,  for any $\tau \in \RR$,
we have $\xi(\tau,u) \in A$ for $u \in A $ 
and $\xi(\tau,u) \in B$ for $u \in B $.
In Fig.\,\ref{fig:1int-line-xi-v0},
three examples of modular evolutions along the chiral direction given by (\ref{xi-map-fund}) are displayed
in the plane  parameterised by $(\xi, \tau)$, where the light grey strip corresponds to $\xi \in A$.
They are  the solid lines and their initial points are the dots having the corresponding color.

While the map (\ref{xi-map-fund}) is regular for any $u \in A$, it is singular for $u \in B$.
Indeed, the denominator in (\ref{xi-map-fund}) 
is strictly positive for any $\tau \in \RR$ when $u \in A$, 
while for  $u \in B$ it has a first order zero at $\tau = \tau_{B}(u)$ given by 
\be
\label{tauB_def}
\tau_{B}(u)
\equiv 
\frac{1}{2\pi} \,
\log \! \left(\frac{u-b}{u-a}\right)  
\;\;\;\;\qquad\;\;\;\;
u \in B\,.
\ee
The horizontal dashed lines in Fig.\,\ref{fig:1int-line-xi-v0}
indicate the asymptotic value (\ref{tauB_def}) corresponding to the initial points $u \in B$.
We remark that  (\ref{xi-map-fund}) solves the differential equations (\ref{der-xi});
hence, while it holds for any $\tau \in \RR$ when $u \in A$,  
it cannot be extended to any $\tau \in \RR$ when $u \in B$ 
because a divergence occurs for $\tau = \tau_{B}(u)$ given by (\ref{tauB_def}).

By introducing $u_0 \equiv -\,\mathsf{d} /\mathsf{c} = (a\, \e^{2\pi \tau} - b)/( \e^{2\pi \tau} - 1)$
to denote the singular point of (\ref{xi-map-fund}) for an assigned $\tau \neq 0$,
we have that $a- u_0 = (b-a)/( \e^{2\pi \tau} - 1)$ and $u_0 -b = (b-a)/( \e^{-2\pi \tau} - 1)$;
hence $u_0 < a$ for $\tau>0$, while $u_0 > b$ for $\tau<0$; hence $u_0 \in B$.

For later use, let us write the second expression in (\ref{der-xi}) more explicitly as follows
\be
\label{pA-qA-def}
\partial_u \xi(\tau,u) =q(\tau,u)^2
\;\;\;\;\;\qquad\;\;\;\;\;
q(\tau,u) \equiv \frac{b-a}{ (b-u)\,\e^{-\pi \tau}  +(u-a)\, \e^{\pi \tau}  } \,.
\ee
Notice that $q(\tau,u) >0$ for any $\tau \in \RR$ when $u \in A$,
while for $u \in B$
we have that $q(\tau,u) $ diverges 
and  changes sign at the finite value $\tau=\tau_{B}(u)$ given by (\ref{tauB_def}),
where the denominator of $q(\tau,u)$ in (\ref{pA-qA-def}) vanishes. 
%
%
In the limit $b \to +\infty$ where $A$ becomes the half line,
we have that $q(\tau,u) \to \e^{\pi \tau}$;
hence in this limit $q(\tau,u) $ is not singular anymore.

The fact that $\xi(\tau,u)$ for $u \in B$ has the horizontal asymptote at $\tau = \tau_B(u)$
tells us that the proper setup to consider 
to observe the same qualitative behaviour for the modular evolutions in $A$ and $B$ 
is a chiral CFT compactified on the circle,
which is discussed in Appendix\;\ref{app-1int-circle-vacuum}.

By applying (\ref{phi-mod-evo}) to the primary fields $\phi_\pm$ in the two different chiral directions,
one obtains
\be
\label{phi-mod-evo-pm}
\phi_\pm(\tau, u_\pm) = \big[ \partial_{u_\pm} \xi_\pm(\tau, u_\pm) \big]^{h_\pm} \phi_\pm\big(\xi_\pm(\tau, u_\pm) \big)
\;\;\;\qquad\;\;\;
\xi_\pm(\tau,u) = \xi(\pm\tau,u) 
\ee
where the notation described at the beginning of this section has been employed.


In the Tomita-Takesaki modular theory \cite{takesaki-book-70, Haag:1992hx}
the fundamental modular objects  are 
the self-adjoint and positive modular operator $\Delta \equiv \e^{-K}$, 
where $K$ is the full modular Hamiltonian,
and the antiunitary modular conjugation operator $J$.
In the setup explored throughout this section, 
$J$ has a geometric action implemented by the real function 
$\mathsf{j} :\mathbb{R}\rightarrow\mathbb{R}$ 
that can be obtained by setting $\tau=\pm\, \ri/2$ in (\ref{xi-map-fund})
and reads \cite{Haag:1992hx}\footnote{The two-dimensional case of Eq.\,(V.4.34) in \cite{Haag:1992hx}\ corresponds to (\ref{j0-map-def}) for $b=-a=1$.} 
\be
\label{j0-map-def}
\mathsf{j}(u) \equiv\frac{a+b}{2}+\frac{\left(\frac{b-a}{2}\right)^{2}}{u-\frac{a+b}{2}}
\ee
which is a bijective and idempotent function sending $A$ onto $B$
(see \cite{Mintchev:2022fcp, Abate:2022dyw} for recent discussions).
From (\ref{j0-map-def}), 
we have that $\mathsf{j}(u) < a$ for $u \in (a, \tfrac{a+b}{2})$, 
while $\mathsf{j}(u) > b$ for $u \in (\tfrac{a+b}{2}, b)$.

The map (\ref{j0-map-def}) is invariant under $a \leftrightarrow b$ and satisfies $\mathsf{j}'(u) < 0$.
Notice also that  (\ref{velocity_fund}) and (\ref{j0-map-def}) are related as follows
\be
\mathsf{j}'(u) \, V(u) = V(\, \mathsf{j}(u)) \,.
\ee
Moreover, (\ref{xi-map-fund}) and (\ref{j0-map-def}) commute, namely
\be
\label{id-xi-j-line}
\mathsf{j}\big( \xi(\tau, u) \big)
=
\xi\big(\tau,\mathsf{j}(u)\big)   
\ee
that will be used in Sec.\,\ref{sec-chiral-distance-mod-evo} and Sec.\,\ref{sec-mod-traject-1int}.

The action of the modular conjugation $J$ on the primary fields of a chiral CFT
(see e.g. \cite{Casini:2010bf}) 
is written in terms of (\ref{j0-map-def}) and  reads
\be
\label{J-on-phi}
J\,\phi(u)\,J 
\,=\, 
\mathsf{j}'(u)^{h} 
\, \phi^\ast (\,\mathsf{j}(u))\,.
\ee

A remarkable simplification occurs in (\ref{j0-map-def}) in the limit $b \to +\infty$, 
where $A$ becomes the right Rindler wedge;
indeed $\mathsf{j}(u) - a \to a - u$ as $b \to +\infty$,
i.e. a reflection with respect to the position of the entangling point.
Consequently, $\mathsf{j}'(u) \to -1$ as either $b \to +\infty$ or $a \to -\infty$.

\subsection{Chiral modular correlators } 
\label{sec-chiral-correlators}


A crucial role in our analysis is played by the modular two-point functions. 

The two-point function of the chiral primaries $\phi_\pm$ with dimension $h_\pm >0$ read
\be 
\label{2pt-chiral-primaries}
\langle \phi_\pm^*(u_1) \, \phi_\pm (u_2)\rangle
=
\langle \phi_\pm(u_1) \, \phi_\pm^* (u_2)\rangle
= 
\frac{ \e^{\mp \ri \pi h_\pm}  }{2\pi\, (u_1-u_2 \mp \ri \varepsilon )^{2h_\pm}}   
\ee 
where the normalisation constant adopted in \cite{Rehren:1987iy, Hollands:2019hje} has been employed.

The modular two-point function for the chiral primaries can be written by
combining (\ref{phi-mod-evo}) and (\ref{2pt-chiral-primaries}) \cite{Longo:2009mn, Hollands:2019hje}.
In order to write its expression in a convenient form, 
let us introduce the following function
\be
\label{W-function-def}
W(\tau ; u_1, u_2) 
\equiv  \frac{1}{(u_1 - u_2)\; \mathcal{R}(\tau ; u_1, u_2) }
\ee
where 
\be
\label{R-fact-def}
\mathcal{R}(\tau ; u_1, u_2) 
\equiv 
\frac{  \e^{2\pi w(u_1) +\pi \tau} - \e^{2\pi w(u_2) - \pi \tau} }{ \e^{2\pi w(u_1)} - \e^{2\pi w(u_2)}  } \,.
\ee
We remark that (\ref{W-function-def}) is well defined for $u_1 \neq u_2$ and 
that (\ref{R-fact-def}) satisfies $\mathcal{R}(0 ; u_1, u_2) = 1$.
Notice also that the limit $\tau \to 0$ and the limit $u_2 \to u_1$ of $\mathcal{R}(\tau ; u_1, u_2) $ in (\ref{R-fact-def}) do not commute;
hence this function is not jointly continuous.
By introducing  $\tau_{12} \equiv \tau_1 -\tau_2$,
the crucial property we need is that,
for $\tau_{12} \neq 0$ and $u_1 \neq u_2$,
the expressions in (\ref{xi-map-fund}) and (\ref{W-function-def}) are related as follows
\be
\label{rel-xi-W-1int}
\frac{\partial_{u_1} \xi(\tau_1, u_1)\, \partial_{u_2} \xi(\tau_2, u_2) }{\big[\xi(\tau_1, u_1) - \xi(\tau_2, u_2) \big]^2} = W(\tau_{12} ; u_1, u_2)^2\,.
\ee
It is worth remarking that  the dependence on $\tau_{12}$ in the l.h.s. of (\ref{rel-xi-W-1int}) is not obvious
and also that the singular behaviour of $\partial_{u_j} \xi(\tau_j, u_j)$ at $\tau_j = \tau_{B}(u_j)$ for $j \in \{1,2\}$ 
discussed in Sec.\,\ref{sec-1int-chiral-dir-mod-evo} (see (\ref{tauB_def}) and (\ref{pA-qA-def})) 
does not occur because their denominators simplify in the ratio.

Combining (\ref{rel-xi-W-1int}) with (\ref{phi-mod-evo}) and (\ref{2pt-chiral-primaries}),
one finds that the modular two-point function of the chiral primaries reads
\be
\label{mod-corr-phi-mu}
\langle \phi_\pm^*(\tau_1, u_1) \,\phi_\pm (\tau_2, u_2)\rangle
\, =\,
\frac{ \e^{\mp \ri \pi h_\pm} }{2\pi }\;
W_\pm(\pm \tau_{12}; u_1 , u_2 )^{2h_{\pm}}
\ee
where we have introduced the following distributions
\be
\label{cap-W-def}
W_\pm(\tau; u_1, u_2)
\,\equiv \,
\frac{ \e^{2\pi w(u_1)} - \e^{2\pi w(u_2)}  }{ u_1 - u_2  }\;
\frac{1}{  \e^{2\pi w(u_1) +\pi \tau} - \e^{2\pi w(u_2) - \pi \tau} \mp \ri \varepsilon}  \,.
\ee

We remark that the following property 
\be
W_\pm(\tau \pm \ri \, ; u_1, u_2) = W_\pm(- \tau; u_2, u_1)
\ee
which is satisfied by (\ref{cap-W-def}), 
guarantees that (\ref{mod-corr-phi-mu}) fulfils the  KMS condition,
a fundamental characteristic feature of any modular flow \cite{Haag:1992hx}.

The distributions in (\ref{cap-W-def}) provide also the modular two-point function 
of  the chiral components of a conserved current $j_\pm$.
The chiral fields $j_\pm$,
which have dimension $h_{j_\pm}=1$ and generate the $U(1)$  transformations,
satisfy the following commutation relations \cite{Luscher}
\be 
\big[ \, j_\pm(u)\, ,\, j_\pm (v) \, \big]_{-} = \mp \, \ri \, \frac{\kappa}{2\pi} \, \delta^\prime(u-v) 
\label{cft7}
\ee
where $\kappa $ is a real constant and the r.h.s. is known as Schwinger term.
The modular correlators of $j_\pm$ can be expressed in terms of (\ref{cap-W-def}) as follows  \cite{Hollands:2019hje}
\be
\label{mod-corr-j}
\langle\, j_\pm(\tau_1, u_1) \,j_\pm (\tau_2, u_2) \,\rangle
=
\frac{\kappa}{4\pi^2} \,W_\pm(\pm \tau_{12}; u_1, u_2)^{2}   
\ee
which is proportional to (\ref{mod-corr-phi-mu}) specialised to $h_\pm =1$.


A crucial role in the subsequent discussion is played by the finite value $\tau = \tilde{\tau}_0 $ 
where the function $\mathcal{R}(\tau ; u_1, u_2) $  in (\ref{R-fact-def}) with $u_1 \neq u_2$
vanishes, which is given by 
\be
\label{tau0tilde-def}
\tilde{\tau}_0(u_1, u_2) \equiv w(u_2) - w(u_1) \,.
 \ee
This expression satisfies also the following equivalent conditions
 \be
 \label{tau0def-u1u2}
 u_2 = \xi(\tilde{\tau}_0,u_1) 
 \;\;\;\qquad\;\;\;
  u_1 = \xi(-\tilde{\tau}_0,u_2) \,.
 \ee
Let us also observe that (\ref{xi-map-fund}) and (\ref{R-fact-def}) are related through the following condition
\be
\label{R-zero-xi-1int}
\mathcal{R}\big(\tau ; u, \xi(\tau,u)  \big) = 0 \,.
\ee
An important limiting regime to consider is given by $b\to + \infty$,
where (\ref{R-fact-def}) simplifies to 
\be
\label{Omega12-rindler}
\lim_{b \to +\infty} \mathcal{R}(\tau ; u_1, u_2) = \frac{(u_1 - a)\, \e^{\pi \tau}  - (u_2 - a) \, \e^{-\pi \tau} }{u_1 - u_2} \,.
\ee

\subsection{Commutator of the currents  } 
\label{sec-1int-commutators}


A universal quantity that we find it worth considering 
is the commutator of the modular evolution of the chiral components of the conserved current.
Under the reasonable assumption that this commutator gives a complex number
(whose validity is checked in Sec.\,\ref{sec-chiral-distance-mod-evo},
where this commutator is obtained through the modular flows)
it can be obtained from the modular correlators (\ref{mod-corr-j}) as follows
\bea
\label{comm-j-curr-v1-1int}
\big[ \, j_{\pm} (\tau_1 , u_1) \, , \, j_{\pm} (\tau_2 , u_2)  \,\big]_{-}
&=&
\langle \,j_{\pm} (\tau_1 , u_1) \,j_{\pm} (\tau_2 , u_2) \,\rangle 
-
\langle \,j_{\pm} (\tau_2 , u_2)\, j_{\pm} (\tau_1 , u_1)  \,\rangle 
\\
\rule{0pt}{.9cm}
& & \hspace{-2.2cm}
=\,
\frac{\kappa}{4\pi^2} \, \frac{ \big(\e^{2\pi w(u_1)} - \e^{2\pi w(u_2)} \big)^2 }{ (u_1 - u_2)^2  }\;
\left\{\,
\frac{1}{  \big( \e^{2\pi w(u_1) \pm \pi \tau_{12}} - \e^{2\pi w(u_2) \mp  \pi \tau_{12}} \mp \ri \varepsilon \big)^2}  
\right.
\nn
\\
\rule{0pt}{.7cm}
& & \hspace{3.3cm}
\left.
- \,\frac{1}{ \big( \e^{2\pi w(u_1) \pm \pi \tau_{12}} - \e^{2\pi w(u_2) \mp  \pi \tau_{12}} \pm \ri \varepsilon \big)^2}  
\right\} \,.
\nn
\eea
Taking the limit $\varepsilon \to 0^+$ of this expression in the sense of the distributions 
and employing  the identity $\frac{1}{(u\pm \ri \varepsilon)^2} = \frac{1}{u^2} \pm \ri \pi \,\delta^\prime (u) $,
we find
\be
\label{comm-j-curr-v3-1int}
\big[ \, j_{\pm} (\tau_1 , u_1) \, , \, j_{\pm} (\tau_2 , u_2)  \,\big]_{-}
=\,
\mp \; \frac{\ri \, \kappa}{2\pi} \,
\Bigg(
\frac{ \e^{2\pi w(u_1)} - \e^{2\pi w(u_2)}  }{ u_1 - u_2  }
\Bigg)^2
\, \delta'\big(  \e^{2\pi w(u_1) \pm \pi \tau_{12}} - \e^{2\pi w(u_2) \mp  \pi \tau_{12}}  \big)
\ee
where the derivative of the Dirac delta with respect to its argument occurs
and  the function multiplying $\delta' $ 
is independent of $\tau_{12}$ and non vanishing for $u_1 \neq u_2$.


The derivative of the Dirac delta in (\ref{comm-j-curr-v3-1int}) can be written by employing the following formula 
\be
\label{delta-prime-res}
G(x)\,
\frac{\partial \,\delta\big(f(x)\big) }{\partial f(x)}
\,=
\sum_j 
\frac{ G(x_j) }{ f'(x_j) \, \big| f'(x_j) \big| } 
\left[\,
\partial_x \delta(x-x_j)
- 
\left(
\frac{G'(x_j)}{G(x_j)} - \frac{f''(x_j)}{f'(x_j) }
\right)
\delta(x-x_j)
\right]
\ee
where the sum is performed over the zeros $x_j$ of $f(x)$ such that $f'(x_j)  \neq 0$.
The derivation of (\ref{delta-prime-res}) 
is reported in Appendix\;\ref{app-derivative-Dirac-delta}.


The identity (\ref{delta-prime-res}) can be applied to the r.h.s. of (\ref{comm-j-curr-v3-1int}),
by identifying  $f$ and $G$ respectively 
with the argument of $\delta'$ 
and the function multiplying $\delta'$ in (\ref{comm-j-curr-v3-1int}),
with $w(u)$ given by (\ref{w_fund}).
Then, considering $u_2$ as the variable $x$ in (\ref{delta-prime-res}),
from (\ref{R-fact-def}) and (\ref{R-zero-xi-1int}) we have  that the sum in the r.h.s. of (\ref{delta-prime-res})
contains only one term corresponding to the zero of the argument of $\delta'$ 
in (\ref{comm-j-curr-v3-1int}), i.e. $u_2 = \xi(\pm\tau_{12},u_1) $.
As for  the expression within the round brackets 
multiplying $\delta(x-x_j)$ in the r.h.s. of (\ref{delta-prime-res}),
it vanishes identically in this case.
Thus,  the identity (\ref{delta-prime-res}) leads us to write
the commutator (\ref{comm-j-curr-v3-1int}) as 
\be
\label{commutator-j-single-int}
\big[ \, j_{\pm} (\tau_1 , u_1) \, , \, j_{\pm} (\tau_2 , u_2)  \,\big]_{-}
=\,
\pm \; \ri\, \frac{\kappa}{2\pi} \,
\partial_{u_1} \xi(\pm\tau_{12},u_1) 
\;\partial_{u_2} \delta\big(  u_2 - \xi(\pm\tau_{12},u_1) \big) \,.
\ee
When $\tau_1 = \tau_2\equiv \tau$, this commutator becomes
\be
\label{commutator-j-single-int-equal-tau}
\big[ \, j_{\pm} (\tau , u_1) \, , \, j_{\pm} (\tau , u_2)  \,\big]_{-}
=\,
\mp \, \ri \, \frac{\kappa}{2\pi} \, \delta^\prime(u_1-u_2) 
\ee
which extends (\ref{cft7}) to any allowed value of the modular parameter. 
In Sec.\,\ref{sec-chiral-distance-mod-evo}, 
the commutator (\ref{commutator-j-single-int-equal-tau}) is obtained by combining 
$j_{\pm} (\tau, u) \equiv \e^{\pm  \ri \tau K }\, j_{\pm} (u)\, \e^{\mp \ri \tau K}  $ and (\ref{cft7}),
without the initial assumption on the commutator made at the beginning of this derivation. 
%


The l.h.s. of (\ref{commutator-j-single-int}) is antisymmetric under exchange $1 \leftrightarrow 2$ by definition,
while this feature is not manifest in the r.h.s. of the same equation. 
This is because $u_2$ as been chosen as the variable $x$ in the application of (\ref{delta-prime-res});
indeed, the antisymmetry under exchange $1 \leftrightarrow 2$ is still manifest in both the sides of (\ref{comm-j-curr-v3-1int}).
In order to write the r.h.s. of (\ref{commutator-j-single-int}) in a form where the 
antisymmetry under exchange $1 \leftrightarrow 2$ is manifest,
first the procedure after (\ref{comm-j-curr-v3-1int}) must be repeated 
by choosing $u_1$ as the variable $x$ in the application of (\ref{delta-prime-res})
and then the resulting expression has to be properly combined with the l.h.s. of (\ref{commutator-j-single-int}).
This leads to 
\bea
\label{commutator-j-single-int-symmetrised}
\big[ \, j_{\pm} (\tau_1 , u_1) \, , \, j_{\pm} (\tau_2 , u_2)  \,\big]_{-}
&=&
\pm \; \ri\, \frac{\kappa}{4\pi} \,
\bigg\{\,
\partial_{u_1} \xi(\pm\tau_{12},u_1) 
\;\partial_{u_2} \delta\big(  u_2 - \xi(\pm\tau_{12},u_1) \big)
\\
& &\hspace{1.6cm}
-\, \partial_{u_2} \xi(\pm\tau_{21},u_2) 
\;\partial_{u_1} \delta\big(  u_1 - \xi(\pm\tau_{21},u_2) \big)
\, \bigg\}
\nn
\eea
which is manifestly antisymmetric under exchange $1 \leftrightarrow 2$ in both its sides. 
Notice that $u_2 = \xi(\pm\tau_{12},u_1)$ and $u_1 = \xi(\pm\tau_{21},u_2) $ 
are the same equation because (\ref{xi-map-fund}) satisfies $\xi( -\tau , \xi(\tau,u) ) = u$,
as already remarked in the text below (\ref{der-xi}).

\subsection{Chiral distance along the modular evolutions  } 
\label{sec-chiral-distance-mod-evo}

The results discussed in Sec.\,\ref{sec-1int-chiral-dir-mod-evo} and Sec.\,\ref{sec-chiral-correlators}
allow us to explore $\xi(\tau_1 ,u_1)  - \xi(\tau_2 ,u_2) $,
which can be considered as the chiral distance between 
the modular evolutions of two generic points along the chiral direction
(with a slight abuse of terminology because this quantity can be negative),
for independent values of the modular parameter in the two distinct evolutions. 
By employing (\ref{W-function-def}) in  (\ref{rel-xi-W-1int}), 
we find that  this chiral distance can be written as follows
\be
\label{xi12-tau-chiral}
\xi(\tau_1 ,u_1)  - \xi(\tau_2 ,u_2) 
\,=\,
 \mathcal{R}(\tau_{12} ; u_1, u_2)\; q(\tau_1 ,u_1) \, q(\tau_2 ,u_2)  \; (u_1 - u_2)
\ee
in terms of the functions introduced in (\ref{pA-qA-def}) and in (\ref{R-fact-def}).

It is instructive to consider the limiting regimes of (\ref{xi12-tau-chiral}) 
given by $u_1 = u_2$ and $\tau_1 = \tau_2$. 

For $u_1 = u_2 \equiv u$, 
since from (\ref{R-fact-def}) we have
\be
\label{R-u12-same-u}
\lim_{v \to u} \mathcal{R}(\tau ; u, v) \, (u - v)
=
\frac{\sinh(\pi \tau)}{\pi\, w'(u)}
\ee
the chiral distance (\ref{xi12-tau-chiral}) in this limiting regime simplifies to 
\be
\label{xi-12-equal-u1u2}
\xi(\tau_1 ,u)  - \xi(\tau_2 ,u) =  \frac{1}{\pi}\, q(\tau_1 ,u) \, q(\tau_2 ,u) \, V(u) \,\sinh(\pi \tau_{12})
\ee
which vanishes when $\tau_1 = \tau_2$, as expected.

\begin{figure}[t!]
\vspace{-.2cm}
\hspace{-1.1cm}
\includegraphics[width=1.15\textwidth]{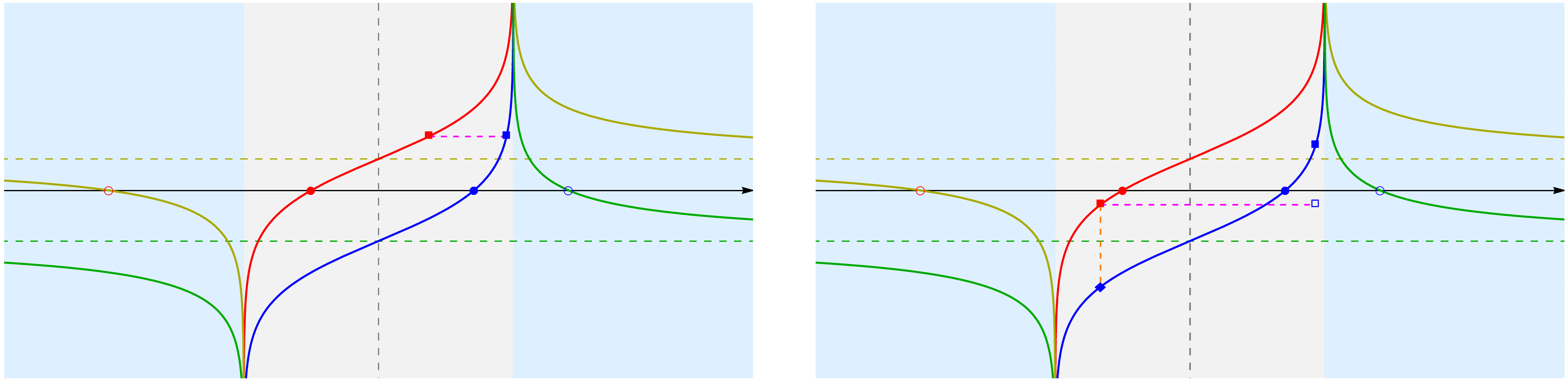}
\vspace{-.3cm}
\caption{
The chiral distance (\ref{xi12-tau-chiral}) between the  points  (red and blue squares)
belonging to two distinct chiral modular evolutions whose initial points (red and blue dots) 
are in $A$ (see also (\ref{xi12-tau-chiral-2})),
for either $\tau_1 = \tau_2$ (left panel) or $\tau_1 \neq \tau_2$ (right panel).
}
\label{fig:1int-line-xi-distance}
\end{figure}

For $\tau_1 = \tau_2 \equiv \tau$, 
since $\mathcal{R}(0 ; u_1, u_2) =1$ 
the chiral distance (\ref{xi12-tau-chiral}) becomes
\be
 \label{xi12-tau-chiral-equal-tau}
\xi(\tau ,u_1)  - \xi(\tau ,u_2) =  q(\tau ,u_1) \, q(\tau ,u_2)  \; (u_1 - u_2)
\ee
which vanishes as $u_1 = u_2$, as expected. 
Since $q(\tau ,u) >0$ for $u \in A$ (see (\ref{pA-qA-def})), 
when $u_1 \in A$ and $u_2 \in A$ we have that 
$\xi(\tau ,u_1)  - \xi(\tau ,u_2) $ and $u_1 - u_2$ have the same sign.
An example of this case is illustrated in the left panel of Fig.\,\ref{fig:1int-line-xi-distance},
where each coloured square indicates the modular evolution of the corresponding initial point 
(denoted by the dot having the same colour)
and the horizontal dashed magenta segment denotes the chiral distance between them.
Instead, in the mixed case where for instance $u_1 \in A$ and $u_2 \in B$, 
the factor $q(\tau ,u_1) \, q(\tau ,u_2) $ diverges at $\tau = \tau_{B}(u_2) $ (see (\ref{tauB_def}))
and the l.h.s. of (\ref{xi12-tau-chiral-equal-tau}) changes sign at this value of $\tau$.
When both  $u_1 \in B$ and $u_2 \in B$, 
 the factor $q(\tau ,u_1) \, q(\tau ,u_2) $ diverges at $\tau = \tau_{B}(u_1) $ and $\tau = \tau_{B}(u_2) $;
and  a change of sign occurs in the l.h.s. of (\ref{xi12-tau-chiral-equal-tau}) at both these values of $\tau$.

When $u_1  \in A$ and $u_2 \in A$,
from (\ref{pA-qA-def}) and the fact that $q(\tau ,u) >0$ for  $u \in A$, 
we have that (\ref{xi12-tau-chiral}) can be written as follows
\be
 \label{xi12-tau-chiral-2}
 \xi(\tau_1 ,u_1)  - \xi(\tau_2 ,u_2) 
\,=\,
 \mathcal{R}(\tau_{12} ; u_1, u_2)\; \sqrt{\partial_{u_1} \xi(\tau_1,u_1)\, \partial_{u_2} \xi(\tau_2,u_2)}   \; (u_1 - u_2)
\ee
hence the sign of $ \xi(\tau_1 ,u_1)  - \xi(\tau_2 ,u_2) $ coincides with
the sign of  $\mathcal{R}(\tau_{12} ; u_1, u_2)\, (u_1 - u_2)$.

In the right panel of Fig.\,\ref{fig:1int-line-xi-distance} an  example 
illustrates the geometrical meaning of the l.h.s. of (\ref{xi12-tau-chiral-2}) when $\tau_1 \neq \tau_2$
and both $u_1$ and $u_2$ belong to $A$.
The blue and red solid lines within the light grey vertical strip describe the modular evolutions
whose initial points $u_1$ and $u_2$ are the blue and the red dots respectively. 
The blue and the red squares denote the generic points along the corresponding modular evolution,
at some values $\tau_1$ and $\tau_2$ of the modular parameter respectively. 
The difference in the l.h.s. of (\ref{xi12-tau-chiral-2}) corresponds to the horizontal dashed magenta segment,
whose sign depends on the horizontal projection of the segment connecting the two squares. 
In particular, it vanishes when the blue square coincides with the blue rhombus
(see the vertical dashed orange segment).


In Fig.\,\ref{fig:1int-line-xi-distance}
(where the vertical dashed grey line indicates the center of $A$, given by $u = \tfrac{a+b}{2}$),
the images of the two initial points in $A$ (blue and red dots) through the map (\ref{j0-map-def})
are the empty circles having  the same colour.
The corresponding modular evolutions (see the yellow and green solid curves)
are obtained from (\ref{xi-map-fund}) and (\ref{id-xi-j-line}) and belong entirely to $B$. 
The horizontal dashed straight lines having the same colour 
indicate their asymptotic values (\ref{tauB_def}).

When $u\in A$, from (\ref{tauB_def}) and (\ref{j0-map-def}), for later use (see (\ref{tauA-minmax-def}))
we find it worth introducing 
\be
\label{tauA_def}
\tau_{A}(u)
\,\equiv\, 
\tau_{B}\big(\,\mathsf{j}(u)\big)
=
\frac{1}{2\pi} \,
\log \! \left(\frac{b-u}{u-a}\right)
=  
- \, w(u)
\;\;\;\;\qquad\;\;\;\;
u \in A
\ee
which satisfies
\be
\xi\big(\tau_{A}(u) , u\big) = \frac{a+b}{2}
\;\;\;\;\qquad\;\;\;\;
u \in A
\ee
hence in Fig.\,\ref{fig:1int-line-xi-distance}
the value (\ref{tauA_def}) corresponds to the intersections between the vertical dashed grey line
and the solid curves within the light grey strip.

The derivation of the commutator (\ref{commutator-j-single-int}) through the modular flow of the currents $j_\pm$ is an instructive application of (\ref{xi12-tau-chiral}).
From  (\ref{phi-mod-evo-pm}) with $h_\pm=1$, the modular flow of $j_\pm$ reads
\be 
j_\pm (\tau,u) 
= 
\partial_u \xi_\pm (\tau,u)  \, j _\pm\big(\xi_\pm (\tau,u)\big) \, .
\ee
This allows to study the commutator (\ref{commutator-j-single-int}) without employing the corresponding modular correlators
(as done in Sec.\,\ref{sec-1int-commutators}), finding 
\bea
\label{commutator-j-single-int-xi}
\big[ \, j_{\pm} (\tau_1 , u_1) \, , \, j_{\pm} (\tau_2 , u_2)  \,\big]_{-}
&=&
\nn
\\
& &
\hspace{-2cm}
=\;
\partial_{u_1} \xi_\pm (\tau_1,u_1) \, \partial_{u_2} \xi_\pm (\tau_2,u_2) \,
\big[ \,  j _\pm\big(\xi_\pm (\tau_1,u_1)\big)  \, , \,  j _\pm\big(\xi_\pm (\tau_2,u_2)\big)  \,\big]_{-}
\\
\rule{0pt}{.55cm}
& &
\label{commutator-j-single-int-xi-s1}
\hspace{-2cm}
=\;
\pm\, \ri \, \frac{\kappa}{2\pi}\;
 \partial_{u_1} \xi_\pm (\tau_1,u_1) \, \partial_{u_2} \xi_\pm (\tau_2,u_2) \;
\delta' \big( \xi_\pm (\tau_2,u_2)- \xi_\pm (\tau_1,u_1) \big)
\hspace{1cm}
\eea
where the argument of $\delta '$ is given by (\ref{xi12-tau-chiral})
and  $\xi_\pm(\tau,u) = \xi(\pm\tau,u) $ (see (\ref{phi-mod-evo-pm})).
The argument of $\delta '$ in (\ref{commutator-j-single-int-xi-s1}) is 
the chiral distance (\ref{xi12-tau-chiral});
hence its vanishing condition can be studied by employing (\ref{R-zero-xi-1int}) and this leads to 
\be
\label{xi-inv-rel-12}
\xi(\tau_1 ,u)  - \xi(\tau_2 , \xi(\tau_{12},u) ) = 0 \,.
\ee
This condition is the modular counterpart of the corresponding result for the standard evolution
without the bipartition of the line, where $\xi(t ,x) = x + v \,t$ for some velocity $v$.

The expression in (\ref{commutator-j-single-int-xi-s1}) can be rewritten 
by applying (\ref{delta-prime-res}) to this case.
Setting $x =u_2$ as the independent variable, we have
 $f(x) =  \xi_\pm (\tau_2,u_2)- \xi_\pm (\tau_1,u_1) $ and $G(x) = \partial_{u_2} \xi_\pm (\tau_2,u_2) $.
Then, (\ref{xi-inv-rel-12}) implies that $f(x)$ has a single zero given by $x_j = \xi(\tau_{12},u_1) $.
In this case it is straightforward to realise that the expression within the round brackets 
multiplying the Dirac delta in the r.h.s. of (\ref{delta-prime-res}) vanishes identically,
as already mentioned in Sec.\,\ref{sec-1int-commutators}.
As for the ratio multiplying the square brackets in the r.h.s. of (\ref{delta-prime-res}),
its expression specialised to this case combined with the fact that (\ref{xi-map-fund}) satisfies 
\be
\label{der-xi-ratio-tau12}
\partial_{u_1} \xi (\tau_{12},u_1)
=
\frac{\partial_{u_1} \xi (\tau_1,u_1) }{ \partial_{u_2} \xi (\tau_2,u_2)  } \, \bigg|_{u_2 \,=\, \xi (\tau_{12},u_1) } 
\ee
hence  (\ref{commutator-j-single-int}) is obtained. 
We remark that the finiteness of $\partial_{u} \xi (\tau,u)$ (see (\ref{pA-qA-def}))
is an important assumption throughout the above analysis.

\subsection{Modular trajectories } 
\label{sec-mod-traject-1int}


In the Minkowski spacetime ${\mathbb M} = {\mathbb R}_x  \times  {\mathbb R}_t$ 
parameterised through the  light-cone coordinates (\ref{light-cone coordinates}),
consider the spacetime domains $\mathcal{D}_A $ and $\mathcal{B}_A $ 
introduced at the beginning of Sec.\,\ref{sec-1int-chiral-dir-mod-evo}.

The modular trajectory associated to a point $P \in \mathcal{D}_A \cup \mathcal{B}_A$ 
having light-cone coordinates $(u_+, u_-)$
is the curve in $\mathcal{D}_A \cup \mathcal{B}_A$ 
made by the points $P(\tau)$ parameterised by $\tau \in \RR$, 
with light-cone coordinates obtained from (\ref{xi-map-fund}) 
and whose spacetime coordinates are
\be
\label{mod-traj-tau-line}
x(\tau) = \frac{\xi(\tau, u_+) + \xi(-\tau, u_-) }{2}
\;\;\;\qquad\;\;\;
t(\tau) = \frac{\xi(\tau, u_+) - \xi(-\tau, u_-) }{2}
\ee
which satisfy the initial condition $P = P(\tau=0)$ because $\xi(\tau =0, u) = u$.
Since $\xi(\tau, u) \in A$ for $u \in A$ and $\xi(\tau, u) \in B$ for $u \in B$,
as already mentioned in Sec.\,\ref{sec-1int-chiral-dir-mod-evo} (see Fig.\,\ref{fig:1int-line-xi-v0}),
we have  that $P(\tau) \in \mathcal{D}_A$ when $P \in \mathcal{D}_A$ 
and that $P(\tau) \in \mathcal{B}_A$ when $P \in \mathcal{B}_A$.

\begin{figure}[t!]
\vspace{-.5cm}
\hspace{.5cm}
\includegraphics[width=.94\textwidth]{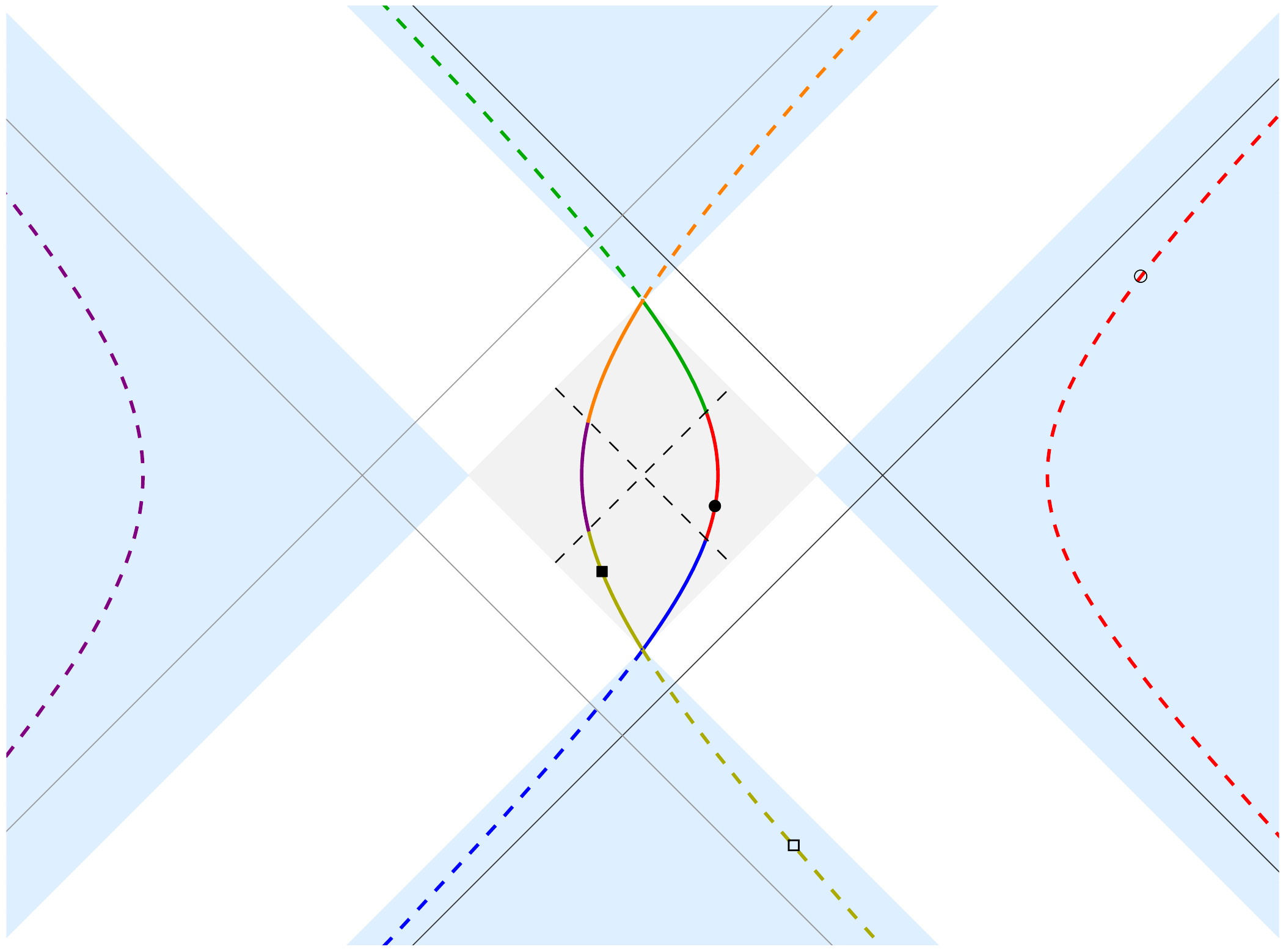}
\vspace{.2cm}
\caption{
Modular trajectories (solid curves) in the diamond $\mathcal{D}_A$ (light grey region),
whose initial points correspond to the black dot and the black square,
and the corresponding modular trajectories (dashed curves)  in $\mathcal{B}_A$ (light blue region),
obtained through the geometric action of the modular conjugation,
which relates dashed and solid arcs having the same colour.
The union of each modular trajectory and of the corresponding modular trajectory
obtained through the modular conjugation map 
(see (\ref{mod-conj-traj-tau-line-x})-(\ref{mod-conj-traj-tau-line-t}))  
gives the hyperbolae (\ref{mod-hyper})-(\ref{mod-hyper-parameters}).
}
\label{fig:diamond-mod-hyper}
\end{figure}

The spacetime region $\mathcal{B}_A $ is naturally partitioned as
$\mathcal{B}_A = \mathcal{W}_{\textrm{\tiny R}} \cup \mathcal{W}_{\textrm{\tiny L}} \cup \mathcal{V}_{\textrm{\tiny F}} \cup \mathcal{V}_{\textrm{\tiny P}} $,
where 
$\mathcal{W}_{\textrm{\tiny R}} = \big\{ (u_+, u_-) \,|\, u_\pm > b \big\} $ 
and 
$\mathcal{W}_{\textrm{\tiny L}} = \big\{ (u_+, u_-) \,|\,  u_\pm < a  \big\} $ 
are right and left Rindler wedges respectively,
while 
$\mathcal{V}_{\textrm{\tiny F}} = \big\{ (u_+, u_-) \,|\, u_{+} > b \,, u_{-} < a  \big\} $ 
and 
$\mathcal{V}_{\textrm{\tiny P}} = \big\{ (u_+, u_-) \,|\,  u_{+} < a \,, u_{-} > b  \big\} $
are a future and past cone respectively.
The subindices R, L, F and P correspond to right, left, future and past respectively.
By employing the modular conjugation map (\ref{j0-map-def}) for both the  light-cone coordinates,
it is straightforward to observe that this partition of $\mathcal{B}_A $ uniquely induces the partition 
$\mathcal{D}_A = \mathcal{D}_{\textrm{\tiny R}} \cup \mathcal{D}_{\textrm{\tiny L}} \cup \mathcal{D}_{\textrm{\tiny F}} \cup \mathcal{D}_{\textrm{\tiny P}} $
of the diamond $\mathcal{D}_A $ made  by four smaller diamonds,
where each $\mathcal{D}_k$ with $k \in \{ \textrm{R} , \textrm{L} , \textrm{F} , \textrm{P}  \}$ 
is the image through the modular conjugation map (\ref{j0-map-def}) of 
the subset of $\mathcal{B}_A $ labelled by the same subindex.
Each $\mathcal{D}_k$ contains only one vertex of $\mathcal{D}_A$ 
which is shared with the corresponding subset of $\mathcal{B}_A $ 
having  the same subindex.
This partition of $\mathcal{D}_A $ is indicated by the dashed black segments in Fig.\,\ref{fig:diamond-mod-hyper},
which intersect in the center of the diamond, whose light-cone coordinates are $u_+ = u_- = \tfrac{a+b}{2}$.

Each modular trajectory belongs either in $\mathcal{D}_A $ or in $\mathcal{B}_A $,
depending on whether its initial point is either in $\mathcal{D}_A $ or in $\mathcal{B}_A $ respectively,
and these partitions of $\mathcal{D}_A $ and $\mathcal{B}_A $ naturally induce on any modular trajectory
a partition made by at most three arcs. 
Considering a point $P \in \mathcal{D}_A$ as initial point, 
the corresponding modular trajectory belongs to $\mathcal{D}_A $,
namely $P(\tau) \in \mathcal{D}_A$ for any $\tau \in \RR$.
If the modular trajectory $P(\tau)$ does not coincide with the vertical segment at $x =\tfrac{a+b}{2}$
passing through the center of $\mathcal{D}_A$ and joining its top and bottom vertices, 
it is convenient to introduce the partition 
$\RR= (-\infty, \tau_A^{<} ) \cup \big[\tau_A^{<}  , \tau_A^{>} \big] \cup (\tau_A^{>}  , +\infty)$ for the values of $\tau$
such that
$P(\tau)  \in \mathcal{D}_{\textrm{\tiny P}}$ for $\tau \in (-\infty, \tau_A^{<}) $,
$P(\tau)  \in \mathcal{D}_{\textrm{\tiny F}}$ for $\tau \in (\tau_A^{>}, +\infty)$
and either 
$P(\tau)  \in \mathcal{D}_{\textrm{\tiny R}}$ or $P(\tau)  \in \mathcal{D}_{\textrm{\tiny L}}$  for $\tau \in \big[\tau_A^{<}  , \tau_A^{>} \big]$,
depending on whether $P$ belongs to either the right half or the left half of the diamond $\mathcal{D}_A $
bipartite by the vertical segment joining its top and bottom vertices 
(this bipartition of $\mathcal{D}_A $ is employed also in Appendix\;\ref{app-new-inversion},
as indicated by the dashed brown segment in Fig.\,\ref{fig:new-inversion-non-chiral}).
By specifying (\ref{tauA_def}) to the light-cone coordinates of the initial point $P$, 
we obtain $\tau_A^\gtrless$ as follows
\be
\label{tauA-minmax-def}
\tau_A^{<} \,\equiv\, \textrm{min} \big\{ \tau_A(u_{+}) \,, -\,\tau_A(u_{-}) \big\}
\;\;\;\qquad\;\;\;
\tau_A^{>} \,\equiv\, \textrm{max} \big\{ \tau_A(u_{+}) \,, -\,\tau_A(u_{-}) \big\} \,.
\ee

In the special case where the initial point $P$ is on the vertical line $x =\tfrac{a+b}{2}$,
the corresponding modular trajectory is a vertical segment belonging to the same line
(see the dashed brown segment in Fig.\,\ref{fig:new-inversion-non-chiral}).
This segment contains the center of $\mathcal{D}_A$
and is bipartite into two segments lying in $\mathcal{D}_{\textrm{\tiny P}}$ and $\mathcal{D}_{\textrm{\tiny F}}$.
In this case $\tau_A(u_{+}) = - \,\tau_A(u_{-})$ and therefore, from (\ref{tauA-minmax-def}),
we have that $\tau_A^{<} = \tau_A^{>} $, 
which characterises the point where the bipartition of the modular trajectory occurs. 
In Fig.\,\ref{fig:diamond-mod-hyper}, 
we show two modular trajectories in $\mathcal{D}_A$ (see the solid curves) 
whose initial points are indicated through the black dot and the black square. 
Since these initial points do not have $x =\tfrac{a+b}{2}$, 
the corresponding modular trajectories are partitioned in three arcs denoted by different colours
and obtained by specifying (\ref{tauA-minmax-def}) 
to the light-cone coordinates of the initial point of each modular trajectory.

Similarly, the modular trajectory with an initial point $P \in \mathcal{B}_A$ 
entirely belongs to $\mathcal{B}_A$, i.e. $P(\tau) \in \mathcal{B}_A$ for any $\tau \in \RR$,
and the above analysis can be adapted to this case.
In particular, 
if $P$ does not belong to the vertical line at $x =\tfrac{a+b}{2}$,
then $P(\tau)$ has a brach either in $\mathcal{W}_{\textrm{\tiny L}}$ or in $\mathcal{W}_{\textrm{\tiny R}}$
and therefore one can define the partition 
$\RR= (-\infty, \tau_B^{<} ) \cup \big[\tau_B^{<}  , \tau_B^{>} \big] \cup (\tau_B^{>}  , +\infty)$ for the values of $\tau$
such that 
$P(\tau)  \in \mathcal{V}_{\textrm{\tiny P}}$ for $\tau \in (-\infty, \tau_B^{<}) $,
$P(\tau)  \in \mathcal{V}_{\textrm{\tiny F}}$ for $\tau \in (\tau_B^{>}, +\infty)$,
and either 
$P(\tau)  \in \mathcal{W}_{\textrm{\tiny R}}$ or $P(\tau)  \in \mathcal{W}_{\textrm{\tiny L}}$  for $\tau \in \big[\tau_B^{<}  , \tau_B^{>} \big]$,
depending on the position of initial point $P \in \mathcal{B}_A$.
%
By specifying (\ref{tauB_def}) to the light-cone coordinates of the initial point $P$,
one introduces
\be
\label{tauB-minmax-def}
\tau_B^{<} \,\equiv\, \textrm{min} \big\{ \tau_B(u_{+}) \,, -\,\tau_B(u_{-}) \big\}
\;\;\;\qquad\;\;\;
\tau_B^{>} \,\equiv\, \textrm{max} \big\{ \tau_B(u_{+}) \,, -\,\tau_B(u_{-}) \big\}
\ee
which characterise the three infinite arcs occurring in the partition of a modular trajectory with initial point $P \in \mathcal{B}_A$ 
induced by the partition $\mathcal{B}_A = \mathcal{W}_{\textrm{\tiny R}} \cup \mathcal{W}_{\textrm{\tiny L}} \cup \mathcal{V}_{\textrm{\tiny F}} \cup \mathcal{V}_{\textrm{\tiny P}} $.


The modular conjugation map (\ref{j0-map-def}) sends
 the modular trajectory with initial point $P\in \mathcal{B}_A \cup \mathcal{D}_A$ 
 with light-cone coordinates $(u_+, u_{-})$ (see (\ref{mod-traj-tau-line}))
into the modular trajectory  whose initial point has  light-cone coordinates $( \, \mathsf{j}(u_+) \,, \, \mathsf{j}(u_{-}) )$
and whose generic point has spacetime coordinates $(\tilde{x}(\tau) , \tilde{t}(\tau) )$ for $\tau \in \RR$ given by 
\cite{Mintchev:2022fcp}
\bea
\label{mod-conj-traj-tau-line-x}
\tilde{x}(\tau) &=& 
\frac{\xi(\tau, \mathsf{j}(u_+)) + \xi(-\tau, \mathsf{j}(u_-)) }{2}
\,=\,
\frac{\mathsf{j} \big( \xi(\tau, u_+) \big) + \mathsf{j} \big( \xi(-\tau, u_-) \big) }{2}
\\
\rule{0pt}{.8cm}
\label{mod-conj-traj-tau-line-t}
\tilde{t}(\tau) &=& 
\frac{\xi(\tau, \mathsf{j}(u_+)) - \xi(-\tau, \mathsf{j}(u_-)) }{2}
\,=\,
\frac{\mathsf{j} \big( \xi(\tau, u_+) \big) - \mathsf{j} \big( \xi(-\tau, u_-) \big) }{2}
\eea
in terms of (\ref{xi-map-fund}) and (\ref{j0-map-def}).
Notice that (\ref{id-xi-j-line}) has been employed to get the last expressions in (\ref{mod-conj-traj-tau-line-x})-(\ref{mod-conj-traj-tau-line-t}).
Thus, any modular trajectory in $\mathcal{D}_A$ (in $\mathcal{B}_A$) with initial point $P$ in $\mathcal{D}_A$ (in $\mathcal{B}_A$)
can be obtained 
either by applying the modular conjugation map to a proper original modular trajectory in $\mathcal{B}_A$ (in $\mathcal{D}_A$)
or as the modular trajectory whose initial point is the image of $P$ through the modular conjugation map (\ref{j0-map-def}).

The initial points of the modular trajectories in $\mathcal{B}_A$ displayed in Fig.\,\ref{fig:diamond-mod-hyper}
are marked by the empty dot and empty square
and have been found by applying the modular conjugation map (\ref{j0-map-def})
to the points in $\mathcal{D}_A$ marked by the black dot and the black square respectively. 
The modular trajectories corresponding to these two initial points
(see the dashed curves in Fig.\,\ref{fig:diamond-mod-hyper}) 
can be found 
either by applying the modular conjugation map (\ref{j0-map-def})
to the solid curves in Fig.\,\ref{fig:diamond-mod-hyper}
(the arcs related through this map are indicated with the same colour)
or by specialising (\ref{mod-conj-traj-tau-line-x})-(\ref{mod-conj-traj-tau-line-t}) to the initial points 
given by  the black dot and the black square in $\mathcal{D}_A$.
This implies that the values of $\tau$ in (\ref{tauB-minmax-def})
identifying the three different arcs building a modular trajectory in $\mathcal{B}_A$ 
are the same ones that determine the partition of the corresponding modular trajectories in $\mathcal{D}_A$,
namely $\tau_B^{<}  = \tau_A^{<} $ and $\tau_B^{>}  = \tau_A^{>} $.


Following \cite{Mintchev:2022fcp}, 
we find it worth observing  that the curve made by the union
of a modular trajectory (\ref{mod-traj-tau-line})  
with initial point $P$ (either in $\mathcal{D}_A$ or in $\mathcal{B}_A$)
and of the corresponding modular trajectory (\ref{mod-conj-traj-tau-line-x})-(\ref{mod-conj-traj-tau-line-t})
obtained through the modular conjugation map (\ref{j0-map-def})
is the hyperbolae $\mathcal{I}_{_P}$ defined by 
\be
\label{mod-hyper}
\big[ x(\tau) - x_0 \big]^2 - t(\tau)^2 = \kappa_0^2
\;\;\;\qquad\;\;\;
\big[ \tilde{x}(\tau) - x_0 \big]^2 -\tilde{t}(\tau)^2 = \kappa_0^2
\ee
where the parameters $x_0$ and $\kappa_0$ 
depend on the light-cone coordinates $(u_+, u_-)$ of the initial point $P$ as follows
\be
\label{mod-hyper-parameters}
x_0 \equiv \frac{u_+ \, u_- - a\, b}{ u_+ + u_- - (a + b) }
\;\;\;\qquad\;\;\;
\kappa_0 \equiv
\frac{\sqrt{ (b-u_+)(u_+ - a) \, (b-u_-)(u_- - a)  } }{ u_+ + u_- - (a + b)} \, .
\ee
%
The asymptotes of $\mathcal{I}_{_P}$ are $t= x - x_0$ and $t=  -x + x_0$.
In Fig.\,\ref{fig:diamond-mod-hyper} 
the two hyperbolae $\mathcal{I}_{_P}$ corresponding to the black dot and the black square are displayed
and their asymptotes are the black and light grey solid straight thin lines respectively. 
As $b \to +\infty$, 
from (\ref{mod-hyper-parameters}) 
we have that $x_0 \to a$ and $\kappa_0 \to - \sqrt{(u_+ - a) \,(u_- - a) }$;
hence the asymptotes of the hyperbolae coincide with the 
boundaries of $\mathcal{D}_A$ and $\mathcal{B}_A$ in this limit. 
In \cite{Mintchev:2022fcp}, 
only the hyperbolae $\mathcal{I}_{_P}$ for initial points $P$ at $t=0$ have been considered,
namely the special case $u_{+} =u_{-}$ in (\ref{mod-hyper-parameters}).

\subsection{Modular properties of the $K$-evolution  } 
\label{sec-tau-evolution}

The evolution generated by the operator $K$ defined in (\ref{K-chiral-1int-line}), 
that will be called $K$-evolution in the following, 
is well defined for any chiral field $\phi(u)$ with arbitrary dimension $h$ on the entire real line. 
In the following we explore the relation between the $K$-evolution
and the modular evolution associated with the Tomita-Takesaki modular theory for a CFT
\cite{Borchers:2000pv}.
This analysis can be carried out for the fields $\phi (u)$ with a single chirality propagating on a light ray 
or for combinations $\Phi(u_+,u_-)$ of right and left movers defining quantum fields in Minkowski space. 
In the former case the modular theory applies to the algebra $\A_A$ of chiral fields localised in the interval $A$ on the light ray,
while in the latter one it applies to the algebra $\A_{\D_A}$ of the fields localised in the diamond $\D_A$. 
Since a fundamental ingredient of the Tomita-Takesaki construction are the commutants $\A^\prime_A$ and $\A^\prime_{\D_A} $, 
it is natural to expect that the locality properties of $\phi (u)$ and $\Phi(u_+,u_-)$ are crucial in this context. 

As for  the chiral fields $\phi (u)$ in unitary CFT,
they obey either Bose or Fermi statistics for $h\in {\mathbb N}$ or $h\in {\mathbb N} + \frac{1}{2}$ respectively, 
being $\mathbb N$ the set of positive integers.
Accordingly, bosonic primaries commute and fermionic primaries anti-commute 
for $u_1 \not= u_2$. The graded commutant $\A^\prime_A$ has therefore its support 
in the complement $B$ of $A$ on the light ray
(see also the blue domain in Fig.\,\ref{fig:1int-line-xi-v0}).
The primary fields with real $h>0$ which is neither integer nor half integer, obey anyon (braid) statistics 
and are nonlocal. 
Indeed, by using (\ref{2pt-chiral-primaries})  and the following formula for distributions \cite{GelfandShilov}
\be
\frac{1}{(u \pm \ri \epsilon)^\lambda} 
\,= \,
\theta(u) \,u^{-\lambda} + \e^{\mp \ri \pi \lambda} \, \theta(-u) \,(- u)^{-\lambda}
\;\;\;\; \qquad \;\;\;
\lambda \notin \mathbb{N}
\label{c1}
\ee
where $\lambda >0$ and $\theta(u) $ is the Heaviside step function, we find the following mean value for the commutator 
\be
\label{d1-d2}
\big\langle \big[\,\phi_\pm^*(u_{1,\pm})\, , \, \phi_\pm(u_{2,\pm}) \, \big]_- \big\rangle 
\,=\,
\mp \, \ri \; \frac{ \sin (\pi h) }{\pi} \; 
\left[ \, \frac{ \theta (u_{12,\pm}) }{ u_{12,\pm}^{2h} } - \frac{ \theta (-u_{12,\pm}) }{ (-u_{12,\pm})^{2h} }\, \right] .
\ee
In this case the absence of locality prevents the application of the Tomita-Takesaki theory. 
For this reason the quantum entanglement features of anyon excitations are usually studied
in the reduced density matrix approach \cite{Bonderson:2017osr, Kirchner:2024ugq}. 
Summarising, at the chiral level the modular theory applies only for primaries with 
$h\in {\mathbb N}$ and $h\in {\mathbb N} + \frac{1}{2}$ \cite{Hollands:2019hje}. 

In Minkowski space $\mathbb{M}$, the situation is more complicated. 
The primary fields $\Phi(u_+,u_-)$ stemming from bosonic or fermionic primaries are local, as expected. 
The corresponding commutators and anti-commutators have the additional property that their support belongs 
to the light-cone $u_+ u_- = 0$. Therefore the graded 
commutant $\A^\prime_{\D_A}$ associated to these fields is localized in
$\mathcal{B}_A = \mathcal{W}_{\textrm{\tiny R}} \cup \mathcal{W}_{\textrm{\tiny L}} \cup \mathcal{V}_{\textrm{\tiny F}} \cup \mathcal{V}_{\textrm{\tiny P}} $
(see the blue domain in Fig.\,\ref{fig:diamond-mod-hyper})
and the $K$-evolution agrees with the modular evolution.

However, in Minkowski space this is not the whole story. 
Indeed, in addition to the local fields generated by bosonic and fermionic primaries, 
in $\mathbb{M}$ there are local fields which originate from nonlocal chiral fields. 
A simple example is the tensor product 
\be 
\label{local-field-v1-0}
\Phi(u_+,u_-) = \phi_+ (u_+) \otimes  \phi_- (u_-) 
\;\;\;\;\;\;\qquad\;\;\;\;\;
h_+ = h_- \equiv h
\qquad
h \neq \frac{k}{2} 
\qquad
k \in \mathbb{N} \,.
\ee 
In order to explore the spacetime support of the commutator of the field (\ref{local-field-v1-0}), 
it is instructive to study its mean value 
\bea 
\label{comm-mean-value-1-0}
\big\langle \, \big[\, \Phi^*(u_{1,+},u_{1,-})\, , \, \Phi(u_{2,+},u_{2,-}) \,\big]_-  \big\rangle 
&=&
\langle \, \phi_+^*(u_{1,+}) \, \phi_+(u_{2,+}) \,  \rangle 
\; \langle \, \phi_-^*(u_{1,-}) \, \phi_-(u_{2,-}) \, \rangle 
\hspace{1cm}
\\
& & 
- \, 
\langle  \, \phi_+(u_{2,+})  \, \phi_+^*(u_{1,+}) \, \rangle \;
\langle  \, \phi_-(u_{2,-})  \, \phi_-^*(u_{1,-}) \, \rangle \,.
\nn
\eea 
By employing the familiar expression (\ref{2pt-chiral-primaries}) for the correlators and (\ref{c1}),
one finds 
\bea 
\label{comm-mean-value-1-0-a}
\big\langle \, \big[\, \Phi^*(u_{1,+},u_{1,-})\, , \, \Phi(u_{2,+},u_{2,-}) \,\big]_-  \big\rangle 
&=&
\frac{\sin (2\pi h) }{2\pi\, \ri}  \;  \frac{  \theta(u_{12,+}) \, \theta(- u_{12,-}) -  \theta(- u_{12,+}) \, \theta(u_{12,-}) }{ \big( -u_{12,+} \, u_{12,-} \big)^{2h} }
\nn
\\
\rule{0pt}{.7cm}
&=&
\frac{\sin (2\pi h) }{4\pi\, \ri}  \;  \frac{  \varepsilon (u_{12,+}) + \varepsilon (- u_{12,-}) }{ \big( - u_{12,+} \, u_{12,-} \big)^{2h} }
\eea
where $\varepsilon (u) $ is the sign function. 
In terms of the spacetime variables, this commutator reads
\be
\label{comm-mean-value-1-0-b}
\big\langle \big[ \, \Phi^*(x_1+t_1,x_1-t_1)\, , \, \Phi(x_2+t_2,x_2-t_2) \, \big]_- \big\rangle 
\,=\, 
\frac{\sin (2\pi h) }{4\pi \ri}  \;
\frac{ \varepsilon (t_{12}-x_{12}) + \varepsilon (t_{12}+x_{12}) }{  \big( t_{12}^2 - x_{12}^2 \big)^{2h}  } \, .
\;\;\;
\ee
Finally, by applying the identity
\be 
\frac{ \varepsilon (t-x) + \varepsilon (t+x) }{2} \,=\, \varepsilon (t) \; \theta \big(t^2 -x^2\big)
\ee
the commutator (\ref{comm-mean-value-1-0-b}) can be written as follows
\bea 
\label{comm-mean-value-1-0-c}
\big\langle \big[ \, \Phi^*(x_1+t_1,x_1-t_1)\, , \, \Phi(x_2+t_2,x_2-t_2) \, \big]_- \big\rangle 
\,=\, 
\frac{ \sin (2\pi h) }{2\pi \ri} \; 
\frac{ \varepsilon (t_{12})\; \theta \big(t_{12}^2 -x_{12}^2 \big) }{   \big(  t_{12}^2 - x_{12}^2 \big)^{2h}  }
\eea
where the r.h.s.  is antisymmetric under the exchange $1 \leftrightarrow 2$ and, 
consistently with locality, vanishes at spacelike distances. 
The result (\ref{comm-mean-value-1-0-c}) implies that the commutator of $\Phi(u_+,u_-)$ has 
a non vanishing support in the past and future cones given by $\V_{\textrm{\tiny P}} $ and $\V_{\textrm{\tiny F}} $ respectively. 
Since the mean value of this commutator is preserved by a unitary evolution,
 in this case the modular evolution outside $\D_A$ remains localised in $\W_{\textrm{\tiny L}}  \cup \W_{\textrm{\tiny R}} $ 
and does not involve $\V_{\textrm{\tiny P}} \cup \V_{\textrm{\tiny F}} $,
in contrast with the modular trajectories outside $\D_A$ discussed in Sec.\,\ref{sec-mod-traject-1int}.
Consequently, the modular evolution of  the local fields defined in (\ref{local-field-v1-0}) differs from their $K$-evolution. 
Understanding the geometric action of the modular evolution of the local fields 
in (\ref{local-field-v1-0}) in $\W_{\textrm{\tiny L}}  \cup \W_{\textrm{\tiny R}} $ 
still represents an open problem to our knowledge.
A first attempt in this direction is discussed in Appendix\;\ref{app-new-inversion},
where non conformal maps sending 
$\mathcal{D}_A $ onto $\mathcal{W}_{\textrm{\tiny R}} \cup \mathcal{W}_{\textrm{\tiny L}}$
in a  bijective way are constructed.

\subsection{Spacetime distance along the modular trajectories  } 
\label{sec-spacetime-distance-1int}


The CFT that we are exploring is a relativistic theory in the two-dimensional Minkowski spacetime;
hence causality between two events is ruled by the sign of their spacetime distance.
In the following we study the Lorentzian spacetime distance 
between two events as they evolve along two different modular trajectories.


Consider the 
Given two points (events) $P_1$ and $P_2$ in $\mathcal{D}_A \cup \mathcal{B}_A$,
consider the Lorentzian spacetime distance $d(P_1, P_2)$ separating them 
and also the modular trajectories having these points as initial points.
Moving along these two modular trajectories,
after the same value of $\tau$ we arrive to the points $P_1(\tau)$ and $P_2(\tau)$,
whose spacetime distance $d\big(P_1(\tau), P_2(\tau)\big)$ is
\bea
\label{s12-tau-def}
d\big(P_1(\tau), P_2(\tau)\big)
&\equiv &
-\; t_{12}(\tau)^2 + x_{12}(\tau)^2 
\\
\label{s12-tau-def-1}
&=&
\big[ \xi(\tau, u_{1,+}) - \xi(\tau, u_{2,+}) \big] \big[ \xi(-\tau, u_{1,-}) - \xi(-\tau, u_{2,-}) \big]
\nn
\eea
i.e. the product of the chiral distances along the two chiral directions, discussed in Sec.\,\ref{sec-chiral-distance-mod-evo}.
Thus, by employing  (\ref{xi12-tau-chiral-equal-tau}),
the spacetime distance (\ref{s12-tau-def}) can be written as 
\be
\label{s12-tau-evolution}
d\big(P_1(\tau), P_2(\tau)\big)
= \omega(\tau;P_1,P_2)  \, 
d(P_1, P_2)
\ee
where $\omega(\tau;P_1,P_2) $ can be written in terms of $q(\tau, u)$ introduced in  (\ref{pA-qA-def})
as follows
\be
\label{OmegaA-def}
\omega(\tau;P_1,P_2) 
\equiv
q(\tau,u_{1,+}) \, q(-\tau,u_{1,-})\; q(\tau,u_{2,+})  \, q(-\tau,u_{2,-})  \,.
\ee
Since $q(0,u) = 1$, we have that $\omega(\tau =0 ;P_1,P_2)  =1 $;
hence 
the  initial condition on (\ref{s12-tau-evolution}) is satisfied, as expected. 

In the limit $b \to +\infty$, 
the diamond $\mathcal{D}_A$ becomes the right Rindler wedge, 
while $ \mathcal{W}_{\textrm{\tiny L}}$ remains the left Rindler wedge.
In this regime, $q(\tau, u) \to \e^{\pi \tau}$
and therefore (\ref{OmegaA-def}) simplifies to 
\be
\lim_{b \to +\infty} \omega(\tau;P_1,P_2) =1 \,.
\ee
This result is expected because 
the modular evolution is a Lorentz boost in the Rindler wedge,
which is an isometry of the Lorentzian spacetime distance.
Hence, causality is preserved along the modular evolution in this limiting regime. 
A recent related analysis based on the result of \cite{Peleska:1984aa} can be found \cite{Sorce:2024zme}.


Causality along the modular evolution 
can be studied by considering the sign of the spacetime distance $d\big(P_1(\tau), P_2(\tau)\big)$
with respect to the sign of the spacetime distance $d(P_1, P_2)$ as $\tau \in \RR$. 
From (\ref{s12-tau-evolution}), this means to explore the sign of (\ref{OmegaA-def}).

\begin{figure}[t!]
\vspace{-.5cm}
\hspace{-1.1cm}
\includegraphics[width=1.15\textwidth]{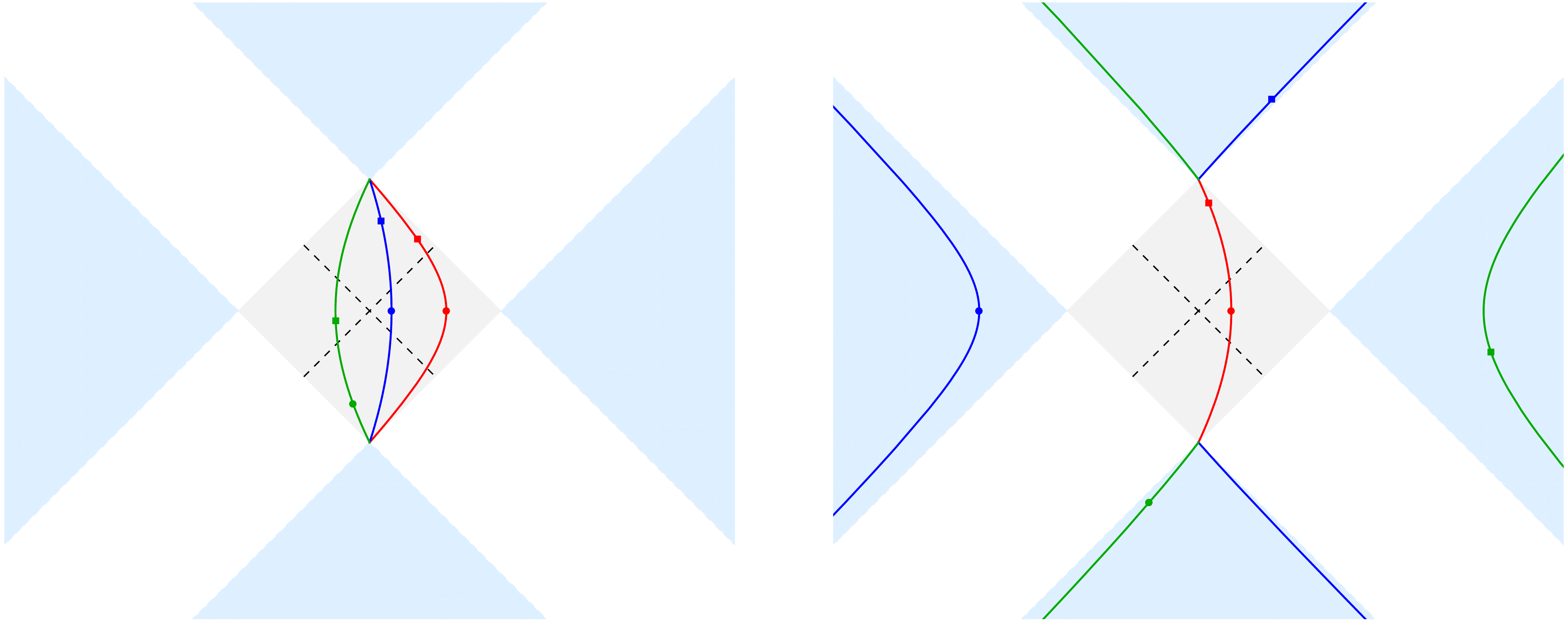}
\vspace{-.4cm}
\caption{
Spacetime distances along the modular evolution 
either for two events in $\mathcal{D}_A$ (left panel)
or for one event in $\mathcal{D}_A$ and the other one in $\mathcal{B}_A$ (right panel).
}
\label{fig:1int-mod-traject-1}
\end{figure}

First we observe that,
given a pair of lightlike separated initial points $P_1$ and $P_2$ in $\mathcal{D}_A \cup \mathcal{B}_A$,
from (\ref{s12-tau-evolution}), their modular evolutions $P_1(\tau)$ and $P_2(\tau)$ remain lightlike separated for all values $\tau \in \RR$.
Hereafter we can focus on initial points $P_1$ and $P_2$ that are either
timelike or spacelike separated, unless stated otherwise.
For any pair of points $P_1$ and $P_2$ providing the initial points of two different modular trajectories, 
three possible cases occur:
both the points belong to $\mathcal{D}_A$,
one point is in $\mathcal{D}_A$ while the other one belongs to $\mathcal{B}_A$
and both the points are in $\mathcal{B}_A$.
In each panel of Fig.\,\ref{fig:1int-mod-traject-1} we show three modular trajectories whose initial points
correspond to the dots while the filled squares of the same colour 
indicate their corresponding modular evolution after an assigned (positive)
value of the modular parameter $\tau$ (the same one for all the initial points).
In the left panel, all the initial points belong to $\mathcal{D}_A$,
while in the right panel only one initial point belongs to $\mathcal{D}_A$.
In the left panel of Fig.\,\ref{fig:1int-mod-traject-1},
the initial points represent all the three possible types of spacetime distance  
(spacelike, timelike and lightlike).
In the following we argue that the relativistic causality is preserved for any $\tau \in \RR$ 
only when both $P_1$ and $P_2$ belong to $\mathcal{D}_A$,
while in the other two cases the modular evolution does not preserve the relativistic causality for any $\tau \in \RR$
because a non empty domain of $\tau$ occurs where $d\big(P_1(\tau), P_2(\tau)\big)$ and $d(P_1, P_2)$ have opposite signs. 

Let us first remind that, 
from (\ref{pA-qA-def}), we have $ q(\tau, u) >0$ for $u \in A$, 
while $ q(\tau, u)$ does not have a defined sign for $u \in B$
(indeed, it vanishes at $\tau=\tau_{B}(u)$ defined in (\ref{tauB_def})).

When both $P_1$ and $P_2$ are in $\mathcal{D}_A$ 
(see e.g. a pair of curves in the left panel of Fig.\,\ref{fig:1int-mod-traject-1}),
all the fours factors in the r.h.s. of (\ref{OmegaA-def}) are separately positive for any $\tau\in \RR$;
hence $\omega(\tau;P_1,P_2) >0$ for any $\tau \in \RR$
and therefore relativistic causality is preserved along this modular evolution.

When an initial point $P_1$ (with coordinates $(u_+ , u_-)$)
belongs to $\mathcal{D}_A$ and the other one $P_2$ (with coordinates $(v_+ , v_-)$)
to  $\mathcal{B}_A$ (see the right panel of Fig.\,\ref{fig:1int-mod-traject-1}),
the sign of $d\big(P_1(\tau), P_2(\tau)\big)$
and the sign of the initial distance $d(P_1, P_2)$ 
are not the same for any value of $\tau \in \RR$.
Indeed, from (\ref{s12-tau-evolution})-(\ref{OmegaA-def}), 
one oberves that in the r.h.s. of (\ref{OmegaA-def}) we have that
$q(\tau,u_{+}) \, q(-\tau,u_{-}) >0$ for any value of $\tau\in \RR$, 
while the sign of the remaining factor $q(\tau,v_{+}) \, q(-\tau,v_{-})$
changes at $\tau=\tau_{B}(v_+)$ and $\tau=\tau_{B}(v_-)$.
Hence, starting from the initial points and evolving along either $\tau>0$ or $\tau<0$, 
the sign of $d\big(P_1(\tau), P_2(\tau)\big)$ changes with respect to the sign of $d(P_1, P_2)$ 
whenever the point in $\mathcal{B}_A$ 
moves from a subset of the partition
$\mathcal{B}_A = \mathcal{W}_{\textrm{\tiny R}} \cup \mathcal{W}_{\textrm{\tiny L}} \cup \mathcal{V}_{\textrm{\tiny F}} \cup \mathcal{V}_{\textrm{\tiny P}} $
to a different one. 
Also in the right panel of Fig.\,\ref{fig:1int-mod-traject-1},
where the same notation employed in the left panel of the same figure has been adopted, 
the initial points  represent a case of either spacelike (red and blue dots) or timelike (red and green dots) 
or lightlike (blue and green dots) spacetime distance.
Moreover, the choice of the value of $\tau$ providing  the positions of the three squares in the right panel of Fig.\,\ref{fig:1int-mod-traject-1}
is such that $d\big(P_1(\tau), P_2(\tau)\big)$ and $d(P_1, P_2)$  have opposite signs
for the cases of spacelike and timelike distance.

\begin{figure}[t!]
\vspace{-.5cm}
\hspace{-1.1cm}
\includegraphics[width=1.15\textwidth]{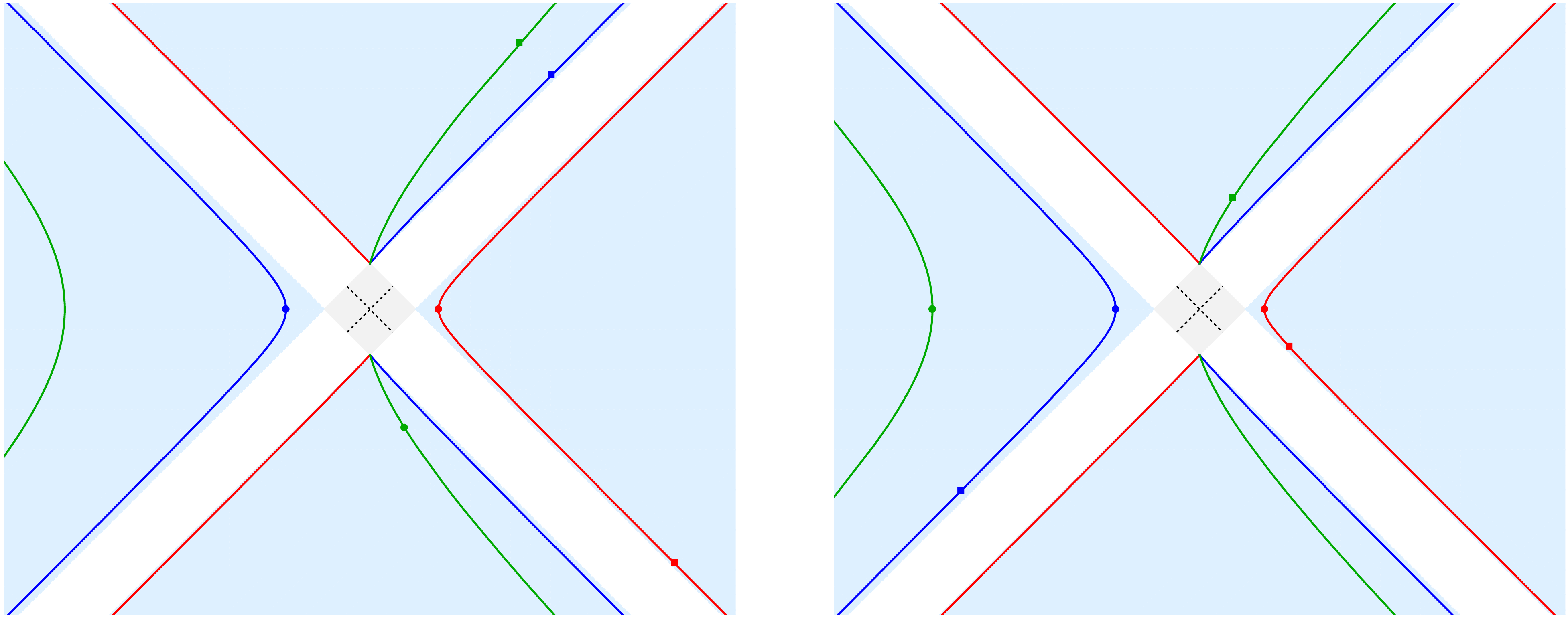}
\vspace{-.4cm}
\caption{
Spacetime distances along the modular evolution for two events in $\mathcal{B}_A$.
}
\label{fig:1int-mod-traject-2}
\end{figure}

In each panel of Fig.\,\ref{fig:1int-mod-traject-2} we show
three distinct modular trajectories with initial points in $\mathcal{B}_A$,
adopting the same notation employed in Fig.\,\ref{fig:1int-mod-traject-1}.
Given two points $P_1$ and $P_2$ in  $\mathcal{B}_A$,
the sign of $d\big(P_1(\tau), P_2(\tau)\big)$
and the sign of the initial distance $d(P_1, P_2)$ 
do not coincide for all  $\tau \in \RR$
because all the four factors in the r.h.s. of (\ref{OmegaA-def}) do not have a definite sign. 
In the left panel of Fig.\,\ref{fig:1int-mod-traject-2} 
the choice of the initial points covers all the three possible kinds of spacetime distance,
while in the right panel of the same figure all the initial points are spacelike separated.
The choice of $\tau$ in both panels of Fig.\,\ref{fig:1int-mod-traject-2} 
displays the fact that $d\big(P_1(\tau), P_2(\tau)\big)$ changes sign with respect to $d(P_1, P_2)$  
whenever one of the two points changes domain among the subsets providing the partition
$\mathcal{B}_A = \mathcal{W}_{\textrm{\tiny R}} \cup \mathcal{W}_{\textrm{\tiny L}} \cup \mathcal{V}_{\textrm{\tiny F}} \cup \mathcal{V}_{\textrm{\tiny P}} $.
Thus, relativistic causality is preserved as long as $P_1(\tau)$ and $P_2(\tau)$ at $\tau \neq 0$
remain in the same domain of $P_1$ and $P_2$ respectively.


The analytic continuation of (\ref{s12-tau-evolution}) to $\tau = \pm \ri/2$ provides a relation involving 
the spacetime distance between $P_1$ and $P_2$ in $\mathcal{D}_A \cup \mathcal{B}_A$
and the one between the corresponding events 
$P_{1, \,\mathsf{j}}$ and $P_{2, \,\mathsf{j}}$ in $\mathcal{D}_A \cup \mathcal{B}_A$,
obtained through  the modular conjugation map (\ref{j0-map-def}) for both the light-cone coordinates.
In terms of (\ref{OmegaA-def}), this relation reads
\be
\label{dist-P1P2-conjugate}
d(P_{1, \,\mathsf{j}}\, , P_{2, \,\mathsf{j}} )
=
 \omega( \pm \ri / 2 \, ;P_1,P_2)  \, 
d(P_1, P_2)
\ee
where
\be
\label{omega-P1P2-conjugate}
 \omega( \pm \ri / 2 \, ;P_1,P_2) = \left( \frac{b-a}{2} \right)^4 \frac{1}{
 \big[\big( u_{1,+} - \tfrac{a+b}{2}\big) \big( u_{1,-} - \tfrac{a+b}{2}\big) \big] \, \big[\big( u_{2,+} - \tfrac{a+b}{2}\big)  \big( u_{2,-} - \tfrac{a+b}{2}\big)\big]  } \,.
\ee
The limit $b\to +\infty$ of (\ref{dist-P1P2-conjugate}) provides also a consistency check between (\ref{j0-map-def}) and (\ref{s12-tau-evolution}).
Indeed, for (\ref{omega-P1P2-conjugate}) in this limit we find $\omega( \pm \ri / 2 \, ;P_1,P_2)  \to 1$;
hence the modular conjugation becomes an isometry, 
as expected from the fact that the geometric action of the modular conjugation 
when the subsystem $A$ is half line is simply given by the reflection with respect to the entangling point. 
%


The most general case to explore is the spacetime distance between two points $P_1(\tau_1)$ and $P_2(\tau_2)$ 
evolving independently along two assigned modular trajectories whose initial points are 
$P_1$ and $P_2$ respectively.
This spacetime distance can be obtained from the chiral distance discussed in Sec.\,\ref{sec-chiral-distance-mod-evo};
indeed
\bea
\label{s12-tau12-def}
d\big(P_1(\tau_1), P_2(\tau_2)\big)
& \equiv &
-\,\big[ t_{1}(\tau_1) - t_{2}(\tau_2) \big]^2 + \big[ x_{1}(\tau_1) - x_{2}(\tau_2) \big]^2
\\
\label{s12-tau12-def-1}
& = &
\big[ \xi(\tau_1, u_{1,+}) - \xi(\tau_2, u_{2,+}) \big] \big[ \xi(-\tau_1, u_{1,-}) - \xi(-\tau_2, u_{2,-}) \big] \,.
\eea
By employing  (\ref{xi12-tau-chiral}), this spacetime distance can be written as 
\be
\label{s12-gen-tau-evolution}
d\big(P_1(\tau_1), P_2(\tau_2)\big)
= \Omega(\tau_1, \tau_2  ;P_1,P_2)  \, 
d(P_1, P_2)
\ee
where $\Omega(\tau_1, \tau_2  ;P_1,P_2) $ is defined in terms of (\ref{pA-qA-def}) and (\ref{R-fact-def}) as follows
\bea
\label{omega-12-def}
\Omega(\tau_1, \tau_2  ;P_1,P_2)  
& \equiv &
\mathcal{R}(\tau_{12} ; u_{1,+}, u_{2,+}) \, \mathcal{R}(-\tau_{12} ; u_{1,-}, u_{2,-}) 
\\
& &
\times\, q(\tau_1,u_{1,+}) \, q(-\tau_1,u_{1,-})\; q(\tau_2,u_{2,+})  \, q(-\tau_2,u_{2,-}) \,.
\nonumber
\eea
When $\tau_1 = \tau_2 \equiv \tau$,
since $\mathcal{R}(0 ; u_1, u_2) = 1$ for $u_1 \neq u_2$ identically, 
(\ref{omega-12-def}) simplifies to  (\ref{OmegaA-def})
and therefore (\ref{s12-tau-evolution}) is recovered from (\ref{s12-gen-tau-evolution})  in this limiting regime, as expected.

In the Rindler wedge limit $b \to +\infty$, from (\ref{pA-qA-def}) and (\ref{Omega12-rindler}) 
we find that (\ref{omega-12-def}) becomes
\bea
\label{Omega-b-large}
\lim_{b \to +\infty}
\Omega(\tau_1, \tau_2  ;P_1,P_2)  
&=&
\\
& & \hspace{-1.3cm}
=
 \frac{(u_{1,+} - a)\, \e^{\pi \tau_{12}}  - (u_{2,+} - a) \, \e^{-\pi \tau_{12}} }{u_{1,+} - u_{2,+}}
 \; 
 \frac{(u_{1,-} - a)\, \e^{- \pi \tau_{12}}  - (u_{2,-} - a) \, \e^{\pi \tau_{12}} }{u_{1,-} - u_{2,-}}
 \nn
\eea
where in the denominator of the r.h.s. we recognise the distance $d(P_1, P_2)$.

When the initial points $P_1$ and $P_2$ are lightlike separated,
by employing (\ref{xi-12-equal-u1u2}) in (\ref{s12-tau12-def-1})
along the chiral direction that makes the distance $d(P_1, P_2)$ vanishing, 
one finds that $d\big(P_1(\tau_1), P_2(\tau_2)\big)$ 
is non vanishing when $\tau_{12} \neq 0$.


When the both the initial points $P_1$ and $P_2$ are in $\mathcal{D}_A$,
since $q(\tau,u) > 0$ for $u \in A$,
from (\ref{s12-gen-tau-evolution})-(\ref{omega-12-def}) it is straightforward to realise 
that the sign of $d\big(P_1(\tau_1), P_2(\tau_2)\big)/ d(P_1, P_2)$
is given by the sign of 
$\mathcal{R}(\tau_{12} ; u_{1,+}, u_{2,+}) \, \mathcal{R}(-\tau_{12} ; u_{1,-}, u_{2,-})$.
Then, since $\mathcal{R}( \tilde{\tau}_{0} ; u_{1}, u_{2}) = 0$ for $\tilde{\tau}_{0}$ defined in (\ref{tau0tilde-def}),
we conclude that $d\big(P_1(\tau_1), P_2(\tau_2)\big)/ d(P_1, P_2)$
changes sign at $\tau_{12} \in \{ \tau_{<} \, , \tau_{>}  \}$, 
where the finite values $ \tau_{<}   < \tau_{>} $  are defined as follows
\bea
\label{tau-min-12-def}
\tau_{<}  
& \equiv & 
\textrm{min} \big\{ \tilde{\tau}_0(u_{1,+}, u_{2,+}) \,, - \,\tilde{\tau}_0(u_{1,-}, u_{2,-})) \big\}
\\
\label{tau-max-12-def}
\rule{0pt}{.5cm}
\tau_{>} 
& \equiv & 
 \textrm{max} \big\{ \tilde{\tau}_0(u_{1,+}, u_{2,+}) \,, - \,\tilde{\tau}_0(u_{1,-}, u_{2,-})) \big\} \,.
\eea

\begin{figure}[t!]
\vspace{-.6cm}
\hspace{-.6cm}
\includegraphics[width=1.07\textwidth]{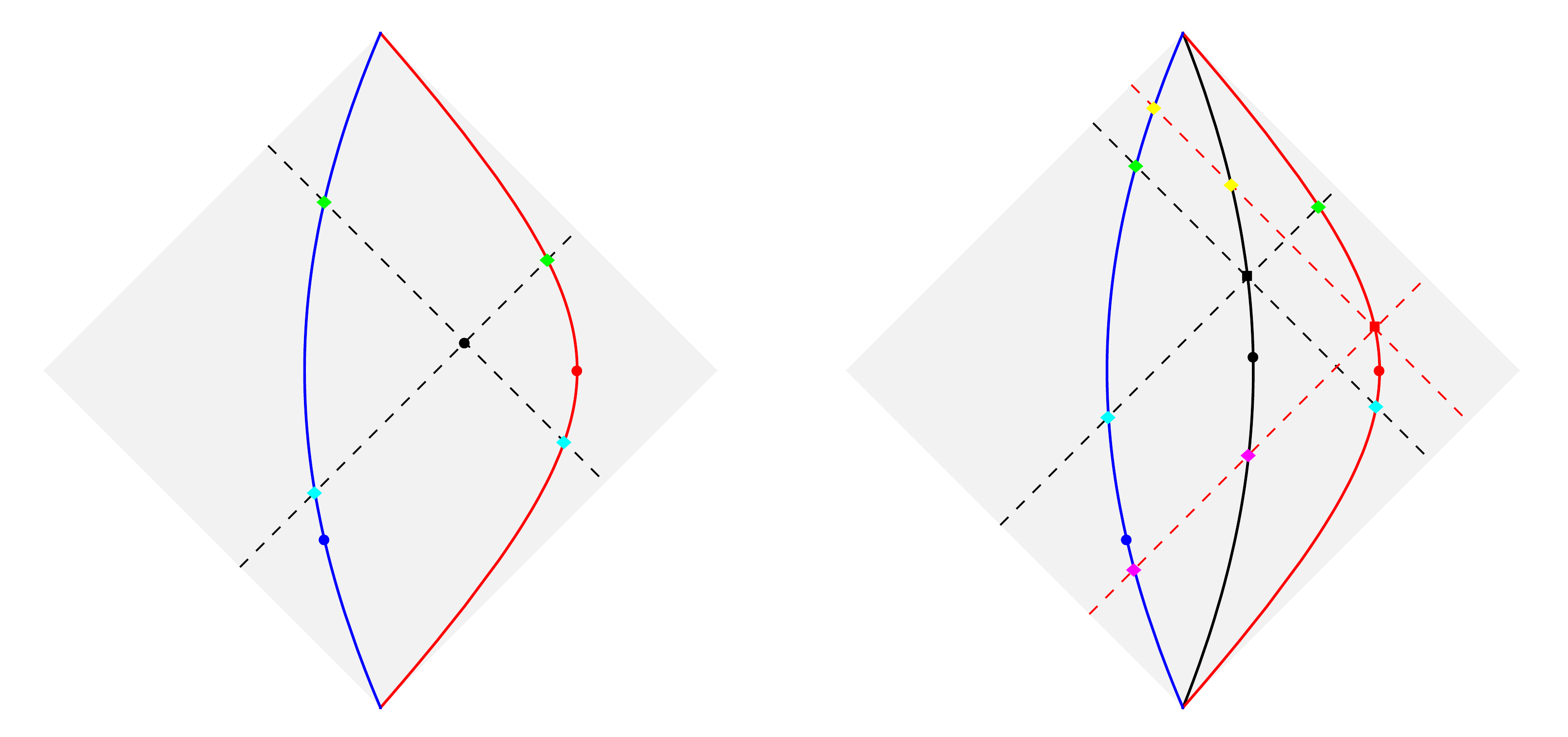}
\vspace{-.3cm}
\caption{
Spacetime distances (see (\ref{s12-gen-tau-evolution}))
between a reference point in $\mathcal{D}_A$ 
(black dot in the left panel and black square in the right panel)
and a point moving along a modular trajectory in $\mathcal{D}_A$
(either the red or the blue solid curve).
In the left panel the reference point is fixed,
while in the right panel it moves along a modular trajectory (solid black curve). 
The green and cyan points are obtained from (\ref{tau-min-12-def})-(\ref{tau-max-12-def}).
}
\label{fig:1int-mod-traject-3}
\end{figure}

In both the panels of Fig.\,\ref{fig:1int-mod-traject-3}, 
the initial points are denoted by dots and the corresponding modular trajectories (\ref{mod-traj-tau-line})
by solid lines having the same colour. 
In the left panel of Fig.\,\ref{fig:1int-mod-traject-3} we consider the case where a point, 
e.g. $P_1$, is kept fixed (see the black dot), 
while the other one moves along a modular trajectory (either the red or the blue solid line).
In this case, the ratio $d\big(P_1 , P_2(\tau_2)\big)/ d(P_1, P_2)$ changes sign 
when $\tau_2 =  \tau_{<}  $ and $\tau_2 =  \tau_{>}  $ given by (\ref{tau-min-12-def})-(\ref{tau-max-12-def}).
The points $P_2(\tau_{<} )$ and $P_2(\tau_{>} )$
are denoted by the cyan and green marker respectively. 

In the right panel of Fig.\,\ref{fig:1int-mod-traject-3}, 
by adopting the notation of the left panel, 
we employ (\ref{s12-gen-tau-evolution}) to study
the spacetime distance between two points $P_1(\tau_1)$ and $P_2(\tau_2)$ 
moving along two distinct modular trajectories.
Consider e.g. the black and red modular trajectories as a generic pair of modular trajectories;
hence $P_1$ and $P_2$ are the black and the red dot respectively.
The points $P_1(\tau_1)$ and $P_2(\tau_2)$ for some $\tau_1 \neq 0$ and $\tau_2 \neq 0$
correspond to the black and the red square respectively. 
For a given value of $\tau_1$ corresponding to the black square, 
the expressions in (\ref{tau-min-12-def})-(\ref{tau-max-12-def}) 
provide the values of the modular evolution parameters $\tau_2$
corresponding to the green and the cyan points on the red  solid curve,
where $d\big(P_1(\tau_1), P_2(\tau_2)\big)/ d(P_1, P_2)$ changes its sign. 
It is straightforward to adapt this procedure also to the red square,
finding through (\ref{tau-min-12-def})-(\ref{tau-max-12-def}) the values of the modular parameter
corresponding to the magenta and yellow points on the black curve,
where $d\big(P_1(\tau_1), P_2(\tau_2)\big)/ d(P_1, P_2)$ changes its sign. 
These changes of sign occur where the Dirac delta functions in the commutators
obtained by applying (\ref{commutator-j-single-int-symmetrised}) in both the chiral directions
are non vanishing.


It is insightful to observe that
the expressions discussed above for the spacetime distance along the modular trajectories 
allow to discuss an interesting geometrical property of the hyperbola $\mathcal{I}_{_P}$ 
in (\ref{mod-hyper})-(\ref{mod-hyper-parameters}), describing the modular trajectories. 
Consider the points  $P_a$ and $P_b$
with  light-cone coordinates $(u_+ , u_-) = (a,a)$ and $(u_+ , u_-) = (b,b)$ respectively,
that are associated to the entangling points of $\mathcal{D}_A$
and do not move under the modular evolution. 
For a generic point on the hyperbola  $\mathcal{I}_{_P}$, 
the ratio between its spacetime distances from $P_a$ and $P_b$ is independent of $\tau$.
Indeed, from (\ref{s12-gen-tau-evolution}) we find that
\be
\label{apo-hyperbola}
\frac{d\big(P(\tau), P_a \big) }{ d\big(P(\tau), P_b \big)  }
=
\frac{d\big(P_{\,\mathsf{j}}(\tau), P_a \big) }{ d\big(P_{\,\mathsf{j}}(\tau), P_b \big)  }
=
\frac{ (u_{+} - a)\, (u_{-} - a) }{ (b- u_{+} )\, (b- u_{-} ) }
=
\frac{ \big(\, \mathsf{j}(u_{+}) - a \big)\, \big(\, \mathsf{j}(u_{-}) - a \big) }{ \big(b- \mathsf{j}(u_{+}) \big)\, \big(b- \mathsf{j}(u_{-}) \big) }
\ee
where $P_{\,\mathsf{j}}(\tau)$ is obtained from $P(\tau)$,
by applying the modular conjugation map (\ref{j0-map-def}) to both the light-cone coordinates.
The relations in (\ref{apo-hyperbola}) imply  that $\mathcal{I}_{_P}$ is a hyperbola of Apollonius,
namely  the Lorentzian counterpart of the circle of Apollonius, 
defined through the corresponding property in the Euclidean signature,
as discussed below, in Sec.\,\ref{sec-mod-flow-1int-Euclid}.

\subsection{Distance along a modular trajectories in Euclidean signature }
\label{sec-mod-flow-1int-Euclid}


We find it worth providing a brief discussion about 
the counterpart of some results described above in Euclidean signature.
In this setup, while the spatial direction parameterised by $x \in \RR$ 
is bipartite as $A \cup B$ with $A =[a,b]$ like in the previous analyses,
 the diamond $\mathcal{D}_A$ cannot be defined anymore
and the spacetime distances are always positive.


In Euclidean signature,
the geometric action of the modular evolution has been explored e.g. in \cite{Cardy:2016fqc, Mintchev:2022fcp} and, 
for a two-dimensional CFT in its ground state and on the line bipartite by the interval $A$,
the modular trajectories are described by the following complex function 
\be
\label{xi-map-euc}
\xi_{\textrm{\tiny E}}(\theta,p) 
\equiv 
\frac{(b-p)\,a+(p-a)\,b\,\mathrm{e}^{\ri \theta}}{(b-p)+(p-a)\,\mathrm{e}^{\ri \theta}}  
\ee
where $p \in A$ corresponds to the initial point and the angle $\theta \in [0, 2\pi)$ is the modular evolution parameter.
The expression (\ref{xi-map-euc}) can be obtained from (\ref{xi-map-fund}), by replacing $2\pi \tau$ with $\ri \theta$. 
The entangling points do not evolve as expected;
indeed, from (\ref{xi-map-euc}),  we have that $\xi_{\textrm{\tiny E}}(\theta,a) = a$ 
and $\xi_{\textrm{\tiny E}}(\theta,b) =b$ for any $\theta$.

Considering the real and the imaginary part of (\ref{xi-map-euc}), 
one finds that the modular trajectories are given by the following circles
\be
\label{mod-circle-euc}
\left( \textrm{Re} \big[\xi_{\textrm{\tiny E}}(\theta,p)  \big] - \frac{p^2-a\, b}{2\,p-(a+b)} \right)^2
+
\Big(
\textrm{Im} \big[\xi_{\textrm{\tiny E}}(\theta,p)  \big] 
\Big)^2
=
\left( \frac{(b-p)(p-a)}{ 2\, p-(a+b) } \right)^2
\ee
which provide the Euclidean counterpart $\mathcal{C}_{_P}$
to the hyperbolae $\mathcal{I}_{_P}$ in (\ref{mod-hyper})-(\ref{mod-hyper-parameters}).
Indeed the center and the radius of  $\mathcal{C}_{_P}$
coincide with the corresponding parameters for $\mathcal{I}_{_P}$
when its initial point $P$ has $u_{+} = u_{-} \equiv p$, i.e. $(x,t)=(p,0)$.

\begin{figure}[t!]
\vspace{-.5cm}
\hspace{.0cm}
\includegraphics[width=1.\textwidth]{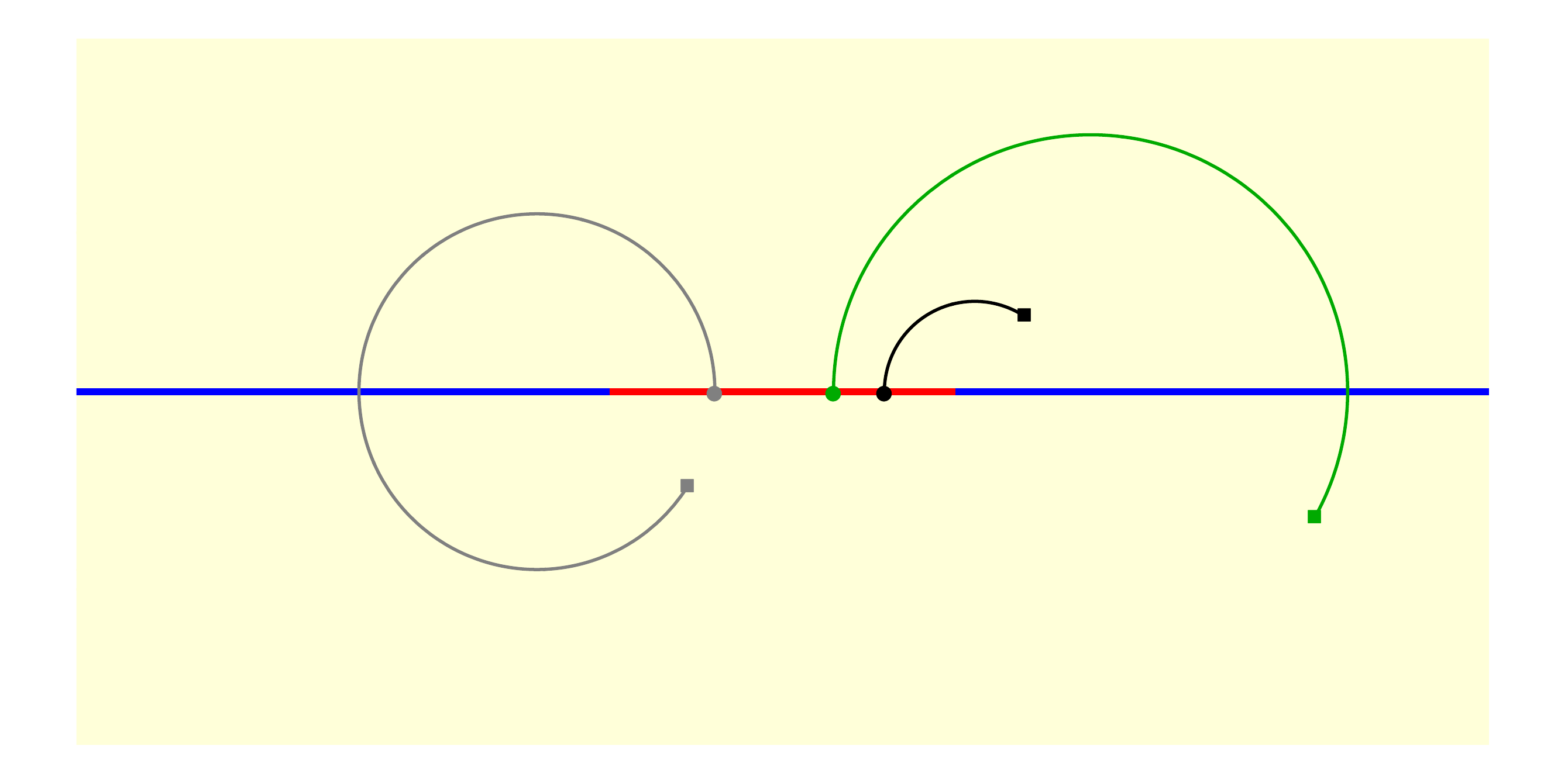}
\vspace{-.6cm}
\caption{Euclidean modular evolutions in the complex plane, 
that are described by arcs of circles of Apollonius. 
}
\label{fig:1int-line-Euclid-evolution}
\end{figure}

In Fig.\,\ref{fig:1int-line-Euclid-evolution} we consider the complex plane 
where the horizontal line is the spatial direction 
bipartite by the interval $A$ (red segment) and its complements (blue half lines),
showing three distinct modular evolutions whose initial points correspond to the coloured dots in $A$.
These modular evolutions are circular arcs belonging to the corresponding modular trajectories in Euclidean signature 
and they have been obtained from (\ref{xi-map-euc}), or equivalently from (\ref{mod-circle-euc}), for $\theta \in [0, \theta_0]$, 
where the ending point $P(\theta_0)$ is marked by a filled square having the same colour.


The complex function (\ref{xi-map-euc}) satisfies the following differential equations
\be
\label{der-xi-euc}
\partial_\theta \xi_{\textrm{\tiny E}}(\theta,p) \,=\, \frac{\ri}{2\pi}\, V(\xi_{\textrm{\tiny E}})
\;\;\;\qquad\;\;\;
\partial_p \xi_{\textrm{\tiny E}}(\theta,p) 
\,=\, \frac{V(\xi_{\textrm{\tiny E}})}{V(p)}
\,=\, q_{\textrm{\tiny E}}(\theta, p)^2
\ee
in terms of (\ref{velocity_fund}),
which are the Euclidean counterparts of the differential equations in (\ref{der-xi});
indeed we have introduced 
\be
q_{\textrm{\tiny E}}(\theta, p)\equiv \frac{b-a}{(b-p) \, \e^{-\ri \theta/2} + (p-a) \, \e^{\ri \theta/2} }
\ee
which corresponds to (\ref{pA-qA-def}) with $\pi \tau$ replaced by $\ri \theta /2$
and whose complex conjugate is $q_{\textrm{\tiny E}}(-\theta, p)$. 
Notice that, while in Euclidean signature the modular trajectories are closed curves in the complex plane, 
in Minkowski space they are open arcs of hyperbola either in $\mathcal{D}_A$ or in $\mathcal{B}_A$,
connecting the top and bottom vertices of $\mathcal{D}_A$.

Given two distinct modular trajectories with initial points $p_1$ and $p_2$ in $A$,
the Euclidean distance between the points $P_1(\theta_1)$ and $P_2(\theta_2)$ along these curves 
can be expressed in terms of (\ref{xi-map-euc}), finding 
\be
\label{s12-theta12-def}
d_{\textrm{\tiny E}}\big(P_1(\theta_1), P_2(\theta_2)\big)
 \equiv 
\big| \, \xi_{\textrm{\tiny E}}(\theta_1 ,p_1)  - \xi_{\textrm{\tiny E}}(\theta_2 ,p_2) \, \big|^2
\ee
which is the Euclidean counterpart of (\ref{s12-tau12-def}).
This Euclidean distance can be rewritten as 
\be
\label{euclid-mod-distance}
d_{\textrm{\tiny E}}\big(P_1(\theta_1), P_2(\theta_2)\big)
=\,
\Omega_{\textrm{\tiny E}}(\theta_1, \theta_2  ;p_1,p_2)  \; 
( p_1 - p_2 )^2
\ee
where we have introduced
\bea
\label{Omega-euc-gen}
\Omega_{\textrm{\tiny E}}(\theta_1, \theta_2  ;p_1,p_2) 
& \equiv &
\mathcal{R}_{\textrm{\tiny E}}(\theta_{12} ; p_1, p_2) \, \mathcal{R}_{\textrm{\tiny E}}(-\theta_{12} ; p_1, p_2) 
\\
& &
\times\, q_{\textrm{\tiny E}}(\theta_1, p_1)\, q_{\textrm{\tiny E}}(-\theta_1, p_1)
\; q_{\textrm{\tiny E}}(\theta_2, p_2)\, q_{\textrm{\tiny E}}(-\theta_2, p_2)
\nn
\eea
in terms of $\theta_{12} \equiv \theta_1 - \theta_2$ and of the function $\mathcal{R}_{\textrm{\tiny E}}(\theta ; p_1, p_2) $ defined as
\be
\label{R-fact-euc-def}
\mathcal{R}_{\textrm{\tiny E}}(\theta ; p_1, p_2) 
\equiv
\frac{  \e^{2\pi w(p_1) + \ri \theta/2} - \e^{2\pi w(p_2) - \ri \theta/2} }{ \e^{2\pi w(p_1)} - \e^{2\pi w(p_2)}  } \, .
\ee
Notice that (\ref{euclid-mod-distance}) and (\ref{Omega-euc-gen}) are the Euclidean counterparts 
of  (\ref{s12-gen-tau-evolution}) and (\ref{omega-12-def})  respectively. 
Moreover, the expression (\ref{R-fact-euc-def}),
which can be obtained from its Lorentzian  counterpart (\ref{R-fact-def}) by replacing $\pi \tau$ with $\ri \theta /2$,
satisfies $|\mathcal{R}_{\textrm{\tiny E}}(\theta ; p_1, p_2) |^2 = \mathcal{R}_{\textrm{\tiny E}}(\theta ; p_1, p_2) \, \mathcal{R}_{\textrm{\tiny E}}(-\theta ; p_1, p_2) $.
The crucial difference between (\ref{R-fact-def}) and (\ref{R-fact-euc-def})
is that $\mathcal{R}_{\textrm{\tiny E}}(\theta ; p_1, p_2) $ never vanishes;
indeed, the equation $\mathcal{R}_{\textrm{\tiny E}}(\theta ; p_1, p_2) =0$ is solved by
$\e^{\ri \theta} = \e^{2\pi w(p_2) - 2\pi w(p_1)} = \tfrac{(p_2-a)(b-p_1) }{(b-p_2)(p_1-a)}$, 
which does not have solutions for $\theta \in \RR$  when $p_1 \neq p_2$.


The circles defined by (\ref{mod-circle-euc}) are circles of Apollonius.
Indeed, denoting by $P_a$ and $P_b$ the points having $z=a$ and $z=b$ respectively
in the complex plane parameterised by the complex coordinate $z$
(which correspond to the entangling points of the bipartition of the spatial direction parameterised by $x$
introduced at the beginning of this subsection),
the ratio of their distances from the
generic point of the circle (\ref{mod-circle-euc}) is independent of $\theta$ and reads
\be
\label{apo-circle}
\frac{ d_{\textrm{\tiny E}}\big(P(\theta), P_a \big) }{ d_{\textrm{\tiny E}}\big(P(\theta), P_b \big) }
 = \left( \frac{p-a}{p-b} \right)^2 . 
\ee
This result can be found from (\ref{s12-theta12-def}) (or equivalently from (\ref{euclid-mod-distance}))
by using that $P_a$ and $P_b$ do not evolve, as already remarked above,
and correspond to the counterpart in Euclidean signature of the observation made 
in the final paragraph of Sec.\,\ref{sec-spacetime-distance-1int} (see (\ref{apo-hyperbola}))
for the Minkowski spacetime.

\section{Interval in the line, thermal state}
\label{sec-1int-line-thermal}


In this section we extend the results reported in Sec.\,\ref{sec-1int-line-vacuum}
about the spacetime distance between two events belonging to different modular trajectories
to the case where the CFT is in a thermal state characterised 
by different temperatures for the right and left moving excitations. 
For the sake of simplicity, we focus only on the modular trajectories in $\mathcal{D}_A$.

\subsection{Chiral distance along the modular evolutions  } 
\label{sec-1int-chiral-dir-temp}


We are interested in a chiral CFT on the line and in the thermal state at finite inverse temperature $\beta$.
Consider the bipartition of the line given by the finite interval $A = [a,b] \subset \RR$ and its complement $B$.
The modular Hamiltonian of $A$ reads \cite{Wong:2013gua, Cardy:2016fqc}
\be
\label{KA-chiral-1int-line-temp}
K_A =
\int_{a}^{b} V_\beta(u) \, T(u) \, \rd u
\ee
where the weight function  multiplying the chiral component of the energy density reads
\be
\label{velocity_fund_temp}
V_\beta(u) =2\beta\, \frac{ \sinh [\pi(b-u)/\beta] \, \sinh [\pi(u-a)/\beta]}{\sinh [\pi(b-a)/\beta]}=\frac{1}{w_\beta'(u)} 
\ee
being $w_\beta(u)$ defined as follows
\be
\label{w_thermal}
w_\beta(u) \equiv \frac{1}{2\pi}\,\log \!\left(\! - \frac{ \sinh [\pi(u-a)/ \beta]}{ \sinh [\pi(u-b)/\beta]} \right) 
\;\;\;\qquad\;\;\;
u \in A
\ee
which becomes (\ref{w_fund}) in the zero temperature limit $\beta \to +\infty$, as expected.

%

By adopting the notation introduced in Sec.\,\ref{sec-1int-chiral-dir-mod-evo},
the modular evolution of a chiral primary generated by (\ref{KA-chiral-1int-line-temp}) reads (see e.g. \cite{Mintchev:2020uom})
\be
\label{phi-mod-evo-thermal}
\phi(\tau, u)
\equiv \e^{\ri \tau K_A }\, \phi(u)\, \e^{-\ri \tau K_A}  
= \big[ \partial_{u} \xi_\beta(\tau, u) \big]^h \,\phi\big(\xi_\beta(\tau, u) \big)
\ee
where 
\be
\label{xi-map-thermal}
\xi_\beta(\tau,u) 
\equiv 
w_\beta^{-1}\big(w_\beta(u) + \tau\big)
=
\frac{\beta}{2\pi} \, \log\! \bigg( \frac{ \e^{\pi  (b+a)/\beta} + \e^{ 2\pi  b/\beta}\, \e^{2\pi w_\beta(u)+2\pi \tau} }{ \e^{\pi (b-a)/\beta} +  \e^{2\pi w_\beta(u)+2\pi \tau}}   \bigg)
\ee
which satisfies the initial condition $\xi_\beta(\tau=0,u) = u$ and also 
\be
\label{der-xi-thermal}
\partial_\tau \xi_\beta(\tau,u) = V_\beta(\xi)
\;\;\;\qquad\;\;\;
\partial_u \xi_\beta(\tau,u) = \frac{V_\beta(\xi)}{V_\beta(u)}
\ee
Notice that $\xi_\beta(\tau,u) \in A$ for any $\tau \in \RR$  when $u \in A $.

\begin{figure}[t!]
\vspace{-.5cm}
\hspace{1.3cm}
\includegraphics[width=.8\textwidth]{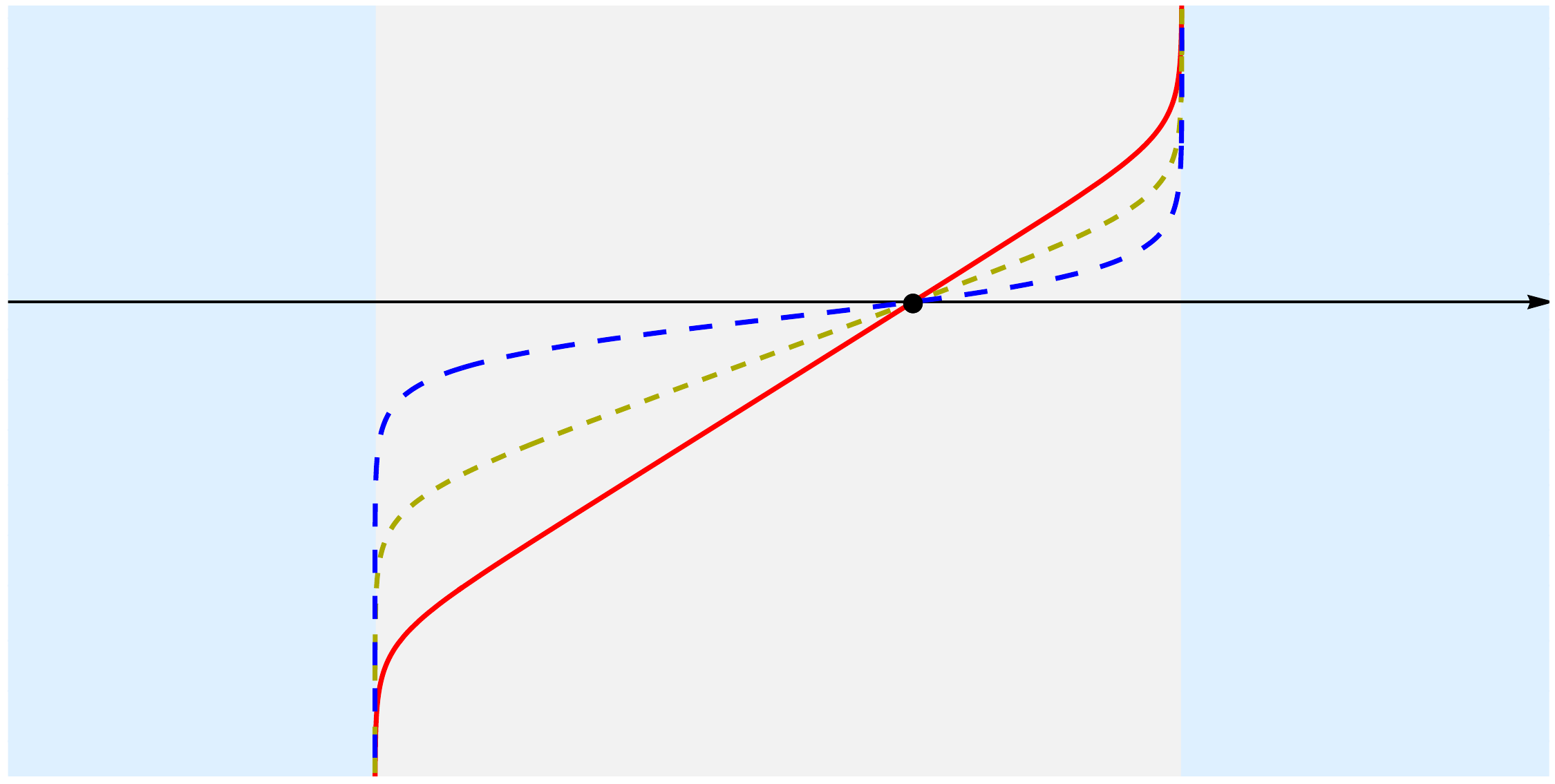}
\vspace{.2cm}
\caption{Modular evolutions along the chiral direction  in the plane $(\xi, \tau)$, given by (\ref{xi-map-thermal}).
The dashed blue line indicates the zero temperature case (see the solid red line in Fig.\,\ref{fig:1int-line-xi-v0}),
while the dashed dark yellow line and the solid red line 
correspond to increasing values of temperature. 
}
\label{fig:1int-line-temp-xi}
\end{figure}

In Fig.\,\ref{fig:1int-line-temp-xi}, by adopting the same notation of Fig.\,\ref{fig:1int-line-xi-v0}, 
we show three examples of chiral modular evolutions given by (\ref{xi-map-thermal}) 
in the plane parameterised by $(\xi, \tau)$ having the same initial point (see the black dot)
and different temperatures 
(the dashed blue curve and the solid red correspond to zero and the highest temperature respectively).
In the regime of high temperature, 
the curve (\ref{xi-map-thermal})  becomes a straight segment in the internal part of $A$.

The two-point function of the chiral primary fields $\phi_\pm$  at finite temperature  read
\be 
\label{2pt-chiral-primaries-temp}
\langle \phi_\pm^*(u_1) \, \phi_\pm (u_2)\rangle_{_\beta}
=
\langle \phi_\pm(u_1) \, \phi_\pm^* (u_2)\rangle_{_\beta}
= 
\frac{ \e^{\mp \ri \pi h_\pm}  }{2\pi\, \big[ \tfrac{\beta}{\pi} \sinh\! \big( \tfrac{\pi}{\beta} (u_1-u_2 \mp \ri \varepsilon) \big)\big]^{2h_\pm}}   
\ee 
where $h_\pm$ are the conformal dimensions. 
In the zero temperature limit $\beta \to +\infty$, the expression (\ref{2pt-chiral-primaries-temp}) becomes (\ref{2pt-chiral-primaries}), as expected. 
From (\ref{phi-mod-evo-thermal}) and (\ref{2pt-chiral-primaries-temp}),
one finds that the modular two-point functions of the chiral primary fields $\phi_\pm$  at finite temperature
are \cite{Mintchev:2020uom}
\be
\label{mod-corr-phi-mu-temp}
\langle \phi_\pm^*(\tau_1, u_1) \,\phi_\pm (\tau_2, u_2)\rangle_{_\beta}
\, =\,
\frac{ \e^{\mp \ri \pi h_\pm} }{2\pi }\;
W_{\beta,\pm}(\pm \tau_{12}; u_1 , u_2 )^{2h_{\pm}}
\ee
in terms of the distributions defined as follows
\be
\label{cap-W-def-temp}
W_{\beta,\pm}(\tau; u_1, u_2)
\,\equiv \,
\frac{ \e^{2\pi w_\beta(u_1)} - \e^{2\pi w_\beta(u_2)}  }{ \tfrac{\beta}{\pi} \sinh\! \big[ \tfrac{\pi}{\beta} (u_1-u_2) \big]  }\;
\frac{1}{  \e^{2\pi w_\beta(u_1) +\pi \tau} - \e^{2\pi w_\beta(u_2) - \pi \tau} \mp \ri \varepsilon}  
\ee
that simplify to (\ref{cap-W-def}) in the zero temperature limit, as expected.

In order to adapt the steps discussed in Sec.\,\ref{sec-chiral-correlators}  to this finite temperature case,
from (\ref{mod-corr-phi-mu-temp}) we find it worth introducing 
\be
\label{W-function-def-temp}
W_\beta(\tau_{12} ; u_1, u_2) 
\equiv  \frac{1}{\tfrac{\beta}{\pi} \sinh\! \big[ \tfrac{\pi}{\beta} (u_1-u_2) \big] \; \mathcal{R}_\beta(\tau_{12} ; u_1, u_2) }
\ee
where 
\be
\label{R-fact-def-temp}
\mathcal{R}_\beta(\tau ; u_1, u_2) 
\equiv 
\frac{  \e^{2\pi w_\beta(u_1) +\pi \tau} - \e^{2\pi w_\beta(u_2) - \pi \tau} }{ \e^{2\pi w_\beta(u_1)} - \e^{2\pi w_\beta(u_2)}  }
\ee
in terms of (\ref{w_thermal}),
which is well defined for $u_1 \neq u_2$ and   satisfies $\mathcal{R}(0 ; u_1, u_2) = 1$ identically.
The function $\mathcal{R}_\beta (\tau ; u_1, u_2) $ in (\ref{R-fact-def-temp}) is not jointly continuous
because its limits $\tau \to 0$ and $u_2 \to u_1$ do not commute.
In the zero temperature limit,
(\ref{W-function-def-temp}) and (\ref{R-fact-def-temp}) become (\ref{W-function-def}) and (\ref{R-fact-def}) respectively, 
as expected.
When $u_1 \neq u_2$, the function  in (\ref{R-fact-def-temp}) vanishes at the finite value $\tau = \tilde{\tau}_{\beta,0}(u_1, u_2) $ given by  
\be
\label{tau0tilde-temp-def}
\tilde{\tau}_{\beta,0}(u_1, u_2) \equiv w_\beta(u_2) - w_\beta(u_1)
 \ee
and such vanishing condition is equivalent to
 \be
 \label{tau0def-u1u2-temp}
 u_2 = \xi_\beta(\tilde{\tau}_{\beta,0} \, ,u_1) 
 \;\;\;\qquad\;\;\;
  u_1 = \xi_\beta(-\tilde{\tau}_{\beta,0} \, ,u_2) \,.
 \ee
We also find that (\ref{xi-map-thermal}) and (\ref{R-fact-def-temp})  satisfy 
\be
\label{R-zero-xi-1int-temp}
\mathcal{R}_\beta\big(\tau ; u, \xi_\beta(\tau,u)  \big) = 0 \,.
\ee
In the limit $b\to + \infty$, when $A$ becomes the half line $(a, +\infty)$,
the function (\ref{R-fact-def-temp})  simplifies to
\be
\label{Omega12-rindler-temp}
\lim_{b \to +\infty} \mathcal{R}_\beta(\tau ; u_1, u_2) = 
\frac{ \e^{\pi \tau} \, \e^{\pi (u_1-a)/\beta} \sinh[\pi(u_1-a)/\beta] -  \e^{-\pi \tau} \, \e^{\pi (u_2-a)/\beta} \sinh[\pi(u_2-a)/\beta] 
}{
\e^{\pi (u_1+ u_2 -2a)/\beta} \sinh[\pi(u_1-u_2)/\beta] 
}
\ee
that becomes (\ref{Omega12-rindler}) in the zero temperature limit, as expected.


The previous results allow to explore the chiral distance $\xi_\beta(\tau_1 ,u_1)  - \xi_\beta(\tau_2 ,u_2) $,
extending to finite temperature the zero temperature analysis reported in Sec.\,\ref{sec-chiral-distance-mod-evo}.
First we observe that, when $\tau_{12} \neq 0$ and $u_1 \neq u_2$, 
the expressions in (\ref{xi-map-thermal}) and (\ref{W-function-def-temp}) satisfy 
\be
\label{rel-xi-W-1int-temp}
\frac{\partial_{u_1} \xi_\beta(\tau_1, u_1)\, \partial_{u_2} \xi_\beta(\tau_2, u_2) }{\big\{ \tfrac{\beta}{\pi}  \sinh \! \big[ \tfrac{\pi}{\beta} \big( \xi_\beta(\tau_1 ,u_1)  - \xi_\beta(\tau_2 ,u_2) \big)\big] \big\}^2} 
\,=\, 
W_\beta(\tau_{12} ; u_1, u_2)^2
\ee
which leads to 
\be
\label{xi12-tau-chiral-thermal}
\sinh \! \big[ \tfrac{\pi}{\beta} \big( \xi_\beta(\tau_1 ,u_1)  - \xi_\beta(\tau_2 ,u_2) \big)\big]
=
\mathcal{R}_\beta(\tau_{12} ; u_1, u_2)\,
\sqrt{\partial_{u_1}\xi_\beta(\tau_1 ,u_1) \; \partial_{u_2}\xi_\beta(\tau_2 ,u_2)  } \; \sinh[ \tfrac{\pi}{\beta} (u_1 - u_2)\big]
\ee
where (\ref{R-fact-def-temp}) occurs. 
The relation (\ref{xi12-tau-chiral-thermal}), 
which becomes (\ref{xi12-tau-chiral-2}) in the  zero temperature limit as expected,
provides the chiral distance at finite temperature
\be
\label{xi-12-temp-F}
\xi_\beta(\tau_1 ,u_1)  - \xi_\beta(\tau_2 ,u_2)  \,=\, \mathcal{F}_\beta(\tau_{1}, \tau_{2} ; u_1, u_2)\,
\big( u_1 - u_2\big)
\ee
where
\be
\label{cal-F-beta-def}
\mathcal{F}_\beta(\tau_{1}, \tau_{2} ; u_1, u_2)
\equiv 
\frac{
\textrm{arcsinh} \!
\left(
\mathcal{R}_\beta(\tau_{12} ; u_1, u_2)\,
\sqrt{\partial_{u_1}\xi_\beta(\tau_1 ,u_1) \; \partial_{u_2}\xi_\beta(\tau_2 ,u_2)  } \; \sinh[ \tfrac{\pi}{\beta} (u_1 - u_2)\big]
\right)
}{
\tfrac{\pi}{\beta} (u_1 - u_2)
}\,.
\ee

It is worth considering (\ref{xi12-tau-chiral-thermal}) in the limiting regimes where either $\tau_1 = \tau_2$ or $u_1 = u_2$.
\\
When $\tau_1 = \tau_2 \equiv \tau$,  since $\mathcal{R}(0 ; u_1, u_2) =1$,
the relation (\ref{xi12-tau-chiral-thermal}) simplifies to 
\be
 \label{xi12-tau-chiral-equal-tau-temp}
\sinh \! \big[ \tfrac{\pi}{\beta} \big( \xi_\beta(\tau ,u_1)  - \xi_\beta(\tau ,u_2) \big)\big]
=
\sqrt{\partial_{u_1}\xi_\beta(\tau ,u_1) \; \partial_{u_2}\xi_\beta(\tau ,u_2)  } \; \sinh[ \tfrac{\pi}{\beta} (u_1 - u_2)\big]
\ee
which vanishes as $u_1 = u_2$ and whose zero temperature limit provides (\ref{xi12-tau-chiral-equal-tau}), as expected. 
By employing $\mathcal{R}(0 ; u_1, u_2) =1$, 
one gets $\mathcal{F}_\beta(\tau , \tau  ; u_1, u_2)$
and the resulting expression does not take a factorised form, 
like in the zero temperature limit (see (\ref{xi12-tau-chiral-equal-tau})).

In order to explore the limit of  (\ref{xi12-tau-chiral-thermal}) when $u_1 = u_2$,
we first notice that for (\ref{R-fact-def-temp}) we have 
\be
\label{R-u12-same-u-temp}
\lim_{v \to u} \mathcal{R}_\beta(\tau ; u, v) \, \frac{ \sinh[ \pi (u - v) / \beta] }{\pi / \beta}
=
\frac{\sinh(\pi \tau)}{\pi\; w_\beta'(u)}
\ee
which gives (\ref{R-u12-same-u}) at zero temperature.
Hence, for $u_1 = u_2 \equiv u$ the relation (\ref{xi12-tau-chiral-thermal})  becomes 
\be
\label{xi-12-equal-u1u2-temp}
\sinh \! \big[ \tfrac{\pi}{\beta} \big( \xi_\beta(\tau_1 ,u)  - \xi_\beta(\tau_2 ,u) \big)\big]
=  
\frac{1}{\beta}\, \sqrt{\partial_{u_1}\xi_\beta(\tau_1 ,u) \; \partial_{u_2}\xi_\beta(\tau_2 ,u)  } \;  V_\beta(u) \,\sinh(\pi \tau_{12})
\ee
which vanishes for $\tau_1 = \tau_2$, as expected.

\subsection{Modular trajectories and spacetime distance between their points  }


In the setup that we are considering throughout this section, 
the generic point of a modular trajectories in  $\mathcal{D}_A$
has  spacetime coordinates $\big( x(\tau) , t(\tau)\big)$ given by 
\be
\label{mod-traj-tau-line-temp}
x(\tau) = \frac{\xi_{\beta_{+}}(\tau, u_+) + \xi_{\beta_{-}}(-\tau, u_-) }{2}
\;\;\;\qquad\;\;\;
t(\tau) = \frac{\xi_{\beta_{+}}(\tau, u_+) - \xi_{\beta_{-}}(-\tau, u_-) }{2}
\ee
in terms of (\ref{xi-map-thermal}),
where $\tau \in \RR$ and 
$(u_+, u_{-})$ are the light-cone coordinates of the initial point.

\begin{figure}[t!]
\vspace{-.6cm}
\hspace{-.25cm}
\includegraphics[width=1.04\textwidth]{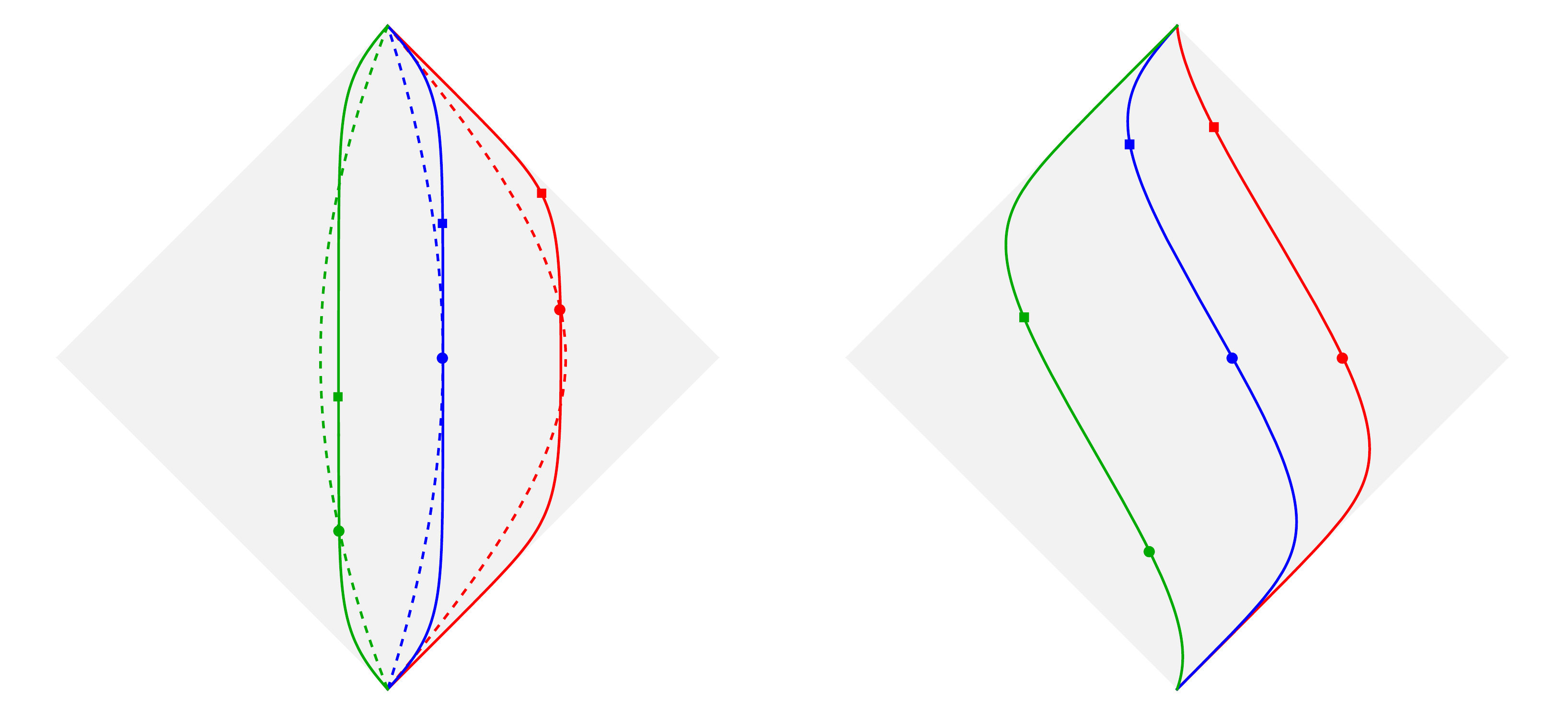}
\vspace{-.3cm}
\caption{
Modular trajectories (solid lines) for
either $\beta_{+} = \beta_{-}$  (left panel) or $\beta_{+} < \beta_{-}$ (right panel).
The dashed curves in the left panel 
are the modular trajectories at zero temperature having the same initial points
(see also the solid lines in Fig.\,\ref{fig:diamond-mod-hyper}).
}
\label{fig:1int-mod-traject-Temp-1}
\end{figure}

In Fig.\,\ref{fig:1int-mod-traject-Temp-1} three distinct modular trajectories are shown
whose initial points are denoted by dots, according to the notation adopted in the previous figures
for either $\beta_+ = \beta_{-}$ (left panel) or $\beta_+ < \beta_{-}$ (right panel).
In order to highlight the deformation of a modular trajectory induced by a non vanishing temperature,
in the left panel of Fig.\,\ref{fig:1int-mod-traject-Temp-1} we show through dashed curves also 
the modular trajectories at zero temperature having the same initial points
(see also the curves in $\mathcal{D}_A$ in Fig.\,\ref{fig:diamond-mod-hyper} or in the left panel of Fig.\,\ref{fig:1int-mod-traject-1}).
At non vanishing finite temperature, the modular trajectories become vertical in the central part of $\mathcal{D}_A$
and this effect is more evident as the temperature increases,
as already observed in \cite{Borchers:1998ye}, 
where the modular trajectories  in the limit $b \to +\infty$ have been considered.  
The right panel of Fig.\,\ref{fig:1int-mod-traject-Temp-1} illustrates 
the characteristic behaviour of the modular trajectories when $\beta_{+} < \beta_{-}$ 
(it is straightforward to figure out the qualitative behaviour of the modular trajectories when $\beta_{+} > \beta_{-}$).


By adopting the same notation of Sec.\,\ref{sec-spacetime-distance-1int},
let us consider two distinct modular trajectories in $\mathcal{D}_A$ with initial points $P_1$ and $P_2$.
The spacetime distance $d\big(P_1(\tau), P_2(\tau)\big)$ between 
the modular evolutions $P_1(\tau)$ and $P_2(\tau)$ of the initial points along the corresponding modular trajectories
after the same value of the modular parameter $\tau$ reads
\bea
\label{s12-tau-thermal-def}
d\big(P_1(\tau), P_2(\tau)\big)
&\equiv &
-\; t_{12}(\tau)^2 + x_{12}(\tau)^2 
\\
&=&
\big[ \xi_{\beta_{+}}(\tau, u_{1,+}) - \xi_{\beta_{+}}(\tau, u_{2,+}) \big] \big[ \xi_{\beta_{-}}(-\tau, u_{1,-}) - \xi_{\beta_{-}}(-\tau, u_{2,-}) \big]\,.
\eea
By employing (\ref{xi-12-temp-F}) in each chiral direction, 
this spacetime distance can be written as follows
\be
\label{s12-tau-evolution-thermal}
d_{\boldsymbol{\beta} }  \big(P_1(\tau), P_2(\tau)\big)
= 
\omega_{\boldsymbol{\beta} }( \tau;P_1,P_2) 
\, 
d (P_1, P_2)
\ee
where $d(P_1, P_2)= u_{12,+} \, u_{12,-}$ is the spacetime distance between the initial points $P_1$ and $P_2$,
that is independent of the temperatures
$\boldsymbol{\beta} \equiv (\beta_+ , \beta_{-})$ 
and  $\omega_{\boldsymbol{\beta} } (\tau;P_1,P_2) $ is defined as 
\be
\label{OmegaA-def-thermal}
\omega_{\boldsymbol{\beta} } ( \tau;P_1,P_2) 
\equiv
\mathcal{F}_{\beta_{+}} (\tau, \tau ; u_{1,+}, u_{2,+}) \,
\mathcal{F}_{\beta_{-}} (\tau, \tau ; u_{1,-}, u_{2,-})
\ee
in terms of (\ref{cal-F-beta-def}),
which takes the factorised form (\ref{OmegaA-def})
in the limiting regime of zero temperatures, i.e. as $\beta_\pm \to +\infty$.

\begin{figure}[t!]
\vspace{-.6cm}
\hspace{-.55cm}
\includegraphics[width=1.05\textwidth]{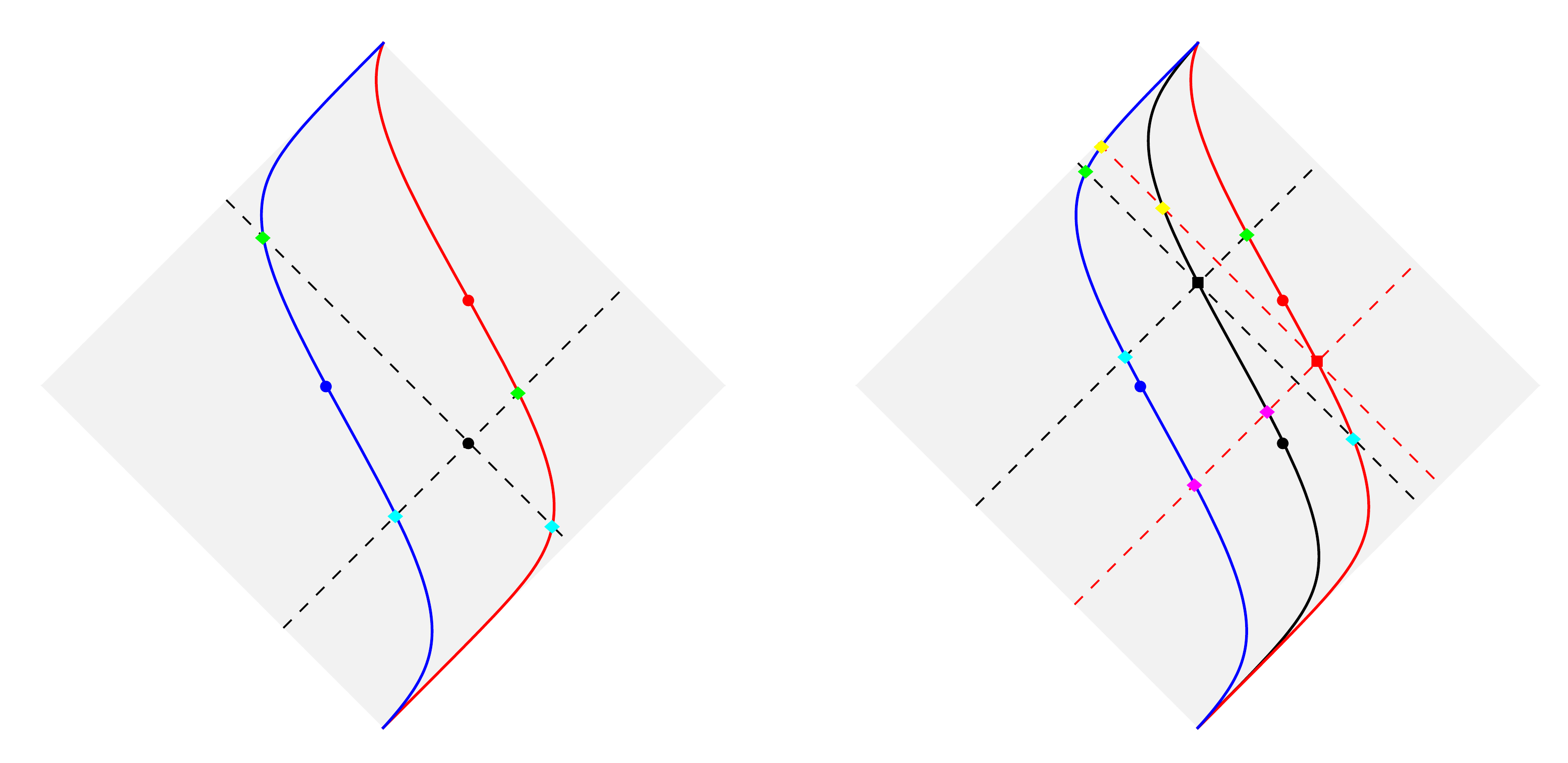}
\vspace{-.3cm}
\caption{
Spacetime distances for  $\beta_{+} < \beta_{-}$ (see (\ref{s12-gen-Temp-tau-evolution}))
between a reference point 
(black dot in the left panel and black square in the right panel)
and a point moving along a modular trajectory in $\mathcal{D}_A$
(either the red or the blue solid curve).
}
\label{fig:1int-mod-traject-Temp-2}
\end{figure}


The most important feature to highlight about (\ref{s12-tau-thermal-def})-(\ref{s12-tau-evolution-thermal}) is that
relativistic causality is preserved along the modular evolution 
for any values of the inverse temperatures $\beta_{-}$ and $\beta_{+}$.
This fact can be easily realised already from (\ref{xi12-tau-chiral-equal-tau-temp}) and (\ref{s12-tau-thermal-def}),
by employing the fact that $x$ and $\sinh(x)$ have the same sign,
and  is illustrated in Fig.\,\ref{fig:1int-mod-traject-Temp-1},
where the coloured squares indicate the modular evolutions of the corresponding dot having the same colour 
after the same value of the modular parameter. 
The choice of the initial points in each panel of Fig.\,\ref{fig:1int-mod-traject-Temp-1}
is such that an example is shown for each type of spacetime distance 
(i.e. timelike, spacelike and lightlike).


The most general case corresponds to the spacetime distance 
between $P_1(\tau_1)$ and $P_2(\tau_2)$ associated to two different values $\tau_1$ and $\tau_2$
of the modular parameter along the corresponding modular trajectories.
This distance reads
\bea
\label{s12-tau12-Temp-def}
d_{\boldsymbol{\beta} } \big(P_1(\tau_1), P_2(\tau_2)\big)
& \equiv &
-\,\big[ t_{1}(\tau_1) - t_{2}(\tau_2) \big]^2 + \big[ x_{1}(\tau_1) - x_{2}(\tau_2) \big]^2
\\
\label{s12-tau12-Temp-def-1}
& = &
\big[ \xi_{\beta_{+}}(\tau_1, u_{1,+}) - \xi_{\beta_{+}}(\tau_2, u_{2,+}) \big] \big[ \xi_{\beta_{-}}(-\tau_1, u_{1,-}) - \xi_{\beta_{-}}(-\tau_2, u_{2,-}) \big]\,.
\hspace{1cm}
\eea
By employing (\ref{xi-12-temp-F}) for both the chiral distances, this distance can be written as follows
\be
\label{s12-gen-Temp-tau-evolution}
d_{\boldsymbol{\beta} }  \big(P_1(\tau_1), P_2(\tau_2)\big)
= \Omega_{\boldsymbol{\beta} }  (\tau_1, \tau_2  ;P_1,P_2)  \, 
d(P_1, P_2)
\ee
where we have introduced 
\be
\label{omega-12-Temp-def}
\Omega_{\boldsymbol{\beta} }  (\tau_1, \tau_2  ;P_1,P_2)  
\equiv
\mathcal{F}_{\beta_{+}} (\tau_{1}, \tau_{2} ; u_{1,+}, u_{2,+}) \,
\mathcal{F}_{\beta_{-}} (\tau_{1}, \tau_{2} ; u_{1,-}, u_{2,-})
\ee
which becomes (\ref{OmegaA-def-thermal}) when $\tau_1 = \tau_2$, as expected.


Extending the analysis reported in Sec.\,\ref{sec-spacetime-distance-1int} for the ground state 
to this finite temperature case, 
we find that $d_{\boldsymbol{\beta} } \big(P_1(\tau_1), P_2(\tau_2)\big)/ d(P_1, P_2)$
changes sign at $\tau_{12} \in \{ \tau_{\boldsymbol{\beta} , <} \, , \tau_{\boldsymbol{\beta} , >}  \}$, 
where the values $ \tau_{\boldsymbol{\beta} , <}    < \tau_{\boldsymbol{\beta} , >}  $  
are defined in terms of (\ref{tau0tilde-temp-def}) as follows
\bea
\label{tau-min-12-Temp-def}
\tau_{\boldsymbol{\beta} , <}  
& \equiv & 
\textrm{min} \big\{ \tilde{\tau}_{\beta_{+},0}(u_{1,+}, u_{2,+}) \,, - \,\tilde{\tau}_{\beta_{-},0}(u_{1,-}, u_{2,-})) \big\}
\\
\label{tau-max-12-Temp-def}
\rule{0pt}{.5cm}
\tau_{\boldsymbol{\beta} , >} 
& \equiv & 
 \textrm{max} \big\{ \tilde{\tau}_{\beta_{+},0}(u_{1,+}, u_{2,+}) \,, - \,\tilde{\tau}_{\beta_{-},0}(u_{1,-}, u_{2,-})) \big\} \,.
\eea

In Fig.\,\ref{fig:1int-mod-traject-Temp-2} 
we show some explicit cases to explore the sign of $d_{\boldsymbol{\beta} } \big(P_1(\tau_1), P_2(\tau_2)\big)/ d(P_1, P_2)$,
as done in Fig.\,\ref{fig:1int-mod-traject-3} at zero temperature, by adopting the same notation.
Comparing these two figures, we observe that 
the same properties discussed for Fig.\,\ref{fig:1int-mod-traject-3} at zero temperature
hold also at finite temperature for Fig.\,\ref{fig:1int-mod-traject-Temp-2}, at least at qualitative level.

\section{Dirac field in the vacuum: Two disjoint intervals in the line}
\label{sec-2int-Dirac}


In this section we consider the free massless Dirac field 
(a prototypical example of two-dimensional CFT with central charge $c=1$)
on the line in its ground state.
The free massless Dirac field is a doublet made by two chiral complex fields $\psi_\pm(u_\pm)$.
While  in Sec.\,\ref{sec-1int-line-vacuum} and Sec.\,\ref{sec-1int-line-thermal}
the bipartition of the line is given by a finite interval, 
in the following the bipartition $A \cup B =\RR$ of the line is characterised by the union of two disjoint intervals, 
i.e. $A \equiv A_1 \cup A_2$, where $A_j \equiv [a_j, b_j]$ with $j \in \{1,2\}$.
The corresponding modular Hamiltonian and modular flows for the chiral fields $\psi_\pm$
have been found by Casini and Huerta \cite{Casini:2009vk},
while the modular correlators of $\psi_\pm$
satisfying the proper KMS condition
have been obtained by Longo, Martinetti and Rehren \cite{Longo:2009mn}.
In this section we explore further the modular evolution generated by the modular Hamiltonian of $A$.
In Appendix\;\ref{app-mod-flow-2int} the derivations 
of the modular flows of $\psi_\pm$ \cite{Casini:2009vk} 
and of their modular correlators \cite{Longo:2009mn} are revisited, 
finding also the modular correlators  of the chiral density fields
$\varrho_\pm \equiv \; : \!\psi_\pm^\ast \psi_\pm \!:\! $\;.
It is often insightful  to employ the lengths $\ell_j = b_j - a_j$ of $A_j$ with $j \in \{1,2\}$
and of their distance $d = a_2 - b_1$, which are positive parameters.
Thus, the bipartition of the line is equivalently specified either by $\{a_1, b_1, a_2, b_2\}$ or by $\{a_1, \ell_1, d, \ell_2\}$.
The symmetric configuration $A=A_{\textrm{\tiny sym}} \equiv [-b, -a] \cup [a, b]$, where $0<a<b$,
is a special case that is worth considering because various expressions take a simpler form.

\subsection{Modular Hamiltonian  }


In the setup described above, the modular Hamiltonian of $A$ is the sum of a local term and a bilocal term.
Along each chiral direction, the modular Hamiltonian found in \cite{Casini:2009vk} reads
\be
\label{mod-ham-2int-KA}
K_A  = K_{A, \textrm{\tiny loc}} + K_{A, \textrm{\tiny biloc}}
\ee
The local term is
\be
\label{K_A-2int-terms-local}
K_{A, \textrm{\tiny loc}}
=
\int_A V_{\textrm{\tiny loc}}(u) \, T(u)\, \rd u
\ee
where 
\be
\label{T00-lambda-def}
T(u) \equiv 
\,\frac{\textrm{i}}{2}
\Big (\! 
:\!  \psi^\ast\, (\partial_u \psi) \! : 
-
:\! (\partial_u \psi^\ast)\, \psi \! :
\! \Big) 
(u)
\ee
which stands for $T_{\pm}(u_\pm) $ 
and we recall that the energy density  of the massless Dirac field is
given by $T_{tt}(t,x)  \equiv T_{+}(u_+) -  T_{-}(u_-)$.
The bilocal term in (\ref{mod-ham-2int-KA}) is
\be
\label{K_A-2int-terms-bi-local}
K_{A, \textrm{\tiny biloc}}
=
\int_A 
V_{\textrm{\tiny biloc}}(u) \, T_{\textrm{\tiny biloc}}(u, u_{\textrm{\tiny c}} ) \, \rd u
\ee
being  $T_{\textrm{\tiny biloc}}(u, v) $ the chiral bilocal quadratic operator defined as follows
\be
\label{T-bilocal-def}
T_{\textrm{\tiny biloc}}(u, v) 
\equiv
\frac{\textrm{i}}{2}\,
\!:\!\!\Big[\, \psi^\ast(u) \,  \psi(v) - \psi^\ast(v) \,  \psi(u) \, \Big]\!\!: 
\ee
and $u_{\textrm{\tiny c}} = \mathcal{C} (u)$ in (\ref{K_A-2int-terms-bi-local}) is introduced below (see (\ref{u-conj-def})).


The weight functions occurring in (\ref{K_A-2int-terms-local}) and (\ref{K_A-2int-terms-bi-local})
can be defined by introducing
\be
\label{w_fund_2int}
w(u) \equiv \frac{1}{2\pi}\,\log \!\left(\! - \frac{(u-a_1)(u-a_2)}{(u-b_1)(u-b_2)} \right) 
\;\;\;\qquad\;\;\;
u \in A \,.
\ee
In the large distance limit $d \to +\infty$, this expression evaluated in $u \in A_j$ for $j \in \{1,2\}$
becomes the corresponding result for the single interval associated to $A_j$ (see (\ref{w_fund})),
while in the opposite limit of adjacent intervals $a_2 \to b_1$, 
the function in (\ref{w_fund_2int}) simplifies to 
the corresponding expression for the resulting single interval, 
namely (\ref{w_fund}) associated to $(a_1, b_2)$.

The weight functions in (\ref{K_A-2int-terms-local}) and (\ref{K_A-2int-terms-bi-local}) 
can be written through (\ref{w_fund_2int}) as follows
\be
\label{velocity_fund-2int}
V_{\textrm{\tiny loc}}(u) =\frac{1}{w'(u)} 
\;\;\;\qquad\;\;\;
V_{\textrm{\tiny biloc}}(u) =
\frac{V_{\textrm{\tiny loc}} (u_{\textrm{c}} )}{ u - u_{\textrm{c}} } 
\ee
where the position $u_{\textrm{c}} = \mathcal{C}(u)$ conjugate\footnote{
The conjugate point $u_{\textrm{c}}$ in (\ref{u-conj-def}) must not be confused with the 
image of the geometric action of the modular conjugation defined in (\ref{j0-map-def}).
} 
to $u$ is  (see also \cite{Mintchev:2022fcp})
\be
\label{u-conj-def}
u_{\textrm{c}} = \mathcal{C}(u) \equiv q_0 - \frac{r_0^2}{u - q_0}
\ee
in terms of the following parameters 
\bea
\label{q0-def}
q_0 
&\equiv &
\frac{b_1\, b_2 - a_1\, a_2}{ b_1 - a_1 + b_2 - a_2 } 
\,=\, 
a_1 +\frac{(\ell_1 + d +\ell_2) \, \ell_1}{\ell_1+\ell_2}
\,=\,
b_1 + \frac{d\, \ell_1}{ \ell_1+\ell_2 }
\,=\,
a_2 - \frac{d\, \ell_2}{ \ell_1+\ell_2 }
\,\in\,(b_1, a_2)
\nn
\\
& &
\\
\label{r0-def}
r_0 & \equiv & 
\frac{ \sqrt{ (b_1 - a_1) (b_2 - a_2) (b_2 - a_1) (a_2 - b_1)} }{ b_1 - a_1 + b_2 - a_2  } 
\,=\,
\frac{\sqrt{(\ell_1 + d +\ell_2) \, d\, \ell_1\, \ell_2}}{\ell_1+\ell_2}\, >\, 0 \,.
\eea
Writing (\ref{u-conj-def}) as $\mathcal{C}(u) - q_0 = - r_0^2 /(u - q_0)$,
one recognises an inversion centered at $q_0$, with radius $r_0$ combined with a reflection. 
The characteristic property defining (\ref{u-conj-def}) is 
\be
\label{w-uc-w-u}
w(u_{\textrm{c}} ) = w(u)
\ee
and we remark that  $u_{\textrm{c}}  \in A_j$ when $u \in A_k$, with $k \neq j$.
An insightful geometric interpretation of (\ref{u-conj-def}) is obtained through the form
$\big(\mathcal{C}(u) - q_0 \big)  ( q_0 - u ) = r_0^2$.
Let us also highlight  the suggestive similarity between 
the expressions providing $q_0$ and $r_0$ in (\ref{q0-def})-(\ref{r0-def})
with the ones for $x_0$ and $\kappa_0$ in  (\ref{mod-hyper-parameters}) 
respectively.

In the symmetric configuration $A=A_{\textrm{\tiny sym}}$,
the parameters in (\ref{q0-def}) and (\ref{r0-def}) simplify to $q_0=0$ and $r_0 = \sqrt{a\,b}$ respectively. 
In the adjacent intervals limit $d \to 0$ for the generic configuration,
we have $r_0 \to 0$ and therefor $ u_{\textrm{c}}(u) \to a_2$ (i.e. the merging point)
for any $u \in A$.

Combining (\ref{velocity_fund-2int}) and (\ref{u-conj-def}), we find the following relations
\be
\label{V-uc-rels}
V_{\textrm{\tiny loc}}(u_{\textrm{c}}) \,=\, \mathcal{C}'(u) \, V_{\textrm{\tiny loc}}(u) 
\;\;\;\;\;\qquad\;\;\;\;\;
V_{\textrm{\tiny biloc}}(u_{\textrm{c}}) \,=\, -\, \frac{1}{ \mathcal{C}'(u) } \; V_{\textrm{\tiny biloc}}(u) \,.
\ee

The weight functions (\ref{velocity_fund-2int}) can be written explicitly in terms of (\ref{u-conj-def}) and (\ref{q0-def})
as follows
\bea
\label{velocity_fund-2int-explicit}
V_{\textrm{\tiny loc}}(u) 
&=& 
\frac{2\pi\, (b_1 - u) (u - a_1) \, (b_2 - u) (u - a_2) }{ (b_1 - a_1 + b_2 - a_2)\, (u - q_0 ) \, (u_{\textrm{c}} - u  ) }
\\
\rule{0pt}{.8cm}
V_{\textrm{\tiny biloc}}(u) 
&=&
\frac{2\pi\, (b_1 - u_{\textrm{c}}) (u_{\textrm{c}} - a_1) \, (b_2 - u_{\textrm{c}}) (u_{\textrm{c}} - a_2) }{ (b_1 - a_1 + b_2 - a_2)\, (u_{\textrm{c}} - q_0 ) \, (u_{\textrm{c}} - u )^2 }
\eea
that are regular functions for $u \in A$;
indeed $q_0 \notin A$ and $u_{\textrm{c}}  \in A_j$ when $u \in A_k$ with $k \neq j$.

\subsection{Modular flow of the chiral fermion  }
\label{sec-mod-flow-psi-2int}


The modular flow $ \psi(\tau, u) \equiv \e^{\ri \tau K_A }\, \psi(u)\, \e^{-\ri \tau K_A} $ of the chiral field $\psi(u)$, 
for $u \in A$ and $\tau \in \RR$,
is generated by (\ref{mod-ham-2int-KA}) 
and it has been studied first by Casini and Huerta in \cite{Casini:2009vk}.
In  Appendix\;\ref{app-mod-flow-2int-psi} we have  revisited its derivation 
by extending to a generic configuration of two disjoint intervals on the line
the analysis in Appendix C of \cite{Mintchev:2020uom} for $A_{\textrm{\tiny sym}}$ 
and the main result is (see (\ref{Psi-tilde-def-app-v6}))
\be
\label{psi-flow-text}
\psi(\tau, u) 
\,=\,
\frac{\sqrt{  \partial_u \xi } }{\tilde{\eta}(\xi, u)} \; \psi(\xi) 
-
\frac{ \sqrt{  \partial_u \xi_{\textrm{c}} } }{ \tilde{\eta}(\xi_{\textrm{c}} , u)} \; \psi(\xi_{\textrm{c}}) 
\ee
where $\xi = \xi(\tau, u)$ (whose explicit expression is reported below, in (\ref{xi-2int-A1-A2}))
satisfies the initial condition $\xi(\tau=0, u) =u$
and we have introduced
\be
\label{xi-conj-def-text}
\xi_{\textrm{c}} \equiv \mathcal{C}(\xi) = q_0 - \frac{r_0^2}{\xi - q_0}
\ee
and 
\be
\label{tilde-eta-def}
\tilde{\eta}(\xi,u) 
\equiv
\frac{ \sqrt{ \big[ (\xi-q_0)^2+r_0^2 \big] \big[ (u-q_0)^2+r_0^2\big] } }{ (\xi -q_0) \, (u-q_0) + r_0^2  }
\ee
that is strictly positive when both $\xi$ and $u$ belong to the same interval.

We find it worth introducing also the following harmonic ratio (see also (\ref{eta-def-app}))
\be
\label{eta-ratio-def}
\eta(u_1, u_2) 
\equiv 
\frac{ (u_1 - u_{1,\textrm{c}} ) \, ( u_2 - u_{2,\textrm{c}} ) }{ (u_1 - u_{2,\textrm{c}} ) \, ( u_2 - u_{1,\textrm{c}} ) }
\,=\,
\frac{ \big[(u_1 - q_0)^2 +r_0^2 \big] \, \big[ (u_2 - q_0)^2 +r_0^2 \big] }{ \big[\,(u_1 - q_0) (u_2 - q_0) + r_0^2\,\big]^2} \,>\, 0
\ee
in terms of the two points $u_j$ and their conjugate points $u_{j,\textrm{c}} \equiv \mathcal{C}(u_j) $ (see  (\ref{u-conj-def})),
which is related to (\ref{tilde-eta-def})  as  follows
\be
\label{tilde-eta-vs-eta}
\tilde{\eta}(\xi,u) 
=
\textrm{sign}\big[ (\xi -q_0) \, (u-q_0) + r_0^2  \big]
\sqrt{ \eta(\xi,u)}
=
\textrm{sign}\big[  (u-q_0) \, \tilde{u}\big]
\sqrt{ \eta(u_{\textrm{c}} +\tilde{u}  ,u)}
\ee
where the last expression has been found by setting $\xi = u_{\textrm{c}} + \tilde{u}$.
Thus, the sign in the r.h.s. of (\ref{tilde-eta-vs-eta})
can be negative only when $\xi$ and $u$ belong to different intervals. 

Since  $\tilde{\eta}(u,u) = 1$ identically and we also have that
$\xi  \to u$ and $\xi_{\textrm{c}} \to u_{\textrm{c}}$ as $\tau \to 0$,
implying (from (\ref{eta-ratio-def})) that $\eta(\xi_{\textrm{c}} , u) \to +\infty$,
the modular flow (\ref{psi-flow-text})
satisfies the initial condition $\psi(\tau=0, u) = \psi(u) $, as expected.

The main feature to highlight in the modular flow (\ref{psi-flow-text}) is its bilocal nature. 
Indeed, the r.h.s. of (\ref{psi-flow-text}) is a mixing between one field localised in $A_j$
and another one in $A_k$ with $k \neq j$.
This originates from the bilocal term (\ref{K_A-2int-terms-bi-local}) 
in the modular Hamiltonian (\ref{mod-ham-2int-KA}).


The harmonic ratio (\ref{eta-ratio-def})
satisfies $\eta(u_1, u_2)  > 1$ for any $u_1$ and $u_2$ in $A$, as shown in (\ref{eta-min-one-app}).
Moreover, it  is invariant under the conjugation relation (\ref{u-conj-def}), namely
(see also (\ref{eta-rels-app1}))
\be
\eta(u_{1,\textrm{c}} , u_{2,\textrm{c}}) = \eta(u_1, u_2) 
\ee
and becomes identically equal to $1$ 
in the limits of adjacent intervals and of large distance,
i.e. 
\be
\label{eta-special-regimes-A}
\lim_{d \,\to\, 0} \eta(u_1, u_2)  = 1
\;\;\;\qquad\;\;\;
\lim_{d \,\to\, +\infty} \eta(u_1, u_2)  = 1
\ee
Notice that the harmonic ratio (\ref{eta-ratio-def})
is not well defined when $u_1$ and $u_2$ are related through the conjugation relation (\ref{u-conj-def});
indeed as $u_2 \to u_{1, \textrm{c}}$ we have 
\be
\label{eta-u2-to-u1c}
\eta ( u, u_{\textrm{c}} + \epsilon) 
=
\left( \frac{ u - u_{\textrm{c}} }{ u - q_0 } \right)^2  
\frac{r_0^2}{\epsilon^2} 
+ O(1/\epsilon)
\ee
in terms of (\ref{q0-def})-(\ref{r0-def}),
that  will be employed  in Sec.\,\ref{sec-2int-distance-chiral}.

\begin{figure}[t!]
\vspace{-.5cm}
\hspace{1.3cm}
\includegraphics[width=.8\textwidth]{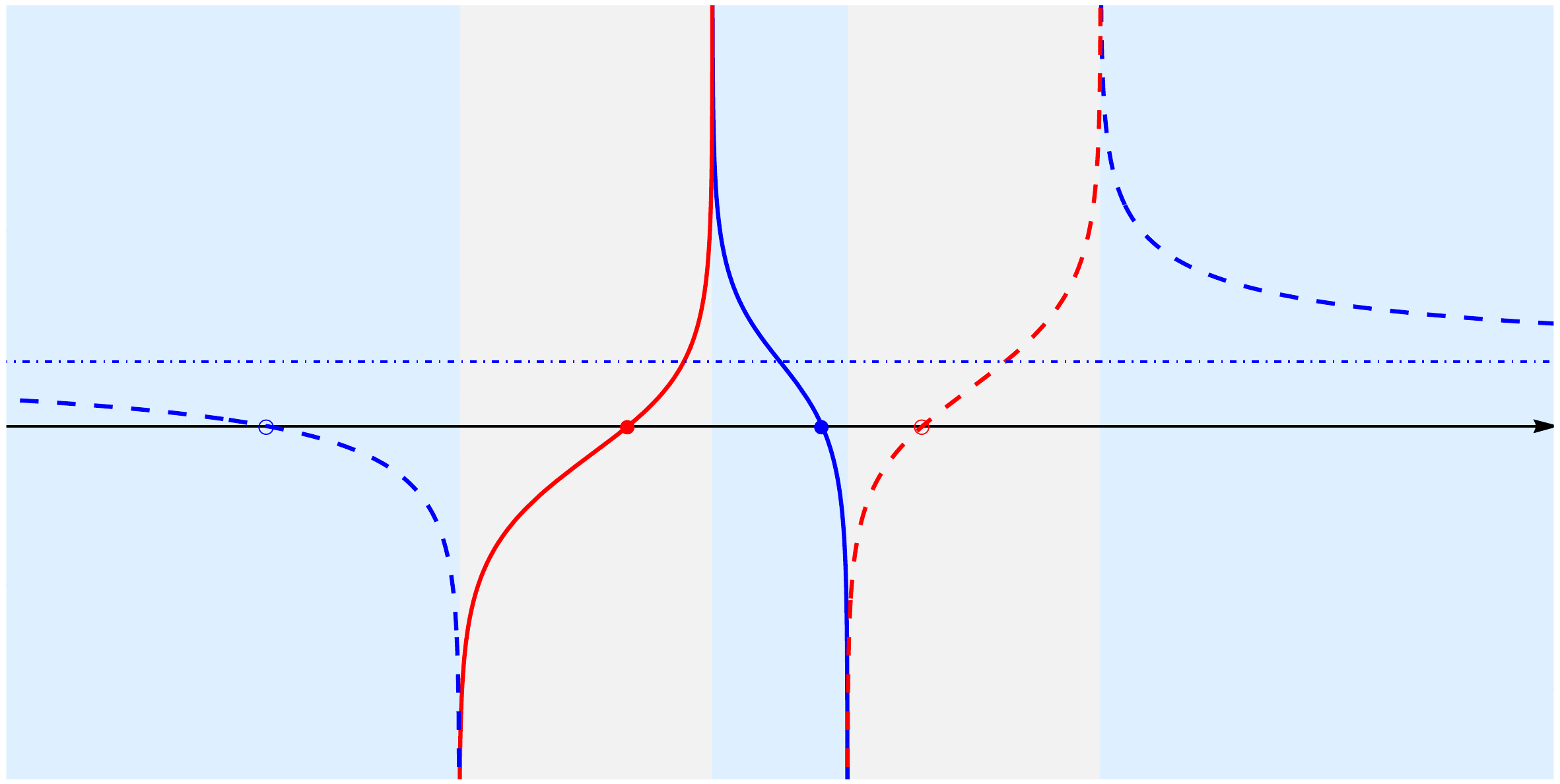}
\vspace{.2cm}
\caption{Modular evolutions $\xi(\tau, u)$ along the chiral direction  in the plane $(\xi, \tau)$,
given by  (\ref{xi-2int-A1-A2}) when $u \in A$ (red dot)
and by (\ref{xi-2int-B1-B2}) when $u \in B$ (blue dot).
}
\label{fig:2int-line-xi-v0}
\end{figure}

The function $\xi = \xi(\tau, u)$ occurring in the modular flow (\ref{psi-flow-text})
provides the geometric action of the modular group of automorphisms induced by the vacuum 
for this setup. 
Its explicit expression reads
\be
\label{xi-2int-A1-A2}
\xi(\tau,u)  \,\equiv \, \Theta_{A_1}  (u) \, \xi_1(\tau,u) + \Theta_{A_2} (u) \, \xi_2(\tau,u)
\hspace{2cm}
u \in A
\ee
where 
$\Theta_{A_j}(u)$ is the characteristic function associated to the interval $A_j$ for $j \in \{1,2\}$,
that is equal to $1$ for $u \in A_j$ and vanishes identically for $u \notin A_j$,
and we have introduced
\be
\label{xi-k-def}
\xi_k(\tau,u) 
\,\equiv \,
w_k^{-1}\big(w(u) + \tau\big)
\;\;\qquad\;\;
u \in A_k
\;\;\;\;\qquad\;\;\;\;
k \in \big\{1,2 \big\}
\ee
in terms of the two solutions $w_k^{-1}(y) $ obtained by inverting (\ref{w_fund_2int}).
Indeed, the function $w(u)$ in (\ref{w_fund_2int}) can be inverted if $u\in A_k$ 
(hence $w_k^{-1}(w(u)) = u$ when $u \in A_k$)
for $k \in \{1,2\}$, finding 
\bea
\label{wk-inverse}
w_k^{-1}(y)  & \equiv & \frac{a_1 + a_2 +(b_1 + b_2)\, \e^{2\pi y} }{2(1+ \e^{2\pi y} )}
\\
\rule{0pt}{1cm}
& & 
+ \, (-1)^k\,
\frac{ \sqrt{\big[ a_1 + a_2 +(b_1 + b_2)\,  \e^{2\pi y}\big]^2 -4 (1+ \e^{2\pi y} ) \,(a_1  a_2 + b_1 b_2\, \e^{2\pi y} )} }{ 2( 1 + \e^{2\pi y} ) }
\;\qquad\;
y \in \RR
\nonumber
\eea
which is well defined because
the expression under the square root is quadratic in the variable $ \e^{2\pi y}$ 
with discriminant equal to $-16\, \ell_1 \ell_2\, d\, (\ell_1 + d + \ell_2)<0$.
The condition $u \in A_k$ in (\ref{xi-k-def}) comes from the initial condition $\xi_k(\tau=0,u) = u$
because $w_k^{-1}(w(u)) = u$ when $u \in A_k$.
In Fig.\,\ref{fig:2int-line-xi-v0}, where the grey strips correspond to $A$ in the plane $(\xi, \tau)$,
the solid red curve is obtained from (\ref{xi-2int-A1-A2}) with $u \in A_1$ (red dot),
while the dashed red curve is given by (\ref{xi-2int-A1-A2}) again with initial point $u_{\textrm{c}}  \in A_2$ 
(red empty circle).

The PDE's defining the functions in (\ref{xi-k-def}) are 
\be
\label{der-xi-2int}
\partial_\tau \xi_k(\tau,u) = V_{\textrm{\tiny loc}}(\xi_k)
\;\;\;\qquad\;\;\;
\partial_u \xi_k(\tau,u) = \frac{V_{\textrm{\tiny loc}}(\xi_k)}{V_{\textrm{\tiny loc}}(u)}
\;\;\;\;\qquad\;\;\;\;
k \in \big\{1,2 \big\}\,.
\ee

The two functions in (\ref{xi-k-def}) having different value of $k$
are related through the following conjugation relation
\be
\label{xi-k-uc}
\xi_k(\tau,u) 
\,=\,
q_0 - \frac{r_0^2}{\xi_j(\tau,u_{\textrm{c}} )  - q_0}
\,=\,
\mathcal{C} \big( \xi_j (\tau, u_{\textrm{c}} ) \big)
\;\;\qquad\;\;
u \in A_k
\;\;\qquad\;\;
k \neq j
\ee
where $j, k \in \{1,2\}$.
Notice that the first equality of (\ref{xi-k-uc})
provides a generalisation to $\tau \neq 0$ of the conjugation relation (\ref{u-conj-def}), 
which is recovered at $\tau=0$.
Moreover, (\ref{xi-k-uc}) implies that the conjugation (\ref{u-conj-def}) and the modular evolution commute, namely
\be
\label{conj-xi-k}
\mathcal{C}\big(\xi_k(\tau,u) \big)
\,=\,
 \xi_j (\tau, u_{\textrm{c}} ) 
 \;\;\qquad\;\;
u \in A_k
\;\;\qquad\;\;
k \neq j \,.
\ee

We remark that, since  (\ref{xi-2int-A1-A2}) implies that $\xi(\tau,u) \in A_k$ when $u \in A_k$ for $k\in \{1,2\}$,
we have $\tilde{\eta}(\xi, u) > 0$ in (\ref{psi-flow-text}).
Instead, the sign of $ \tilde{\eta}(\xi_{\textrm{c}} , u)$ in (\ref{psi-flow-text}) is not well defined for any $\tau \in \RR$.


The domain $B \equiv \RR \setminus A$, complementary to $A$ on the line,
is partitioned as $B = B_1 \cup B_2$,
where $B_1 \equiv (-\infty , a_1) \cup (b_2, +\infty)$ and $B_2 \equiv (a_1, b_2)$.
For  $u\in B$,
the function $\xi = \xi(\tau, u)$ characterising the geometric action of the modular group of automorphisms induced by the vacuum 
 can be written  as follows
\be
\label{xi-2int-B1-B2}
\xi(\tau,u)  \,\equiv \, \Theta_{B_1}  (u) \, \xi_1(\tau,u) + \Theta_{B_2} (u) \, \xi_2(\tau,u)
\hspace{2cm}
u \in B
\ee
in terms of the expressions reported in (\ref{xi-k-def}), whose domain of validity can be extended. 
Since $u\in B$, 
a divergence occurs in (\ref{xi-2int-B1-B2}) when $1 + \e^{2\pi [w(u) +\tau]} = 0$
(see the denominator in (\ref{wk-inverse})),
whose solution is
\be
\label{tauB_def_2int}
\tau_{B}(u)
\equiv 
\frac{1}{2\pi} \,
\log \! \left(\frac{(u-b_1)(u-b_2)}{(a_1-u)(a_2-u)}\right)  
\;\;\;\;\qquad\;\;\;\;
u \in B\,.
\ee
Since this expression becomes (\ref{tauB_def}) in the limit $b_1 \to a_2$ of adjacent intervals, as expected,
considerations similar to the ones reported in Sec.\,\ref{sec-1int-chiral-dir-mod-evo} apply also in this setup. 
In Fig.\,\ref{fig:2int-line-xi-v0},
where the union of the light blue regions corresponds to $B$ in the plane $(\xi, \tau)$,
the solid blue curve is given by (\ref{xi-2int-B1-B2}) with $u \in B_2$ (blue dot).
The corresponding dashed blue curve is obtained from (\ref{xi-2int-B1-B2}) 
with initial point $u_{\textrm{c}} \in B_1$ (blue empty circle)
and the associated horizontal blue dot-dashed line 
comes from (\ref{tauB_def_2int}) specialised to $u_{\textrm{c}}  \in B_1$.
As for the limiting case $d \to 0$ of (\ref{xi-2int-A1-A2}) and (\ref{xi-2int-B1-B2}),
we refer the reader to Fig.\,10 of \cite{Mintchev:2022fcp}.

\subsection{Chiral modular correlators of the fermionic field and of the density }
\label{sec-2int-mod-corr}


The two-point function of the modular flow $\psi_\pm (\tau , u)$ 
of the chiral fermionic fields $\psi_\pm$ (discussed in  Sec.\,\ref{sec-mod-flow-psi-2int})
satisfying the KMS condition reads \cite{Longo:2009mn}
\be
\label{mod-corr-psi-2int-pm}
\langle \,\psi_{\pm}^\ast (\tau_1 , u_1) \,\psi_{\pm} (\tau_2 , u_2) \,\rangle 
=
\frac{ \e^{2\pi w(u_1)} - \e^{2\pi w(u_2)}  }{ 2\pi (\pm \ri)\,(u_1 - u_2)  }\;
\frac{1}{  \e^{2\pi w(u_1) \pm \pi \tau_{12}} - \e^{2\pi w(u_2) \mp  \pi \tau_{12}} \mp \ri \varepsilon}  
\ee
in terms of (\ref{w_fund_2int}) 
(see Appendix\;\ref{app-mod-corr-psi} for the derivation of (\ref{mod-corr-psi-2int-pm}) 
and (\ref{2pt-chiral-primaries}) for its normalisation),
which corresponds to (\ref{mod-corr-phi-mu}) specialised to $h_\pm =1/2$ at formal level;
indeed, the function $w(u)$ is given by  (\ref{w_fund_2int}) in this case.


We find it worth exploring also the modular two-point function 
of the chiral bosonic density operators 
$\varrho_\pm ( u) \equiv \;  :\! \psi_\pm^\ast \psi_\pm \!: \! (u) $,
which are hermitian operators with chiral dimension $h_{\pm} =1$
providing the charge and helicity densities of the massless Dirac field. 
They have been studied in Appendix\;\ref{app-mod-flow-density}, finding 
\be
\label{mod-corr-rho-2int-pm}
\langle \,\varrho_\pm (\tau_1 , u_1) \,\varrho_\pm (\tau_2 , u_2) \,\rangle 
=
 - \frac{ \big( \e^{2\pi w(u_1)} - \e^{2\pi w(u_2)} \big)^2 }{ 4\pi^2 \,(u_1 - u_2)^2  }\;
\frac{1}{  \big( \e^{2\pi w(u_1) \pm \pi \tau_{12}} - \e^{2\pi w(u_2) \mp  \pi \tau_{12}} \mp \ri \varepsilon \big)^2}  
\ee
which satisfy the KMS condition.
This modular correlator is proportional to 
(\ref{mod-corr-phi-mu}) specialised to $h_\pm =1$ only at formal level 
because, like in (\ref{mod-corr-psi-2int-pm}), the function $w(u)$ is  (\ref{w_fund_2int}).


By adapting the analysis performed in Sec.\,\ref{sec-chiral-correlators}
(see (\ref{2pt-chiral-primaries})-(\ref{R-fact-def})) to this setup, 
we observe that the modular correlators (\ref{mod-corr-psi-2int-pm}) and (\ref{mod-corr-rho-2int-pm})
naturally suggest to introduce the function 
\be
\label{R-fact-def-2int}
\mathcal{R}(\tau ; u_1, u_2) \equiv 
\frac{  \e^{2\pi w(u_1) +\pi \tau} - \e^{2\pi w(u_2) - \pi \tau} }{ \e^{2\pi w(u_1)} - \e^{2\pi w(u_2)}  }
\ee
and
\be
\label{W-function-def-2int}
W(\tau ; u_1, u_2) 
\equiv  \frac{1}{(u_1 - u_2)\; \mathcal{R}(\tau ; u_1, u_2) }
\ee
where $w(u)$ is (\ref{w_fund_2int});
hence (\ref{R-fact-def-2int}) and (\ref{W-function-def-2int}) are just formally equal to (\ref{R-fact-def}) and (\ref{W-function-def}) respectively. 
However, like (\ref{R-fact-def}), also (\ref{R-fact-def-2int}) is not jointly continuous because
the limits $\tau \to 0$ and $u_2 \to u_1$ do not commute.

Given $u_1 \neq u_2$ in $A$,
for (\ref{R-fact-def-2int}) we have $\mathcal{R}(\tilde{\tau}_0 ; u_1, u_2) = 0$ 
at the finite value of $\tau$ given by 
\be
\label{tau0tilde-def-2int}
\tilde{\tau}_0(u_1, u_2) \equiv w(u_2) - w(u_1)
\ee
which is just formally equal to (\ref{tau0tilde-def})
because $w(u)$ is (\ref{w_fund_2int}), as above. 
The special value of $\tau$ introduced in (\ref{tau0tilde-def-2int})
provides an interesting property of $\xi(\tau, u)$ in (\ref{xi-2int-A1-A2});
indeed, for $u_1$ and $u_2$ in the same interval we have 
  \be
 \label{tau0def-2int-u1u2-same-int}
 u_2 = \xi(\tilde{\tau}_0,u_1) 
 \;\;\;\qquad\;\;\;
  u_1 = \xi(-\tilde{\tau}_0,u_2) 
   \;\;\;\;\;\qquad\;\;\;\;\;
   u_1, u_2 \in A_k
 \ee
while for $u_1$ and $u_2$ in different intervals we find 
  \be
 \label{tau0def-2int-u1u2-diff-int}
 u_2 = \xi(\tilde{\tau}_0, u_{1,\textrm{c}}) 
 \;\;\;\qquad\;\;\;
  u_1 = \xi(-\tilde{\tau}_0,u_{2,\textrm{c}}) 
     \;\;\;\;\;\qquad\;\;\;\;\;
   u_1 \in A_k
    \qquad 
       u_2 \in A_j
       \qquad 
       j \neq k
 \ee
 where $j, k \in \{1,2\}$ label the two intervals in $A = A_1 \cup A_2$.

For the expressions introduced in (\ref{xi-k-def}) and (\ref{R-fact-def-2int})
we also find that 
\be
\label{R-zero-xi-2int}
\mathcal{R}\big(\tau ; u, \xi_k (\tau,u)  \big) = 0
\;\;\;\;\qquad\;\;\;
k \in \big\{ 1,2  \big\}
\ee
which will be employed in Sec.\,\ref{sec-anti-comm-dirac} 
(see (\ref{argument-delta-v0})).

Let us conclude this discussion by highlighting  that the limit $u_2 \to u_{1, \textrm{c}}$ of (\ref{R-fact-def-2int}) gives
\be
\label{R-u2-to-u1c}
\mathcal{R}(\tau ; u, u_{\textrm{c}} + \epsilon) 
=
- \frac{ \sinh(\pi \tau) \; \mathcal{C}'(u) \, V_{\textrm{\tiny loc}}(u)  }{\pi \, \epsilon} 
+ 
O(1)
\;\;\;\;\qquad\;\;\;
\tau \neq 0
\ee
which can be slightly simplified through  the first relation in (\ref{V-uc-rels})
and will be used  in Sec.\,\ref{sec-2int-distance-chiral}.

\subsection{Anti-commutator of the field and commutator of the density}
\label{sec-anti-comm-dirac}


The anti-commutator of the fermionic chiral field $\psi_{\pm} (\tau , u) $ along the modular flow
provides useful insights about the causal properties of this modular evolution.
Since this quantity is a complex number, 
it can be evaluated by employing the modular correlator (\ref{mod-corr-psi-2int-pm})
as follows
\bea
& & \hspace{-.5cm}
\big[ \, \psi_{\pm}^\ast (\tau_1 , u_1) \, , \, \psi_{\pm} (\tau_2 , u_2)  \,\big]_+
=\,
\langle \,\psi_{\pm}^\ast (\tau_1 , u_1) \,\psi_{\pm} (\tau_2 , u_2) \,\rangle 
+
\langle \,\psi_{\pm} (\tau_2 , u_2)\, \psi_{\pm}^\ast (\tau_1 , u_1)  \,\rangle 
\\
\rule{0pt}{.9cm}
& & \hspace{3.cm}
=\,
\frac{ \e^{2\pi w(u_1)} - \e^{2\pi w(u_2)}  }{ 2\pi (\pm \ri)\,(u_1 - u_2)  }\;
\left\{\,
\frac{1}{  \e^{2\pi w(u_1) \pm \pi \tau_{12}} - \e^{2\pi w(u_2) \mp  \pi \tau_{12}} \mp \ri \varepsilon}  
\right.
\\
\rule{0pt}{.7cm}
& & \hspace{7.5cm}
\left.
- \,\frac{1}{  \e^{2\pi w(u_1) \pm \pi \tau_{12}} - \e^{2\pi w(u_2) \mp  \pi \tau_{12}} \pm \ri \varepsilon}  
\right\}
\nn
\\
\label{comm-psi-v3}
\rule{0pt}{.7cm}
& & \hspace{3.cm}
=\,\frac{ \e^{2\pi w(u_1)} - \e^{2\pi w(u_2)}  }{ u_1 - u_2  }\;\,
\delta\big(  \e^{2\pi w(u_1) \pm \pi \tau_{12}} - \e^{2\pi w(u_2) \mp  \pi \tau_{12} }  \big)
\eea 
where the last expression has been obtained by taking the limit $\varepsilon \to 0^+$ in the sense of the distributions
and by using $\frac{1}{u\pm \ri \varepsilon} = \frac{1}{u} \mp \ri \pi \delta (u) $.

It is well known that the standard time evolution of the chiral fields $\psi_\pm$ respects relativistic causality, 
which implies that $\psi^\ast_\pm(u_1)$ and $\psi_\pm(u_2)$ anti-commute for $u_1 \not=u_2$. 
From (\ref{comm-psi-v3}) we 
conclude that this property holds also for the fields $\psi^\ast_\pm(\tau_1,u_1)$ and 
$\psi_\pm(\tau_2,u_2)$, except in the case when 
the relative modular time $\tau_{12}$ satisfies 
\be
\label{argument-delta}
 \e^{2\pi w(u_1) \pm \pi \tau_{12}} - \e^{2\pi w(u_2) \mp  \pi \tau_{12}} =0 \,.
\ee 
From (\ref{tau0tilde-def-2int}), the solution of this equation reads
\be 
\label{argument-delta-bis}
\tau_{12} = \pm \,\tilde{\tau}_0(u_1, u_2) \,.
\ee 
When this condition is fulfilled, the delta function in the r.h.s. of (\ref{comm-psi-v3}) provides a non vanishing contribution.
For $u_1$ and $u_2$ belonging to different components of $A_1 \cup A_2$, 
such a contribution signals a violation of relativistic causality along the modular flow. 
This violation is generated by the 
bilocal term (\ref{K_A-2int-terms-bi-local}) in the modular Hamiltonian, which implies that 
the modular evolution (\ref{psi-flow-text}) of $\psi_\pm$ is a superposition 
of the initial value of the field along the trajectory $\xi$ and its conjugate $\xi_{\textrm{c}} $,
that live in different intervals. 
In order to clarify this aspect it is useful to rewrite (\ref{comm-psi-v3}) in a form 
that involve $\xi$ and $\xi_{\textrm{c}} $ explicitly. 
For this purpose, we observe that, 
from (\ref{w_fund_2int}), (\ref{xi-k-def}) and (\ref{w-uc-w-u}), 
for $k \in \{1, 2\}$ one obtains
\be
\label{argument-delta-v0}
\e^{2\pi w(u) + \pi \tau } - \e^{2\pi w (\xi_k (\tau ,u) ) -  \pi \tau } 
=\,
\e^{2\pi w(\xi_k (-\tau ,u)) + \pi \tau } - \e^{2\pi w ( u ) -  \pi \tau } 
=\, 0
\;\;\qquad\;\;
u \in A_k\,.
\ee

Combining (\ref{w-uc-w-u}) and (\ref{argument-delta-v0}),
we find  that (\ref{argument-delta}) is fulfilled when
\be
\label{argument-delta-zeros}
u_2 = \xi_k (\pm \tau_{12},u_1) 
\;\;\;\qquad\;\;\;
u_2 = \mathcal{C} \big(  \xi_k (\pm \tau_{12},u_1) \big)
\;\;\;\;\;\qquad\;\;\;\;\;
u_1 \in A_k
\ee
in terms of $u_1$, or, equivalently, in terms of $u_2$, when
\be
\label{argument-delta-zeros-bis}
u_1 = \xi_j (\pm \tau_{21},u_2) 
\;\;\;\qquad\;\;\;
u_1 = \mathcal{C} \big(  \xi_j (\pm \tau_{21},u_2) \big)
\;\;\;\;\;\qquad\;\;\;\;\;
u_2 \in A_j\,.
\ee
It is straightforward to check e.g. that when both $u_1$ and $u_2$ belong to $A_k$
(hence $j=k$ in (\ref{argument-delta-zeros-bis})),
the first expression in (\ref{argument-delta-zeros-bis}) is equivalent to the first expression in (\ref{argument-delta-zeros}).
Thus, all the possibilities have been considered because
either both $u_1$ and $u_2$ belong to the same interval $A_k$ 
or they are in different intervals.

The limit of large separation distance for the two intervals can be explored by 
first setting $a_1 = -d/2 -\ell_1$, $b_1 = -d/2$, $a_2 = d/2 $, $b_2 = d/2 + \ell_2 $
and then sending $d \to +\infty$.
In this limiting regime, when $u_1 \neq u_2$ the ratio multiplying the Dirac delta in (\ref{comm-psi-v3}) falls off as follows 
\be
\frac{ \e^{2\pi w(u_1)} - \e^{2\pi w(u_2)}  }{ u_1 - u_2  } = \frac{4(\ell_1 +\ell_2)}{d^2} +O(1/d^3) \,.
\ee
The same power law decay is observed for (\ref{argument-delta-bis});
indeed, in this limiting regime we find that $\tau_{12} = -2(\ell_1 +\ell_2)(u_1 - u_2)/d^2 +O(1/d^3)$ as $d \to +\infty$.

The r.h.s. of (\ref{comm-psi-v3}) can be written more explicitly
by using for  its Dirac delta the well known identity
$\delta\big(f(x)\big) = \sum_j \tfrac{1}{ | f'(x_j) | } \,\delta(x-x_j)$,
where the sum is over the zeros $x_j$ of $f(x)$ such that $f'(x_j)  \neq 0$.
From this formula, the first relation in  (\ref{velocity_fund-2int}) and the fact that $ V_{\textrm{\tiny loc}}(u) > 0$ when $u \in A$,
assuming $u_1 \in A_k$ without loss of generality we find that (\ref{comm-psi-v3}) becomes
\bea
\label{comm-psi-T0}
\big[ \, \psi_{\pm}^\ast (\tau_1 , u_1) \, , \, \psi_{\pm} (\tau_2 , u_2)  \,\big]_+
& = &
\mathcal{F}\big(u_1 , \xi_k (\pm \tau_{12},u_1)  ; \pm \tau_{12} \big) \;
\delta\big( u_2 - \xi_k (\pm \tau_{12},u_1)  \big)
\\
\rule{0pt}{.5cm}
& &
+\;
\mathcal{F}\big(u_1 ,  \mathcal{C} ( \xi_k (\pm \tau_{12},u_1) )  ; \pm \tau_{12} \big) 
\; \delta\big( u_2 - \mathcal{C} ( \xi_k (\pm \tau_{12},u_1)  ) \big)
\nn
\eea
where we have introduced the following function 
\be
\label{cal-F-def-2int}
\mathcal{F}(u,v; \tau)
\equiv
\frac{  \e^{2\pi w(u)} -  \e^{2\pi w(v)}  }{  u - v }\;
\frac{ V_{\textrm{\tiny loc}}( v )  }{2\pi \, \e^{2\pi w( v ) -  \pi \tau } }
\ee
and the second term in the r.h.s. of (\ref{comm-psi-T0}) corresponds to $ u_{2,\textrm{c}} = \xi_k (\pm \tau_{12},u_1)$.
From (\ref{xi-k-def}) in (\ref{cal-F-def-2int}), we find 
\be
\label{calF-on-xi}
\mathcal{F}\big(u_1,\xi_k (\pm \tau_{12},u_1) ; \pm \tau_{12} \big)
=
\frac{ \sinh(\pm \pi \tau_{12}) \; V_{\textrm{\tiny loc}}( \xi_k (\pm \tau_{12},u_1) )  }{ \pi \big[  \xi_k (\pm \tau_{12},u_1) - u_1 \big] }
\ee
which can be used also for $\mathcal{F}(u_1, \mathcal{C} ( \xi_k (\pm \tau_{12},u_1)) ; \pm \tau_{12} )$ in (\ref{comm-psi-T0}).
Thus, (\ref{comm-psi-T0}) becomes
\bea
\label{comm-psi-T1}
\big[ \, \psi_{\pm}^\ast (\tau_1 , u_1) \, , \, \psi_{\pm} (\tau_2 , u_2)  \,\big]_+
& = &
\mathcal{F}\big(u_1,\xi_k (\pm \tau_{12},u_1) ; \pm \tau_{12} \big)
\; \delta\big( u_2 - \xi_k (\pm \tau_{12},u_1)  \big)
\\
\rule{0pt}{.5cm}
& & \hspace{-0.cm}
 +\;
 \mathcal{F}\big(u_1, \mathcal{C} ( \xi_k (\pm \tau_{12},u_1) ) ; \pm \tau_{12} \big)
\; \delta\big( u_2 - \mathcal{C} ( \xi_k (\pm \tau_{12},u_1)  ) \big)
\nn
\\
\label{comm-psi-T1-bis}
\rule{0pt}{1.cm}
& & \hspace{-3cm}
=\;
\frac{ V_{\textrm{\tiny loc}}( \xi_k (\pm \tau_{12},u_1) )}{\pi} \;
\bigg\{\,
\frac{  \sinh(\pm \pi \tau_{12})   }{ \xi_k (\pm \tau_{12},u_1) - u_1 }
\; \delta\big( u_2 - \xi_k (\pm \tau_{12},u_1)  \big)
\\
\rule{0pt}{.8cm}
& & \hspace{1.cm}
 +\;
\frac{  \sinh(\pm \pi \tau_{12}) \; \mathcal{C}' ( \xi_k (\pm \tau_{12},u_1) )    }{   \mathcal{C} \big(  \xi_k (\pm \tau_{12},u_1) \big) - u_1  }
\; \delta\big( u_2 - \mathcal{C} ( \xi_k (\pm \tau_{12},u_1)  ) \big)
\bigg\}
\nn
\eea
where the last expression is obtained by applying the first relation in (\ref{V-uc-rels}).
Finally, since (\ref{xi-k-uc}) holds, the anti-commutator (\ref{comm-psi-T1}) can be written as follows
\bea
\label{comm-psi-T2}
\big[ \, \psi_{\pm}^\ast (\tau_1 , u_1) \, , \, \psi_{\pm} (\tau_2 , u_2)  \,\big]_+
& = &
\frac{ \sinh(\pm \pi \tau_{12}) \; V_{\textrm{\tiny loc}}( \xi_k (\pm \tau_{12},u_1) )  }{ \pi \big[ \, \xi_k (\pm \tau_{12},u_1) - u_1 \big] } 
\; \delta\big( u_2 - \xi_k (\pm \tau_{12},u_1)  \big)
\\
\rule{0pt}{.8cm}
& & 
\hspace{-0.cm}
 +\,
\frac{ \sinh(\pm \pi \tau_{12}) \; V_{\textrm{\tiny loc}} ( \xi_j (\pm \tau_{12},u_{1, \textrm{c}}) )  }{  \pi \big[ \, \xi_j (\pm \tau_{12},u_{1, \textrm{c}})  - u_1 \big]}
\; \delta\big( u_2 -  \xi_j (\pm \tau_{12},u_{1, \textrm{c}})   \big)
\hspace{.7cm}
\nn
\eea
where $j \neq k$ and we remind that $u_1 \in A_k$.
We remark that  only one Dirac delta occurs in the r.h.s. of (\ref{comm-psi-T2});
indeed, either its first or its second term is non vanishing,
 depending on whether $u_2 \in A_k$ or $u_2 \notin  A_k$
 respectively.

It is important to study the anti-commutator (\ref{comm-psi-T2})  when $\tau_{1} = \tau_{2}  \equiv \tau$.
Since for $u_1 \in A_k$  and $u_{1, \textrm{c}} \in A_j$ with $j \neq k$,
the second term in the r.h.s of (\ref{comm-psi-T2}) vanishes identically in this limit
because $ \mathcal{F} (u_1, \mathcal{C} ( \xi_k (\pm \tau_{12},u_1) ) ; \pm \tau_{12} ) \to 0$ as $\tau_{12} \to 0$.
As for the first term in the r.h.s of (\ref{comm-psi-T2}),
since $\xi_k(\tau,u) = u + V_{\textrm{\tiny loc}}( u)\,\tau +O(\tau^2)$ as $\tau \to 0$
(see the first relation in  (\ref{der-xi-2int})), 
we have that 
$ \mathcal{F} (u_1, \xi_k (\pm \tau_{12},u_1)  ; \pm \tau_{12} ) \to 1$ as $\tau_{12} \to 0$.
Thus, (\ref{comm-psi-T2}) provides the expected result for  $\tau_{1} = \tau_{2}  \equiv \tau$,
namely
\be
\label{anti-comm-psi-zero}
\big[ \, \psi_{\pm}^\ast (\tau , u_1) \, , \, \psi_{\pm} (\tau , u_2)  \,\big]_+
=\, \delta(u_1 - u_2)\,.
\ee


While the l.h.s. of (\ref{comm-psi-T2}) is manifestly symmetric under the exchange $(u_1, \tau_1) \leftrightarrow (u_2, \tau_2)$,
in the r.h.s. this feature is not manifest.
In order to make this symmetry evident also in the r.h.s. of (\ref{comm-psi-T2}),
one first observes that the argument $F_{1,2}$ of the Dirac delta $\delta(F_{1,2})$ in (\ref{comm-psi-v3}) is antisymmetric under $(u_1, \tau_1) \leftrightarrow (u_2, \tau_2)$,
namely $F_{1,2} = - F_{2,1}$.
Thus, by writing $\delta(F_{1,2}) =\tfrac{1}{2} [\delta(F_{1,2})+ \delta(F_{2,1})]$ and repeating the steps described above, 
the anti-commutator (\ref{comm-psi-T2}) can be expressed as follows
\bea
\label{comm-psi-T2-symm}
\big[ \, \psi_{\pm}^\ast (\tau_1 , u_1) \, , \, \psi_{\pm} (\tau_2 , u_2)  \,\big]_+
& = &
\frac{ \sinh(\pm \pi \tau_{12})  }{ 2\pi  }
\\
\rule{0pt}{.8cm}
& &
\hspace{-4.8cm}
\times\,
\Bigg\{\,
\frac{ V_{\textrm{\tiny loc}}( \xi_k (\pm \tau_{12},u_1) ) }{  \xi_k (\pm \tau_{12},u_1) - u_1 }
\; \delta\big( u_2 - \xi_k (\pm \tau_{12},u_1)  \big)
+
\frac{ V_{\textrm{\tiny loc}} ( \xi_j (\pm \tau_{12},u_{1, \textrm{c}}) )   }{   \xi_j (\pm \tau_{12},u_{1, \textrm{c}})  - u_1 }
\; \delta\big( u_2 -  \xi_j (\pm \tau_{12},u_{1, \textrm{c}})   \big)
\nn
\\
\rule{0pt}{.9cm}
& &
\hspace{-4.2cm}
-\,
\frac{ V_{\textrm{\tiny loc}}( \xi_m (\pm \tau_{21},u_2) ) }{  \xi_m (\pm \tau_{21},u_2) - u_2 }
\; \delta\big( u_1 - \xi_m (\pm \tau_{21},u_2)  \big)
-
\frac{ V_{\textrm{\tiny loc}} ( \xi_n (\pm \tau_{21},u_{2, \textrm{c}}) )   }{   \xi_n (\pm \tau_{21},u_{2, \textrm{c}})  - u_2 }
\; \delta\big( u_1 -  \xi_n (\pm \tau_{21},u_{2, \textrm{c}})   \big)
\Bigg\}
\nn
\eea
where $u_1 \in A_k$ and $u_2 \in A_m$ with $j \neq k$  and $n \neq m$.
In (\ref{comm-psi-T2-symm}) the symmetry under $(u_1, \tau_1) \leftrightarrow (u_2, \tau_2)$ 
is evident in both sides.


In order to check the consistency between (\ref{comm-psi-T2}) and (\ref{psi-flow-text}),
let us first set $\tau_2 = 0$ in the l.h.s. of (\ref{comm-psi-T2}), 
where $u_1 \in A_k$ for a given $k \in \{1,2\}$,
and then employ the explicit expression (\ref{psi-flow-text}) for the modular flow.
Consistency between the resulting anti-commutator 
and the r.h.s. of (\ref{comm-psi-T2}) specified to $\tau_2=0$
requires the following relations (where $j \neq k$)
\be
\label{consistency-modflow-psi-anticomm}
 \frac{ \sqrt{\partial_u \xi_k }}{ \tilde{\eta}(\xi_k , u)} 
\,=\,
\frac{ \sinh(\pi \tau) \; V_{\textrm{\tiny loc}}( \xi_k (\tau,u) )  }{ \pi \big[ \, \xi_k (\tau,u) - u \big] }
\;\;\qquad\;\;
 \frac{ \sqrt{ \partial_u \xi_{k,\textrm{c}} } }{ \tilde{\eta}(\xi_{k,\textrm{c}} , u)} 
\,=\,-
\frac{ \sinh(\pi \tau) \; V_{\textrm{\tiny loc}} ( \xi_j (\tau ,u_{\textrm{c}}) )  }{  \pi \big[ \, \xi_j (\tau,u_{\textrm{c}})  - u \big]} \,.
\ee

\begin{figure}[t!]
\vspace{-.2cm}
\hspace{-1.4cm}
\includegraphics[width=1.2\textwidth]{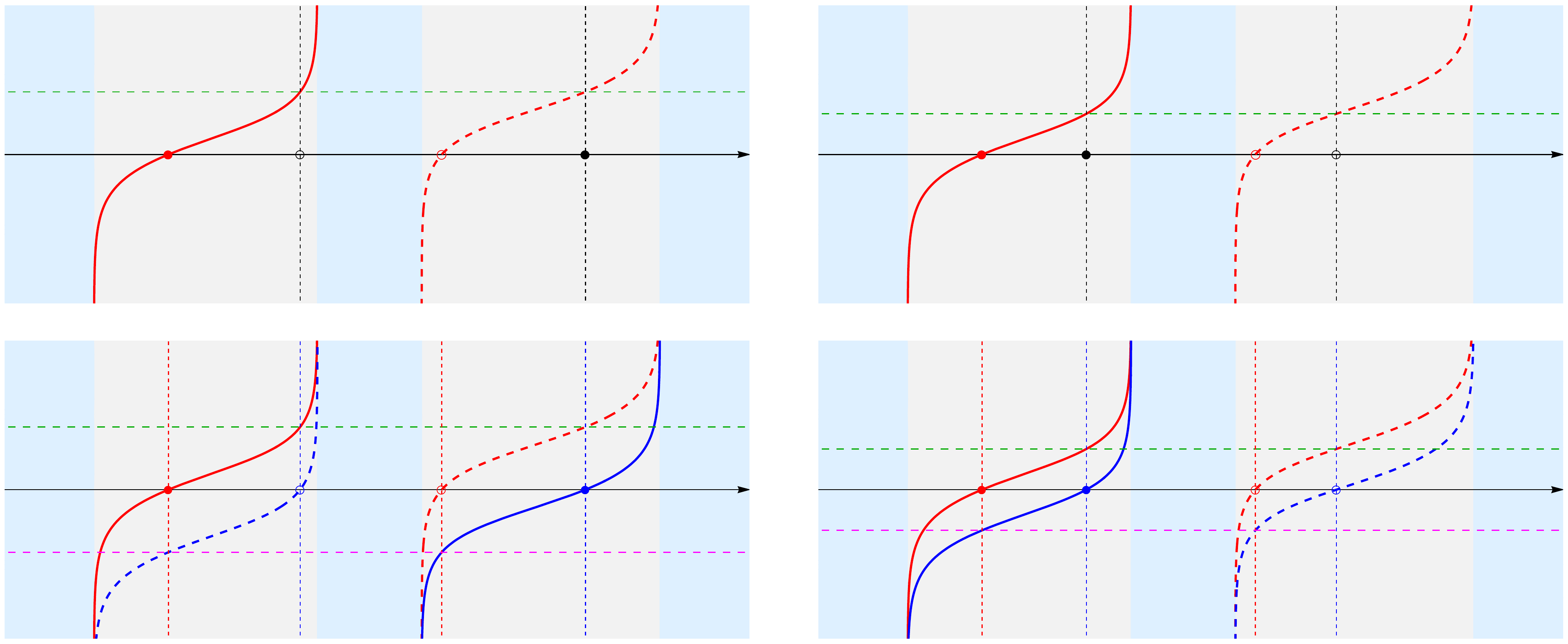}
\vspace{-.3cm}
\caption{Chiral modular evolutions illustrating the non vanishing terms
in the r.h.s. of the anti-commutator (\ref{comm-psi-T2}) 
and in the r.h.s. of the commutator (\ref{comm-rho-T2}).
In the top panels one point does not evolve ($\tau_2=0$, see the black dot),
while in the bottom panels both the initial points evolve.
In the right panels both the initial points belong to $A_1$,
while in the left panels they are in different intervals. 
}
\label{fig:2int-commutators}
\end{figure}

 
The above analysis about the anti-commutator of the chiral fermionic fields
can be straightforwardly adapted to the case 
where the bipartition of the line is given by a single interval (see Sec.\,\ref{sec-1int-line-vacuum}),
finding only the first term in the r.h.s. of (\ref{comm-psi-T2}),
with $\xi_k (\pm \tau_{12},u_1) $ and $V_{\textrm{\tiny loc}}( u_1 )$ 
replaced  respectively by  $\xi (\pm \tau_{12},u_1) $ in (\ref{xi-map-fund})
and  by $V( u_1 )$ in (\ref{velocity_fund}).


Considering e.g. the anti-commutator for $\psi_+$ in (\ref{comm-psi-T2}), 
in the top panels Fig.\,\ref{fig:2int-commutators}
we show an example for each one of the two possible configurations of initial points 
(see the red and black dot in each panel), 
where they belong 
either to the same interval (top right panel)
or to different intervals (top left panel).
At $\tau_1 = \tau_2 =0$, 
the red dot corresponds e.g. to $u_1 \in A_1$ (hence $k =1$ in the r.h.s. of (\ref{comm-psi-T2})) in both the top panels,
while $u_2$, denoted by the black dot, 
is either in $A_1$ (top right panel) or in $A_2$ (top left panel).
While $\tau_2 =0$ is kept fixed,
$\tau_1$ takes all the possible real values, providing the solid red curves.
In each panel of the top line, the empty circle indicates $u_{1,\textrm{c} } $ at $\tau_1 =0$
and the dashed curve describes the corresponding modular evolution. 
Thus, in the r.h.s. of the anti-commutator (\ref{comm-psi-T2}), 
where only one term occurs,
the first (second) term is non vanishing for the setup of the top right (left) panel.
In both the top panels of Fig.\,\ref{fig:2int-commutators}, 
the horizontal dashed green line denotes the value of $\tau_1=\tilde{\tau}_0(u_1, u_2) $ (see (\ref{tau0tilde-def-2int}))
where the argument of the Dirac delta occurring in the r.h.s. of the anti-commutator (\ref{comm-psi-T2}) vanishes.

In the  bottom panels of Fig.\,\ref{fig:2int-commutators}, generalising the case just considered in the corresponding top panels, 
we show an example where both the points $u_1$ and $u_2$ move along their chiral modular trajectories. 
In these bottom panels the horizontal dashed green and magenta straight lines
are at the same distance from the horizontal black line at $\tau = 0$, given by (\ref{argument-delta-bis}).


The above analysis for the anti-commutator of the chiral fermionic fields $\psi_\pm (\tau , u) $
can be adapted to explore 
the commutation relation of the chiral bosonic operators $\varrho_\pm (\tau , u) $.
The corresponding analysis in the case where the bipartition of the line is characterised by a single interval 
can be performed for a generic CFT and it has been reported in Sec.\,\ref{sec-1int-commutators}.

In the case that we are exploring throughout this section, 
by adapting (\ref{comm-j-curr-v1-1int})-(\ref{comm-j-curr-v3-1int}) to this case
and employing (\ref{2pt-mod-rho-app}) for the modular correlators, we find 
\bea
& & \hspace{-.8cm}
\big[ \, \varrho_{\pm} (\tau_1 , u_1) \, , \, \varrho_{\pm} (\tau_2 , u_2)  \,\big]_{-}
=\,
\langle \,\varrho_{\pm} (\tau_1 , u_1) \,\varrho_{\pm} (\tau_2 , u_2) \,\rangle 
-
\langle \,\varrho_{\pm} (\tau_2 , u_2)\, \varrho_{\pm} (\tau_1 , u_1)  \,\rangle 
\\
\rule{0pt}{.9cm}
& & \hspace{2.2cm}
=\,
\frac{ \big(\e^{2\pi w(u_1)} - \e^{2\pi w(u_2)} \big)^2 }{ 4\pi^2 \,(u_1 - u_2)^2  }\;
\left\{\,
\frac{1}{  \big( \e^{2\pi w(u_1) \pm \pi \tau_{12}} - \e^{2\pi w(u_2) \mp  \pi \tau_{12}} \mp \ri \varepsilon \big)^2}  
\right.
\nn
\\
\rule{0pt}{.7cm}
& & \hspace{7.cm}
\left.
- \,\frac{1}{ \big( \e^{2\pi w(u_1) \pm \pi \tau_{12}} - \e^{2\pi w(u_2) \mp  \pi \tau_{12}} \pm \ri \varepsilon \big)^2}  
\right\}
\nn
\eea
and 
\be
\label{comm-rho-v3}
\big[ \, \varrho_{\pm} (\tau_1 , u_1) \, , \, \varrho_{\pm} (\tau_2 , u_2)  \,\big]_{-}
=\,
\mp \, \frac{ \ri }{2\pi}\;
\frac{ \big(\e^{2\pi w(u_1)} - \e^{2\pi w(u_2)} \big)^2 }{ (u_1 - u_2)^2  }
\;\,
\delta'\big(  \e^{2\pi w(u_1) \pm \pi \tau_{12}} - \e^{2\pi w(u_2) \mp  \pi \tau_{12}}  \big)
\ee
(see also (\ref{comm-j-curr-v3-1int}) for $\kappa=1$),
where $w(u)$ is  (\ref{w_fund_2int}) 
and whose r.h.s. contains the derivative of the Dirac delta with respect to its argument.

The expression in (\ref{comm-rho-v3}) can be treated through  (\ref{delta-prime-res})
and in this case one gets  a sum made by two terms 
(see either (\ref{argument-delta-zeros}) or (\ref{argument-delta-zeros-bis})).
In particular, setting $u_1=u$, $u_2 =v$ and $\tau_{12} = \tau$ 
in the r.h.s. of (\ref{comm-rho-v3}) to enlighten the notation and
choosing $v$ as the variable playing the role of $x$ in (\ref{delta-prime-res}),
the ratio multiplying the square brackets in  (\ref{delta-prime-res})
becomes $  \pm  \tfrac{ \ri }{2\pi} \,\mathcal{F}(u,v; \tau)^2$, in terms of (\ref{cal-F-def-2int})
(notice that $f'(x_j) < 0$ in this calculation),
and for the expression within the round brackets multiplying the Dirac delta in (\ref{delta-prime-res})
we obtain
\bea
\label{cal-G-def-2int-v0}
\mathcal{G}(u,v)
& \equiv &
- \,\frac{w''(v) }{w'(v) }
-   \frac{  \e^{2\pi w(u)} + \e^{2\pi w(v)}  }{  \e^{2\pi w(u)} -  \e^{2\pi w(v)}  } \; 2\pi \,  w'(v)
+ \frac{2}{u-v} 
\\
\label{cal-G-def-2int-v1}
\rule{0pt}{.7cm}
&=&
\frac{V'_{\textrm{\tiny loc}}(v)  }{ V_{\textrm{\tiny loc}}(v)  }
-   \frac{  \e^{2\pi w(u)} + \e^{2\pi w(v)}  }{  \e^{2\pi w(u)} -  \e^{2\pi w(v)}  } \;  \frac{2\pi}{ V_{\textrm{\tiny loc}}(v) } 
+ \frac{2}{u-v} 
\\
\label{cal-G-def-2int}
\rule{0pt}{.7cm}
&=&
\frac{2(b_1 - a_1)(b_2 - a_2) (a_2 - b_1)(b_2 - a_1)}{ (b_1 - a_1 + b_2 - a_2)^2 \, (u- q_0)( v - q_0)\, (v - v_{\textrm{c}} ) } \;\,
\frac{u-v}{ v - u_{\textrm{c}} }
\eea
in terms of (\ref{w_fund_2int})-(\ref{u-conj-def}),
which is independent of $\tau$  and singular only at $v = u_{\textrm{c}}$ for any choice of 
$u,v \in A_1 \cup A_2$ because $q_0 \notin A$ (see (\ref{q0-def})).
When the line is bipartite by a single interval, 
by employing $w(u)$ in (\ref{w_fund}) into (\ref{cal-G-def-2int-v0}), one finds $\mathcal{G}(u,v) = 0$ identically,
as already highlighted in Sec.\,\ref{sec-1int-commutators}.
Consistently, this result can be found 
also by taking the limit $a_2 \to b_1$ in (\ref{cal-G-def-2int}).

As for the case where the bipartition of the line is given by the union of two disjoint intervals, 
by restoring the notation occurring in the r.h.s. of (\ref{comm-rho-v3}),
let us observe that the zeros of the argument of $\delta'$ in (\ref{comm-rho-v3})
have been already discussed above 
(see (\ref{argument-delta})-(\ref{argument-delta-zeros-bis})).
Combining these results, we find that 
(\ref{delta-prime-res}) allows to write (\ref{comm-rho-v3}) as follows
\bea
\label{comm-rho-final}
\big[ \, \varrho_{\pm} (\tau_1 , u_1) \, , \, \varrho_{\pm} (\tau_2 , u_2)  \,\big]_{-}
&=&
  \pm\, \frac{\ri}{2\pi} \;
  \bigg\{ \,
  \mathcal{F}\big(u_1,\xi_k (\pm \tau_{12},u_1) ; \pm \tau_{12} \big)^2
  \\
  & & \hspace{-.9cm}
  \times
  \Big[\,
 \delta' \big( u_2 - \xi_k (\pm \tau_{12},u_1)  \big)
 -
  \mathcal{G}\big(u_1,\xi_k (\pm \tau_{12},u_1)  \big)\,
  \delta \big( u_2 - \xi_k (\pm \tau_{12},u_1)  \big)
 \Big]
\nn
\\
\rule{0pt}{.6cm}
& & \hspace{-2.9cm}
+\; \mathcal{F}\big(u_1, \mathcal{C} ( \xi_k (\pm \tau_{12},u_1)) ; \pm \tau_{12} \big)^2
  \nn
  \\
  & & \hspace{-2.5cm}
  \times
  \Big[\,
 \delta' \big( u_2 - \mathcal{C} ( \xi_k (\pm \tau_{12},u_1) )  \big)
 -
  \mathcal{G}\big(u_1, \mathcal{C} ( \xi_k (\pm \tau_{12},u_1) )  \big)\,
  \delta \big( u_2 - \mathcal{C} ( \xi_k (\pm \tau_{12},u_1) )  \big)
 \Big]
 \bigg\}
\nn
\eea
in terms of (\ref{cal-F-def-2int}) and (\ref{cal-G-def-2int-v0}),
which can be expressed more explicitly by employing (\ref{calF-on-xi}) and
\be
\label{calG-on-xi}
  \mathcal{G}\big(u_1,\xi_k (\pm \tau_{12},u_1)  \big)
= 
 \frac{V'_{\textrm{\tiny loc}}\big( \xi_k (\pm \tau_{12},u_1) \big)  }{ V_{\textrm{\tiny loc}} \big( \xi_k (\pm \tau_{12},u_1)  \big)}
- \frac{2}{ \xi_k (\pm \tau_{12},u_1) - u_1 } 
\pm  \frac{2\pi}{ \tanh(\pi \tau_{12})\, V_{\textrm{\tiny loc}} \big( \xi_k (\pm \tau_{12},u_1)  \big) } 
\ee
that has been obtained from (\ref{cal-G-def-2int-v1})
through steps similar to the ones providing (\ref{calF-on-xi}).
Finally,  by using (\ref{xi-k-uc}),
we have that (\ref{comm-rho-final}) can be written as 
\bea
\label{comm-rho-T2}
\big[ \, \varrho_{\pm} (\tau_1 , u_1) \, , \, \varrho_{\pm} (\tau_2 , u_2)  \,\big]_{-}
&=&
  \pm\, \frac{\ri}{2\pi} \;
  \bigg\{ \,
  \mathcal{F}\big(u_1,\xi_k (\pm \tau_{12},u_1) ; \pm \tau_{12} \big)^2
  \\
  & & \hspace{-1.2cm}
  \times
  \Big[\,
 \delta' \big( u_2 - \xi_k (\pm \tau_{12},u_1)  \big)
 -
  \mathcal{G}\big(u_1,\xi_k (\pm \tau_{12},u_1)  \big)\,
  \delta \big( u_2 - \xi_k (\pm \tau_{12},u_1)  \big)
 \Big]
\nn
\\
\rule{0pt}{.6cm}
& & \hspace{-2.2cm}
+\; \mathcal{F}\big(u_1, \xi_j (\pm \tau_{12},u_{1, \textrm{c} }) ; \pm \tau_{12} \big)^2
  \nn
  \\
  & & \hspace{-1.8cm}
  \times
  \Big[\,
 \delta' \big( u_2 - \xi_j (\pm \tau_{12},u_{1, \textrm{c} } )   \big)
 -
  \mathcal{G}\big(u_1,  \xi_j (\pm \tau_{12},u_{1, \textrm{c} } )  \big)\,
  \delta \big( u_2 -  \xi_j (\pm \tau_{12},u_{1, \textrm{c} } )   \big)
 \Big]
 \bigg\}
\nn
\eea
where $j \neq k$, being $u_1 \in A_k$.
Similarly to (\ref{comm-psi-T2}), 
in the r.h.s. of (\ref{comm-rho-T2}) only 
two terms are non vanishing because
only one argument of the Dirac delta functions  vanishes
and the same considerations made above for Fig.\,\ref{fig:2int-commutators} 
can be repeated here.

In the limit $\tau_{1} = \tau_{2} \equiv \tau$, the r.h.s. of (\ref{comm-rho-T2}) drastically simplifies.
Indeed, 
by employing that $\xi_k(\tau,u) = u + V_{\textrm{\tiny loc}}( u)\,\tau +O(\tau^2)$ as $\tau \to 0$ when $u \in A_k$
(from the first relation in  (\ref{der-xi-2int})),
we have that
$\mathcal{G}(u_1, u_1) = 0$ and
$\mathcal{F}\big(u_1 ,u_{1, \textrm{c} } ; 0 \big) = 0$,
from (\ref{cal-G-def-2int}) and (\ref{calF-on-xi}) respectively.
Moreover, we remark that 
$\mathcal{F} (u_1, \xi_j (\tau,u_{1, \textrm{c} }) ;  \tau)^2\, \mathcal{G}(u_1,  \xi_k (\tau,u_{1, \textrm{c} } )  ) \to 0$ 
and 
$ \mathcal{F}\big(u_1,\xi_k (\tau,u_1) ; \tau ) \to 1$
as $\tau \to 0$.
Thus, the commutator (\ref{comm-rho-T2}) provides the expected result when $\tau_{1} = \tau_{2} \equiv \tau$, namely
\be
\label{rho-comm-tau12-zero}
\big[ \, \varrho_{\pm} (\tau , u_1) \, , \, \varrho_{\pm} (\tau , u_2)  \,\big]_{-}
=\, \mp\, \frac{\ri}{2\pi}\, \delta'(u_1 - u_2) \,.
\ee
which can be compared with (\ref{commutator-j-single-int-equal-tau}), 
that holds for the single interval case and a generic CFT.


The argument $F_{1,2}$ of the derivative of the Dirac delta $\delta'(F_{1,2})$ in (\ref{comm-rho-v3}) 
is the same argument of the Dirac delta in (\ref{comm-psi-v3}) 
and it is antisymmetric under exchange $(u_1, \tau_1) \leftrightarrow (u_2, \tau_2)$.
Since the derivative of the Dirac delta is antisymmetric, 
we can write $\delta'(F_{1,2}) =\tfrac{1}{2} [\delta'(F_{1,2}) - \delta'(F_{2,1})]$ and 
follow the steps discussed above (below (\ref{anti-comm-psi-zero})), 
finding a form for the r.h.s. of (\ref{comm-rho-T2}) where 
the antisymmetry under under exchange $(u_1, \tau_1) \leftrightarrow (u_2, \tau_2)$ of the commutator is manifest. 
Since this expression is lengthy and not very illuminating, it has not been reported here.

\subsection{Chiral distance along the modular evolutions  }
\label{sec-2int-distance-chiral}


The chiral distance $\xi(\tau_1 ,u_1)  - \xi(\tau_2 ,u_2) $
for any choice of $u_1$ and $u_2$ in $A = A_1 \cup A_2$
and any real value of $\tau_1$ and $\tau_2$
can be investigated by employing the expressions discussed in Sec.\,\ref{sec-mod-flow-psi-2int}
to describe the modular flow of the chiral fermions,
following the analysis for the single interval case 
performed in Sec.\,\ref{sec-chiral-distance-mod-evo}.

The crucial observation is that (\ref{tilde-eta-def}), (\ref{xi-2int-A1-A2}),
(\ref{R-fact-def-2int}) and (\ref{W-function-def-2int}) are related as follows
\be
\label{W-eta-tilde-2int}
W(\tau_{12} ; u_1, u_2) =  \frac{ \tilde{\eta}\big(  \xi(\tau_1,u_1), \xi(\tau_2,u_2)\big) }{ \tilde{\eta}(u_1, u_2) } \; 
\frac{ \sqrt{\partial_{u_1} \xi(\tau_1,u_1) \; \partial_{u_2} \xi(\tau_2,u_2)}  }{ \xi(\tau_1 ,u_1)  - \xi(\tau_2 ,u_2)  }
\ee
%
whose derivation has been reported in Appendix\;\ref{app-W-tilde-relation}.
This relation implies that
\bea
\label{xi12-tau-chiral-2int}
\xi(\tau_1 ,u_1)  - \xi(\tau_2 ,u_2) 
=  
& &
\\
\rule{0pt}{.9cm}
& & \hspace{-1.7cm}
=\,
 \frac{ \tilde{\eta}\big(  \xi(\tau_1,u_1), \xi(\tau_2,u_2)\big) }{ \tilde{\eta}(u_1, u_2) } \;
\mathcal{R}(\tau_{12} ; u_1, u_2) \;
\sqrt{\partial_{u_1} \xi(\tau_1,u_1) \; \partial_{u_2} \xi(\tau_2,u_2)} 
 \; (u_1 - u_2)
\nonumber
\eea
As consistency check, 
we consider the limit of adjacent intervals, finding that,
by using  (\ref{tilde-eta-vs-eta}) and the first relation in (\ref{eta-special-regimes-A}),
the expressions in (\ref{rel-xi-W-1int}) and (\ref{xi12-tau-chiral-2}) are recovered
from (\ref{W-eta-tilde-2int}) and (\ref{xi12-tau-chiral-2int}) respectively in this limiting regime.

Comparing (\ref{xi12-tau-chiral-2int}) with (\ref{xi12-tau-chiral-2}),
one observes that the main qualitative feature 
of the two intervals case with respect to the single interval one 
is the occurrence of the ratio involving $\tilde{\eta}$,
which makes evident the role of $u_{1,\textrm{c}}$ and $u_{2,\textrm{c}}$ 
(through (\ref{tilde-eta-vs-eta})), 
even when $u_1$ and $u_2$ belong to the same interval. 
Moreover, from (\ref{xi12-tau-chiral-2int}), notice that 
\be
\label{xi12-tau-chiral-2int-v2}
\frac{ \xi(\tau_1 ,u_1)  - \xi(\tau_2 ,u_2) }{  \tilde{\eta}\big(  \xi(\tau_1,u_1), \xi(\tau_2,u_2)\big)  }
\,=\,
\mathcal{R}(\tau_{12} ; u_1, u_2) \;
\sqrt{\partial_{u_1} \xi(\tau_1,u_1) \; \partial_{u_2} \xi(\tau_2,u_2)} 
 \; \frac{ u_1 - u_2 }{ \tilde{\eta}(u_1, u_2) }
\ee
hence, when  the l.h.s. of this relation is well defined,
its behaviour is formally similar to the one of  $\xi(\tau_1 ,u_1)  - \xi(\tau_2 ,u_2) $ 
for the bipartition of the line given by a single interval (see (\ref{xi12-tau-chiral-2})).

From (\ref{tilde-eta-def}), (\ref{xi-2int-A1-A2}) and (\ref{R-fact-def-2int})
it is straightforward to find  that, 
 since $\xi(\tau ,u) \in A_k$ when $u \in A_k$ with $k \in \{1,2\}$ for any $\tau$,
in the case where  $u_1 \neq u_2$ are in the same interval  (hence $u_2 \neq u_{1, \textrm{c}}$)
 we have that 
 $\tilde{\eta}(u_1, u_2)  > 0$ and $\tilde{\eta}\big(  \xi(\tau_1,u_1), \xi(\tau_2,u_2)\big)  >0$ 
 in (\ref{xi12-tau-chiral-2int}), for any value of $\tau_1$ and $\tau_2$.
 Hence, the chiral distance $\xi(\tau_1 ,u_1)  - \xi(\tau_2 ,u_2) $ 
 may have a different sign with respect to $u_1 - u_2$ (depending on $\mathcal{R}(\tau_{12} ; u_1, u_2) $)
and it vanishes for $\mathcal{R}(\tau_{12} ; u_1, u_2) =0$.

 Instead, when $u_1$ and $u_2$ are in different intervals,
with the additional assumption $u_2 \neq u_{1, \textrm{c}}$, 
by employing again that $\xi(\tau ,u) \in A_k$ when $u \in A_k$ for any $\tau$,
the sign of $\xi(\tau_1 ,u_1)  - \xi(\tau_2 ,u_2)$ 
coincides with the sign of $u_1 - u_2$ for any real values of $\tau_{1}$ and $\tau_2$
(see e.g. the red and blue solid lines in the bottom left panel of Fig.\,\ref{fig:2int-commutators})
and $\xi(\tau_1 ,u_1)  - \xi(\tau_2 ,u_2) $ never vanishes. 
For $u_1$ and $u_2$ in different intervals,
by adopting the shorthand notation $\xi_j \equiv \xi(\tau_j ,u_j) $ with $j \in \{1,2\}$, 
from (\ref{tilde-eta-def}) and (\ref{tilde-eta-vs-eta}) we obtain
\be
\label{sign-eta-eta-ratio}
\textrm{sign} \bigg( \frac{ \tilde{\eta}(\xi_1, \xi_2)  }{ \tilde{\eta}(u_1, u_2) } \bigg)
=\,
\textrm{sign} \bigg( \frac{  (u_1 -q_0) \, ( u_2 - q_0) + r_0^2  }{  (\xi_1 -q_0) \, ( \xi_2 - q_0) + r_0^2 }\bigg)
=\,
\textrm{sign} \bigg( \frac{  (u_1 -q_0)  \, \tilde{u}_{1,2}  }{  (\xi_1 -q_0) \, \tilde{\xi}_{1,2} }\bigg)
=\,
\textrm{sign} \bigg( \frac{   \tilde{u}_{1,2}  }{ \tilde{\xi}_{1,2} }\bigg)
\ee
where we have defined $u_2 = u_{1, \textrm{c}} + \tilde{u}_{1,2} $ and $\xi_2 = \xi_{1, \textrm{c}} + \tilde{\xi}_{1,2} $.

%

The most relevant special case of (\ref{xi12-tau-chiral-2int}) to consider 
corresponds to the limit $\tau_2 =\tau_1 \equiv \tau$.
Since $\mathcal{R}(0 ; u_1, u_2)  =1$ identically,
when $u_2 \neq u_1$ and $u_2 \neq u_{1, \textrm{c}}$, 
independently of the position of $u_1$ and $u_2$ in $A$,
we have 
\be
\label{xi12-2int-same-tau}
\xi(\tau ,u_1)  - \xi(\tau ,u_2)
 =  
 \frac{ \tilde{\eta}\big(  \xi(\tau,u_1), \xi(\tau,u_2)\big) }{ \tilde{\eta}(u_1, u_2) } \;
\sqrt{\partial_{u_1} \xi(\tau,u_1) \; \partial_{u_2} \xi(\tau,u_2)} 
 \; (u_1 - u_2)
\ee
hence  the sign $\xi(\tau ,u_1)  - \xi(\tau ,u_2)$ 
and the sign of $u_1 - u_2$ are the same, for any $\tau \in \RR$.
In the case of $u_1$ and $u_2$ in different intervals,
this is obtained also by using (\ref{sign-eta-eta-ratio}) and observing that 
$\textrm{sign} (\tilde{\xi}_{1,2}) = \textrm{sign} (\tilde{u}_{1,2} )$.


Another interesting limiting regime for (\ref{xi12-tau-chiral-2int}) occurs when $u_1$ and $u_2$
are related through (\ref{u-conj-def}), namely $u_2 = u_{1, \textrm{c}}$.
In this limit and for $\tau_{12} \neq 0$,
the divergencies in $ \tilde{\eta}(u_1, u_2)$ and $\mathcal{R}(\tau_{12} ; u_1, u_2) $
(see (\ref{tilde-eta-def}), (\ref{eta-u2-to-u1c}) and (\ref{R-u2-to-u1c}) respectively) cancel
in (\ref{xi12-tau-chiral-2int}), leading to 
\bea
\label{xi12-tau-chiral-2int-different-u2u1c}
\xi(\tau_1 ,u)  - \xi(\tau_2 ,u_{\textrm{c}} )
\, =  
& &
\\
\rule{0pt}{.8cm}
& & \hspace{-3.6cm}
=\,
\left| \frac{  (u - q_0)\, V_{\textrm{\tiny loc}}(u_{\textrm{c}})  \, \sinh(\pi \tau_{12}) }{ \pi\, r_0\, ( u - u_{\textrm{c}} )} \right|  \,
\sqrt{ \eta\big(  \xi(\tau_1,u), \xi(\tau_2,u_{\textrm{c}})\big) \,\partial_{u} \xi(\tau_1,u) \; \partial_{u_{\textrm{c}}} \xi(\tau_2,u_{\textrm{c}})} 
 \; (u - u_{\textrm{c}})
\nonumber
\\
\rule{0pt}{.9cm}
& & \hspace{-3.6cm}
=
\frac{ (u - q_0)^2 \, V_{\textrm{\tiny loc}}(u_{\textrm{c}}) \, \big|  \sinh(\pi \tau_{12})  \big|    }{ \pi\, r_0 \big[ ( u - q_0)^2 + r_0^2 \big]}\;
\sqrt{ \eta\big(  \xi(\tau_1,u), \xi(\tau_2,u_{\textrm{c}})\big) \, \frac{ V_{\textrm{\tiny loc}}\big(\xi(\tau_1,u) \big) }{ V_{\textrm{\tiny loc}}(u) } \; \frac{ V_{\textrm{\tiny loc}}\big( \xi(\tau_2,u_{\textrm{c}} ) \big) }{ V_{\textrm{\tiny loc}}(u_{\textrm{c}} ) } } 
 \; (u - u_{\textrm{c}})
\nonumber
\eea
where (\ref{u-conj-def}) and the second relation in (\ref{der-xi-2int}) have been employed to write the last expression.


Finally, the  limiting case defined by $u_2 = u_{1, \textrm{c}}$ and $\tau_{1} = \tau_2 \equiv \tau$ 
in (\ref{xi12-tau-chiral-2int}) deserves some care. 
Indeed, one could try to take $\tau_1 \to \tau_2$ in (\ref{xi12-tau-chiral-2int-different-u2u1c}),
by using the fact that 
(from (\ref{conj-xi-k}), the first relation in (\ref{der-xi-2int}) and (\ref{eta-u2-to-u1c})) 
we have 
\be
\label{lim-eta-1}
\lim_{\epsilon \,\to\,0} \;
[  \sinh(\pi \epsilon ) ]^2 \;
 \eta\big(  \xi(\tau+\epsilon ,u), \xi(\tau,u_{\textrm{c}})\big) 
 =
\frac{\pi^2 \,\big[ \xi_k(\tau,u) - \xi_j(\tau,u_{ \textrm{c}} )  \big]^2}{ V_{\textrm{\tiny loc}} \big( \xi_k(\tau,u)  \big) \, V_{\textrm{\tiny loc}} \big( \xi_j(\tau,u_{ \textrm{c}})  \big)  }
\ee
where $u\in A_k$, with $j \neq k$ and $j,k \in \{1,2\}$.
However, once (\ref{lim-eta-1}) is employed to find the $\tau_{12} \to 0$ limit of (\ref{xi12-tau-chiral-2int}),
an identity is obtained. 
Another way to proceed could be 
first setting  $\tau_{1} = \tau_2 \equiv \tau$ with $\tau \neq 0$ in (\ref{xi12-tau-chiral-2int}),
which implies $\mathcal{R}(\tau_{12} =0 ; u_1, u_2) =1 $ identically,
and then taking $u_2 \to u_{1, \textrm{c}}$,
which can be evaluated by using that for $u \in A_k$ and $j \neq k$ 
the following result holds
\be
\label{limit-eta-over-eta}
\lim_{v  \to u_{ \textrm{c}} } \! \frac{\eta\big(  \xi(\tau,u), \xi(\tau,v)\big) }{ \eta(u, v) }
\,=\,
\frac{\big[ \xi_k(\tau,u) - \xi_j(\tau,u_{ \textrm{c}} )  \big]^2}{ V_{\textrm{\tiny loc}} \big( \xi_k(\tau,u)  \big) \, V_{\textrm{\tiny loc}} \big( \xi_j(\tau,u_{ \textrm{c}})  \big)  }\;
\frac{ V_{\textrm{\tiny loc}}(u) \, V_{\textrm{\tiny loc}}(u_{\textrm{c}})  }{ (u - u_{\textrm{c}} )^2}
\ee
obtained from (\ref{conj-xi-k}), the second relation in (\ref{der-xi-2int}) and (\ref{eta-u2-to-u1c}).
After employing (\ref{limit-eta-over-eta}) into the r.h.s. of (\ref{xi12-tau-chiral-2int}) evaluated for  $\tau_{1} = \tau_2 \equiv \tau \neq 0$,
an identity is found again. 

The difference $\xi(\tau ,u)  - \xi(\tau ,u_{\textrm{c}} )$ can be studied by using 
(\ref{xi-k-def})-(\ref{wk-inverse}) and (\ref{w-uc-w-u}).
This gives 
\bea
\label{diff-xi-uc-same-tau}
\xi(\tau ,u)  - \xi(\tau ,u_{\textrm{c}} )
& = &
\\
\rule{0pt}{1.cm}
& & \hspace{-2.8cm}
=\,
(-1)^k\,
\frac{ \sqrt{\big( a_1 + a_2 +(b_1 + b_2)\,  \e^{2\pi [w(u) +\tau] }\big)^2 -4 \big(1+ \e^{2\pi [w(u) +\tau] } \big) \,\big(a_1  a_2 + b_1 b_2\, \e^{2\pi [w(u) +\tau] } \big)} }{ 1 + \e^{2\pi [w(u) +\tau] }  }
\nn
\eea
where $u \in A_k$,
which tells that $\xi(\tau ,u)  - \xi(\tau ,u_{\textrm{c}} )$ and $ u - u_{\textrm{c}}$ have the same sign, 
as expected.

These results tell us that the relativistic causality is preserved along the modular evolution 
also in this setup where the line is bipartite by the union of two disjoint intervals.

\subsection{Spacetime distance along the modular trajectories} 
\label{sec-2int-distance-full}

Following the analysis of Sec.\,\ref{sec-spacetime-distance-1int} for the bipartition characterised by a single interval,
it is insightful to explore the relativistic spacetime distance between two points along two distinct modular trajectories 
by employing the results reported in Sec.\,\ref{sec-2int-distance-chiral} for the chiral distance.


The charge density $\varrho \equiv \varrho_{+}  + \varrho_{-} $ and current density $j \equiv \varrho_{+}  - \varrho_{-} $ 
for the massless Dirac field
are important bosonic operators
and their modular evolutions read respectively
\be
\label{rho-j-def-dirac}
\varrho (\tau ; x,t) \equiv \varrho_{+} (\tau , u_+) + \varrho_{-} (\tau , u_{-})
\;\;\;\;\qquad\;\;\;\;
j(\tau ; x,t) \equiv \varrho_{+} (\tau , u_+) - \varrho_{-} (\tau , u_{-}) \,.
\ee
These quantities satisfy the following commutation relations 
\bea
& &
\big[ \, \varrho (\tau_1 ; x_1 ,t_1 ) \, , \, \varrho (\tau_2 ; x_2 ,t_2)  \,\big] _{-}
=\,
\big[ \, j (\tau_1 ; x_1 ,t_1 ) \, , \, j (\tau_2 ; x_2 ,t_2)  \,\big] _{-}
\nn
\\
\label{commutators-rho-j-mod-2}
\rule{0pt}{.4cm}
& &
=\,
\big[ \, \varrho_{+} (\tau_1 , u_{1,+}) \, , \, \varrho_{+} (\tau_2 , u_{2,+})  \,\big]_{-}
+
\big[ \, \varrho_{-} (\tau_1 , u_{1,-}) \, , \, \varrho_{-} (\tau_2 , u_{2,-})  \,\big]_{-}
\eea
where $(x_1 ,t_1)$ and $(x_2 ,t_2)$ are the spacetime coordinates of the initial points 
for the corresponding the modular flow 
and the commutation relation $\big[ \, \varrho_{+} (\tau_1 , u) \, , \, \varrho_{-} (\tau_2 , v)  \,\big]_{-} = 0$ has been employed. 
The final expression (\ref{commutators-rho-j-mod-2})
is written through the chiral commutators (\ref{comm-rho-T2}).

\begin{figure}[t!]
\vspace{-1.2cm}
\hspace{-.6cm}
\includegraphics[width=1.07\textwidth]{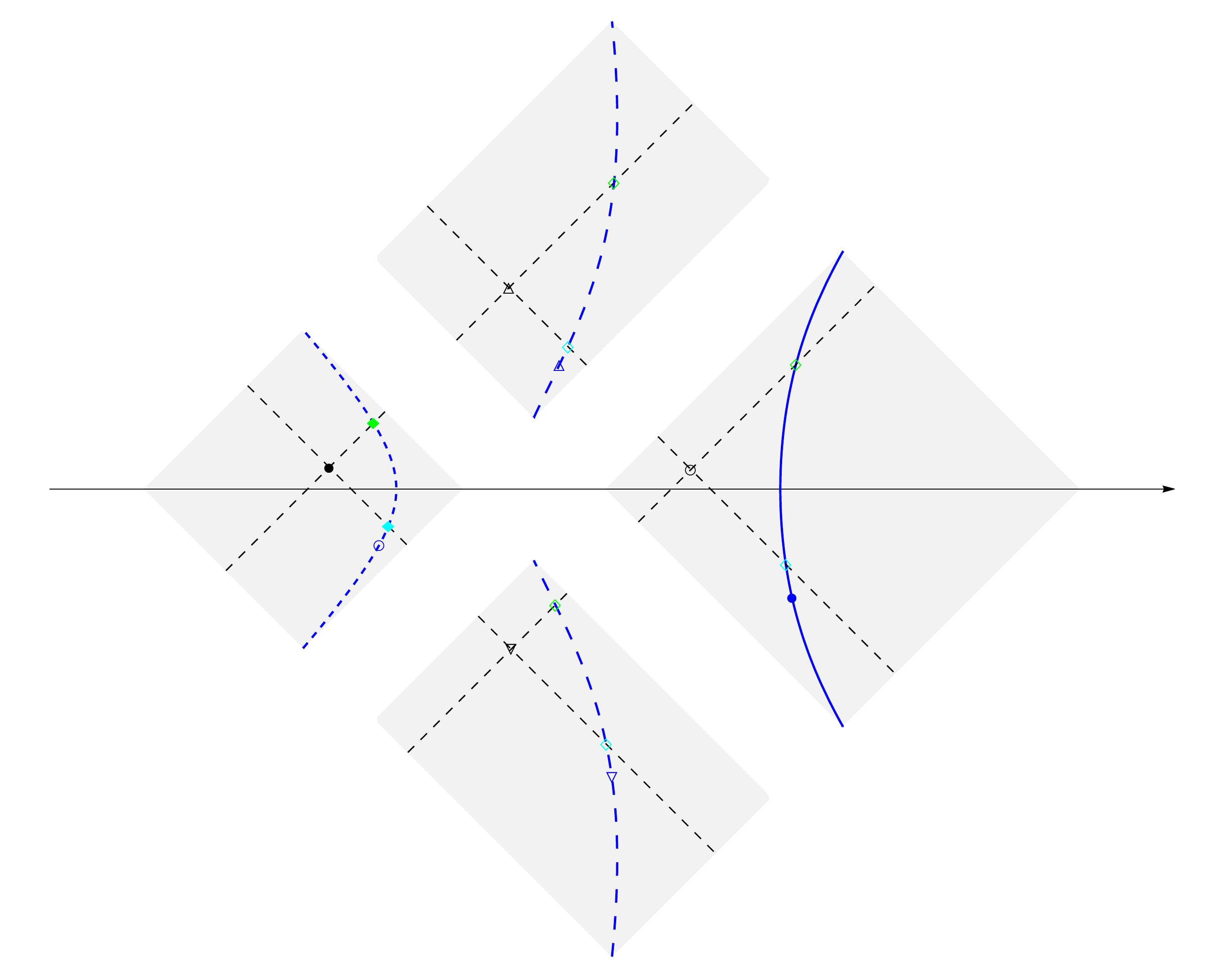}
\vspace{-.3cm}
\caption{
Modular trajectories in the spacetime region $\widetilde{\mathcal{D}}_{A} $ (grey domain)
whose initial points are denoted by the blue markers, whose coordinates are related through (\ref{u-conj-def}).
Also the reference point (black dot) is related to the points denoted by the empty markers through (\ref{u-conj-def}).
The spacetime distance between the reference point and a generic point along the modular trajectories is 
given by (\ref{s12-gen-tau-evolution-2int-v0})-(\ref{tilde-Omega-def-2int}) with e.g. $\tau_1 =0$.
This setup for the solid curves is obtained by combining the one
in the  top left panel of Fig.\,\ref{fig:2int-commutators} along both chiral directions.
}
\label{fig:2int-mod-traject-four-diamonds}
\end{figure}

\begin{figure}[t!]
\vspace{-1.2cm}
\hspace{-.6cm}
\includegraphics[width=1.07\textwidth]{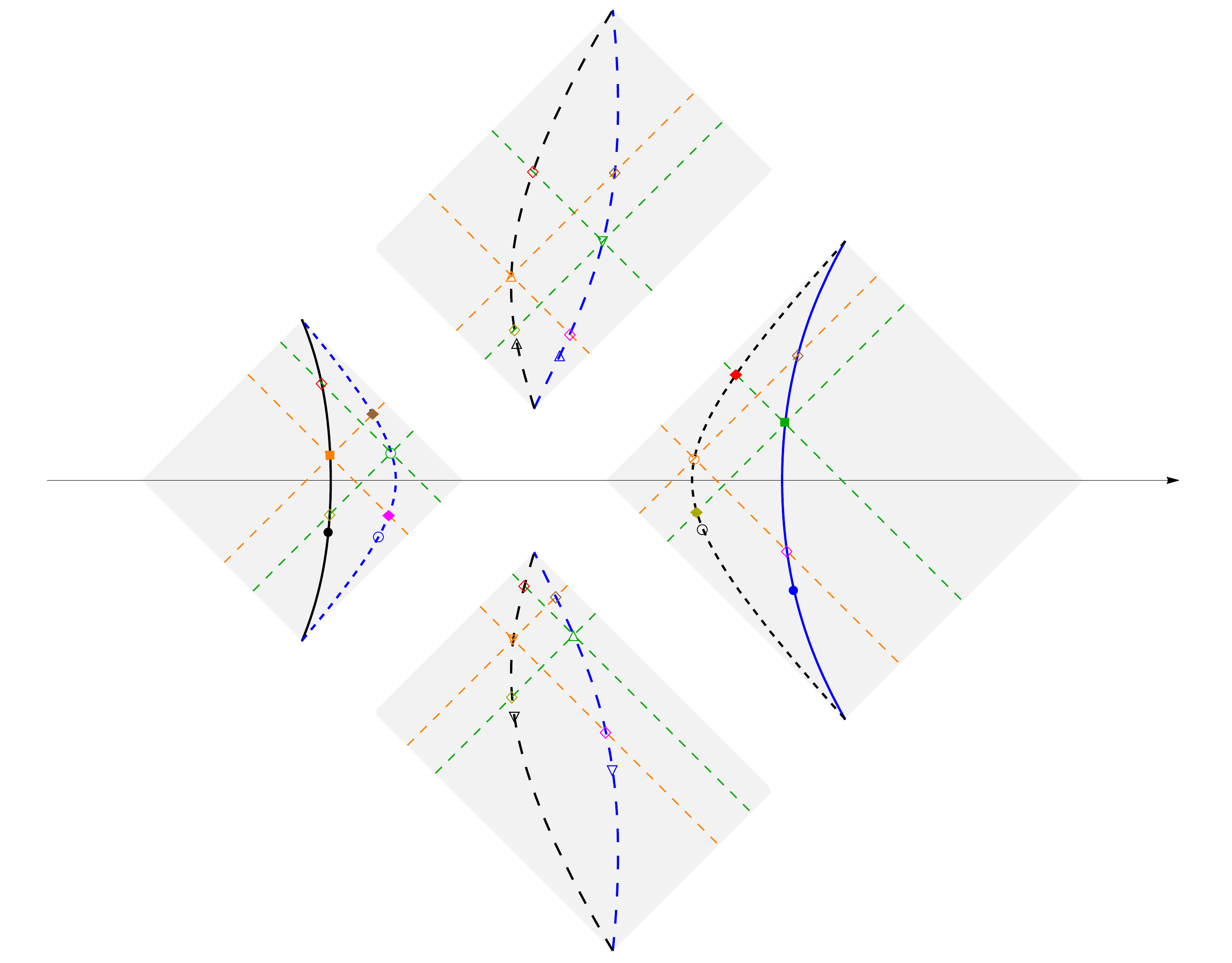}
\vspace{-.3cm}
\caption{
Two sets of modular trajectories  (blue and black curves), 
whose initial points are related through (\ref{u-conj-def}) in each set,
according to the notation adopted in Fig.\,\ref{fig:2int-mod-traject-four-diamonds}.
The spacetime distance between a generic point along a modular trajectory 
(red dot, related to the points denoted by red empty markers through (\ref{u-conj-def})) 
and a generic point along another modular trajectory 
is given by (\ref{s12-gen-tau-evolution-2int-v0})-(\ref{tilde-Omega-def-2int}).
This setup for the solid curves is obtained 
by combining the one illustrated in the bottom left panel of Fig.\,\ref{fig:2int-commutators}
along both chiral directions.
}
\label{fig:2int-mod-traject-four-diamonds-bis}
\end{figure}

The result (\ref{commutators-rho-j-mod-2}) implies that
the spacetime region involved in the modular evolution 
corresponding to the subsystem $ A_{1,x} \cup A_{2,x}$ in the spatial direction parameterised by $x$
is obtained by taking the subsystem $A_1 \cup A_2$ along both the chiral directions parameterised by $u_{\pm}$.
This provides 
$\widetilde{\mathcal{D}}_{A}  \equiv \mathcal{D}_{A_1} \cup \mathcal{D}_{A_2} \cup \mathcal{D}_{A, \textrm{\tiny F}} \cup  \mathcal{D}_{A, \textrm{\tiny P}}$
(see the grey domain in Fig.\,\ref{fig:2int-mod-traject-four-diamonds} and Fig.\,\ref{fig:2int-mod-traject-four-diamonds-bis}),
where $\mathcal{D}_{A_1} =\{(u_+ , u_{-}) ; \, u_+ \in A_1 , u_{-} \in A_1\}$ 
and $\mathcal{D}_{A_2} =\{(u_+ , u_{-}) ; \, u_+ \in A_2 , u_{-} \in A_2\}$ 
are the causal diamonds associated to $A_{1,x}$ and $A_{2,x}$ respectively along the $x$-direction,
while $\mathcal{D}_{A, \textrm{\tiny F}} =\{(u_+ , u_{-}) ; \, u_+ \in A_2 , u_{-} \in A_1\}$ 
and $\mathcal{D}_{A, \textrm{\tiny P}} =\{(u_+ , u_{-}) ; \, u_+ \in A_1 , u_{-} \in A_2\}$.
Given a point $P$ in $\widetilde{\mathcal{D}}_{A} $ with light-cone coordinates $(u_{+} , u_{-})$
(see e.g. the blue, black and red dots in Fig.\,\ref{fig:2int-mod-traject-four-diamonds} and Fig.\,\ref{fig:2int-mod-traject-four-diamonds-bis}),
three other points in $\widetilde{\mathcal{D}}_{A} $ are naturally associated to $P$ through the map (\ref{u-conj-def}),
namely $P_{\textrm{c}, +} \equiv (u_{+,\textrm{c}} \, , u_{-})$, 
$P_{\textrm{c}, -} \equiv(u_{+} \, , u_{-,\textrm{c}})$ and $P_{\textrm{c}} \equiv(u_{+,\textrm{c}} \, , u_{-,\textrm{c}})$
(in Fig.\,\ref{fig:2int-mod-traject-four-diamonds} and Fig.\,\ref{fig:2int-mod-traject-four-diamonds-bis}, 
these are points denoted through the empty markers, obtained from the dot having the same colour).
These four points belong to the four different subsets of $\widetilde{\mathcal{D}}_{A} $
and provide the initial points of four different modular trajectories 
obtained by employing (\ref{xi-2int-A1-A2}) into  (\ref{mod-traj-tau-line})
(see the blue and black curves in Fig.\,\ref{fig:2int-mod-traject-four-diamonds} and Fig.\,\ref{fig:2int-mod-traject-four-diamonds-bis},
whose initial points are marked with the same colour);
hence $\tau \in \RR$ is the modular parameter along each one of these trajectories.
All these four curves are naturally involved in the modular flow of the massless Dirac field.
Indeed, all of them must be considered to obtain a consistent result in the  limit of adjacent intervals,
as discussed in \cite{Mintchev:2022fcp}, whose Fig.\,11 illustrates this limiting procedure in the spacetime.


Considering the modular evolution along the two distinct modular trajectories with initial points $P_1$ and $P_2$,
the spacetime distance between the points $P_1(\tau)$ and $P_2(\tau)$ 
after the same value of the modular parameter $\tau$
along the corresponding modular trajectories can be written by using 
(\ref{s12-tau-def}) and (\ref{xi12-2int-same-tau}), finding 
\be
\label{dist-spacetime-2int-same-tau}
d\big(P_1(\tau), P_2(\tau)\big)
= \,
\tilde{\omega}(\tau;P_1,P_2)  \, 
d(P_1, P_2)
\ee
where $\tilde{\omega}(\tau;P_1,P_2) $ is defined as follows 
\bea
\label{omega-factor-tau-2int}
\tilde{\omega}(\tau;P_1,P_2)  
& \equiv &
 \frac{ \tilde{\eta}\big(  \xi(\tau,u_{1,+}), \xi(\tau,u_{2,+})\big) }{ \tilde{\eta}(u_{1,+}, u_{2,+}) }  \; 
 \frac{ \tilde{\eta}\big(  \xi( - \tau,u_{1,-}), \xi( - \tau,u_{2,-})\big) }{ \tilde{\eta}(u_{1,-}, u_{2,-}) } 
 \\
 \rule{0pt}{.6cm}
 & &
 \times\,
\sqrt{
\partial_{u_{1,+}} \xi(\tau,u_{1,+}) \; \partial_{u_{1,-}} \xi(- \tau,u_{1,-})  \; \,
\partial_{u_{2,+}} \xi(\tau,u_{2,+}) \; \partial_{u_{2,-}} \xi(- \tau,u_{2,-})
} 
 \nn
\eea
in terms of  (\ref{tilde-eta-def}), (\ref{xi-2int-A1-A2}) and (\ref{der-xi-2int}).
The above considerations imply that 
$\tilde{\omega}(\tau=0;P_1,P_2)  = 1$, as expected,
and that $\tilde{\omega}(\tau;P_1,P_2) > 0$ for any $\tau$.
We remark that,
since (\ref{dist-spacetime-2int-same-tau})-(\ref{omega-factor-tau-2int}) imply that 
$d\big(P_1(\tau), P_2(\tau)\big)$ and $d(P_1, P_2)$ have the same sign for any  $\tau\in \RR$,
the modular evolution 
along the modular trajectories containing the initial points $P_1$ and $P_2$
(indeed, the modular evolution in this case is non local and involves four modular trajectories, 
as shown in Fig.\,\ref{fig:2int-mod-traject-four-diamonds} and Fig.\,\ref{fig:2int-mod-traject-four-diamonds-bis})
preserves the relativistic causality also in this case. 
As consistency check of (\ref{omega-factor-tau-2int}), 
we observe that, by using (\ref{tilde-eta-vs-eta}) and (\ref{eta-special-regimes-A}),
the corresponding expression for the single interval case
in (\ref{OmegaA-def}) is recovered in the adjacent intervals limit.


It is worth studying  the spacetime distance between $P_1(\tau_1)$ and $P_2(\tau_2)$
at generic real values $\tau_1$ and $\tau_2$ of the modular parameter 
along the corresponding modular trajectories, 
as done in Sec.\,\ref{sec-spacetime-distance-1int} for the bipartition of the line induced by the single  interval 
(see (\ref{s12-gen-tau-evolution})-(\ref{omega-12-def}) and Fig.\,\ref{fig:1int-mod-traject-3}).
Assuming that $d(P_1, P_2) \neq 0$ for the initial distance, 
by employing (\ref{xi12-tau-chiral-2int}) into (\ref{s12-tau12-def})-(\ref{s12-tau12-def-1}),
we find that this quantity reads
\bea
\label{s12-gen-tau-evolution-2int-v0}
d\big(P_1(\tau_1), P_2(\tau_2)\big)
&=&
\big[ \xi(\tau_1, u_{1,+}) - \xi(\tau_2, u_{2,+}) \big] \big[ \xi(-\tau_1, u_{1,-}) - \xi(-\tau_2, u_{2,-}) \big]
\\
\rule{0pt}{.5cm}
\label{s12-gen-tau-evolution-2int}
&=& 
\widetilde{\Omega}(\tau_1, \tau_2  ;P_1,P_2)  \, 
d(P_1, P_2)
\eea
where we have introduced 
\bea
\label{tilde-Omega-def-2int}
\widetilde{\Omega}(\tau_1, \tau_2  ;P_1,P_2)  
& \equiv &
 \frac{ \tilde{\eta}\big(  \xi(\tau_1 ,u_{1,+}), \xi(\tau_2 ,u_{2,+})\big) }{ \tilde{\eta}(u_{1,+}, u_{2,+}) }  \; 
 \frac{ \tilde{\eta} \big(  \xi( - \tau_1 ,u_{1,-}), \xi( - \tau_2 ,u_{2,-})\big) }{ \tilde{\eta}(u_{1,-}, u_{2,-}) } 
  \\
 \rule{0pt}{.5cm}
 & &
 \times\,
\mathcal{R}(\tau_{12} ; u_{1,+}, u_{2,+}) \; \mathcal{R}(-\tau_{12} ; u_{1,-}, u_{2,-}) 
 \nn
 \\
 \rule{0pt}{.6cm}
 & &
 \times\,
\sqrt{
\partial_{u_{1,+}} \xi(\tau_1,u_{1,+}) \; \partial_{u_{1,-}} \xi(- \tau_1,u_{1,-})  \; \,
\partial_{u_{2,+}} \xi(\tau_2,u_{2,+}) \; \partial_{u_{2,-}} \xi(- \tau_2,u_{2,-})
} 
 \nn
\eea
which satisfies $\widetilde{\Omega}(\tau, \tau  ;P_1,P_2)  = \tilde{\omega}(\tau;P_1,P_2) $, as expected. 
From (\ref{s12-gen-tau-evolution-2int-v0})-(\ref{tilde-Omega-def-2int}), 
the sign of $d\big(P_1(\tau_1), P_2(\tau_2)\big)$
can be different from the sign of $d(P_1, P_2)$ as $\tau_{12}$ changes.

Combining (\ref{s12-gen-tau-evolution-2int-v0})-(\ref{tilde-Omega-def-2int}) 
with the discussion made for  (\ref{xi12-tau-chiral-2int}),
we conclude  that the sign of the ratio $d\big(P_1(\tau_1), P_2(\tau_2)\big)/ d(P_1, P_2)$ 
remains positive for every value of $\tau_1$ and $\tau_2$ only when 
either $P_1 \in \mathcal{D}_{A_1}$ and $P_2 \in \mathcal{D}_{A_2}$ (or viceversa)
or $P_1 \in \mathcal{D}_{A, \textrm{\tiny F}} $ and $P_2 \in \mathcal{D}_{A, \textrm{\tiny P}} $  (or viceversa).
%
For all the other choices of initial points in $\widetilde{\mathcal{D}}_{A} $,
the sign of the ratio $d\big(P_1(\tau_1), P_2(\tau_2)\big)/ d(P_1, P_2)$ 
changes at $\tau_{12} = \tau_{>} $ and $\tau_{12} = \tau_{>} $,
whose explicit expressions can be obtained by using that
$\mathcal{R}(\tau ; u_1, u_2) $ in (\ref{xi12-tau-chiral-2int}) 
has a first order zero at $\tau = \tilde{\tau}_0(u_{1}, u_{2})$ given by (\ref{tau0tilde-def-2int}).
This leads to 
\bea
\label{tau-min-12-def-2int}
\tau_{<}  
& \equiv & 
\textrm{min} \big\{ \tilde{\tau}_0(u_{1,+}, u_{2,+}) \, , \;  - \,\tilde{\tau}_0(u_{1,-}, u_{2,-})) \big\}
\\
\label{tau-max-12-def-2int}
\rule{0pt}{.5cm}
\tau_{>} 
& \equiv & 
 \textrm{max} \big\{ \tilde{\tau}_0(u_{1,+}, u_{2,+}) \,, \, - \,\tilde{\tau}_0(u_{1,-}, u_{2,-})) \big\} \,.
\eea


In the limiting cases where 
$u_{2,+} = \mathcal{C}(u_{1,+})$ or $u_{2,-} = \mathcal{C}(u_{1,-})$ (see (\ref{u-conj-def})),
the spacetime distance $d\big(P_1(\tau_1), P_2(\tau_2)\big)$  
can be written as in (\ref{s12-tau12-def})-(\ref{s12-tau12-def-1})
by employing for the chirality where this condition is fulfilled 
the expression in (\ref{xi12-tau-chiral-2int-different-u2u1c}) when $\tau_{1} \neq \tau_2 $
and the one in (\ref{diff-xi-uc-same-tau}) when $\tau_{1} = \tau_2 $.
The resulting formulas for $d\big(P_1(\tau_1), P_2(\tau_2)\big)$  
can be written straightforwardly 
and we do not find it worth reporting them here. 
In particular, 
in the special case of $P_2 = P_{1, \textrm{c}}$ (i.e. $u_{2,\pm} = \mathcal{C}(u_{1,\pm})$),
this tells us that $d\big(P(\tau_1), P_{\textrm{c}}(\tau_2)\big) > 0$
for all real values of $\tau_1$ and $\tau_2$.
Indeed, the factor multiplying $u - u_{\textrm{c}}$ in the r.h.s. of (\ref{xi12-tau-chiral-2int-different-u2u1c})
is always positive and the sign $(-1)^k$ in (\ref{diff-xi-uc-same-tau})
does not affect the sign of $d\big(P(\tau), P_{\textrm{c}}(\tau)\big)$ 
because two of them must be multiplied, one for each chirality.


In Fig.\,\ref{fig:2int-mod-traject-four-diamonds}
(that should be compared with the left panel of Fig.\,\ref{fig:1int-mod-traject-3} for the single interval case)
we show an example where one point (see the black dot), e.g. $P_1$, is kept fixed, i.e. $\tau_1=0$,
and its spacetime distance from  another point $P_2(\tau_2)$ moving along its modular trajectory is considered.  
This modular trajectory is one of the blue arcs in the different parts of $\widetilde{\mathcal{D}}_{A} $ (grey domain),
whose initial point $P_2$ is indicated by a blue marker.

Taking, for instance, as the initial point $P_2$ the blue dot in $\mathcal{D}_{A_2} $ 
the corresponding modular trajectory for $P_2(\tau_2)$ is the blue solid line.
The initial distance $d(P_1, P_2) >0$ is spacelike and, 
from (\ref{s12-gen-tau-evolution-2int-v0})-(\ref{tilde-Omega-def-2int}),
it remains spacelike all over the entire modular trajectory, 
namely  $d\big(P_1, P_2(\tau_2)\big) >0$ for every $\tau_2 \in \RR$.
Indeed, all the points of $ \mathcal{D}_{A_2}$ are spacelike separated from any point of $ \mathcal{D}_{A_1}$.
However, considering the anti-commutator of  a massless Dirac field localised in $P_1$
and the modular flow of a massless Dirac field localised in $P_2$ at the beginning of the flow, 
by applying (\ref{comm-psi-T2}) or (\ref{comm-psi-T2-symm}) on both the chiralities, 
we find that Dirac deltas occur when  $\tau_2 =  - \tau_{<} $ and $\tau_2 =  - \tau_{>} $
(see (\ref{tau-min-12-def-2int}) and (\ref{tau-max-12-def-2int})).
This is due to the fact that the modular flow (\ref{psi-flow-text}) is bilocal
and these Dirac deltas originate from the term containing $ \psi(\xi_{\textrm{c}}) $.
Such bilocal term makes the geometric action of the modular flow non-local (in particular, bilocal) \cite{Casini:2009vk, Longo:2009mn},
and therefore all the blue curves in Fig.\,\ref{fig:2int-mod-traject-four-diamonds}
must be considered for a complete geometrical description of the modular flow of a massless Dirac field initially localised in $P_2$
(see e.g. \cite{Mintchev:2022fcp}).
This implies that, in $\widetilde{\mathcal{D}}_{A} $, the above mentioned Dirac deltas are supported on all the points 
denoted by the cyan and green markers in Fig.\,\ref{fig:2int-mod-traject-four-diamonds}.

Instead, taking as initial point $P_2$ one of the other empty blue markers
(whose coordinates are related to the blue dot through (\ref{u-conj-def})), 
the corresponding modular trajectory is either in $\mathcal{D}_{A_1} $
or in $\mathcal{D}_{A, \textrm{\tiny F}}$ or in $\mathcal{D}_{A, \textrm{\tiny P}}$
(for the circle, the up-pointing triangle and down-pointing triangle respectively);
hence the distance $d\big(P_1, P_2(\tau_2)\big)$ changes sign with respect to $d(P_1, P_2)$
when  $\tau_2 =  - \tau_{<} $ and $\tau_2 =  - \tau_{>} \,$, 
given by (\ref{tau-min-12-def-2int}) and (\ref{tau-max-12-def-2int}).
In particular, $P_2(- \tau_{<} )$ and $P_2(- \tau_{>} )$
are the points denoted by the green and cyan markers respectively
along the blue curve passing through $P_2$.
As for  the anti-commutator of a massless Dirac field in $P_1$
and the modular flow of a massless Dirac field in $P_2$ at the beginning of the flow, 
from (\ref{comm-psi-T2}) or (\ref{comm-psi-T2-symm}) 
we reach the same conclusion described for the previous case, 
namely that Dirac delta occurs supported in all the points of $\widetilde{\mathcal{D}}_{A} $
corresponding to the cyan and green markers in Fig.\,\ref{fig:2int-mod-traject-four-diamonds}.


In Fig.\,\ref{fig:2int-mod-traject-four-diamonds-bis}
(to be compared with the right panel of Fig.\,\ref{fig:1int-mod-traject-3} for the single interval case),
by adopting a notation similar to the one of Fig.\,\ref{fig:2int-mod-traject-four-diamonds},
we illustrate an example where both $P_1(\tau_1)$ and $P_2(\tau_2)$ 
move along distinct modular trajectories whose initial points $P_1$ and $P_2$
correspond to a black and blue marker respectively (each of them can be either filled or empty).
The point $P_1(\tau_1)$  at $\tau_1 \neq 0$ is indicated by an orange marker
and the situation in Fig.\,\ref{fig:2int-mod-traject-four-diamonds} is recovered for $\tau_1 =0$,
when any orange marker coincides with the corresponding black marker on its modular trajectory. 
Instead, the point $P_2(\tau_2)$  at $\tau_2 \neq 0$ corresponds to a green marker.
Extending the observations made for Fig.\,\ref{fig:2int-mod-traject-four-diamonds} is straightforward. 
The sign of the ratio $d\big(P_1(\tau_1), P_2(\tau_2)\big)/ d(P_1, P_2)$ has been already discussed in the text
above (\ref{tau-min-12-def-2int})-(\ref{tau-max-12-def-2int}), finding that
e.g.  when $P_1$ and $P_2$ are the black dot in $\mathcal{D}_{A_1}$ and the blue dot in $\mathcal{D}_{A_2}$ respectively,
 $d(P_1, P_2)>0$ and $d\big(P_1(\tau_1), P_2(\tau_2)\big) > 0$ for any $\tau_1$ and $\tau_2$.
 Indeed, any point in $ \mathcal{D}_{A_2}$ is spacelike separated from any other point in $ \mathcal{D}_{A_1}$.
However, 
considering the anti-commutator of the modular flows of two massless Dirac fields (see  (\ref{psi-flow-text})), 
initially localised in $P_1$ and $P_2$,
from (\ref{comm-psi-T2}) or (\ref{comm-psi-T2-symm}) 
we find that Dirac deltas contributions occur
when  $\tau_{12} =   \tau_{<} $ and $\tau_{12} =  \tau_{>} $, 
given by (\ref{tau-min-12-def-2int}) and (\ref{tau-max-12-def-2int}) respectively. 
In Fig.\,\ref{fig:2int-mod-traject-four-diamonds-bis} the points where one of these conditions 
is realised have been highlighted through rhombi of various colours.

Instead, choosing as $P_1$ and $P_2$ e.g. the black dot in $\mathcal{D}_{A_1}$ 
and the blue empty  up-pointing triangle in $\mathcal{D}_{A, \textrm{\tiny F}}$ respectively,
the signs of $d\big(P_1(\tau_1), P_2(\tau_2)\big)$ and $d(P_1, P_2)$ might be different (depending on $\tau_1$ and $\tau_2$)
and, as for the anti-commutator of the modular flows of two massless Dirac fields initially in $P_1$ and $P_2$,
the same conclusion holds.



We find it instructive to compare the above results about 
the signs of $d\big(P_1(\tau_1), P_2(\tau_2)\big)$ and $d(P_1, P_2)$ 
when $P_1 \in \mathcal{D}_{A_1}$ and $P_2 \in \mathcal{D}_{A_2}$
with the ones for the same quantities in the case where the bipartition is determined by a single interval
when $P_1 \in \mathcal{D}_{A}$ and $P_2 \in \mathcal{W}_{\textrm{\tiny R}} $,
discussed in Sec.\,\ref{sec-spacetime-distance-1int} (see the right panel of Fig.\,\ref{fig:1int-mod-traject-1}). 
While in the former case $d\big(P_1(\tau_1), P_2(\tau_2)\big)$ and $d(P_1, P_2)$ have the same sign
and the Dirac delta contributions in the commutators given by (\ref{commutators-rho-j-mod-2}) 
and (\ref{comm-rho-T2}) are due to the bi-local nature of the modular flow, 
in the latter case the Dirac deltas in the commutator of the modular flows of the currents 
(see Sec.\,\ref{sec-1int-commutators}) occur when 
$d\big(P_1(\tau_1), P_2(\tau_2)\big)$ changes sign with respect to  $d(P_1, P_2)$, 
consistently with the fact that the modular flow is local.

\section{Conclusions}
\label{sec-conclusions}


The modular evolution is an interesting unitary evolution to study and
understanding its the relation with causality could provide new insights 
in the entanglement structure of quantum field theories. 
This is usually a difficult task because the explicit expressions of the modular Hamiltonian 
and of the corresponding  modular flows are not known. 

Since causality and spacetime distance are deeply related in Minkowski spacetime, 
in this manuscript we have studied the spacetime distance between points 
moving along two distinct modular trajectories for some known modular flows 
in two-dimensional CFT in the Minkowski spacetime.
When the bipartition of the spatial line is characterised by a single interval,
we have explored first the case of the ground state (see Sec.\,\ref{sec-1int-line-vacuum}) 
and then its thermal generalisation
where the CFT have different temperatures for the two chiral components (see Sec.\,\ref{sec-1int-line-thermal}).
Instead, in the more complicated case where the bipartition of the spatial line 
is determined by the union of two disjoint intervals, 
we have considered the free massless Dirac field in its ground state,
by employing the known expressions for 
the modular Hamiltonian, the corresponding modular flows \cite{Casini:2009vk} 
and the modular correlators for the field \cite{Longo:2009mn, Hollands:2019hje} 
(see Sec.\,\ref{sec-2int-Dirac}).
While the modular flows related to the finite spatial subsystem $A$ have been mainly explored, 
in the case of the interval $A$ and of the ground state (see Sec.\,\ref{sec-1int-line-vacuum})
some results about its complement $B$ have been also obtained.


As for the subsystem $A$, 
analytic expressions for the spacetime distance $d\big(P_1(\tau_1), P_2(\tau_2)\big)$
between two points along distinct modular trajectories 
for generic values of the modular evolution parameters $\tau_1$ and $\tau_2$ along them
have been obtained 
in terms of the spacetime distance $d(P_1 , P_2)$ between the two initial points $P_1$ and $P_2$
of the modular trajectories. 
When $A$ is a single interval in the line, this distance is given by 
(\ref{s12-tau12-def})-(\ref{omega-12-def}) for the CFT in its ground state (see also Fig.\,\ref{fig:1int-mod-traject-3})
and by (\ref{s12-tau12-Temp-def})-(\ref{omega-12-Temp-def}) 
for the CFT at finite inverse temperatures $\beta_{+}$ and $\beta_{-}$
(see also Fig.\,\ref{fig:1int-mod-traject-Temp-2}).
In the former case, also the corresponding result in the Euclidean signature has been discussed
 (see Sec.\,\ref{sec-mod-flow-1int-Euclid}).
When $A$ is the union of two disjoint intervals 
and the underlying CFT is the free massless Dirac field, 
this distance is given by (\ref{s12-gen-tau-evolution-2int-v0})-(\ref{tilde-Omega-def-2int}) 
(see also Fig.\,\ref{fig:2int-mod-traject-four-diamonds-bis}).
In this case, the harmonic ratio (\ref{eta-ratio-def}) 
allows to write the modular flow of the Dirac field in the form (\ref{psi-flow-text})
through (\ref{tilde-eta-def}) and (\ref{tilde-eta-vs-eta}).
The relativistic causality along these modular flows has been studied by considering the sign of 
$d\big(P_1(\tau), P_2(\tau)\big)$ with respect to the sign of $d(P_1 , P_2)$
in the special case of $\tau_1=\tau_2 \equiv \tau$.
Since these two quantities keep the same sign for any $\tau \in \RR$,
the relativistic causality is preserved by these modular evolutions.
Explicit expressions for the values of the difference $\tau_1 - \tau_2$ 
where $d\big(P_1(\tau_1), P_2(\tau_2)\big)$ changes its sign with respect to $d(P_1 , P_2)$
have been obtained (see (\ref{tau-min-12-def})-(\ref{tau-max-12-def}),
(\ref{tau-min-12-Temp-def})-(\ref{tau-max-12-Temp-def}) 
and (\ref{tau-min-12-def-2int})-(\ref{tau-max-12-def-2int})
in terms of  (\ref{tau0tilde-def}), (\ref{tau0tilde-temp-def}) and (\ref{tau0tilde-def-2int}) respectively),
finding for them a simple form in terms of the function 
determining the weight function of the local term of the modular Hamiltonian
(see (\ref{velocity_fund}), (\ref{velocity_fund_temp}) and (\ref{velocity_fund-2int})).


The commutators  (or anti-commutators in the case of fermionic operators) of certain modular flows
are interesting quantities to consider to explore the relativistic causality along the modular evolution. 
They can be studied by taking their mean value first and then employing the modular correlators
involved in the resulting scalar quantities. 
For a CFT in the ground state and the bipartition of the line given by an interval, 
in Sec.\,\ref{sec-1int-commutators}  the commutators of the modular flow of the chiral currents 
have been considered,  finding (\ref{commutator-j-single-int})
(or (\ref{commutator-j-single-int-symmetrised}) equivalently). 
For the massless Dirac field in the ground state 
and  the bipartition of the line given by the union of two disjoint intervals, 
in Sec.\,\ref{sec-anti-comm-dirac} the anti-commutators of the modular flow of the chiral field
and the commutators of the modular flow of the chiral density have been studied,
finding (\ref{comm-psi-T2}) (or (\ref{comm-psi-T2-symm}) equivalently) and (\ref{comm-rho-T2}) respectively. 
The latter result,
which has been obtained by first writing the modular correlators of the chiral density fields in (\ref{mod-corr-rho-2int-pm})
(see also Appendix\;\ref{app-mod-flow-density}),
provides the commutators of the modular flows of the charge density and of the current density
for the massless Dirac field (see (\ref{rho-j-def-dirac})-(\ref{commutators-rho-j-mod-2})).


The main result of the comparison between the spacetime distance 
of points along two distinct modular trajectories and the commutators of the corresponding modular flows 
is about the massless Dirac field and the bipartition of the line given  by the union of two disjoint intervals;
indeed, in this case the  non-locality (bilocality, in particular) of the modular flow
 induced by the bilocal term of the modular Hamiltonian (see (\ref{K_A-2int-terms-bi-local})-(\ref{T-bilocal-def}))
provides some Dirac delta contributions in the commutators 
also when the all the points of a modular trajectories are spacelike separated from all the points of the other one,
as discussed in the final part of Sec.\,\ref{sec-2int-distance-full}
and represented in Fig.\,\ref{fig:2int-mod-traject-four-diamonds} and Fig.\,\ref{fig:2int-mod-traject-four-diamonds-bis}.


Another interesting case as been discussed in Sec.\,\ref{sec-1int-line-vacuum}
and involves, instead, the local operator $K$ in (\ref{K-chiral-1int-line}),
associated to the ground state of a CFT on the line bipartite by an interval. 
Considering a modular trajectory inside the diamond $\D_A$ 
and another one in $\mathcal{B}_A$
(i.e. in the grey and light blue region  in Fig.\,\ref{fig:diamond-mod-hyper} respectively),
the spacetime distance $d\big(P_1(\tau_1), P_2(\tau_2)\big)$ changes its sign with respect to $d(P_1 , P_2)$
when singularities occur in the evolution in $\mathcal{B}_A$
corresponding to the singularity of (\ref{xi-map-fund}) (see (\ref{tauB_def})).
This feature can be traced back to the singularity of the map employed 
in \cite{Hislop:1981uh, Casini:2011kv} to obtain (\ref{K-chiral-1int-line}) 
from the Bisognano and Wichmann modular Hamiltonian,
as discussed in Appendix\;\ref{app-HL-map}.
Moreover, focussing on primaries having $h_{+} = h_{-} \equiv h$ for simplicity, 
with $h>0$,
in Sec.\,\ref{sec-tau-evolution} we have shown that, 
for $h \neq k/2$ with $k \in \mathbb{N}$,
local fields with conformal dimension $h$ can be constructed whose commutator
has  a non vanishing support in the past and future cones 
made by the points that are timelike separated from the points of $\D_A$.
Hence, their modular evolution outside the diamond $\D_A$  
is not the one generated by $K$ in (\ref{K-chiral-1int-line}) and,
as far as we know, it is not available in the literature. 
%


Our analyses can be extended in various directions. 
It is worth considering other cases whose modular Hamiltonian is known explicitly, 
including the ones where non-local terms occur.
Considering the two-dimensional massless Dirac field in the ground state, 
besides the case investigated in Sec.\,\ref{sec-2int-Dirac},
bilocal terms have been found also 
for the bipartition of the half line given by an interval separated from the boundary \cite{Mintchev:2020uom}
and for the bipartition of the line with a point-like defect characterised by two equal intervals 
placed symmetrically with respect to the defect \cite{Mintchev:2020jhc}.
They also occur for the same model on the circle and at finite temperature,
when the bipartition is given by a single arc \cite{Blanco:2019xwi, Fries:2019ozf}.
Another case that is worth considering is 
the free chiral current of the massless scalar field in the ground state and on the line
bipartite by the union of two disjoint intervals, 
whose modular Hamiltonian contains a genuine non-local term \cite{Arias:2018tmw}.
It is instructive to extend our analyses also in two-dimensional inhomogeneous CFT,
already in the cases of local modular Hamiltonians \cite{Tonni:2017jom}.
Let us also mention prototypical CFT models like the compactified boson and the Ising model
in their ground state on the line bipartite by the union of two disjoint intervals and its complement,
whose entanglement entropies are known 
(see e.g. \cite{Furukawa:2008uk, Caraglio:2008pk, Calabrese:2009ez, Calabrese:2010he, Coser:2013qda,Grava:2021yjp, DeNobili:2015dla, Coser:2015dvp})
but analytic expressions for the corresponding modular Hamiltonians are still not available.


An important arena to explore the relation between causality and modular evolution is provided
by the gauge/gravity correspondence. 
In this context, by employing the geometric prescription on the gravitational side for
the holographic entanglement entropy \cite{Ryu:2006bv, Ryu:2006ef, Hubeny:2007xt},
causality has been studied e.g. in \cite{Headrick:2014cta}.
When the subsystem in the gauge theory side on the boundary is the union of disjoint regions,
interesting transitions occur (see e.g. \cite{Headrick:2010zt, Tonni:2010pv, Fonda:2014cca})
coming from the limit of large degrees of freedom where the classical gravity picture emerges. 
Relevant insights for these analyses could come from the connection with von Neumann algebras 
recently discussed in \cite{Leutheusser:2021qhd, Leutheusser:2021frk, Leutheusser:2022bgi}
or from the reformulation of the holographic entanglement entropy prescription 
through the bit threads \cite{Freedman:2016zud}.
In the latter framework, for instance, an interesting flow provided by the geodesic bit threads has been explored \cite{Agon:2018lwq},
finding a relation with the geometric action of the modular conjugation in the boundary CFT
\cite{Mintchev:2022fcp, Caggioli:2024uza}.


The invariance under the Lorentz transformation plays a crucial role
in the notion of causality and in the determination of the modular evolutions;
hence it would be very insightful to explore the questions addressed in this manuscript
also in non-relativistic models
(see e.g. \cite{Mintchev:2022xqh, Mintchev:2022yuo, Eisler:2023yys} for some results in the free fermionic Schr\"odinger model).
%
Since the continuum limit of lattice models typically provides non-relativistic quantum field theories, 
we remark that it would be interesting to explore the notions of modular evolutions and causality 
also in lattice models, 
where some entanglement Hamiltonians and the corresponding spectra have been studied 
(see e.g. \cite{peschel-03, EislerPeschel:2009review, Casini:2009sr, Lauchli:2013jga, Eisler:2017cqi, Eisler:2018aaa, Eisler:2019rnr, DiGiulio:2019cxv,
Surace:2019mft, Eisler:2022rnp}).

\vskip 20pt 
\centerline{\bf Acknowledgments } 
\vskip 5pt

We thank Domenico Seminara, John Sorce 
and especially Sebastiano Carpi, Stefan Hollands, Roberto Longo and Diego Pontello for insightful discussions and correspondence. 
We are grateful in particular to Hong Liu for collaboration during the initial part of this project
and important discussions throughout its development.

\appendix

\section{Mappings}
\label{app-mappings}

In Appendix\;\ref{app-HL-map} 
we review the map \cite{Hislop:1981uh, Haag:1992hx, Casini:2011kv} employed to obtain 
the modular Hamiltonian (\ref{K-chiral-1int-line})
from the one of Bisognano and Wichmann \cite{Bisognano:1975ih, Bisognano:1976za},
discussing also its singularities.
Inversions mappings sending $\mathcal{D}_A $ onto $\mathcal{W}_{\textrm{\tiny R}} \cup \mathcal{W}_{\textrm{\tiny L}}$
in a bijective way are constructed in Appendix\;\ref{app-new-inversion}.

\subsection{Mapping the wedge into the diamond}
\label{app-HL-map}

\begin{figure}[t!]
\vspace{-.5cm}
\hspace{-.7cm}
\includegraphics[width=1.08\textwidth]{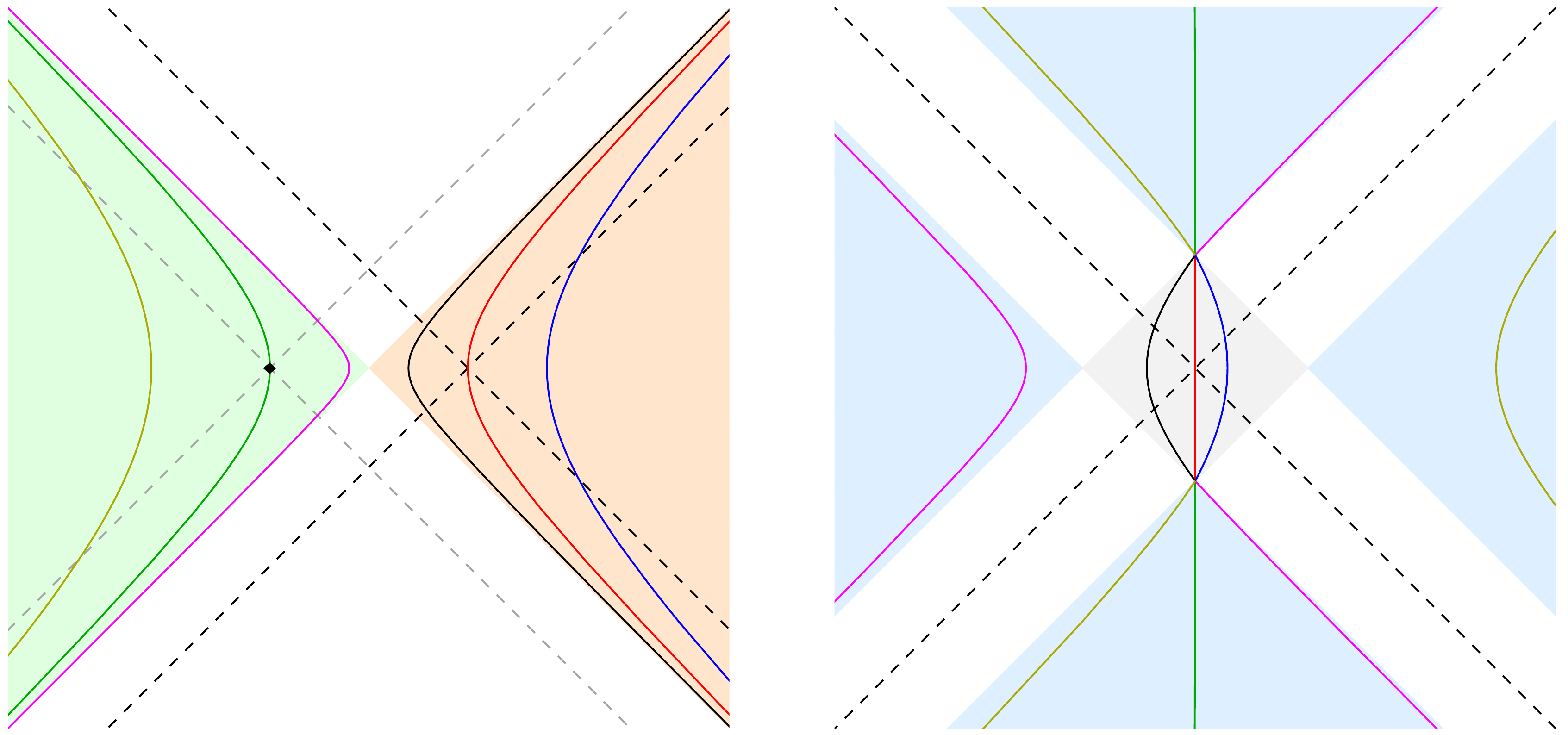}
\vspace{-.2cm}
\caption{Modular trajectories in the double wedge domain
$\widetilde{ \mathcal{W} }_{\textrm{\tiny R}} \cup \widetilde{ \mathcal{W} }_{\textrm{\tiny L}}$ (left panel)
and their images in $\mathcal{D}_A \cup \mathcal{B}_A $ under the map (\ref{HL-map-2dim}) (right panel).
}
\label{fig:HL-map}
\end{figure}

In the Minkowski spacetime parameterised by the coordinates $\widetilde{X} = (\tilde{x} , \tilde{t}\,)$, 
consider the double wedge domain $\widetilde{ \mathcal{W} }_{\textrm{\tiny R}} \cup \widetilde{ \mathcal{W} }_{\textrm{\tiny L}}$
made by the right and left Rindler wedges sharing the origin, 
denoted by $\widetilde{ \mathcal{W} }_{\textrm{\tiny R}}$ and $\widetilde{ \mathcal{W} }_{\textrm{\tiny L}}$ respectively
(in the left panel of Fig.\,\ref{fig:HL-map}, see the green and the orange domains respectively).
We are interested in the following map
\cite{Hislop:1981uh, Haag:1992hx, Casini:2011kv}
\be
\label{HL-map-2dim}
\left\{ 
\begin{array}{l}
\displaystyle
\; t \; =
\frac{ (b-a)\,  \big[\, \tilde{t} - \big( \widetilde{X} \cdot \widetilde{X} \big) \, C_0 \big] }{ 1- 2\, \widetilde{X} \cdot C + \big(\widetilde{X} \cdot \widetilde{X} \big) (C \cdot C ) }
+\frac{b-a}{2}\; C_0
\,=\,
\frac{(b-a)\, \tilde{t}}{ \big( \tilde{x} + \tilde{t} + 1 \big) \big( \tilde{x} - \tilde{t} + 1 \big) }
\\
\rule{0pt}{1.cm}
\displaystyle
\; x\, =
\frac{a+b}{2} +
 \frac{ (b-a)\,  \big[\, \tilde{x} - \big( \widetilde{X} \cdot \widetilde{X} \big) \, C_1 \big] }{ 1- 2\, \widetilde{X} \cdot C + \big(\widetilde{X} \cdot \widetilde{X} \big) (C \cdot C ) }
+\frac{b-a}{2}\; C_1
\,=\,
\frac{a + (a+b)\, \tilde{x} + b\,\big( \tilde{x}^2 - \tilde{t}^{\, 2}\big) }{ \big( \tilde{x} + \tilde{t} + 1 \big) \big( \tilde{x} - \tilde{t} + 1 \big) }
\end{array}
\right.
\ee
where $C = (C_1, C_0) \equiv (-1,0)$ is a constant vector
and the scalar products in the spacetime given by
$\widetilde{X} \cdot \widetilde{X}  \equiv -\,\tilde{t}^{\, 2} + \tilde{x}^2$,
$\widetilde{X} \cdot C  \equiv -\,\tilde{t}\, C_0 + \tilde{x}\, C_1$ and $C\cdot C  \equiv - \,C_0^2 + C_1^2$ occur.

In terms of the light-cone coordinates $ \tilde{u}_\pm \equiv \tilde{x} \pm \tilde{t} $ 
and of the ones defined in (\ref{light-cone coordinates}),
the map (\ref{HL-map-2dim}) takes the following simple form
\be
\label{HL-map-2dim-null-coords}
u_\pm = \frac{a + b\, \tilde{u}_\pm}{ \tilde{u}_\pm + 1} 
\ee
whose inverse reads
\be
\label{HL-map-2dim-null-coords-inv}
\tilde{u}_\pm = \frac{u_\pm - a}{ b - u_\pm  } \,.
\ee

The map (\ref{HL-map-2dim}) sends 
$\widetilde{ \mathcal{W} }_{\textrm{\tiny R}}$ and $\widetilde{ \mathcal{W} }_{\textrm{\tiny L}}$ 
into $\mathcal{D}_A$ and $\mathcal{B}_A$ respectively 
( in the right panel of Fig.\,\ref{fig:HL-map}, see the light grey and light blue region respectively),
described in Sec.\,\ref{sec-mod-traject-1int}
(see Fig.\,\ref{fig:diamond-mod-hyper}).
In particular, (\ref{HL-map-2dim}) relates the 
spatial axes parameterised by $x$ and $\tilde{x}$;
indeed we have 
$t |_{\tilde{t} = 0}=0$ and $x |_{\tilde{t} = 0} = (a +b\, \tilde{x})/( \tilde{x} + 1)$.
The latter expression, which is singular at $\tilde{x} = -1$, 
satisfies $x |_{\tilde{t} = 0} = a $ when $\tilde{x} =0 $
and $x |_{\tilde{t} = 0} \to b^\mp $ as $\tilde{x} \to \pm \infty$.
Moreover, $x |_{\tilde{t} = 0} = (a+b)/2 $ when $\tilde{x} =1 $.
Consider also $\tilde{u}_\pm = 1$ 
(see the dashed black straight lines in the left panel of Fig.\,\ref{fig:HL-map}),
whose images through (\ref{HL-map-2dim-null-coords}) are the 
dashed black straight lines in the right panel of Fig.\,\ref{fig:HL-map}.
These lines identify the partition 
$\mathcal{D}_A = \mathcal{D}_{\textrm{\tiny R}} \cup \mathcal{D}_{\textrm{\tiny L}} \cup \mathcal{D}_{\textrm{\tiny F}} \cup \mathcal{D}_{\textrm{\tiny P}} $
(see Sec.\,\ref{sec-mod-traject-1int}),
that corresponds (through (\ref{HL-map-2dim-null-coords-inv}))
to the partition of $\widetilde{ \mathcal{W} }_{\textrm{\tiny R}}$ 
identified by the dashed black straight lines in the left panel of Fig.\,\ref{fig:HL-map};
indeed, the red curve in $\mathcal{D}_A$ belongs to $\mathcal{D}_{\textrm{\tiny F}} \cup \mathcal{D}_{\textrm{\tiny P}}$.

The singularity in (\ref{HL-map-2dim-null-coords}) corresponds to $\tilde{u}_\pm = -1$,
i.e. to the straight lines $\tilde{t} = \pm \tilde{x} \pm 1$
(see the dashed grey lines in the left panel of Fig.\,\ref{fig:HL-map}, whose intersection is  the black rhombus). 
These lines provide a partition of $\widetilde{ \mathcal{W} }_{\textrm{\tiny L}}$ 
that is mapped through (\ref{HL-map-2dim-null-coords}) into the partition
$\mathcal{B}_A = \mathcal{W}_{\textrm{\tiny R}} \cup \mathcal{W}_{\textrm{\tiny L}} \cup \mathcal{V}_{\textrm{\tiny F}} \cup \mathcal{V}_{\textrm{\tiny P}} $
described in Sec.\,\ref{sec-mod-traject-1int}.
As consistency check, notice  that the green curve in $\mathcal{B}_A$ 
belongs to $\mathcal{V}_{\textrm{\tiny F}} \cup \mathcal{V}_{\textrm{\tiny P}}$.
This picture is  described also in Fig.\,11 of \cite{Leutheusser:2022bgi}.

Combining (\ref{HL-map-2dim-null-coords-inv}) with the Bisognano-Wichmann 
modular Hamiltonian (i.e. the generator of the Lorentz boosts),
one obtained  the parabolic weight function (\ref{velocity_fund}); indeed
\be
V(u_\pm) = \frac{ 2\pi \, \tilde{u}_\pm }{ \partial_{u_{\pm}} \tilde{u}_\pm}
\ee
where the transformation law of $T_\pm(u_\pm)$ under conformal transformations has been employed.


It is well known that  the geometric action of the modular conjugation 
in $\widetilde{ \mathcal{W} }_{\textrm{\tiny R}} \cup \widetilde{ \mathcal{W} }_{\textrm{\tiny L}}$
is given by $(\tilde{x} , \tilde{t}\,) \mapsto ( - \tilde{x} , - \tilde{t}\,)$, namely $\tilde{u}_\pm \mapsto - \tilde{u}_\pm$.
This inversion map commute with (\ref{HL-map-2dim}),
or with (\ref{HL-map-2dim-null-coords}) equivalently. 
Indeed, along each chiral direction, 
from (\ref{HL-map-2dim-null-coords}) and $\mathsf{j}(u)$ in  (\ref{j0-map-def}),
we have 
\be
\frac{a + b\, (- \tilde{u}_\pm)}{ (-\tilde{u}_\pm) + 1} 
=\,
 \mathsf{j}\left( \frac{a + b\, \tilde{u}_\pm}{ \tilde{u}_\pm + 1}  \right) .
\ee


In a Rindler wedge 
(either $\widetilde{ \mathcal{W} }_{\textrm{\tiny R}} $ or $\widetilde{ \mathcal{W} }_{\textrm{\tiny L}} $), 
a modular trajectory whose initial point has light-cone coordinates $(\tilde{u}_{0, +}, \tilde{u}_{0, -})$
is the branch of hyperbola described by 
\be
\label{hyperbolas-in-double-wedge}
\tilde{u}_{\pm} = 
\tilde{\xi}(\pm \tau, \tilde{u}_{0, \pm})
\;\;\;\;\qquad\;\;\;\;
\tilde{\xi}(\pm \tau, \tilde{u}_{0, \pm})
\equiv 
 \tilde{u}_{0, \pm} \, \e^{\pm 2\pi \tau}
\ee
as shown in the left panel of Fig.\,\ref{fig:HL-map}, 
where the various solid curves have  different initial points.

A modular trajectory (\ref{hyperbolas-in-double-wedge}), 
either in $\widetilde{ \mathcal{W} }_{\textrm{\tiny R}}$ or in $\widetilde{ \mathcal{W} }_{\textrm{\tiny L}}$, 
is mapped through (\ref{HL-map-2dim}) into a modular trajectory, 
either in $\mathcal{D}_A$ or in $\mathcal{B}_A$ respectively,
that belongs to the hyperbola $\mathcal{I}_{_P}$ introduced in Sec.\,\ref{sec-mod-traject-1int} 
(see (\ref{mod-traj-tau-line}) and (\ref{mod-hyper})).
Indeed,  plugging (\ref{hyperbolas-in-double-wedge}) into (\ref{HL-map-2dim-null-coords}), 
one obtains (\ref{xi-map-fund}), namely
\be
u_\pm \big|_{\tilde{u}_{\pm} = \tilde{\xi}(\pm \tau, \tilde{u}_{0, \pm})}
= \xi(\pm \tau, u_{0, \pm})
\ee
where the initial values $\tilde{u}_{0, \pm}$ are related to $u_{0, \pm}$ through (\ref{HL-map-2dim-null-coords-inv}).
In Fig.\,\ref{fig:HL-map}
we show various modular trajectories 
either in $\widetilde{ \mathcal{W} }_{\textrm{\tiny R}}$ or in $\widetilde{ \mathcal{W} }_{\textrm{\tiny L}}$ 
and their images under the map (\ref{HL-map-2dim}), 
which are modular trajectories either in $\mathcal{D}_A$ or in $\mathcal{B}_A$ respectively. 
In Fig.\,\ref{fig:HL-map}, a specific colour is associated to any
pair of modular trajectories related through (\ref{HL-map-2dim}).
Notice that the initial points for the red and the green curves 
could be $\widetilde{X} = (1,0) $ and $\widetilde{X} = (-1,0) $ respectively,
in $\widetilde{ \mathcal{W} }_{\textrm{\tiny R}}$ and $\widetilde{ \mathcal{W} }_{\textrm{\tiny L}}$  respectively.


The spacetime distance $d(P_1, P_2) = - \,t_{12}^2 + x_{12}^2 $ between two events $P_1$ and $P_2$
(the notation $y_{12} \equiv y_1 - y_2$ for the differences of certain coordinate $y$ is adopted)
can be written through the spacetime distance $d\big( \widetilde{P}_1, \widetilde{P}_2 \big) \equiv -\, \tilde{t}_{12}^{\, 2} + \tilde{x}_{12}^2$
between their pre-images under (\ref{HL-map-2dim}) as follows
\be
\label{HL-map-conformality}
d(P_1, P_2)
=
\frac{(b-a)^2}{ d\big( \widetilde{P}_1, \widetilde{P}_0 \big)\, d\big( \widetilde{P}_2, \widetilde{P}_0 \big)}\;
d\big( \widetilde{P}_1, \widetilde{P}_2 \big)
\ee
where the  spacetime coordinates of $\widetilde{P}_0$ are  $\widetilde{X} = (-1,0) $.
Since $\widetilde{P}_0 \in \widetilde{ \mathcal{W} }_{\textrm{\tiny L}}$, 
the ratio multiplying $d\big( \widetilde{P}_1, \widetilde{P}_2 \big)$ in the r.h.s. of (\ref{HL-map-conformality})
is strictly positive (and therefore causality is preserved)
for any choice of $ \widetilde{P}_1$ and $ \widetilde{P}_2$ in $\widetilde{ \mathcal{W} }_{\textrm{\tiny R}}$.
However, this ratio is singular when $ \widetilde{P}_1$ or $ \widetilde{P}_2$ is lightlike separated from $\widetilde{P}_0$
(i.e. when at least one of them belongs to the grey dashed lines in the left panel of Fig.\,\ref{fig:HL-map})
and it does not have a definite sign as $ \widetilde{P}_1$ or $ \widetilde{P}_2$ spans $\widetilde{ \mathcal{W} }_{\textrm{\tiny L}}$.
This observation highlights the asymmetric role played by 
$\widetilde{ \mathcal{W} }_{\textrm{\tiny R}}$ or in $\widetilde{ \mathcal{W} }_{\textrm{\tiny L}}$ 
in (\ref{HL-map-2dim}).

\subsection{Alternative inversion maps for the diamond}
\label{app-new-inversion}


Motivated by  the results discussed in Sec.\,\ref{sec-tau-evolution},
in the following we construct two inversion maps sending the diamond $\mathcal{D}_A $ 
onto $\mathcal{W}_{\textrm{\tiny R}} \cup \mathcal{W}_{\textrm{\tiny L}}$
(i.e. the domain of the Minkowski spacetime made by the points that are spacelike separated by all the points in $\mathcal{D}_A $)
in a bijective way. 
%

\begin{figure}[t!]
\vspace{-.5cm}
\hspace{.5cm}
\includegraphics[width=.98\textwidth]{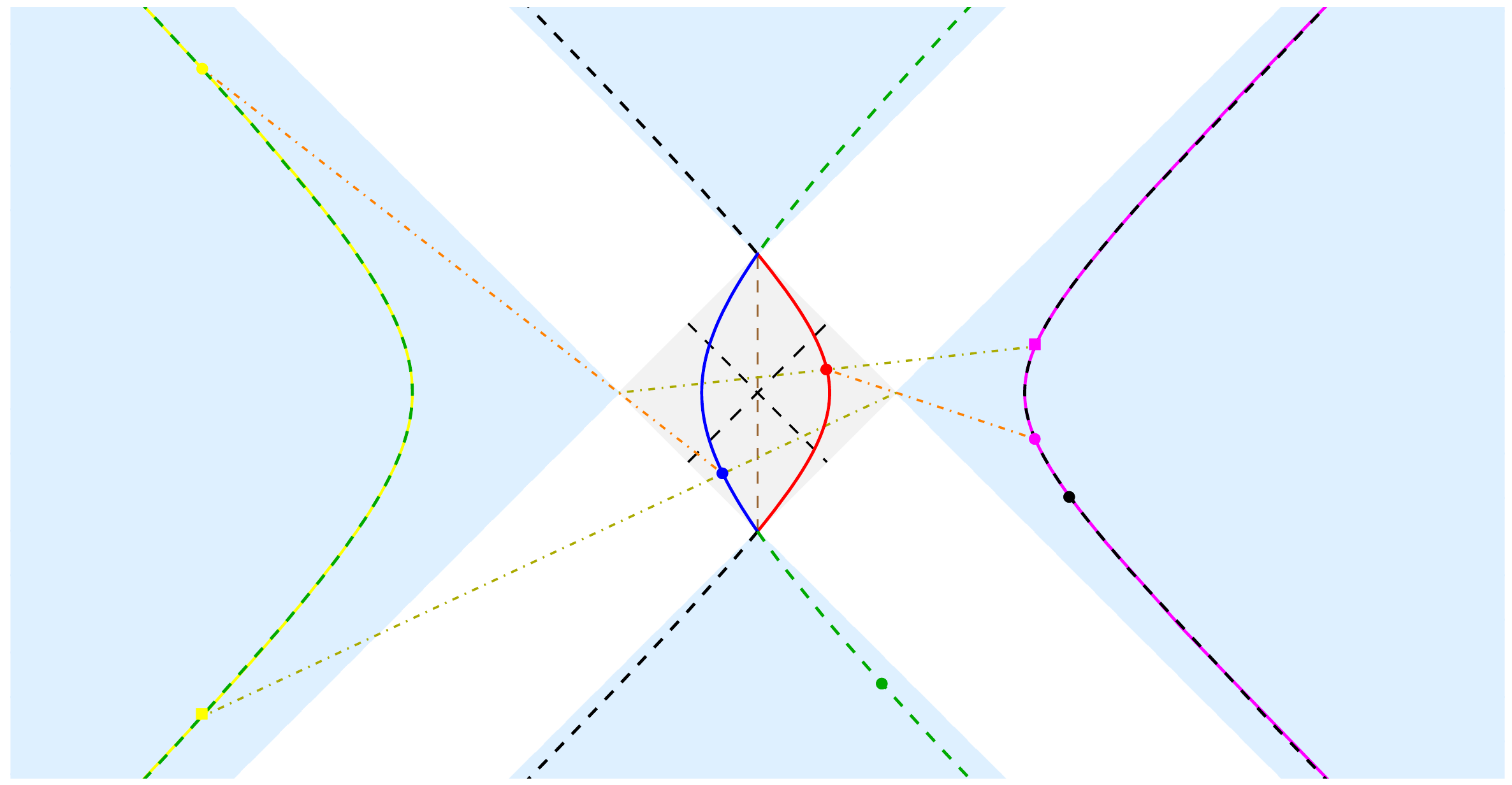}
\vspace{-.2cm}
\caption{
Inversions of the modular trajectories in $\mathcal{D}_A$ (red and blue solid curves)
through either (\ref{inversion-map-V2}) (magenta and yellow solid curves respectively)
or (\ref{mod-conj-traj-tau-line-x})-(\ref{mod-conj-traj-tau-line-t})
(black and green dashed curves respectively).
The latter inversion has been already represented in Fig.\,\ref{fig:diamond-mod-hyper}.
}
\label{fig:new-inversion-non-chiral}
\end{figure}

Consider the partition 
$\mathcal{D}_A = \widetilde{\mathcal{D}}_{\textrm{\tiny R}} \cup \widetilde{\mathcal{D}}_{\textrm{\tiny L}} $
of the diamond $\mathcal{D}_A$ into the triangular spacetime regions 
$\widetilde{\mathcal{D}}_{\textrm{\tiny R}} \equiv \big\{ (x, t) \in \mathcal{D}_A \,;\,  x \geqslant (a+b)/2\big\}$
and $\widetilde{\mathcal{D}}_{\textrm{\tiny L}} \equiv \big\{ (x, t) \in \mathcal{D}_A \,;\,  x \leqslant (a+b)/2\big\}$.
This bipartition of $\mathcal{D}_A $ is identified by the vertical dashed brown segment in Fig.\,\ref{fig:new-inversion-non-chiral}.

We  introduce the mapping $(x,t) \mapsto \big( \bar{x}(x,t) , \bar{t}(x,t) \big)$
sending the generic point $(x,t) \in \mathcal{D}_A$ 
in the point in the spacetime whose spatial and temporal spacetime coordinates are respectively
\be
\label{inversion-map-V2}
\bar{x}(x,t) \equiv \, \mathsf{j}(x)
\;\;\;\;\;\qquad\;\;\;\;\;\;\;
\bar{t}(x,t) \equiv
\left\{ \begin{array}{ll}
\displaystyle
- \frac{(b-a)\, t}{2\big(x -\tfrac{a+b}{2} \big)}
\hspace{1cm}&
x \in \widetilde{\mathcal{D}}_{\textrm{\tiny R}} 
\\
\rule{0pt}{.8cm}
\displaystyle
\frac{(b-a)\, t}{2\big(x -\tfrac{a+b}{2} \big)}
\hspace{1cm}&
x \in \widetilde{\mathcal{D}}_{\textrm{\tiny L}} 
\end{array}
\right.
\ee
where the chiral inversion map (\ref{j0-map-def}) has been employed for the spatial coordinate. 
The map  (\ref{inversion-map-V2}) sends $\widetilde{\mathcal{D}}_{\textrm{\tiny R}} $ onto the right wedge $\mathcal{W}_{\textrm{\tiny R}}$
and $\widetilde{\mathcal{D}}_{\textrm{\tiny L}} $ onto the left wedge $\mathcal{W}_{\textrm{\tiny L}}$ in a bijective way. 
Notice that $\bar{x}(x,t) \to \pm \infty$ as $x \to (\tfrac{a+b}{2})^\pm$ 
and that $\bar{t}(x,0) = 0$ for any $x \in (a,b)$.
Furthermore, the map (\ref{inversion-map-V2}) is idempotent. 
Given a point in $\widetilde{\mathcal{D}}_{\textrm{\tiny R}} $, its image under (\ref{inversion-map-V2}) 
is obtained 
by first imposing the spatial coordinate $\bar{x}(x,t)$ through the chiral inversion map (\ref{j0-map-def})
and then $\bar{t}(x,t)$ is found by constructing 
the two right-angled triangles that share the vertex of $\mathcal{D}_A $ in  $(x,t)= (b,0)$,
with one side along the $x$-axis 
and their hypotenuses along the straight line passing through the initial point in $\widetilde{\mathcal{D}}_{\textrm{\tiny R}} $
and the vertex in $(b,0)$. 
For the red dot in Fig.\,\ref{fig:diamond-mod-hyper},
this construction provides the magenta dot in $\mathcal{W}_{\textrm{\tiny R}}$
and the orange dot-dashed segment connecting them is made by the two hypotenuses 
of the two above mentioned right-angled triangles.
A slight modification of the above geometric construction, 
which consists in just replacing the vertex in  $(x,t)= (b,0)$ with the one in  $(x,t)= (a,0)$,
allows to send $\widetilde{\mathcal{D}}_{\textrm{\tiny L}} $ onto $\mathcal{W}_{\textrm{\tiny L}}$ in a bijective way 
(in Fig.\,\ref{fig:diamond-mod-hyper}, see e.g. the blue dot and its image denoted by the yellow dot,
that are connected by the orange dot-dashed segment).

A different map that also sends
$\widetilde{\mathcal{D}}_{\textrm{\tiny R}} $ onto $\mathcal{W}_{\textrm{\tiny R}}$ in a bijective way 
can be written
by imposing the same spatial coordinate $\bar{x}(x,t)$
and considering, instead, 
the two right-angled triangles that share the vertex of $\mathcal{D}_A $ in  $(a,0)$
and having one side along the $x$-axis again
but their hypotenuses along the straight line passing through the initial point in  $\widetilde{\mathcal{D}}_{\textrm{\tiny R}} $ 
and the vertex in $(a,0)$.
Then, $\widetilde{\mathcal{D}}_{\textrm{\tiny L}} $ is sent onto $\mathcal{W}_{\textrm{\tiny L}}$ in a bijective way
by replacing the vertex in  $(a,0)$ with the one in  $(b,0)$ in the previous construction. 
Two examples correspond to the dark yellow dot-dashed segments in Fig.\,\ref{fig:diamond-mod-hyper}.
This geometric construction provides the map $(x,t) \mapsto \big( \bar{x}(x,t) , -\,\bar{t}(x,t) \big)$.
In Fig.\,\ref{fig:diamond-mod-hyper}, the images of the red and the blue dots under this map
are the magenta and the yellow squares respectively. 

We remark that these inversion maps cannot be written in terms of a chiral map, and therefore are not conformal;
in contrast with the inversion map in the spacetime given by $(u_+, u_{-}) \mapsto ( \, \mathsf{j}(u_+) \,, \, \mathsf{j}(u_{-}) )$,
i.e. by applying (\ref{j0-map-def}) along each chiral direction, 
which has been discussed in Sec.\,\ref{sec-mod-traject-1int}.

It is worth comparing the images  in the spacetime
of a modular trajectory in $\mathcal{D}_A $ (see (\ref{mod-traj-tau-line}))
under (\ref{inversion-map-V2}) and under $(u_+, u_{-}) \mapsto \big( \, \mathsf{j}(u_+) \,, \, \mathsf{j}(u_{-}) \big)$.
In Fig.\,\ref{fig:new-inversion-non-chiral}, two modular trajectories in $\mathcal{D}_A $  are shown 
(see the red and blue solid lines), whose initial points are the dots having the corresponding color. 
The geometric construction discussed above for the inversion map (\ref{inversion-map-V2})
and illustrated by the dot-dashed orange segments in Fig.\,\ref{fig:new-inversion-non-chiral} for the initial points
can be extended to the whole red and blue solid lines, 
finding the solid magenta line and the solid yellow line respectively.
The dashed black and dashed green curves in Fig.\,\ref{fig:new-inversion-non-chiral}
denote the images of the red and blue solid lines respectively 
under (\ref{mod-conj-traj-tau-line-x})-(\ref{mod-conj-traj-tau-line-t})
and their initial points are the black and green dots respectively
(see also Fig.\,\ref{fig:diamond-mod-hyper}).

These examples show that, remarkably, 
the image of  a modular trajectory in $\mathcal{D}_A $
under (\ref{inversion-map-V2}) coincides with the branches of the curve 
obtained from (\ref{mod-conj-traj-tau-line-x})-(\ref{mod-conj-traj-tau-line-t}) 
and contained in $\mathcal{W}_{\textrm{\tiny R}} \cup \mathcal{W}_{\textrm{\tiny L}}$
(see either the red dashed arc or the purple dashed arc in Fig.\,\ref{fig:diamond-mod-hyper}).
Indeed, the image of a modular trajectory in $\mathcal{D}_A $ with initial point $P \in \mathcal{D}_A $
under (\ref{inversion-map-V2}) belongs to the corresponding hyperbola $\mathcal{I}_{_P}$
(see (\ref{mod-hyper})-(\ref{mod-hyper-parameters}));
namely  (\ref{mod-traj-tau-line}) and (\ref{inversion-map-V2}) satisfy 
\be
\label{mod-hyper-bar}
\big[ \bar{x}\big( x(\tau)  , t(\tau) \big) - x_0 \big]^2 - \bar{t}\big( x(\tau)  , t(\tau) \big) ^2 = \kappa_0^2 \,.
\ee

\begin{figure}[t!]
\vspace{-.5cm}
\hspace{-.65cm}
\includegraphics[width=1.075\textwidth]{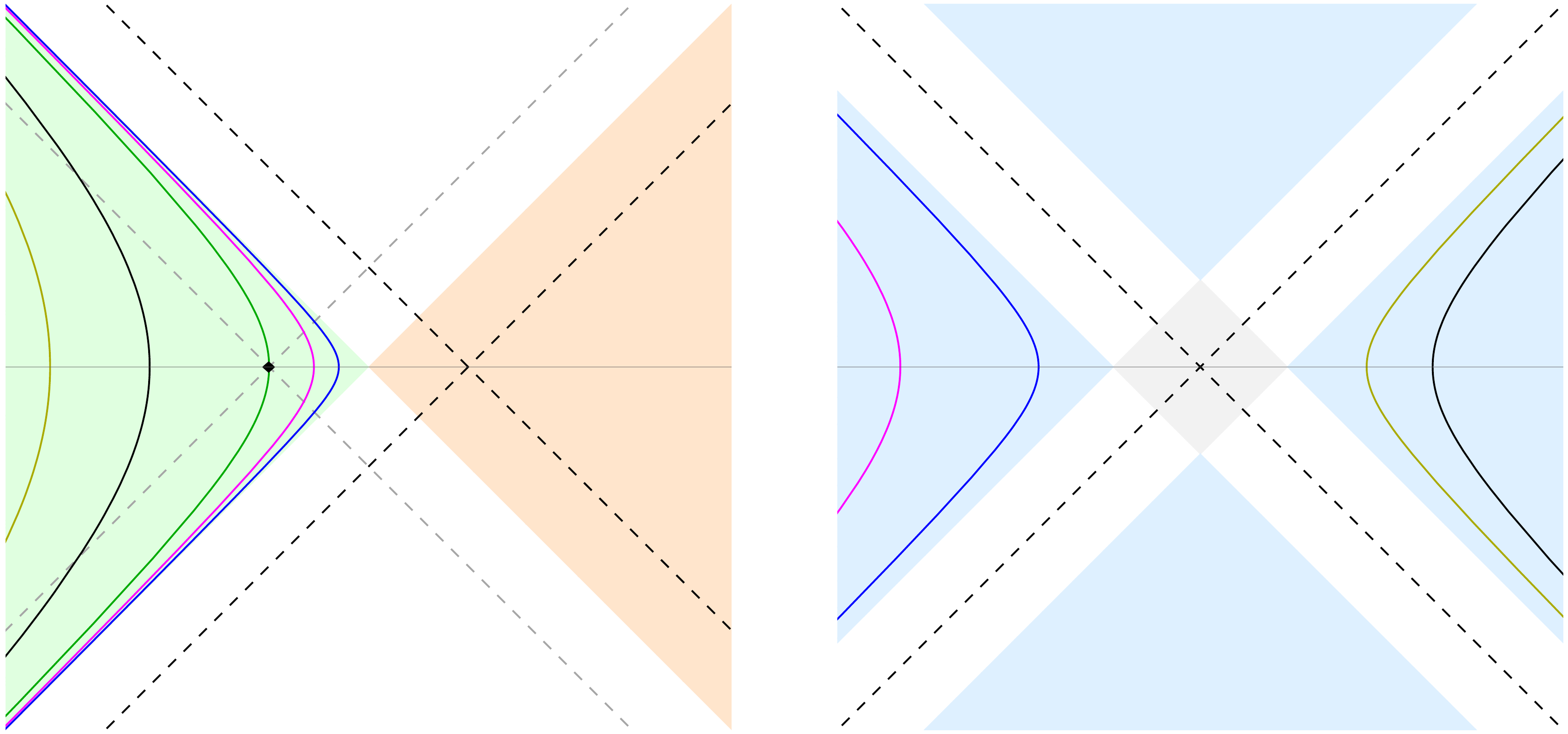}
\vspace{-.2cm}
\caption{
Modular trajectories in the left Rindler wedge $\widetilde{ \mathcal{W} }_{\textrm{\tiny L}} $ (left panel)
and their images in $\mathcal{W}_{\textrm{\tiny R}} \cup  \mathcal{W} _{\textrm{\tiny L}}$ under the map (\ref{inversion-map-out})
(right panel).
}
\label{fig:HL-map-bis}
\end{figure}


The inversion map (\ref{inversion-map-V2}) 
does not preserve the sign of the spacetime distance.
For instance, consider for simplicity the points in $\mathcal{D}_A$ having $t= x- (a+b)/2$,
which are lightlike separated and 
belong to the black dashed segment in Fig.\,\ref{fig:new-inversion-non-chiral} with positive slope. 
Given the points $P_1$ and $P_2$ in this class, 
with spatial coordinate $x_1$ and $x_2$ respectively,
and lying either in $\widetilde{\mathcal{D}}_{\textrm{\tiny R}}$ or in $\widetilde{\mathcal{D}}_{\textrm{\tiny L}}$,
their images under (\ref{inversion-map-V2}), 
denoted by  $\bar{P}_1$ and $\bar{P}_2$ respectively, 
are spacelike separated; 
indeed $d( \bar{P}_{1} , \bar{P}_{2} ) = x_{12}^2$.


The maps discussed in Appendix\;\ref{app-HL-map} and in this Appendix can be employed to construct 
a map sending the left Rindler wedge $\widetilde{ \mathcal{W} }_{\textrm{\tiny L}} $ 
(see the green region in the left panel of Fig.\,\ref{fig:HL-map}) 
onto $ \mathcal{W}_{\textrm{\tiny R}} \cup  \mathcal{W} _{\textrm{\tiny L}}$ in a bijective way.
Consider the bipartition of $\widetilde{ \mathcal{W} }_{\textrm{\tiny L}} $ given by 
$ \widetilde{ \mathcal{W} }_{\textrm{\tiny L}} = \widetilde{ \mathcal{W} }_{\textrm{\tiny L}}^{^{\textrm{\tiny $\,<$}}} \cup \widetilde{ \mathcal{W} }_{\textrm{\tiny L}}^{^{\textrm{\tiny $\,>$}}}$,
where $ \widetilde{ \mathcal{W} }_{\textrm{\tiny L}}^{^{\textrm{\tiny $\,<$}}}$ and $ \widetilde{ \mathcal{W} }_{\textrm{\tiny L}}^{^{\textrm{\tiny $\,>$}}}$ 
are identified by the branch of the hyperbola $\tilde{x}^2 - \tilde{t}^{\, 2}  = 1 $ in $\widetilde{ \mathcal{W} }_{\textrm{\tiny L}} $
(see the solid green curve in the left panels of Fig.\,\ref{fig:HL-map} and Fig.\,\ref{fig:HL-map-bis})
and correspond to the region respectively on the left and on the right of this curve.
A map sending $ \widetilde{ \mathcal{W} }_{\textrm{\tiny L}}^{^{\textrm{\tiny $\,<$}}}$ and $ \widetilde{ \mathcal{W} }_{\textrm{\tiny L}}^{^{\textrm{\tiny $\,>$}}}$  
onto $  \mathcal{W}_{\textrm{\tiny R}}$ and $  \mathcal{W} _{\textrm{\tiny L}}$  respectively in a bijective way
can be constructed by composing (in this order) the reflection $(\tilde{x} , \tilde{t}\,)  \mapsto ( - \tilde{x} , - \tilde{t}\,) $, the map (\ref{HL-map-2dim})  and (\ref{inversion-map-V2});
hence it act as 
$ \widetilde{ \mathcal{W} }_{\textrm{\tiny L}} \to \widetilde{ \mathcal{W} }_{\textrm{\tiny R}} \to \mathcal{D}_A \to  \mathcal{W}_{\textrm{\tiny R}} \cup  \mathcal{W} _{\textrm{\tiny L}}$.
This composition provides a map acting on a given point $(\tilde{x} , \tilde{t}\,) \in \widetilde{ \mathcal{W} }_{\textrm{\tiny L}} $ as follows
\be
\label{inversion-map-out}
x\big( \tilde{x},\tilde{t}\, \big) \equiv \, b -\frac{(b-a)\, ( \tilde{x}-1)}{ \tilde{x}^2 - \tilde{t}^{\, 2} -1}
\;\;\;\qquad\;\;\;\;\;\;
x\big( \tilde{x},\tilde{t}\, \big) 
\equiv
\left\{ \begin{array}{ll}
\displaystyle
- \frac{(b-a)\, \tilde{t} }{ \tilde{x}^2 - \tilde{t}^{\, 2} -1 }
\hspace{1cm}&
x \in \widetilde{ \mathcal{W} }_{\textrm{\tiny L}}^{^{\textrm{\tiny $\,<$}}}
\\
\rule{0pt}{.9cm}
\displaystyle
\frac{(b-a)\, \tilde{t} }{ \tilde{x}^2 - \tilde{t}^{\, 2} -1 }
\hspace{1cm}&
x \in \widetilde{ \mathcal{W} }_{\textrm{\tiny L}}^{^{\textrm{\tiny $\,>$}}} \,.
\end{array}
\right.
\;\;
\ee
In Fig.\,\ref{fig:HL-map-bis} we show some modular trajectories in $\widetilde{ \mathcal{W} }_{\textrm{\tiny L}} $ (left panel)
and their images in $\mathcal{W}_{\textrm{\tiny R}} \cup  \mathcal{W} _{\textrm{\tiny L}}$ 
(having the same colour of the original curve) under the map (\ref{inversion-map-out}).
This mapping  does not preserve the sign of the spacetime distance because  (\ref{inversion-map-V2}) is involved in its construction.

\section{Interval in the circle, vacuum state: Modular inversion}
\label{app-1int-circle-vacuum}


In this Appendix we consider a finite volume case (with periodic boundary conditions),
extending the results discussed in
Sec.\,\ref{sec-1int-chiral-dir-mod-evo} and Sec.\,\ref{sec-mod-traject-1int} for the infinite line.

A well known way to compactify a chiral direction is based on the Cayley map,
which relates the real line to the unit circle with one point removed (see e.g. \cite{Hollands:2019hje}). 
This map can be defined as $u \mapsto z=\tfrac{1+ \ri \, u}{1- \ri \, u}$ where $u \in \RR$ and $z\in \mathbb{S}\setminus \{P_0\}$,
being $P_0$ the point corresponding to $\theta=\pi$ on 
the unit circle $ \mathbb{S}$. Parameterising $ \mathbb{S}$ through the complex number $z=\e^{2\pi \ri v/L} $ with  $v \sim v+L$,
where $L$ corresponds to the compactification parameter, 
the Cayley map and its inverse become respectively
\be
\label{cayley-one-dim}
\e^{2\pi \ri v/L} = \frac{1+ \ri \, u}{1- \ri \, u}
\;\;\;\; \qquad\;\;\;\;
u \,=\, \ri\; \frac{1 - \e^{2\pi \ri v/L} }{ 1 + \e^{2\pi \ri v/L} } \,=\, \tan(\pi v/L)\,\equiv \, \mathcal{C}(v) \,.
\ee
Alternatively, 
following the standard approach to two-dimensional CFT in Euclidean signature \cite{Ginsparg:1988ui},
one first introduces the periodic identification $v \sim v+L$ on the real line 
and then employs the exponential map $v \mapsto \e^{2\pi \ri v/L}$.
%

\begin{figure}[t!]
\vspace{-.6cm}
\hspace{-.85cm}
\includegraphics[width=1.1\textwidth]{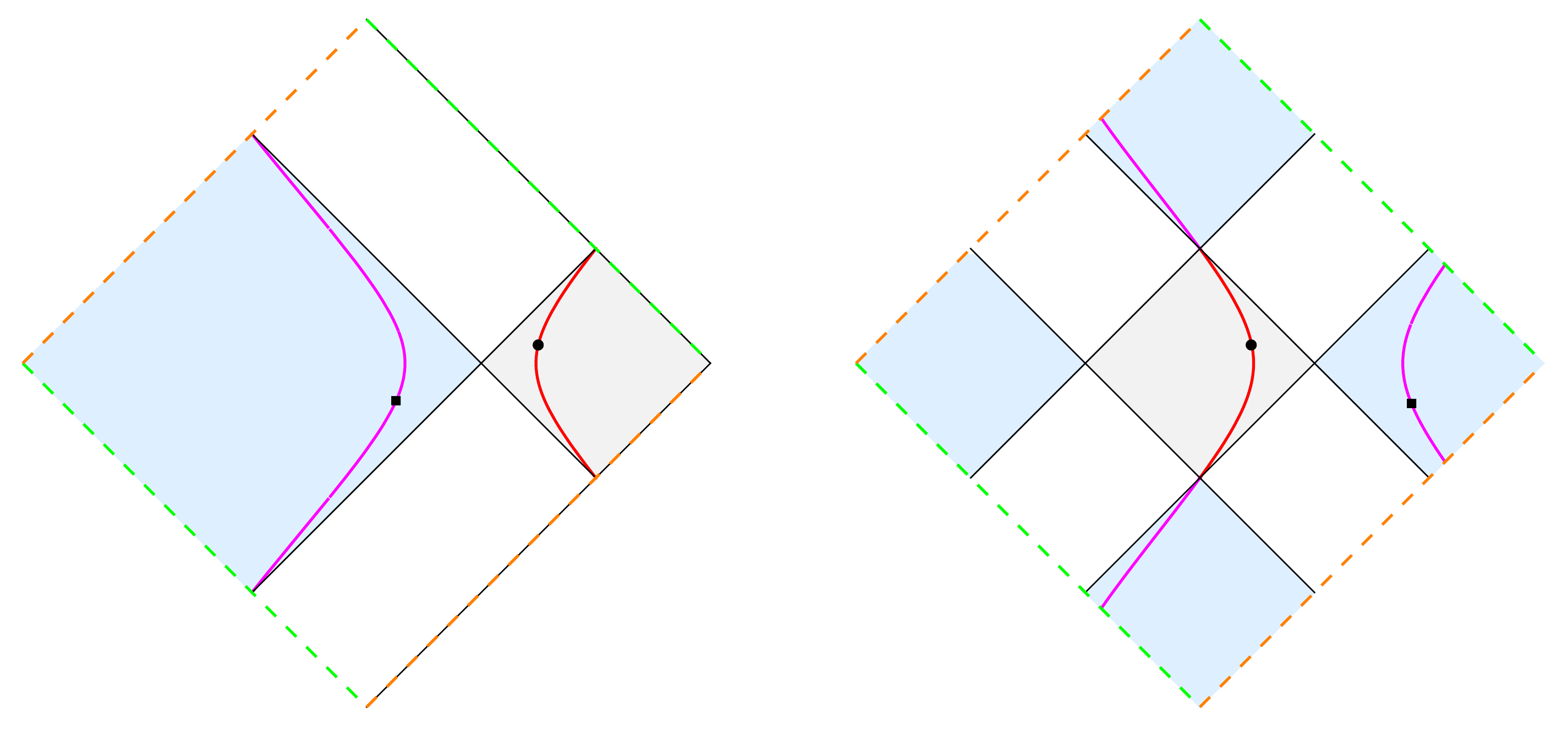}
\vspace{-.2cm}
\caption{
Modular trajectories (red and magenta solid curves)
generated by (\ref{fmh-circle})
and related through the inversion map constructed by 
applying (\ref{j0-map-circle-def}) along each chiral direction.
}
\label{fig:diamond-circle}
\end{figure}

In the following analysis both the chiral directions are compactified in a circle of length $L$.
The rhombi in both the panels of Fig.\,\ref{fig:diamond-circle} represent the entire spacetime and
the dashed edges having the same color are identified.
Considering along each chiral direction the bipartition given by the interval $A$ and its complement $B$ on the circle of length $L$
parameterised by $v_\pm \in (-L/2\,, L/2)$,
the modular Hamiltonian of $A$ and the corresponding full modular Hamiltonian  read respectively  
\cite{Wong:2013gua, Cardy:2016fqc}
\be 
\label{KA-def-u-pm-circle}
K_A  
= 
\int_A  V_L(v_+) \left[\, T_+(v_+) + \frac{\pi c}{12 L^2}\,  \right]  \!\rd v_+
+ 
\int_A  V_L(v_-) \left[\, T_-(v_-)  + \frac{\pi c}{12 L^2}\,  \right]  \! \rd v_-
\ee
and 
\bea
\label{fmh-circle}
K
\equiv
K_{A} \otimes \boldsymbol{1}_B - \boldsymbol{1}_A \otimes K_{B} 
&=&
\int_{-L/2}^{L/2} V_L(v_+) \left[\, T_+(v_+) + \frac{\pi c}{12 L^2}\,  \right]   \rd v_+
\nn
\\
\rule{0pt}{.7cm}
& &
+ 
\int_{-L/2}^{L/2}  V_L(v_-) \left[\, T_-(v_-) + \frac{\pi c}{12 L^2}\,  \right]   \rd v_-
\eea
where the weight function $V_L(v)$ is
\be
\label{velocity_fund-circle}
V_L(v) =2L\, \frac{\sin[\pi(b-v)/L]\, \sin[\pi(v-a)/L]}{\sin[\pi(b-a)/L]}=\frac{1}{w_L'(v)} 
\;\;\;\qquad\;\;\;
v \in A
\ee
being $w_L(v)$ defined as
\be
\label{w-function-def-circle}
w_L(v) \equiv \frac{1}{2\pi}\,\log \!\left(\! - \frac{\sin[\pi(v-a)/L]}{\sin[\pi(v-b)/L]} \right)  .
\ee


The weight function (\ref{velocity_fund-circle}) is related to the corresponding weight function for the interval on the line given in (\ref{velocity_fund}) 
as follows
\be
\label{VL-from-V}
V_L(v) 
= \frac{\widetilde{V}\big( \e^{2\pi \ri v/L}\big)}{\partial_v\big( \e^{2\pi \ri v/L}\big)}
= \frac{\widehat{V}  \big( \mathcal{C}(v)\big)}{ \mathcal{C}'(v) }
\ee
where in the first equality we have employed the exponential map 
and $\widetilde{V}(v)$ is defined as (\ref{velocity_fund}) with $a$ and $b$ replaced by $\e^{2\pi \ri a/L}$ and $\e^{2\pi \ri b/L}$ respectively,
while the last expression is obtained through the Cayley map (\ref{cayley-one-dim})
and $\widehat{V}(v)$ is defined as (\ref{velocity_fund}) with  $a$ and $b$ replaced by $\mathcal{C}(a)$ and $\mathcal{C}(b)$ respectively. 

The bipartitions induced by $A$ in each chiral direction determine the partition of the compact spacetime 
provided by the black solid thin straight lines in each panel of Fig.\,\ref{fig:diamond-circle}.
The light grey domain corresponds to the diamond $\mathcal{D}_A$.
Since periodic boundary conditions are imposed along each chiral direction, 
the two panels in Fig.\,\ref{fig:diamond-circle} are equivalent.

The modular evolution generated by (\ref{fmh-circle}) are  described through the following function \cite{Mintchev:2020uom}
\be
\label{xi-map-fund-circle}
\xi_L(\tau,v) 
\equiv \frac{L}{2\pi \, \ri} \, \log\! \bigg( \frac{ \e^{\pi \ri  (b+a)/L} + \e^{ 2\pi \ri  b/L}\, \e^{2\pi w_L(v)+2\pi \tau} }{ \e^{\pi \ri (b-a)/L} +  \e^{2\pi w_L(v)+2\pi \tau}}   \bigg)
\ee
whose infinite volume limit $L \to +\infty$ gives (\ref{xi-map-fund}), as expected.
Notice that 
\bea
\label{exp-xi-L-1}
\e^{2\pi \ri  \xi_L(\tau,v)  /L} 
&=&
\frac{
\big( \e^{ 2\pi \ri  b/L} - \e^{ 2\pi \ri  v/L}  \big)\,\e^{ 2\pi \ri  a/L} + \big(\e^{ 2\pi \ri  v/L} - \e^{ 2\pi \ri  a/L} \big)\,\e^{ 2\pi \ri  b/L} \,\mathrm{e}^{2\pi \tau}
}{
\big( \e^{ 2\pi \ri  b/L}  - \e^{ 2\pi \ri  v/L} \big)+ \big( \e^{ 2\pi \ri  v/L} - \e^{ 2\pi \ri  a/L} \big)\, \mathrm{e}^{2\pi \tau}} 
\\
\label{exp-xi-L-2}
\rule{0pt}{.75cm}
&=&
\e^{\pi \ri (a+b)/L}\;
\frac{ \e^{\pi \ri a/L} \sin\!\big[\pi(b-v)/L\big] + \e^{\pi \ri b/L} \sin\!\big[\pi(v-a)/L\big]\, \e^{2\pi \tau} 
}{ 
\e^{\pi \ri b/L} \sin\!\big[\pi(b-v)/L\big] + \e^{\pi \ri a/L} \sin\!\big[\pi(v-a)/L\big]\, \e^{2\pi \tau} 
 }
 \hspace{1cm}
\eea
where  the r.h.s. of (\ref{exp-xi-L-1}) 
can be obtained as the r.h.s. of $\xi(\tau,v)$ in  (\ref{xi-map-fund})
by replacing $a$, $b$ and $v$ with  $\e^{ 2\pi \ri  a/L} $, $\e^{ 2\pi \ri  b/L} $ and $\e^{ 2\pi \ri  v/L} $ respectively. 

A modular trajectory in $\mathcal{D}_A$ 
whose initial point at $\tau=0$ has light-cone coordinates $(v_+, v_-)$
can be written through $\xi_L(\tau,v) $ in  (\ref{xi-map-fund-circle})
and its spacetime coordinates are
\be
\label{mod-traj-tau-circle}
x(\tau) = \frac{\xi_L(\tau, v_+) + \xi_L(-\tau, v_-) }{2}
\;\;\;\qquad\;\;\;
t(\tau) = \frac{\xi_L(\tau, v_+) - \xi_L(-\tau, v_-) }{2} \,.
\ee
%
The solid red and magenta curves in both panels of Fig.\,\ref{fig:diamond-circle} are two modular trajectories 
whose  initial points correspond to the black dot and to the square dot respectively.


In this setup, the inversion map along each chiral direction is
defined as follows \cite{Mintchev:2022fcp}
\be
\label{j0-map-circle-def}
\mathsf{j}_L(v)\equiv 
\frac{L}{2\pi \ri} \, \log
\!\left(
\frac{ \e^{2\pi \ri b/L} + \e^{2\pi \ri a/L} }{2} 
+ \frac{  \big[ \big( \e^{2\pi \ri b/L} - \e^{2\pi \ri a/L} \big)/2 \big]^2 }{ \e^{2\pi \ri v/L} - \big( \e^{2\pi \ri b/L} + \e^{2\pi \ri a/L} \big)/2 }
\right)
\ee
which is a bijective and idempotent function sending $A$ onto $B$ with negative derivative
\be
\label{der-j0-map-circle}
\mathsf{j}'_L(v)
=\,
-\, \frac{\sin^2[\pi (b-a)/L] }{ \big| \sin[\pi (b-v)/L] - \e^{\pi \ri (b-a)/L}  \sin[\pi (v-a)/L] \, \big|^2 }  \,.
\ee
The expressions in (\ref{velocity_fund-circle}), (\ref{j0-map-circle-def}) and (\ref{der-j0-map-circle}) are related as follows
\be
\mathsf{j}_L'(v) \, V_L(v) = V_L(\, \mathsf{j}_L(v)) \,.
\ee
We remark that the maps (\ref{xi-map-fund-circle}) and (\ref{j0-map-circle-def}) commute, namely they satisfy
\be
\label{id-xi-j-circle}
\mathsf{j}_L \big( \xi_L(\tau, v) \big)
=
\xi_L\big(\tau,\mathsf{j}_L (v)\big)
\ee
whose infinite volume limit gives (\ref{id-xi-j-line}), as expected.

In the compact spacetime introduced above, the inversion map  
is obtained by applying (\ref{j0-map-circle-def}) along each chiral direction,
i.e. $(v_+, v_{-}) \mapsto ( \, \mathsf{j}_L(v_+) \,, \, \mathsf{j}_L(v_{-}) )$.
This map relates the two modular trajectories 
corresponding to the red and magenta solid curves in Fig.\,\ref{fig:diamond-circle}.

Taking  the infinite volume limit $L \to \infty$ in the right panel of Fig.\,\ref{fig:diamond-circle},
$\mathcal{D}_A$ remains fixed and one obtains the setup described in Fig.\,\ref{fig:diamond-mod-hyper}.
Instead, taking the infinite volume limit $L \to \infty$ in the left panel of Fig.\,\ref{fig:diamond-circle},
$\mathcal{D}_A$ and the light blue diamond become the right and the left Rindler wedges respectively
(see e.g. the left panel of Fig.\,\ref{fig:HL-map}).

We remark that in the above analysis the universal covering discussed in \cite{Segal:1971aa, Luscher:1974ez, Brunetti:1992zf} 
and mentioned in the final part of Sec.\,\ref{sec:intro} has not been considered.

\section{A formula for the derivative of the Dirac delta}
\label{app-derivative-Dirac-delta}

In this Appendix we report a derivation of the following formula about the
derivative of the Dirac delta function
\bea
\label{delta-prime-res-app}
G(x)\,
\frac{\partial }{\partial f(x)}\,\delta\big(f(x)\big) 
&=&
\frac{G(x)}{f'(x)}\,
\sum_j \frac{1}{\big| f'(x_j)  \big|} \,\partial_x \delta(x-x_j)
\\
\rule{0pt}{.6cm}
& & \hspace{-1.7cm}
=\,
\sum_j 
\bigg(\,
\frac{ G(x_j) }{ f'(x_j) \, \big| f'(x_j) \big| } \,\partial_x \delta(x-x_j)
- 
\frac{ G'(x_j)\, f'(x_j) - G(x_j)\, f''(x_j)}{ \big| f'(x_j) \big|^3 } \; \delta(x-x_j)
\bigg)
\nn
\eea
which has been used
to explore the commutation relations of the modular evolutions
of the chiral components of the conserved current in CFT
(see Sec.\,\ref{sec-1int-commutators})
and of the chiral density operators for the free massless Dirac field 
(see Sec.\,\ref{sec-anti-comm-dirac}).

Consider the integral representation of the Dirac delta and its derivative given by 
\be
\label{dirac-delta-int-reps}
\int_{-\infty}^{\infty} \e^{-\ri p x}\, \rd p \,=\, 2\pi \, \delta(x)
\;\;\;\;\;\;\qquad\;\;\;\;\;
\int_{-\infty}^{\infty} p\; \e^{-\ri p x}\, \rd p \,=\, 2\pi \, \ri\, \delta'(x)
\ee
respectively,  where the second expression can be found by taking the derivative with respect to $x$ of the first one. 
By employing the first identity in (\ref{dirac-delta-int-reps})  for $\delta(f(x))$ 
and then taking the derivative with respect to $x$ of the resulting expression, one gets
\be
\label{delta-prime-sequence}
 \frac{\partial }{\partial f(x)} \, \delta \big( f(x) \big)
=
\frac{1}{2\pi  \ri } \int_{-\infty}^{\infty} p\, \e^{-\ri p f(x)}\, \rd p 
=
\frac{1}{f'(x)} \; \partial_x \,\delta \big( f(x) \big)
=
\frac{1}{f'(x)} \;  \sum_j \frac{1}{ \big| f'(x_j) \big| } \, \partial_x\delta(x-x_j)
\ee
where we used the second expression in (\ref{dirac-delta-int-reps}) in the first equality 
and the well known formula $\delta\big(f(x)\big) = \sum_j \tfrac{1}{ | f'(x_j) | } \,\delta(x-x_j)$ 
(where the sum is over the zeros $x_j$ of $f(x)$ such that $f'(x_j)  \neq 0$) in the last equality.
The last expression in (\ref{delta-prime-sequence}) 
provides the first equality in (\ref{delta-prime-res-app}).

For a generic test function $g(x)$ we have that 
\be
\label{id-app-der-dirac}
\int_{-\infty}^{\infty} g(x)\,F(x) \, \partial_x  \delta(x-y) \, \rd x
\,=\,
- \int_{-\infty}^{\infty} g'(x)\,F(x) \, \delta(x-y) \, \rd x
- \int_{-\infty}^{\infty} g(x)\,F'(x) \, \delta(x-y) \, \rd x \,.
\ee
The final  expression in (\ref{delta-prime-res-app}) is obtained
by applying (\ref{id-app-der-dirac}) in the special case where  $F(x) = G(x) / f'(x)$ 
to each term of the sum over $j$ in the expression 
found through the first equality of the same equation.

\section{Two intervals: Modular flow and modular correlators}
\label{app-mod-flow-2int}


In the setup described in Sec.\,\ref{sec-2int-Dirac},
we revisit the derivations of 
the modular flow of the chiral fermions (Appendix\;\ref{app-mod-flow-2int-psi}) \cite{Casini:2009vk}
and their modular two-point functions (Appendix\;\ref{app-mod-corr-psi}) \cite{Longo:2009mn},
by adapting the analysis described in Appendix C of \cite{Mintchev:2020uom} 
to the case of a generic configuration of two disjoint intervals on the line.
The  modular two-point functions for the chiral density fields
$\varrho_\pm \equiv \; : \!\psi_\pm^\ast \psi_\pm \!:\, $ are also derived (Appendix\;\ref{app-mod-flow-density}).

\subsection{Modular flow of the chiral fermionic field  }
\label{app-mod-flow-2int-psi}


Consider the massless Dirac field in its ground state
and on the line bipartite by $A =A_1 \cup A_2$,
where $A_j =[a_j, b_j]$  with $j \in \{1,2\}$.
The modular flow of this field is generated by (\ref{mod-ham-2int-KA});
hence  it can be obtained as the solution of the following system of PDE's
\be
\label{Dirac-mod-flow-app}
\frac{d}{d\tau}
\left( \begin{array}{c}
\psi_+ (\tau, u_+) \\
\psi_+ (\tau, u_{+, \textrm{c}} ) \\
\psi_{-}(\tau, u_{-}) \\
\psi_{-}(\tau, u_{-,\textrm{c}}) 
\end{array} \right)
=\,
\big[  \boldsymbol{V}(u_+)  \oplus \big( \! - \boldsymbol{V}(u_{-}) \big)  \big]
\left( \begin{array}{c}
\psi_+ (\tau, u_+) \\
\psi_+ (\tau, u_{+, \textrm{c}} ) \\
\psi_{-}(\tau, u_{-}) \\
\psi_{-}(\tau, u_{-,\textrm{c}}) 
\end{array} \right)
\ee
where the matrix operators in the r.h.s. are 
\be
\boldsymbol{V}(u) \equiv 
\Bigg( \begin{array}{cc}
V_{\textrm{\tiny loc}}(u)\, \partial_u +\tfrac{1}{2} \, \partial_u V_{\textrm{\tiny loc}}(u) \;\;& -  \,V_{\textrm{\tiny biloc}}(u)
\\
\rule{0pt}{.5cm}
 -  V_{\textrm{\tiny biloc}}(u_{\textrm{c}}) \;\;& 
 V_{\textrm{\tiny loc}}(u_{\textrm{c}})\, \partial_{u_{\textrm{c}}} +\tfrac{1}{2} \, \partial_{u_{\textrm{c}}} \!V_{\textrm{\tiny loc}}(u_{\textrm{c}}) 
\end{array} \Bigg)
\ee
in terms of the weight functions (\ref{velocity_fund-2int}) and $u_{\pm, \textrm{c}} \equiv \mathcal{C}(u_\pm)$ (see (\ref{u-conj-def})).


The block diagonal structure of (\ref{Dirac-mod-flow-app}) allows to focus separately on the chiral fields $\psi_\pm$,
whose modular evolutions can be written in terms of the following doublet 
\be
\label{Psi-doublet-def-app}
\Psi(\tau, u) \equiv 
\bigg( \begin{array}{c}
\psi(\tau, u) \\
\psi(\tau, u_{\textrm{c}}) 
\end{array} \bigg)
\ee
where the notation $\psi =\psi_\pm$ is adopted throughout this analysis to enlighten the formulas.
Let us recall that, since $- \boldsymbol{V}(u_{-}) $ occurs in the r.h.s. of (\ref{Dirac-mod-flow-app}),
the modular evolution parameter to employ when  $u=u_{-}$ is $-\tau$.
The doublet (\ref{Psi-doublet-def-app})  solves  the following system of PDE's
\be
\label{system-pde-V}
\frac{d}{d\tau} \Psi(\tau, u) =  \boldsymbol{V}(u) \, \Psi(\tau, u) 
\;\;\;\qquad\;\;\;
 \Psi(\tau =0, u) = \Psi(u) 
\ee
where $ \Psi(u) $ is the initial field configuration.


The first relation in (\ref{V-uc-rels}) implies
$V_{\textrm{\tiny loc}}(u_{\textrm{c}}) \, \partial_{u_{\textrm{c}}} =  V_{\textrm{\tiny loc}}(u) \, \partial_u  $,
that leads us to write (\ref{system-pde-V}) as
\be
\label{system-pde-V2}
\Big[ \,\partial_\tau -  V_{\textrm{\tiny loc}}(u) \, \partial_u \,\Big]
\Psi(\tau, u) 
\,=\,  
\Bigg( \begin{array}{cc}
\tfrac{1}{2} \, \partial_u V_{\textrm{\tiny loc}}(u) \;\;& -  V_{\textrm{\tiny biloc}}(u)
\\
\rule{0pt}{.5cm}
 -  V_{\textrm{\tiny biloc}}(u_{\textrm{c}}) \;\;& 
 \tfrac{1}{2} \, \partial_{u_{\textrm{c}}} \!V_{\textrm{\tiny loc}}(u_{\textrm{c}}) 
\end{array} \Bigg)
\, \Psi(\tau, u) \,.
\ee
This system can be studied by introducing the following redefinition of the field
\be
\label{Psi-tilde-def-app}
\Psi(\tau, u) 
\equiv 
\boldsymbol{M}(u) \, \tilde{\Psi}(\tau, u) 
\equiv 
\boldsymbol{M}(u) 
\bigg( \begin{array}{c}
\tilde{\psi}(\tau, u) \\
\tilde{\psi}(\tau, u_{\textrm{c}}) 
\end{array} \bigg)
\;\;\;\qquad\;\;\;
\boldsymbol{M}(u) 
\equiv 
\bigg( \begin{array}{cc}
G(u) & 0
\\
0 & G(u_{\textrm{c}})
\end{array} \bigg)
\ee
where $G(u)$ is defined by imposing the vanishing of the diagonal terms in (\ref{system-pde-V2}).
This requirement gives
\be
\label{cond-zero-app}
V_{\textrm{\tiny loc}}(u)\, \partial_u G(u)
\,=\,
-\frac{1}{2} \; G(u) \, \partial_u V_{\textrm{\tiny loc}}(u)
\ee
whose solution is
\be
\label{app-G(u)-def}
G(u) = \frac{G_0}{\sqrt{V_{\textrm{\tiny loc}}(u)}}
\ee
for some constant $G_0$.
The function $G(u)$ in (\ref{app-G(u)-def}) satisfies the following properties
\be
\label{G-properties-app}
G(\xi_{\textrm{c}}) =  \frac{ | \xi - q_0 |}{r_0}\; G(\xi)
\;\;\;\;\qquad\;\;\;\;
\frac{G(u) }{ G(\xi) } = \sqrt{\frac{ V_{\textrm{\tiny loc}}(\xi)  }{ V_{\textrm{\tiny loc}}(u)} } \;.
\ee

From (\ref{cond-zero-app}),
the system (\ref{system-pde-V2}) can be equivalently written  in terms of the field  in (\ref{Psi-tilde-def-app}) as 
\be
\label{system-pde-V3}
\Big[ \,\partial_\tau -  V_{\textrm{\tiny loc}}(u) \, \partial_u \,\Big]
\tilde{\Psi}(\tau, u) 
\,=\,  
-\, \bigg( \begin{array}{cc}
0 &\widetilde{V}(u)
\\
\widetilde{V}(u_{\textrm{c}})&  0
\end{array} \bigg)
\, \tilde{\Psi}(\tau, u) 
\ee
where
\be
\widetilde{V}(u) 
\equiv 
V_{\textrm{\tiny biloc}}(u)\, \frac{G(u_{\textrm{c}}) }{G(u)}
=
\frac{\sqrt{V_{\textrm{\tiny loc}}(u)\, V_{\textrm{\tiny loc}}(u_{\textrm{c}}) } }{u - u_{\textrm{c}}}\;.
\ee
Since $\widetilde{V}(u_{\textrm{\tiny c}}) =-\,\widetilde{V}(u)$, 
the system of PDE's in (\ref{system-pde-V3}) becomes 
\be
\label{system-pde-V4}
\Big[ \,\partial_\tau -  V_{\textrm{\tiny loc}}(u) \, \partial_u \,\Big]
\tilde{\Psi}(\tau, u) 
\,=\,  
\widetilde{V}(u) \, \boldsymbol{J} \, \tilde{\Psi}(\tau, u) 
\ee
in terms of the constant symplectic matrix $\boldsymbol{J} $ defined as follows
\be
\label{J-mat-diagonalization}
\boldsymbol{J} \equiv  
\bigg( \begin{array}{cc}
\,0 \,\, & -1
\\
\,1 \,\, &  0 
\end{array} \bigg)
=\,
\boldsymbol{U}^{-1}
\bigg( \begin{array}{cc}
\,\ri \,\, & 0
\\
\, 0 \,\, &  -\ri
\end{array} \bigg)
\; \boldsymbol{U} 
\;\;\qquad\;\;
\boldsymbol{U} \equiv 
\frac{1}{\sqrt{2}}
\bigg( \begin{array}{cc}
-\ri \,\; & 1\,
\\
 \ri \,\; &  1\,
\end{array} \bigg)
\ee
where also the unitary matrix $\boldsymbol{U} $ diagonalising $\boldsymbol{J} $ has been reported.

The diagonalisation of $\boldsymbol{J}$ in the r.h.s. of (\ref{system-pde-V4})
leads to the decoupling of the two PDE's in the system (\ref{system-pde-V4}) 
through the following field redefinition 
\be
\label{Phi-tilde-def-app}
 \tilde{\Phi}(\tau, u) 
\equiv
\bigg( \begin{array}{c}
\varphi_1(\tau, u) \\
\varphi_2(\tau, u_{\textrm{c}}) 
\end{array} \bigg)
\equiv \,
 \boldsymbol{U} \, \tilde{\Psi}(\tau, u) 
\ee
where $\varphi_1$ and $\varphi_2$ are not distinct fields; 
indeed, this definition implies that $\varphi_2(\tau, u)= \ri\,\varphi_1(\tau, u)$.
This is consistent with the fact that the same field $\tilde{\psi}$ occurs in the two components of $\tilde{\Psi}$ (see (\ref{Psi-tilde-def-app})).
The field redefinition (\ref{Phi-tilde-def-app}) leads to write (\ref{system-pde-V4}) as 
\be
\label{system-pde-V4-bis}
\Big[ \,\partial_\tau -  V_{\textrm{\tiny loc}}(u) \, \partial_u \,\Big]
\bigg( \begin{array}{c}
\varphi_1(\tau, u) \\
\varphi_2(\tau, u_{\textrm{c}}) 
\end{array} \bigg)
\,=\,  
\widetilde{V}(u) \, 
\bigg( \begin{array}{cc}
\,\ri \,\, & 0
\\
\, 0 \,\, &  -\ri
\end{array} \bigg)
\bigg( \begin{array}{c}
\varphi_1(\tau, u) \\
\varphi_2(\tau, u_{\textrm{c}}) 
\end{array} \bigg)
\ee
By using 
$V_{\textrm{\tiny loc}}(u_{\textrm{c}}) \, \partial_{u_{\textrm{c}}} =  V_{\textrm{\tiny loc}}(u) \, \partial_u  $ 
and 
$\widetilde{V}(u_{\textrm{\tiny c}}) =-\,\widetilde{V}(u)$ again,  
we find that the two decoupled PDE's 
given by the first and the second component in (\ref{system-pde-V4-bis}) 
take the same form expressed in terms of either $u$ or $u_{\textrm{\tiny c}}$ respectively, 
which is given by the following linear PDE
\be
\label{pde-single-app}
\Big[ \,\partial_\tau -  V_{\textrm{\tiny loc}}(u) \, \partial_u \,\Big]
\varphi(\tau, u) 
\,=\,  
\ri\, \widetilde{V}(u) \, \varphi(\tau, u) \,.
\ee


The solution of the PDE in (\ref{pde-single-app}) 
for the initial field configuration $\varphi(u)$ at $\tau=0$
reads
\be
\label{app-soln-varphi}
\varphi(\tau, u) = \e^{\ri\, \theta(\tau, u) }\, \varphi\big(\xi(\tau, u)\big)
\;\;\;\qquad\;\;\;
\theta(\tau, u) \equiv \Gamma\big(\xi(\tau, u) \big) - \Gamma(u)
\ee
where $\xi(\tau, u) $ has been defined in (\ref{xi-2int-A1-A2}) and 
\be
\Gamma(\xi) 
\equiv
 \int  \frac{\widetilde{V}(\xi) }{ V_{\textrm{\tiny loc}}(\xi) }\, \rd \xi
\,= 
\int \frac{1}{\xi - \xi_{\textrm{c}} } \, \sqrt{\frac{V_{\textrm{\tiny loc}}(\xi_{\textrm{c}} ) }{V_{\textrm{\tiny loc}}(\xi) } } \; \rd \xi 
\,=\,
\textrm{sign}(\xi -q_0) \; \textrm{arctan}\bigg( \frac{\xi - q_0}{r_0} \bigg) +\textrm{const}
\phantom{xxx}
\ee
(we remind that $V_{\textrm{\tiny loc}}(\xi)  \geqslant 0$ for $\xi \in A$)
where the notation $\xi_{\textrm{c}}(\xi) \equiv \mathcal{C}(\xi)$, in terms of (\ref{u-conj-def}), has been adopted.
Since $q_0 \in (b_1, a_2) $ (see (\ref{q0-def})), 
we have that $\textrm{sign}(\xi -q_0) = +1$ for $u \in A_2$
while $\textrm{sign}(\xi -q_0) = -1$ for $u \in A_1$.
Moreover, the relation 
$\partial_{\xi} \xi_{\textrm{c}} = V_{\textrm{\tiny loc}}(\xi_{\textrm{c}} )  / V_{\textrm{\tiny loc}}(\xi) $
implies that $\Gamma(\xi_{\textrm{c}}) = -\,\Gamma(\xi) $.

Combining (\ref{Psi-tilde-def-app}), (\ref{Phi-tilde-def-app}) and  (\ref{app-soln-varphi}),
we find that  (\ref{Psi-doublet-def-app})  becomes
\be
\label{Psi-tilde-def-app-v2}
\bigg( \begin{array}{c}
\psi(\tau, u) \\
\psi(\tau, u_{\textrm{c}}) 
\end{array} \bigg)
\,=\,
\boldsymbol{M}(u) \,
\bigg( \begin{array}{cc}
\cos(\theta) \; & -\sin(\theta) \\
\sin(\theta) \; & \cos(\theta)
\end{array} \bigg) \,
\boldsymbol{M}(\xi)^{-1} \,
\bigg( \begin{array}{c}
\psi(\xi) \\
\psi(\xi_{\textrm{c}}) 
\end{array} \bigg)
\ee
where we used that
\be
\boldsymbol{U}^{-1}
\bigg( \begin{array}{cc}
\e^{\ri \theta}  & 0 \\
0  & \e^{-\ri \theta}
\end{array} \bigg) \;
\boldsymbol{U} 
\,=
\bigg( \begin{array}{cc}
\cos(\theta) \; & -\sin(\theta) \\
\sin(\theta) \; & \cos(\theta)
\end{array} \bigg)\,.
\ee
Since the second component of (\ref{Psi-tilde-def-app-v2}) 
can be obtained from the first one
by replacing $\xi$ with $\xi_{\textrm{c}}$ and $u$ with $u_{\textrm{c}}$;
we can focus only on the first component of (\ref{Psi-tilde-def-app-v2}).


By employing (\ref{G-properties-app}),
the explicit expression of the first component in (\ref{Psi-tilde-def-app-v2}) reads
\be
\label{Psi-tilde-def-app-v3}
\psi(\tau, u) = 
\sqrt{\frac{ V_{\textrm{\tiny loc}}(\xi)  }{ V_{\textrm{\tiny loc}}(u)} }\,
\left[\,
\cos(\theta)\, \psi(\xi) - \frac{r_0}{| \xi - q_0 | } \, \sin(\theta)\, \psi(\xi_{\textrm{c}})
\,\right]
\ee
where, by using that $\textrm{sign}(\xi - q_0) = \textrm{sign}(u - q_0)$
(notice that $\textrm{sign}(u - q_0) = (-1)^j$ for $u \in A_j$ with $j \in \{1,2\}$), 
the functions $\cos(\theta)$ and $\sin(\theta)$ are given respectively by 
\bea
\label{cos-theta-app}
\cos(\theta)
&=&
\cos\! \bigg[
\textrm{arctan}\bigg( \frac{\xi - q_0}{r_0} \bigg) - \textrm{arctan}\bigg( \frac{u - q_0}{r_0} \bigg) 
\bigg]
\,=\,
\frac{ (\xi -q_0) \, (u- q_0) +r_0^2 }{\sqrt{\big[ (\xi-q_0)^2+r_0^2 \big] \big[ (u-q_0)^2+r_0^2\big]}}
\nn
\\
& &
\\
\label{sin-theta-app}
\sin(\theta)
&=&
\textrm{sign}(u - q_0) \;
\sin\! \bigg[
\textrm{arctan}\bigg( \frac{\xi - q_0}{r_0} \bigg) - \textrm{arctan}\bigg( \frac{u - q_0}{r_0} \bigg) 
\bigg]
\nn
\\
\rule{0pt}{.7cm}
&=&
\textrm{sign}(u - q_0) \;
\frac{r_0 \, (\xi - u) }{\sqrt{\big[ (\xi-q_0)^2+r_0^2 \big] \big[ (u-q_0)^2+r_0^2\big]} }
\eea
which give
\bea
\label{tan-theta-app}
\tan(\theta) 
&=&
\textrm{sign}(u - q_0) \;
\tan\! \bigg[
\textrm{arctan}\bigg( \frac{\xi - q_0}{r_0} \bigg) - \textrm{arctan}\bigg( \frac{u - q_0}{r_0} \bigg) 
\bigg]
\nn
\\
\rule{0pt}{.7cm}
&=&
\textrm{sign}(u - q_0) \;
\frac{r_0 \, (\xi - u) }{ (\xi -q_0) \, (u- q_0) +r_0^2  } \,.
\eea
By using the expressions in (\ref{cos-theta-app}) and (\ref{sin-theta-app}),
the modular flow (\ref{Psi-tilde-def-app-v3}) becomes
\bea
\label{Psi-tilde-def-app-v4}
\psi(\tau, u) 
&=&
\sqrt{\frac{ V_{\textrm{\tiny loc}}(\xi)  }{ V_{\textrm{\tiny loc}}(u)} } \,
\left\{
\frac{ (\xi -q_0) \, (u- q_0) +r_0^2 }{ \sqrt{\big[ (\xi-q_0)^2+r_0^2 \big] \big[ (u-q_0)^2+r_0^2\big] } } \, \psi(\xi) 
\right.
\\
& & \hspace{2cm}
\left.
-\, \frac{ r_0^2 }{ \sqrt{\big[ (\xi-q_0)^2+r_0^2 \big] \big[ (u-q_0)^2+r_0^2\big] } } \; \frac{ \xi -u  }{\xi - q_0}\, \psi(\xi_{\textrm{c}})
\right\} \,.
\nonumber
\eea

Longo, Martinetti and Rehren \cite{Longo:2009mn} have introduced the parameters
\be
L_{\textrm{\tiny LMR}} \equiv b_1 - a_1 + b_2 - a_2
\;\qquad
M_{\textrm{\tiny LMR}} \equiv b_1 b_2 - a_1 a_2
\;\qquad
N_{\textrm{\tiny LMR}} \equiv (b_1 - a_1) \, a_2 b_2 + (b_2 - a_2)\, a_1 b_1
\ee
which are related to the parameters defined in (\ref{q0-def})-(\ref{r0-def}) 
as follows
\be
\label{LMR-notation-relation}
\frac{L_{\textrm{\tiny LMR}} }{ \sqrt{ L_{\textrm{\tiny LMR}} \, N_{\textrm{\tiny LMR}} - M_{\textrm{\tiny LMR}}^2 } } = \frac{1}{r_0}
\;\;\;\qquad\;\;\;
\frac{M_{\textrm{\tiny LMR}} }{ \sqrt{ L_{\textrm{\tiny LMR}}  \,N_{\textrm{\tiny LMR}} - M_{\textrm{\tiny LMR}}^2 } } = \frac{q_0}{r_0} \,.
\ee
In particular, the relations in (\ref{LMR-notation-relation}) tell us that the argument of the first arctan function in the r.h.s. of Eq.\,(3.12) in \cite{Longo:2009mn} 
corresponds to $(\xi-q_0)/r_0$ in (\ref{cos-theta-app})-(\ref{tan-theta-app}).


We find it worth expressing the modular flow (\ref{Psi-tilde-def-app-v4}) in terms of 
the four-point harmonic ratio defined as follows
\be
\label{eta-def-app}
\eta(\xi, u) 
\equiv 
\frac{ (\xi - \xi_{\textrm{c}}) \, (u - u_{\textrm{c}}) }{ (\xi - u_{\textrm{c}}) \, (u - \xi_{\textrm{c}}) }
\,=\,
\frac{ \big[ (\xi-q_0)^2+r_0^2 \big] \big[ (u-q_0)^2+r_0^2\big] }{ \big[ (\xi -q_0) \, (u-q_0) + r_0^2 \big]^2 }
\ee
(which has been introduced also in (\ref{eta-ratio-def}) in terms of other variables),
where the last expression, obtained by employing  (\ref{u-conj-def}),
is symmetric under exchange $\xi \leftrightarrow u$.
From (\ref{eta-def-app}) it is straightforward to observe that 
 \be
 \label{eta-min-one-app}
 \eta(\xi, u) - 1
=
 - \frac{ (\xi - u ) \, (\xi_{\textrm{c}}  - u_{\textrm{c}} ) }{ (\xi - u_{\textrm{c}}) \, ( \xi_{\textrm{c}} - u) }
\,=\,
\left[\,
\frac{ r_0 \,(\xi - u) }{ (\xi -q_0) \, (u-q_0) + r_0^2  }
\,\right]^2
\ee
hence $\eta(\xi, u) \geqslant  1$ for any choice of $u$ and $\xi$ in $A$,
which is saturated when $\xi = u$.
Combining (\ref{tan-theta-app}) and (\ref{eta-min-one-app}), one finds
\be
\label{tan-theta-eta-app}
\tan(\theta) 
\,=\, 
\textrm{sign}(u - q_0) \, \sqrt{ \eta(\xi, u) - 1} \,.
\ee
The expression in (\ref{eta-def-app}) satisfies the following relations
\be
\label{eta-rels-app1}
\eta(\xi_{\textrm{c}} , u_{\textrm{c}}) = \eta(\xi, u) 
\;\;\;\;\qquad\;\;\;\;
\eta(\xi_{\textrm{c}} , u) = \frac{\eta(\xi, u) }{ \eta(\xi, u) - 1  }
\ee


It is worth exploring the possible occurrence of an upper bound for (\ref{eta-def-app}) with $u, \xi \in A$.
When $u \in A_i$ and $\xi \in A_j$ with $i \neq j$,
an upper bound does not exist because $\eta(\xi, u)  \to +\infty$ as $\xi \to u_{\textrm{c}}$
(see (\ref{eta-u2-to-u1c})).
Instead, when $u$ and $\xi$ belong to the same interval
$\eta(\xi, u)  \leqslant \eta_{\textrm{\tiny max}} $, where 
\be
\label{eta-max-def}
\eta_{\textrm{\tiny max}} 
\equiv \frac{(a_2-a_1)(b_2-b_1)}{(a_2-b_1)(b_2 - a_1)}
\;\;\;\qquad\;\;\;
\eta_{\textrm{\tiny max}} 
= \eta(a_1, b_1) 
= \eta(a_2, b_2) 
\ee
which can be equivalently written in terms of (\ref{tan-theta-eta-app}) as follows \cite{Chen:2019iro}
\be
\big|\tan(\theta) \big| < \sqrt{\eta_{\textrm{\tiny max}} -1 } \,.
\ee
When both $u \in A_j$ and $\xi \in A_j$ with $j \in \{1,2\}$,
we can focus on  $\xi \geqslant u$ without loss of generality
because (\ref{eta-def-app}) is symmetric under exchange $\xi \leftrightarrow u$.
By introducing $\xi_0 \equiv \xi - q_0$ and $u_0 \equiv u - q_0$, 
from (\ref{eta-min-one-app}) we get
\be
\label{eta-max-bound-step1}
\frac{1}{  \eta(\xi, u) - 1 }
=
\frac{1}{r_0^2}
\left(
\frac{ \xi_0 \,  u_0 + r_0^2  }{ \xi_0 - u_0 }
\right)^2
=
\frac{1}{r_0^2}
\left(
\frac{u_0^2 + r_0^2  }{ \xi_0 - u_0 } +u_0
\right)^2
\geqslant 
\frac{1}{r_0^2}
\left(
\frac{u_0^2 + r_0^2  }{ b_{j,0} - u_0 } +u_0
\right)^2
\ee
where $b_{j,0} \equiv b_j - q_0$ in the last expression, 
which can be bounded as follows
\be
\label{eta-max-bound-step2}
\frac{1}{r_0^2}
\left(
\frac{u_0^2 + r_0^2  }{ b_{j,0} - u_0 } +u_0
\right)^2
=
\frac{1}{r_0^2}
\left(
\frac{b_{j,0}^2 + r_0^2  }{ b_{j,0} - u_0} - b_{j,0}
\right)^2
\geqslant 
\frac{1}{r_0^2}
\left(
\frac{b_{j,0}^2 + r_0^2  }{ b_{j,0} - a_{j,0}} - b_{j,0}
\right)^2
=
\frac{1}{  \eta(b_j , a_j) - 1 }
\ee
in terms of $a_{j,0} \equiv a_j - q_0$.
The upper bound $\eta(\xi, u)  \leqslant \eta_{\textrm{\tiny max}} $ for $\xi$ and $u$ in the same interval
is obtained by combining (\ref{eta-max-def}), (\ref{eta-max-bound-step1}) and (\ref{eta-max-bound-step2}).


In (\ref{tilde-eta-def}) we have introduced a specific notation to denote  the ratio multiplying the field $\psi(\xi)$ 
within the curly brackets in the r.h.s. of (\ref{Psi-tilde-def-app-v4}). 
From (\ref{xi-conj-def-text}), it is straightforward to get
\be
\label{xi-conj-der}
 \partial_u \xi_{\textrm{c}} = \left( \frac{r_0}{\xi -q_0} \right)^2  \partial_u \xi
\ee
and, as for the numerator and the denominator in the r.h.s. of (\ref{tilde-eta-def}),
we find  respectively 
\be
\label{eta-tilde-xi-conj-num}
\big[ (\xi_{\textrm{c}}   -q_0)^2+r_0^2 \big]
=
\left( \frac{r_0}{\xi -q_0} \right)^2 \big[ (\xi   -q_0)^2+r_0^2 \big]
\;\;\;\qquad\;\;\;
(\xi_{\textrm{c}} -q_0) \, (u-q_0) + r_0^2  
=
 \frac{ \xi - u }{ \xi - q_0}\, r_0^2 \,.
 \phantom{xxx}
\ee
Combining (\ref{der-xi-2int}), (\ref{xi-conj-der}) and (\ref{eta-tilde-xi-conj-num}),
we observe that
\be
\label{step-32}
\sqrt{\frac{ V_{\textrm{\tiny loc}}(\xi)  }{ V_{\textrm{\tiny loc}}(u)} } \,\;
\frac{ r_0^2 }{ \sqrt{\big[ (\xi-q_0)^2+r_0^2 \big] \big[ (u-q_0)^2+r_0^2\big] } } \; \frac{ \xi -u  }{\xi - q_0}
\,=\,
\frac{ \sqrt{ \partial_u \xi_{\textrm{c}} } }{ \tilde{\eta}(\xi_{\textrm{c}} ,u)  } \,.
\ee
Thus, by employing (\ref{der-xi-2int}) again, (\ref{tilde-eta-def}) and (\ref{step-32}), 
the modular flow (\ref{Psi-tilde-def-app-v4}) becomes
\be
\label{Psi-tilde-def-app-v6}
\psi(\tau, u) 
=
 \frac{ \sqrt{ \partial_u \xi } }{ \tilde{\eta}(\xi, u)}  \, \psi(\xi) 
-
 \frac{ \sqrt{ \partial_u \xi_{\textrm{c}} } }{ \tilde{\eta}(\xi_{\textrm{c}} , u)} \, \psi(\xi_{\textrm{c}}) 
\ee
which satisfies the initial condition $\psi(\tau=0, u) = \psi(u)$, as expected,
because $\tilde{\eta}(u,u) = 1$ and $\tilde{\eta}(\xi_{\textrm{c}},u) $ diverges as $\xi_{\textrm{c}} \to u_{\textrm{c}}$.
It is useful to write (\ref{Psi-tilde-def-app-v7}) in the following  form
\be
\label{Psi-tilde-def-app-v7}
\psi(\tau, u) 
=
 \frac{ \sqrt{ \partial_u \xi } }{ \tilde{\eta}(\xi, u) }  
 \left[\,  \psi(\xi) 
-
 \frac{ \tilde{\eta}(\xi, u) }{ \tilde{\eta}(\xi_{\textrm{c}} , u)} \; 
 \sqrt{ \frac{ \partial_u \xi_{\textrm{c}} }{ \partial_u \xi  } } 
 \; \psi(\xi_{\textrm{c}}) 
 \, \right]
 \equiv\,
  \frac{ \sqrt{ \partial_u \xi } }{ \tilde{\eta}(\xi, u) }  
 \Big[\,  \psi(\xi)  - M(\xi, u) \, \psi(\xi_{\textrm{c}})  \, \Big]
\ee
where we have introduced
\bea
\label{Mdef-explicit-step1}
M(\xi, u) 
&\equiv &
 \frac{ \tilde{\eta}(\xi, u) }{ \tilde{\eta}(\xi_{\textrm{c}} , u)} \; 
 \sqrt{ \frac{ \partial_u \xi_{\textrm{c}} }{ \partial_u \xi  } } 
 \,=\,
 \frac{ (\xi_{\textrm{c}} -q_0) \, (u-q_0) + r_0^2   }{ (\xi -q_0) \, (u-q_0) + r_0^2  }
 \\
 \rule{0pt}{.8cm}
 \label{Mdef-explicit-step2}
 & = &
  \frac{ r_0^2\, (\xi - u) }{ (\xi - q_0) \big[ (\xi -q_0) \, (u-q_0) + r_0^2 \big]  } 
   =
     \frac{ (\xi_{\textrm{c}} - q_0)  \big[ (\xi_{\textrm{c}} -q_0) \, (u-q_0) + r_0^2  \big] }{ r_0^2\, (\xi_{\textrm{c}} - u)  } 
     \hspace{1cm}
\eea
whose explicit expressions are obtained from (\ref{xi-conj-def-text}), (\ref{tilde-eta-def}), (\ref{xi-conj-der}) and (\ref{eta-tilde-xi-conj-num}).
The expression (\ref{Mdef-explicit-step2}) is employed in a crucial way 
to investigate the modular correlators 
of the chiral fermionic fields and of the chiral density fields
in Appendix\;\ref{app-mod-corr-psi} and Appendix\;\ref{app-mod-flow-density} respectively.

We find it worth concluding this analysis by observing that
the expression (\ref{psi-flow-text}) for the modular flow of the chiral field
provides a useful starting point to explore the flow of the same field generated by the negativity Hamiltonian.
Indeed, since this operator is given by (\ref{mod-ham-2int-KA}) with $a_2$ and $b_2$ interchanged
when the partial transposition of $A_2$ is considered \cite{Murciano:2022vhe, Rottoli:2022plr},
according to the procedure discussed in \cite{Calabrese:2012ew, Calabrese:2012nk, Calabrese:2014yza}
to study the entanglement negativity in quantum field theory (see also \cite{Calabrese:2013mi, Coser:2015eba}),
the corresponding flow can be obtained by interchanging $a_2$ and $b_2$ in (\ref{psi-flow-text}) as well.

\subsection{Modular correlators of the chiral fermionic fields  }
\label{app-mod-corr-psi}

The modular flow of the fermionic chiral fields $\psi_\pm$ derived in Appendix\;\ref{app-mod-flow-2int-psi}
allows to construct the corresponding modular two-point functions satisfying the proper KMS condition, 
which have been first obtained by Longo, Martinetti and Rehren in \cite{Longo:2009mn}.

In the following, to avoid confusion with (\ref{xi-k-def}),
we adopt the notation $\xi_{\tilde{r}} \equiv \xi(\pm \tau_r, u_{r,\pm})$ 
and $\xi_{{\tilde{r}},\textrm{c}} \equiv \mathcal{C}(\xi_{\tilde{r}} )  $ 
for $r, \tilde{r} \in \{1,2\}$,
in terms of $\xi(\tau, u)$ defined in (\ref{xi-2int-A1-A2}).
Combining  (\ref{Psi-tilde-def-app-v7}) and 
the two-point function of the chiral fermionic fields $\psi_\pm$ 
given by $\langle \psi_\pm^\ast(u)\, \psi_\pm(v) \rangle = \tfrac{1}{2\pi (\pm \ri) \,(u-v)}$ 
(see (\ref{2pt-chiral-primaries}) with $h_\pm =1/2$),
for the modular correlator of $\psi_\pm$ we find 
\be
\label{2pt-mod-psi-app-0}
\langle \,\psi_\pm^\ast (\tau_1 , u_1) \,\psi_\pm (\tau_2 , u_2) \,\rangle 
\,=\,
\frac{ \sqrt{ \partial_{u_1} \xi_{\tilde{1}} \; \partial_{u_2} \xi_{\tilde{2}} } }{ \tilde{\eta}(\xi_{\tilde{1}} , u_1) \, \tilde{\eta}(\xi_{\tilde{2}}, u_2) } 
\; \mathcal{M}_{\psi_\pm}(\tau_1 , u_1; \tau_2 , u_2)
\ee
where  $u_j$ means $u_{j,\pm}$ with $j\in \{1,2\}$ and we have introduced
\bea
\label{2pt-mod-psi-app-00}
\mathcal{M}_{\psi_\pm} (\tau_1 , u_1; \tau_2 , u_2)
&\equiv &
\langle \,
\big[ \, \psi_\pm^\ast(\xi_{\tilde{1}}) - M(\xi_{\tilde{1}} ,u_1) \,   \psi_\pm^\ast( \xi_{{\tilde{1}},\textrm{c}} ) \big]
\big[ \, \psi_\pm(\xi_{\tilde{2}}) - M(\xi_{\tilde{2}} ,u_2) \,   \psi_\pm(\xi_{{\tilde{2}},\textrm{c}} ) \big]
\,\rangle 
\\
\rule{0pt}{.8cm}
& & \hspace{-2.3cm}
=\,
N_{\psi_\pm} \!
\left\{\,
\frac{1}{\xi_{\tilde{1}} - \xi_{\tilde{2}}  \mp \ri \varepsilon } 
- \frac{M(\xi_{\tilde{1}} ,u_1) }{\xi_{{\tilde{1}},\textrm{c}} - \xi_{\tilde{2}}  \mp \ri \varepsilon } 
- \frac{M(\xi_{\tilde{2}} ,u_2) }{\xi_{\tilde{1}} - \xi_{{\tilde{2}},\textrm{c}}  \mp \ri \varepsilon } 
+ \frac{M(\xi_{\tilde{1}} ,u_1) \, M(\xi_{\tilde{2}} ,u_2) }{\xi_{{\tilde{1}},\textrm{c}} - \xi_{{\tilde{2}},\textrm{c}}  \mp \ri \varepsilon } 
\,\right\}
\hspace{1cm}
\nn
\eea
in terms of the normalisation constant $N_{\psi_\pm} = 1/[2\pi (\pm \ri)]$, that depends on the chirality of the field. 
By employing (\ref{Mdef-explicit-step2}) into (\ref{2pt-mod-psi-app-00}),  
for the modular correlator (\ref{2pt-mod-psi-app-0}) one obtains 
\be
\label{2pt-mod-psi-app}
\langle \,\psi_\pm^\ast (\tau_1 , u_1) \,\psi_\pm (\tau_2 , u_2) \,\rangle 
=
\sqrt{ \partial_{u_1} \xi_{\tilde{1}} \, \partial_{u_2} \xi_{\tilde{2}} } \;\,
\sqrt{ \frac{ \eta(\xi_{\tilde{1}}, \xi_{\tilde{2}})  }{ \eta(u_1, u_2)  } }\;\,
\frac{N_{\psi_\pm} }{ \xi_{\tilde{1}} - \xi_{\tilde{2}}  \mp \ri \varepsilon }
\,=\,
 N_{\psi_\pm} W_\pm(\tau_{12}; u_1, u_2)
\ee
(we remind that $u_j$ must be replaced with $u_{j,\pm}$)
in terms of the distributions $W_\pm(\tau; u_1, u_2)$ introduced in  (\ref{cap-W-def}).
Further comments about (\ref{2pt-mod-psi-app}) have been reported in Sec.\,\ref{sec-2int-mod-corr},
where the relevant properties of the function provided by this modular correlator
which are relevant for the chiral distance along the modular evolutions (see Sec.\,\ref{sec-2int-distance-chiral})
are discussed.

\subsection{Modular correlators of the chiral density fields }
\label{app-mod-flow-density}


The modular flow of the fermionic chiral fields $\psi_\pm$ 
discussed in Appendix\;\ref{app-mod-flow-2int-psi} 
can be employed to study 
also the modular two-point functions of the chiral density fields
$\varrho_\pm \equiv \; : \!\psi_\pm^\ast \psi_\pm \!:\! $\,,
which provide the modular two-point functions 
of the charge density and of the helicity density 
for the massless Dirac fermion. 
Indeed, from (\ref{Psi-tilde-def-app-v7}) we find 
\be
\label{2pt-mod-rho-app-00}
\langle \,
\varrho_\pm(\tau_1 , u_1) \,
\varrho_\pm(\tau_2 , u_2)
\,\rangle 
\,=\,
\frac{ \partial_{u_1} \xi_{\tilde{1}} \; \partial_{u_2} \xi_{\tilde{2}} }{ \tilde{\eta}(\xi_{\tilde{1}}, u_1)^2 \; \tilde{\eta}(\xi_{\tilde{2}}, u_2)^2 } \; 
\mathcal{M}_{\varrho_\pm}(\tau_1 , u_1; \tau_2 , u_2)
\ee
where $u_j$ must be understood as $u_{j,\pm}$ (like in (\ref{2pt-mod-psi-app-0}))
and  we have introduced
\bea
\mathcal{M}_{\varrho_\pm}(\tau_1 , u_1; \tau_2 , u_2)
& \equiv &
\big\langle\,
\Big\{
\! : \!\psi_\pm^\ast ( \xi_{\tilde{1}} ) \, \psi_\pm( \xi_{\tilde{1}} ) \!: 
+ \,M(\xi_{\tilde{1}} ,u_1)^2
: \!\psi_\pm^\ast ( \xi_{{\tilde{1}},\textrm{c}} ) \, \psi_\pm( \xi_{{\tilde{1}},\textrm{c}} ) \!: 
\\
\rule{0pt}{.4cm}
& & \hspace{2.5cm}
- \,
M(\xi_{\tilde{1}} ,u_1) 
\big[
 : \!\psi_\pm^\ast ( \xi_{{\tilde{1}},\textrm{c}} ) \, \psi_\pm( \xi_{\tilde{1}} ) \!: 
+
 : \!\psi_\pm^\ast ( \xi_{\tilde{1}} ) \, \psi_\pm( \xi_{{\tilde{1}},\textrm{c}} ) \!: 
\big]
 \Big\}
\nn
\\
\rule{0pt}{.5cm}
& & \hspace{.3cm}
\Big\{
\! : \!\psi_\pm^\ast ( \xi_{\tilde{2}} ) \, \psi_\pm( \xi_{\tilde{2}} ) \!: 
+ \,M(\xi_{\tilde{2}} ,u_2)^2
: \!\psi_\pm^\ast ( \xi_{{\tilde{2}},\textrm{c}} ) \, \psi_\pm( \xi_{{\tilde{2}},\textrm{c}} ) \!: 
\nn \\
\rule{0pt}{.4cm}
& & \hspace{2.5cm}
- \,
M(\xi_{\tilde{2}} ,u_2) 
\big[
 : \!\psi_\pm^\ast ( \xi_{{\tilde{2}},\textrm{c}} ) \, \psi_\pm( \xi_{\tilde{2}} ) \!: 
+
 : \!\psi_\pm^\ast ( \xi_{\tilde{2}} ) \, \psi_\pm( \xi_{{\tilde{2}},\textrm{c}} ) \!: 
\big]
 \Big\} \, \big\rangle
 \nn
\eea
which is an algebraic sum containing $16$ terms.
Indeed, performing the various contractions, we obtain 
\bea
\label{M-rho-v1}
\frac{ \mathcal{M}_{\varrho_\pm}(\tau_1 , u_1; \tau_2 , u_2) }{ N_\varrho }
& = &
\\
\rule{0pt}{.8cm}
& & \hspace{-3.8cm}
=\;
\frac{1}{ (\xi_{\tilde{1}} - \xi_{\tilde{2}} \mp \ri \varepsilon )^2 } 
+ \frac{M(\xi_{\tilde{1}} ,u_1)^2 }{ ( \xi_{{\tilde{1}},\textrm{c}} - \xi_{\tilde{2}} \mp \ri \varepsilon )^2 } 
+ \frac{M(\xi_{\tilde{2}} ,u_2)^2 }{ ( \xi_{\tilde{1}} - \xi_{{\tilde{2}},\textrm{c}} \mp \ri \varepsilon )^2 }  
+ \frac{M(\xi_{\tilde{1}} ,u_1)^2 \, M(\xi_{\tilde{2}} ,u_2)^2 }{ ( \xi_{{\tilde{1}},\textrm{c}} - \xi_{{\tilde{2}},\textrm{c}} \mp \ri \varepsilon )^2 } 
\nn
\\
\rule{0pt}{.8cm}
& & \hspace{-3.3cm}
- \,2\, M(\xi_{\tilde{1}} ,u_1)
\left[ \,
\frac{1}{(\xi_{\tilde{1}} - \xi_{\tilde{2}} \mp \ri \varepsilon ) \, (\xi_{\tilde{1},\textrm{c}} - \xi_{{\tilde{2}}} \mp \ri \varepsilon  )}
+
\frac{ M(\xi_{\tilde{2}} ,u_2)^2 }{ (\xi_{\tilde{1}} - \xi_{{\tilde{2},\textrm{c}}} \mp \ri \varepsilon  ) \, (\xi_{\tilde{1},\textrm{c}} - \xi_{\tilde{2},\textrm{c}} \mp \ri \varepsilon )  }
\, \right]
\nn
\\
\rule{0pt}{.9cm}
& & \hspace{-3.3cm}
- \,2\, M(\xi_{\tilde{2}} ,u_2)
\left[ \,
\frac{1}{(\xi_{\tilde{1}} - \xi_{\tilde{2}} \mp \ri \varepsilon ) \, (\xi_{\tilde{1}} - \xi_{{\tilde{2},\textrm{c}}} \mp \ri \varepsilon )}
+
\frac{ M(\xi_{\tilde{1}} ,u_1)^2 }{(\xi_{\tilde{1},\textrm{c}} - \xi_{{\tilde{2} }}  \mp \ri \varepsilon)\, (\xi_{\tilde{1},\textrm{c}} - \xi_{\tilde{2},\textrm{c}} \mp \ri \varepsilon )  }
\, \right]
\nn
\\
\rule{0pt}{.9cm}
& & \hspace{-3.3cm}
+ \,2\, M(\xi_{\tilde{1}} ,u_1)\, M(\xi_{\tilde{2}} ,u_2)
\left[ \,
\frac{1}{(\xi_{\tilde{1}} - \xi_{{\tilde{2}}} \mp \ri \varepsilon )\, (\xi_{\tilde{1},\textrm{c}} - \xi_{\tilde{2},\textrm{c}}\mp \ri \varepsilon ) }
+
\frac{1}{(\xi_{\tilde{1}} - \xi_{{\tilde{2},\textrm{c}}} \mp \ri \varepsilon )\, (\xi_{\tilde{1},\textrm{c}} - \xi_{\tilde{2}} \mp \ri \varepsilon ) }
\, \right]
\nn
\eea
where the normalisation constant is $N_\varrho = 1/(4\pi^2)$.
%
Plugging the expression for $M(\xi, u) $ 
(see (\ref{Mdef-explicit-step2})) into (\ref{M-rho-v1}),  
a remarkable simplification occurs
and the modular correlators (\ref{2pt-mod-rho-app-00}) become
\be
\label{2pt-mod-rho-app}
\langle \,\varrho_\pm (\tau_1 , u_1) \,\varrho_\pm (\tau_2 , u_2) \,\rangle 
\,=\,
\partial_{u_1} \xi_{\tilde{1}} \, \partial_{u_2} \xi_{\tilde{2}}  \;\,
\frac{ \eta(\xi_{\tilde{1}}, \xi_{\tilde{2}})  }{ \eta(u_1, u_2)  } \;\,
\frac{N_\varrho}{ ( \xi_{\tilde{1}} - \xi_{\tilde{2}}  \mp \ri \varepsilon )^2}
\,=\,
N_\varrho \, W_\pm(\tau_{12}; u_1, u_2)^2
\ee
where any $u_j$ means $u_{j,\pm}$,
according to the notation adopted in  (\ref{2pt-mod-psi-app-0}) and (\ref{2pt-mod-rho-app-00}),
and the distributions $W_\pm$ are (\ref{cap-W-def}) with $w(u)$ given by (\ref{w_fund_2int}).
We remark that (\ref{2pt-mod-rho-app}),
which has been discussed further in Sec.\,\ref{sec-2int-mod-corr},
displays an intriguing formal similarity with (\ref{mod-corr-phi-mu}) 
for $h_\pm =1$.

It is tempting to introduce the following ansatz 
for the modular flow of  $\varrho_\pm$
\be
\label{rho-mod-evo-guess}
\tilde{\varrho}_\pm (\tau, u) 
\equiv
\frac{\partial_u \xi }{\eta(\xi, u)}  \; \varrho_\pm(\xi) 
-
\frac{\partial_u \xi_{\textrm{c}} }{\eta(\xi_{\textrm{c}} , u)}  \; \varrho_\pm(\xi_{\textrm{c}}) \,.
\ee
However, 
evaluating $\langle \,\tilde{\varrho}_\pm (\tau_1 , u_1) \,\tilde{\varrho}_\pm (\tau_2 , u_2) \,\rangle  $ 
through this ansatz and 
taking into account the proper subset of terms in (\ref{M-rho-v1}),
we find  that the resulting correlators do not coincide with the modular correlators (\ref{mod-corr-rho-2int-pm});
hence (\ref{rho-mod-evo-guess}) is not the correct modular flow of the chiral density fields.
It would be interesting to find the analytic expression for the modular flow of the chiral density fields $\varrho_\pm$
providing the modular correlators (\ref{mod-corr-rho-2int-pm}).

\subsection{Details on the chiral distance }
\label{app-W-tilde-relation}

In the following we derive the relation (\ref{W-eta-tilde-2int})
that provides the chiral distance (\ref{xi12-tau-chiral-2int}).

Let us adopt the notation $\xi_j \equiv \xi(\tau_j ,u_j) $ for $j \in \{1,2\}$ in order 
to shorten the expressions occurring below. 
From (\ref{W-function-def-2int}), (\ref{R-fact-def-2int}), (\ref{tilde-eta-def})
and also (\ref{xi-k-def}), which implies $w(\xi_j) = w(u_j) + \tau_j$, 
the relation (\ref{W-eta-tilde-2int}) is obtained as follows
\bea
W(\tau_{12} ; u_1, u_2)  
&=& 
\frac{1}{u_1 - u_2} \;\,
\frac{ \e^{2\pi w(u_1)} - \e^{2\pi w(u_2)}  }{  \e^{2\pi w(u_1) +\pi (\tau_1 - \tau_2)} - \e^{2\pi w(u_2) - \pi (\tau_1 - \tau_2)} }
\\
\label{W-eta-tilde-derivation-step2}
\rule{0pt}{.8cm}
& = &
\frac{1}{\xi_1 - \xi_2} \;\,
\frac{ \xi_1 - \xi_2 }{  \e^{2\pi w(\xi_1)} - \e^{2\pi w(\xi_2)} }\;
\frac{  \e^{2\pi w(u_1)} - \e^{2\pi w(u_2)} }{ u_1 - u_2 } \; 
\e^{\pi (\tau_1 + \tau_2)}
\\
\rule{0pt}{1cm}
& = &
\frac{1}{\xi_1 - \xi_2} \;\,
\frac{  (u_1 -q_0) \, ( u_2 - q_0) + r_0^2  }{  (\xi_1 -q_0) \, ( \xi_2 - q_0) + r_0^2 }\; \mathcal{B}_{1,2} \;
\sqrt{ \frac{ \e^{2\pi [w(\xi_1) + w(\xi_2) ]}  }{ \e^{2\pi [w(u_1) + w(u_2) ]}  } }
\hspace{1cm}
\\
\label{W-eta-tilde-derivation-step4}
\rule{0pt}{1cm}
& & \hspace{-2cm}
=\; 
\frac{1}{\xi_1 - \xi_2} \;\,
\frac{  \tilde{\eta} (\xi_1 , \xi_2) }{ \tilde{\eta} (u_1 , u_2)  }\; 
\mathcal{B}_{1,2} \;
\sqrt{ 
\frac{  \big[ (u_1 - q_0)^2+r_0^2 \big] \big[ (u_2 - q_0)^2+r_0^2\big]  }{  \big[ (\xi_1 - q_0)^2+r_0^2 \big] \big[ (\xi_2 - q_0)^2+r_0^2\big] }
\; \frac{ \e^{2\pi [w(\xi_1) + w(\xi_2) ]}  }{ \e^{2\pi [w(u_1) + w(u_2) ]}  } 
}
\hspace{1.2cm}
\\
\label{W-eta-tilde-derivation-step5}
\rule{0pt}{.9cm}
& & \hspace{-2cm}
=\; 
\frac{\sqrt{\partial_{u_1} \xi_1 \; \partial_{u_2} \xi_2 } }{\xi_1 - \xi_2} \;\,
\frac{  \tilde{\eta} (\xi_1 , \xi_2) }{ \tilde{\eta} (u_1 , u_2)  }\; 
\hspace{1cm}
\eea
where we have introduced
\be
\mathcal{B}_{1,2} 
\equiv
\frac{ (b_1 - \xi_1)(b_2 - \xi_1)\, (b_1 - \xi_2)(b_2 - \xi_2) }{ (b_1 - u_1)(b_2 - u_1)\, (b_1 - u_2)(b_2 - u_2)  }
\ee
which is strictly positive  because $(b_j - \xi_k) / (b_j - u_k) > 0$ for any  choice of $j$ and $k$,
with $j,k \in \{1,2\}$.
Notice that in (\ref{W-eta-tilde-derivation-step2}) and (\ref{W-eta-tilde-derivation-step4}) we have used 
\be
\label{x1-x2-from-exp--app}
\frac{ \xi_1 - \xi_2 }{  \e^{2\pi w(\xi_1)} - \e^{2\pi w(\xi_2)} }\;
\frac{  \e^{2\pi w(u_1)} - \e^{2\pi w(u_2)} }{ u_1 - u_2 } 
\,=\,
\frac{  (u_1 -q_0) \, ( u_2 - q_0) + r_0^2  }{  (\xi_1 -q_0) \, ( \xi_2 - q_0) + r_0^2 }\; \mathcal{B}_{1,2}
\ee
and 
\be
\label{exp-w-12-app}
\frac{ \e^{2\pi [w(\xi_1) + w(\xi_2) ]}  }{ \e^{2\pi [w(u_1) + w(u_2) ]}  } 
\,=\,
 \frac{\partial_{u_1} \xi_1 \; \partial_{u_2} \xi_2 }{ \mathcal{B}_{1,2}^{\, 2} }\;
\frac{  \big[ (\xi_1 - q_0)^2+r_0^2 \big] \big[ (\xi_2 - q_0)^2+r_0^2\big] }{  \big[ (u_1 - q_0)^2+r_0^2 \big] \big[ (u_2 - q_0)^2+r_0^2\big]  }
\ee
respectively,
that can be found from (\ref{w_fund_2int}), (\ref{q0-def})-(\ref{r0-def}) and the second relation in (\ref{der-xi-2int}).

\newpage

\bibliographystyle{nb}

\bibliography{refsMT2}

\end{document}